\documentclass[%
    11pt,               
    a4paper,            
    twoside=semi,       
    DIV=12,             
    british,            
    chapterprefix=true, 
]{scrbook}

\usepackage[listings=false]{scrhack}    
\usepackage[british]{babel}             
\usepackage[T1]{fontenc}                
\usepackage{setspace}
\usepackage{silence}                    
\usepackage{setspace}                   

\usepackage[%
    backend=biber,      
    style=numeric-comp, 
    giveninits=true,    
    sorting=nyt,        
    hyperref=true,      
    indexing=cite,      
    maxbibnames=99,     
    isbn=false,         
    url=true,           
    doi=true,           
]{biblatex}
\DeclareNameAlias{author}{family-given}
\preto\fullcite{\AtNextCite{\defcounter{maxnames}{99}}}

\usepackage{csquotes}   

\newcommand{\nocontentsline}[3]{}\newcommand{\notoc}[2]{\bgroup\let\addcontentsline=\nocontentsline#1{#2}\egroup}

\newcommand{\tocAddFauxChapter}[1]{%
    \cleardoublepage
    \phantomsection  
    \addcontentsline{toc}{chapter}{#1}  
}

\DeclareTOCStyleEntry[numwidth=3em]{tocline}{table}
\DeclareTOCStyleEntry[numwidth=3em]{tocline}{figure}

\usepackage[section]{placeins}  
\usepackage{float}              

\usepackage{caption}            
\usepackage{tabularx}           
\usepackage{longtable}          
\usepackage{booktabs}           
\usepackage{multirow}           
\usepackage{makecell}           
\usepackage{array}              
\usepackage{ragged2e}           
\usepackage{afterpage}          

\usepackage[table]{xcolor}      

\usepackage{adjustbox}          
\usepackage{pdfpages}           
\usepackage{subcaption}         

\usepackage{enumitem}           
\setlist{itemsep=2pt}           

\usepackage{dashrule}           
\usepackage{tikz}               
\usepackage[most]{tcolorbox}    

\tcbuselibrary{skins,breakable}
\newtcolorbox{mybox}[2][]{breakable,sharp corners, skin=enhancedmiddle jigsaw,parbox=false,
boxrule=0mm,leftrule=2mm,boxsep=0mm,arc=0mm,outer arc=0mm,attach title to upper,
after title={.\ }, coltitle=black,colback=gray!10,colframe=black, title={#2},
fonttitle=\bfseries,#1}

\usepackage{imakeidx}   
\usepackage{hyperref}   
\usepackage[
    nogroupskip,        
]{glossaries-extra}     

\newglossary*{abbreviations}{Abbreviations}
\newglossary*{glossary}{Glossary}

\newglossarystyle{mystyle}{%
    \setglossarystyle{long}%
}

\newcommand*{\xglsindex}[1]{\index[termindex]{\glsxtrlong{#1}}}
\newcommand*{\xgls}[1]{\gls{#1}\xglsindex{#1}}

\newcommand*{\xglspl}[1]{\glspl{#1}\xglsindex{#1}}

\newcommand*{\xglsfirst}[1]{\glsfirst{#1}\xglsindex{#1}}
\newcommand*{\xglsfirstplural}[1]{\glsfirstplural{#1}\xglsindex{#1}}

\makeglossaries

\makeindex[name=termindex,title={Term Index}]   
\makeindex[title={Author Index}]                
\renewbibmacro*{citeindex}{\ifciteindex{\indexnames[family][1-999]{labelname}}{}}

\hypersetup{pdfborder=0 0 0}    
\usepackage{nameref}            

\usepackage{epigraph}
\newcommand*{\mytilde}{\raise.17ex\hbox{$\scriptstyle\mathtt{\sim}$}}

\newcommand*{\nameWM}{Prof.~Dr.~Maalej}

\newcommand*{\nameCL}{Dr.~L\"uders}

\newcommand*{\nameCR}{Christian Rahe}

\newcommand*{\statementPublication}[1]{%
    \begin{spacing}{1}
        \vspace{5mm}
        \footnotesize
        \noindent
        \textbf{Publications.}
        This chapter contains portions of my collaborative publication, some copied verbatim~#1.
        \vspace{-5mm}
    \end{spacing}
}

\newcommand*{\bp}[2]{\csname bp#1#2\endcsname}

\newcommand*{\fixoverfull}[1]{\emergencystretch 3em #1}  
\newcommand*{\ltexignore}[1]{#1}    
\newcommand*{\ltexdummy}[1]{#1}     

\newcommand*{\refBPTable}[1]{``\bp{#1}{UID}: \bp{#1}{NAM}'' (see Table~\ref{tab:bp_\bp{#1}{REF}})}

\newcommand*{\numItebp}{40}
\newcommand*{\numItebpAlgs}{18}
\newcommand*{\varNumUsedIssueTypes}{352}

\newcommand*{\addtoabbreviations}[5]{
    \newglossaryentry{#1}{
        type        = {abbreviations},
        name        = {#2},
        short       = {#2},
        plural      = {#3},
        long        = {#4},
        longplural  = {#5},
        description = {#4}}}

\newcommand*{\addtoglossary}[3]{
    \newglossaryentry{#1}{
        type={glossary},
        name={#2},
        long={#2},          
        description={#3}}}

\newcommand*{\addtoabbreviationsandglossary}[6]{

    \addtoglossary{#1_gls}{#4}{#6}  

    \newglossaryentry{#1}{
        type        = {abbreviations},
        name        = {#2},
        short       = {#2},
        plural      = {#3},
        long        = {#4},
        longplural  = {#5},
        first       = {#4 (#2)\glsadd{#1_gls}},
        firstplural = {#5 (#3)\glsadd{#1_gls}},
        see         = [Glossary:]{#1_gls},  
        description = {#4},
    }
}

\addtoglossary{are}{Agile Requirements Engineering}
    {The approach to Requirements Engineering whereby iteration is a key principle. Instead of gated phases (as in Traditional Requirements Engineering), requirements are gathered and implemented in short cycles of 2--6 weeks. These cycles allow for quick prototyping, customer interaction, and small increments of delivered software.}

\addtoglossary{cre}{Change-Based Requirements Engineering}
    {The approach to Requirements Engineering whereby the focus is on product maintenance, instead of product creation. The requirements are gathered through various bottom-up requests like Bug Reports and Feature Requests, and very little (if any) roadmap or corporate direction. Instead, it is the users and small group of maintainers who keep the software updated and relevant.}

\addtoabbreviationsandglossary{erd}
    {ERD}{ERDs}
    {Entity Relationship Diagram}{Entity Relationship Diagrams}
    {A type of UML diagram that illustrates how entities, such as people, objects, or concepts, related to each other within a defined scope.}

\addtoglossary{issue}{issue}{An artefact in an Issue Tracking System (ITS), the represents and individual unit of work to be completed. Each issue has a type, which dictates its role within the ITS. Examples of issue types include User Story, Feature Request, and Bug Report.}

\addtoabbreviationsandglossary{ite}
    {ITE}{ITEs}
    {Issue Tracking Ecosystem}{Issue Tracking Ecosystems}
    {The context that surrounds an Issue Tracking System, including the organisation/team/people who use it, the processes defined to interact with it, and the communication channels that interface with it.}

\addtoabbreviationsandglossary{its}
    {ITS}{ITSs}
    {Issue Tracking System}{Issue Tracking Systems}
    {Software that manages and maintains lists of issues. Traditionally used by Software companies to build software, but can also be found within other organisational contexts.}

\addtoabbreviationsandglossary{iqr}
    {IQR}{IQRs}
    {Interquartile Range}{Interquartile Ranges}
    {A measure of statistical distribution across a set of data. It is defined as the difference between the 75th and 25th percentiles of the data. In other words, the central 50\% of the data.}

\addtoglossary{is}{Information Systems}
    {The field within computer science concerned with (and the process of) collecting, processing, storing, and distributing information, usually within large software systems.}

\addtoabbreviationsandglossary{jr}
    {Jira Repo}{Jira Repos}
    {Jira Repository}{Jira Repositories}
    {Jira is an Issue Tracking System designed by the company Atlassian. A Jira Repository is a singular instance of Jira, which was created by an organisation and contains their data.}

\addtoabbreviationsandglossary{nlp}
    {NLP}{}
    {Natural Language Processing}{}
    {The application of computational techniques to the analysis and synthesis of natural language.}

\addtoabbreviationsandglossary{oss}
    {OSS}{}
    {Open-Source Software}{}
    {Software that is released under a licence in which the copyright holder grants users the rights to use, change, and distribute the software and its source code.}

\addtoabbreviations{qa}
    {QA}{}
    {Quality Assurance}{}

\addtoabbreviationsandglossary{re}
    {RE}{}
    {Requirements Engineering}{}
    {The field within Software Engineering concerned with (and the process of) defining, documenting, and maintaining requirements.}

\addtoabbreviationsandglossary{sdd}
    {SDD}{SDDs}
    {Software Design Document}{Software Design Documents}
    {A representation of a software design that is to be used for recording design information, addressing various design concerns, and communicating that information to the design's stakeholders.}

\addtoabbreviationsandglossary{sdm}
    {SDM}{SDMs}
    {Software Development Methodology}{Software Development Methodologies}
    {A process for organising, planning, and controlling the process of developing software systems.}

\addtoabbreviationsandglossary{spm}
    {SPM}{SPMs}
    {Software Process Model}{Software Process Models}
    {An abstraction of the actual processes being conducted when building software.}

\addtoabbreviationsandglossary{se}
    {SE}{}
    {Software Engineering}{}
    {The field within computer science concerned with (and the process of) designing, developing, testing, and maintaining software.}

\addtoabbreviationsandglossary{sms}
    {SMS}{SMSs}
    {Systematic Mapping Study}{Systematic Mapping Studies}
    {A scientific methodology for collecting and summarising a field (or topic within a field) in a repeatable, rigorous way. Not to be confused with a Systematic Literature Review, which also includes the process of synthesising the data in some critical way.}

\addtoabbreviationsandglossary{srs}
    {SRS}{SRSs}
    {Software Requirements Specification}{Software Requirements Specifications}
    {A description of a software system to be developed. Often includes functional and non-functional requirements, use cases, important stakeholders, existing environment, and scope.}

\addtoglossary{tre}{Traditional Requirements Engineering}
    {The approach of Requirements Engineering whereby each phase is gated, requiring the previous phase to be complete before the next phase is started. This approach is also called the Waterfall Model, since you cannot go back up the waterfall. In this approach, the entire Requirements Engineering phase is completed before moving on to implementation. In other words, the Requirements Engineering is conducted only and completely up-front.}

\addtoglossary{ta}{Thematic Analysis}
    {A qualitative data analysis technique for analysing large quantities of textual data, with the goal of producing a taxonomy of information summarising the textual data. The output is a thematic map, which is a set of themes representing the information in the textual data. Each theme in a thematic map also normally has codes that further break down the themes into meaningful insights from the data.}

\addtoabbreviations{uml}
    {UML}{}
    {Unified Modelling Language}{}

\newcommand*{\bpHeaderMain}[1]{\multicolumn{2}{l}{#1}}
\newcommand*{\bpHeaderSecondary}[1]{\footnotesize{\textbf{#1:} }}
\newcommand*{\bpContent}[1]{\footnotesize{#1}}

\newcommand*{\bpTable}[2][]{
    \begin{table}[ht]
    \centering
    \renewcommand*{\arraystretch}{1}
    \setlength{\LTleft}{-20cm plus -1fill}
    \setlength{\LTright}{\LTleft}
    \begin{longtable}{p{.01\textwidth}p{.95\textwidth}}
            \captionsetup{font={Large}}
            \caption{\bp{#2}{UID}: \bp{#2}{NAM}.}\label{tab:#1bp_\bp{#2}{REF}}\\
        \toprule
            \bpHeaderMain{Summary} \\
                & \bpHeaderSecondary{Objective}\bpContent{\bp{#2}{OBJ}} \\
                & \bpHeaderSecondary{Motivation}\bpContent{\bp{#2}{MOT}} \\
        \midrule
            \bpHeaderMain{Recommendation} \\
                & \bpHeaderSecondary{Process}\bpContent{\bp{#2}{RECa}} \\
                & \bpHeaderSecondary{ITS}\bpContent{\bp{#2}{RECb}} \\
        \midrule
            \bpHeaderMain{Context} \\
                & \bpHeaderSecondary{Stakeholder Benefits}\bpContent{\bp{#2}{CONa}} \\
                & \bpHeaderSecondary{Stakeholder Costs}\bpContent{\bp{#2}{CONb}} \\
                & \bpHeaderSecondary{Artefact Scope}\bpContent{\bp{#2}{CONc}} \\
                & \bpHeaderSecondary{Issue Types}\bpContent{\bp{#2}{CONd}} \\
                & \bpHeaderSecondary{Inclusion Factors}\bpContent{\bp{#2}{CONe}} \\
                & \bpHeaderSecondary{Exclusion Factors}\bpContent{\bp{#2}{CONf}} \\
        \midrule
            \bpHeaderMain{Violation} \\
                & \bpHeaderSecondary{Smells}\bpContent{\bp{#2}{VIOa}} \\
                & \bpHeaderSecondary{Consequences}\bpContent{\bp{#2}{VIOb}} \\
                & \bpHeaderSecondary{Causes}\bpContent{\bp{#2}{VIOc}} \\
                & \bpHeaderSecondary{Algorithmic Detection}\bpContent{\bp{#2}{VIOd}} \\
        \bottomrule
            \multicolumn{2}{c}{\color{gray}\scriptsize Extracted and extended from \bp{#2}{AUT}} \\
    \end{longtable}
    \end{table}
}

\newcommand*{\bpTocLine}[1]{
    \bp{#1}{UID} &
    \bp{#1}{NAM} &
    \bp{#1}{OBJ} &
    \cite{\bp{#1}{CIT}} &
    \pageref{tab:bp_\bp{#1}{REF}}  \\
}
\newcommand*{\bpCategoryHeader}[1]{\multicolumn{5}{c}{--- #1 ---}  \\ \midrule}

\newcolumntype{R}[1]{>{\RaggedRight}p{#1}}

\newcommand*{\bpToC}[1]{
\begingroup
  
    \setlength{\LTleft}{-20cm plus -1fill}
    \setlength{\LTright}{\LTleft}

    \footnotesize  

    \renewcommand*{\arraystretch}{1.2}  
    \setlength{\tabcolsep}{2pt}         

    \begin{longtable}{
        l  
        >{\baselineskip=10pt}R{.25\textwidth}  
        >{\baselineskip=10pt}R{.50\textwidth}  
        >{\baselineskip=10pt}R{.13\textwidth}  
        c  
    }

        \caption{Best Practices Catalogue - Table of Contents.}\label{tab:catalogue_toc_#1} \\

        \toprule
        \textbf{ID} & \textbf{Name} & \textbf{Objective} & \textbf{Source} & \textbf{Page} \\ \midrule
        \endfirsthead
        \multicolumn{5}{@{}l}{Table \thetable, continued} \\
        \addlinespace
        \toprule
        ID & Name & Objective & Source & Page \\ \midrule
        \endhead
        \midrule
        \multicolumn{5}{r@{}}{\em\small (continued on next page)} \\
        \endfoot
        \bottomrule
        \endlastfoot
      
        \bpCategoryHeader{Issue Properties}
        \bpTocLine{GoodBugReport}
        \bpTocLine{AtomicFeatureRequests}
        \bpTocLine{AssignBugsToIndividuals}
        \bpTocLine{SufficientDescription}
        \bpTocLine{SuccinctDescription}
        \bpTocLine{AvoidStatusPingPong}
        \bpTocLine{AvoidAssigneePingPong}
        \bpTocLine{SetBugReportAssignee}
        \bpTocLine{SetBugReportPriority}
        \bpTocLine{SetBugReportSeverity}
        \bpTocLine{SetBugReportEnvironment}
        \bpTocLine{AssigneeBugResolution}
        \bpTocLine{AvoidZombieBugs}
        \bpTocLine{ActiveBugReports}
        \bpTocLine{BugReportDiscussion}
        \bpTocLine{RespectfulCommunication}
        \bpTocLine{ConsistentProperties}
        \bpTocLine{GoodFirstAssignee}
        \bpTocLine{StableClosedState}
        \bpTocLine{TimelySevereIssueResolution}
        \bpTocLine{IssueCreationGuidelines}
        \bpTocLine{OnTopicDiscussions}
      
        \midrule
        \bpCategoryHeader{Issue Linking}
        \bpTocLine{BugToCommitLinking}
        \bpTocLine{LinkDuplicates}
        \bpTocLine{MinimalLinkTypes}
        \bpTocLine{RecordLinks}
        \bpTocLine{RealisticDependencies}
        \bpTocLine{SingularRelationships}
        \bpTocLine{ConnectedHierarchies}
        \bpTocLine{HighDependencyBugsFirst}
        \bpTocLine{SearchReminders}

        \midrule
        \bpCategoryHeader{Issue Processes}
        \bpTocLine{OrderedProductBacklog}
        \bpTocLine{TeamProducedWorkEstimates}
        \bpTocLine{StoryPointsOverHours}
        \bpTocLine{EstimateAllItems}
        \bpTocLine{AvoidUnplannedWork}
        \bpTocLine{RecommendedSprintLength}
        \bpTocLine{ConsistentSprintLength}
        \bpTocLine{UseAcceptanceCriteria}
        \bpTocLine{LimitAcceptanceCriteria}

    \end{longtable}
\endgroup
}

\begin{document}

    \frontmatter

\thispagestyle{empty}	

\begin{tikzpicture}[remember picture, overlay]
	\node [anchor=north west, inner sep=0pt]  at (current page.north west)
	   {\includegraphics[height=4cm]{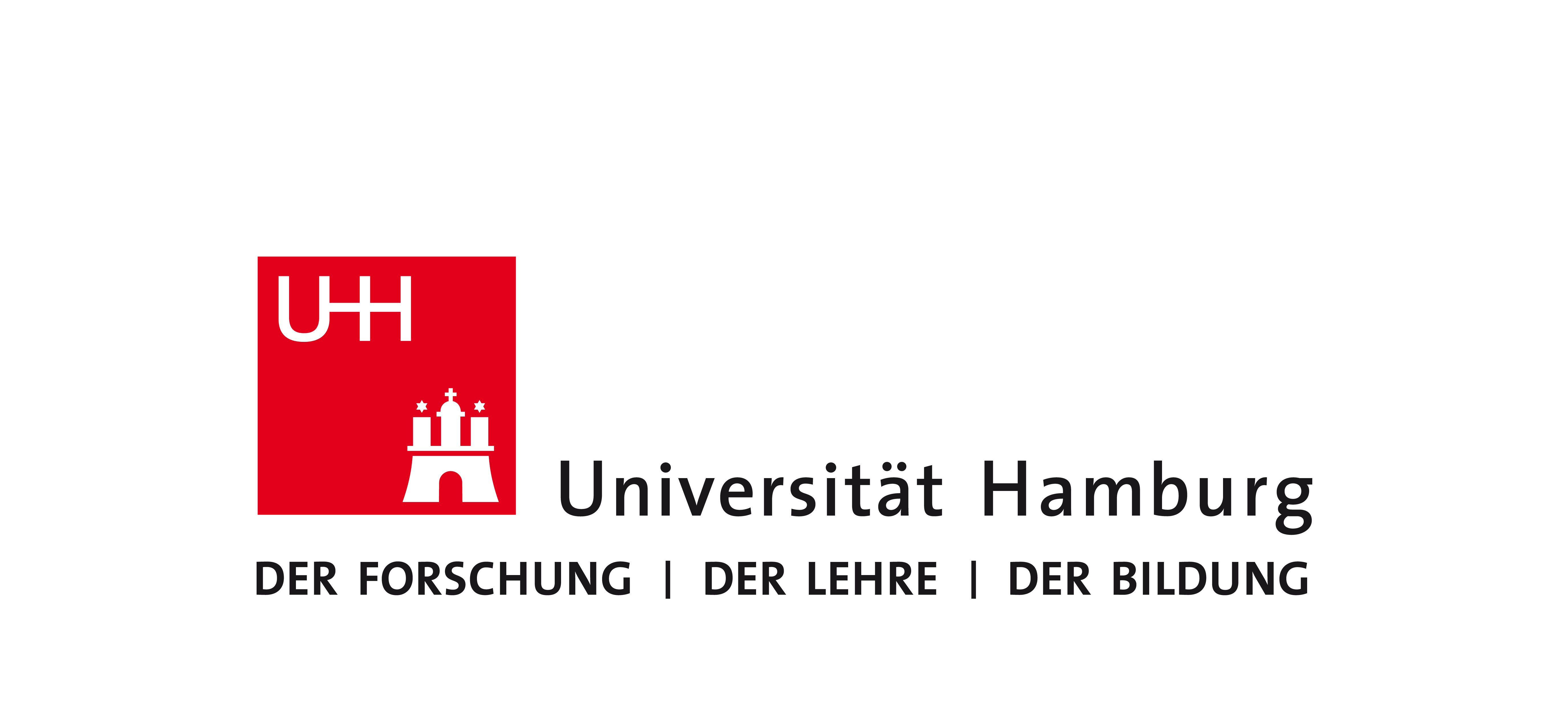}};
\end{tikzpicture}

\begin{center}

	\vspace{20mm}
	{\huge\textbf{Issue Tracking Ecosystems}\\[8pt]Context and Best Practices}
	\vspace{20mm}


	Dissertation for the doctoral degree \textbf{(Dr. rer. nat.)}\\

	at the Faculty of Mathematics, Informatics, and Natural Sciences\\
	Department of Informatics\\
	University of Hamburg\\
    Hamburg, Germany

	\vspace{20mm}

    submitted by

	\vspace{5mm}
	{\Large Lloyd Montgomery}
	\vspace{5mm}

	from White Rock,\\British Columbia, Canada

	\vspace{20mm}

	Hamburg, 2025
\end{center}

\clearpage                  
\thispagestyle{empty}		
\enlargethispage{30mm}		
\vspace*{\fill}             


\begin{table}[b]
    \begin{tabular}{ll}
        Oral Defence Date       & April 11th, 2025 \\
                                & \\
        Examination Commission  & Prof. Dr. Thomas Ludwig (Chair) \\
                                & Prof. Dr. Peter Kling (Vice Chair) \\
        Dissertation Evaluators & 1. Prof. Dr. Walid Maalej \\
                                & 2. Prof. Dr. Andreas Vogelsang \\
                                & 3. Ass. Prof. Dr. Eray Tüzün
    \end{tabular}
\end{table}

\clearpage                  
\thispagestyle{empty}		
\enlargethispage{30mm}		
\vspace*{\fill}             

\begin{center}
    \scriptsize
    This work \textcopyright\ 2025 by Lloyd Montgomery is licenced under Creative Commons Attribution 4.0 International.\\
    To view a copy of this licence, visit https://creativecommons.org/licences/by/4.0/
\end{center}

\cleardoublepage

\includepdf[pages=-]{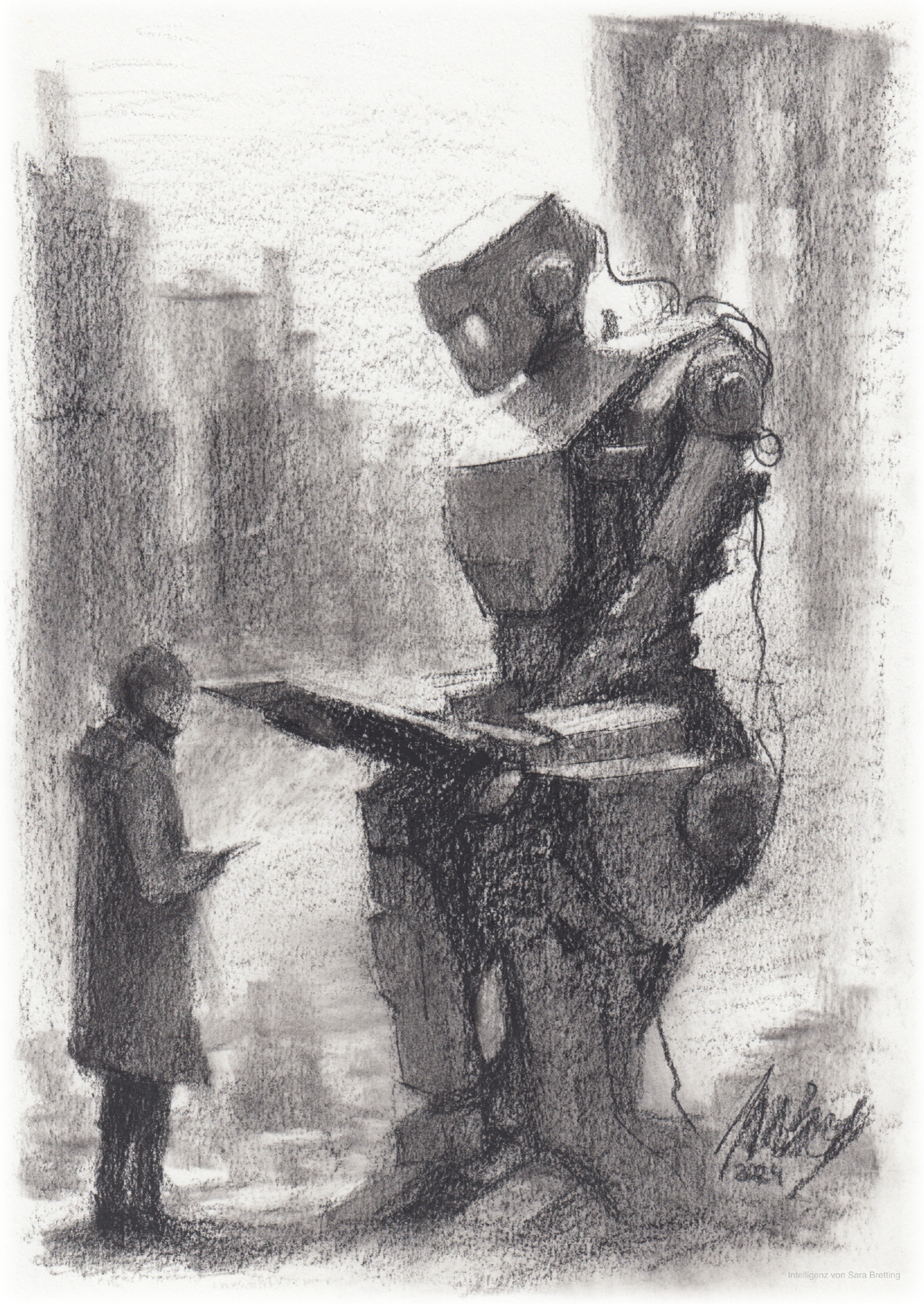}

\cleardoublepage            
\thispagestyle{empty}		

\vspace*{\fill}             
\centerline{\emph{To my wife, Lisa.}}  
\vfill                      

\clearpage                  
\thispagestyle{empty}       
\textcolor{white}{.}        

\chapter*{Acknowledgements}

\thispagestyle{empty}  

The first person I need to acknowledge is my wife, Lisa.
She provided me with both the encouragement to push through difficult times, and the kindness to tell me when to take a break.
Most importantly, she taught me the value of living a full life, a life beyond the PhD.

I am deeply grateful for the opportunities afforded to me by my Doktorvater, Walid Maalej.
Walid's acclaim in the research community brought me to his research group in Germany.
His hard work and dedication provided me a position in an EU-funded project that enabled me to travel Europe and collaborate with internationally renown researchers.
His guidance elevated me into the research community as a respected scientist, reviewer, and peer.
Lastly, his patience and feedback supported my journey towards achieving my PhD.
I am forever grateful.

I'd like to thank my family for their continued support throughout my academic journey.
My sister, Alexis, has provided me with a lifetime of sibling joy and connection.
My mother, Jennifer, has been patient and supportive through my growth as an individual.
My father, Dale, has always been there for me providing both support and entertainment.
My grandmother, Julie, has always been there as a steady, reliable source of wisdom and delicious food.
My stepfather, Mike, has provided stability to our second family and is always there for my mother.
I'd also like to acknowledge my new family who have accepted and embraced me as one of their own.
Thank you to Christoph, Anja, Lea, and Mathis for welcoming me into their home.
Without their feedback and support, this journey would have been much harder.

My friends have brought me untold amounts of joy, acceptance, patience, and support.
Volodymyr has been a good friend ever since I first started my PhD, and has provided much needed entertainment.
Davide was always there when an event was to be had, a real friend you could count on.
Guy is someone I can always go to for brutally honest feedback, which is something we all need every now and then.
Dorothea has wonderfully refreshing perspectives on both academia and life in general.
Thomas and Ela are two beautifully kind souls who brought me into their life and provided me with my first home in Germany.
Svenja warmed my heart, and together we formed many lasting memories together.
In recent years, I am grateful to have many new and meaning connections in my life, including Tim, Aref, and Ariane.
Without such kind and giving people around me, this journey would have been much more difficult.

\clearpage              
\thispagestyle{empty}   

The MAST research group has been my home away from home, filled with people I deeply respect and have formed many strong memories with.
Davide, Daniel, Christoph, and Volodymyr have been great colleagues since I first started my PhD.
They have challenged my understanding of the world, improved my work, and helped shape my academic career.
Marlo, Clara, Yen, and Abir brought many interesting conversations that improved my perspective on many topics.
Christian, Tim Puhlf{\"u}r{\ss}, Tim Pietz, and Aref have all been fantastic colleagues to get to know over the past few years.
Jakob, Pavel, Michael, and Amna have all brought meaningful outside perspectives into the MAST group.
I also want to thank Farnaz, Alexander, and Zijad for their feedback and support on various academic topics.
I'd like to thank those who gave me feedback on my thesis: Tim, Abir, Volodymyr, Pavel, Christian, and Michael.
I'd also like to thank the University of Hamburg for hosting my PhD, and providing a platform for me to teach and mentor many intelligent and talented individuals.

The Software Engineering research community has many characteristics---many positive and some negative.
What is most positive about this community, is the people who stand out as champions of strong scientific principles, act as meaningful mentors, and offer paradigm-shifting ideals---none of which are extrinsically rewarded by the academic system.
In this category of people, I'd like to give a special thank you to Daniel Mendez, Julian Frattini, Marvin Wyrich, Davide Fucci, Alessio Ferrari, Andreas Vogelsang, Margaret-Anne Storey, and Alexander Serebrenik.
Daniel Mendez understands and applies the true value of empathy within our community, and he knows how to support those around him.
If was his words of encouragement, and perspective-altering discussions about the Software Engineering community, that kept me pushing through my PhD.
Julian Frattini and Marvin Wyrich are beacons of excellence in a system of mediocracy; they know where meaningful change is needed, and work to achieve it.
Both Julian and Marvin have given me much hope in my outlook on the Software Engineering research community: that there exists a future generation of scientists who will make a meaningful difference.
Davide Fucci understands the value of community, and works vigorously to keep people connected, invested, and producing the best output they are capable of.
He has had a positive impact on my experience in this community by showcasing just how powerful bringing people together can be.
Alessio Ferrari and Andreas Vogelsang have been champions for young scientists looking to join, integrate, and excel in our community.
While I have not had the privilege of working directly with either of them, they have both shown me kindness as a young member of the community, encouraging me to continue the academic journey.
Margaret-Anne Storey and Alexander Serebrenik are academics who have connected with me along my PhD journey, offering their time, their attention, and their advice.
Each of them is a highly respected prominent member of the community in their respective research areas, giving keynotes at top-tier Software Engineering venues, and yet they immediately drop everything when they sense that you need them.
I received words of wisdom from both of them at critical points in my PhD journey, and I am very thankful for their advice.

\clearpage              
\thispagestyle{empty}   
\vspace*{\fill}         
\begin{center}          
    \begin{minipage}{.8\textwidth}  
        \begin{center}  
            The PhD journey is like competing in the Olympics: you showcase feats of personal intelligence, skill, and endurance, but this is only possible because of the team of coaches, peers, family, and friends who enable them.
        \end{center}
    \end{minipage}
\end{center}
\vspace{5mm}            
\centerline{Thank you team.}
\vfill                  
\clearpage              

\chapter*{Abstract}

Issue Tracking Systems (ITSs), such as GitHub and Jira, are popular tools that support Software Engineering (SE) organisations through the management of ``issues'', which represent different SE artefacts such as requirements, development tasks, and maintenance items.
ITSs also support internal linking between issues, and external linking to other tools and information sources.
This provides SE organisations key forms of documentation, including forwards and backwards traceability (e.g., Feature Requests linked to sprint releases and code commits linked to Bug Reports).
An Issue Tracking Ecosystem (ITE) is the aggregate of the central ITS and the related SE artefacts, stakeholders, and processes---with an emphasis on how these contextual factors interact with the ITS.
The quality of ITEs is central to the success of these organisations and their software products.
There are challenges, however, within ITEs, including complex networks of interlinked artefacts and diverse workflows.
While ITSs have been the subject of study in SE research for decades, ITEs as a whole need further exploration.

In this thesis, I undertake the challenge of understanding ITEs at a broader level, addressing these questions regarding complexity and diversity.
I interviewed practitioners and performed archival analysis on a diverse set of ITSs.
These analyses revealed the context-dependent nature of ITE problems, highlighting the need for context-specific ITE research.
While previous work has produced many solutions to specific ITS problems, these solutions are not consistently framed in a context-rich and comparable way, leading to a desire for more aligned solutions across research and practice.
To address this emergent information and lack of alignment, I created the Best Practice Ontology for ITEs.
Using this ontology, I curated a catalogue of Best Practices from existing literature, including Timely Severe Issue Resolution, Bug-to-Commit Linking, and Avoid Zombie Bugs.
I also collected and created algorithms to automatically detect violations to these Best Practices.
Finally, I proposed and evaluated tooling solutions that describe how to integrate these Best Practices into existing development environments.

The findings from this thesis enable a structured approach to improving the quality of ITEs.
The Best Practice ontology, catalogue, and algorithms are contributions to researchers interested in understanding and improving ITEs.
In practice, the context-aware catalogue and algorithms can be used to identify key areas for improvement, and to automate organisational processes such as maintaining a meaningful backlog and ensuring the completeness of issues.


\chapter*{Zusammenfassung}

Issue Tracking Systems (ITSs) wie GitHub und Jira sind beliebte Tools, die Software Engineering (SE) Organisationen durch die Verwaltung von „Issues“ unterstützen, die verschiedene SE-Artefakte wie Anforderungen, Entwicklungsaufgaben und Wartungselemente darstellen.
ITS unterstützen auch die interne Verknüpfung von Issues und die externe Verknüpfung mit anderen Tools und Informationsquellen.
Dies bietet SE-Organisationen wichtige Formen der Dokumentation, einschließlich vorwärts und rückwärts gerichteter Rückverfolgbarkeit (z. B. Verknüpfung von Feature Requests mit Sprint Releases und Code Commits mit Bug Reports).
Ein Issue Tracking Ecosystem (ITE) ist die Gesamtheit des zentralen ITS und der damit verbundenen SE-Artefakte, Stakeholder und Prozesse---mit dem Schwerpunkt darauf, wie diese kontextuellen Faktoren mit dem ITS interagieren.
Die Qualität der ITEs ist entscheidend für den Erfolg dieser Organisationen und ihrer Softwareprodukte.
Es gibt jedoch Herausforderungen innerhalb von ITEs, einschließlich komplexer Netzwerke miteinander verbundener Artefakte und unterschiedlicher Arbeitsabläufe.
Während ITS seit Jahrzehnten Gegenstand der SE-Forschung sind, müssen ITEs als Ganzes weiter erforscht werden.

In dieser Arbeit stelle ich mich der Herausforderung, ITEs auf einer breiteren Ebene zu verstehen und diese Fragen hinsichtlich Komplexität und Vielfalt zu beantworten.
Ich befragte Praktiker und führte eine Archivanalyse einer Vielzahl von ITS durch.
Diese Analysen haben die kontextabhängige Natur von ITS-Problemen und -Lösungen aufgezeigt und den Bedarf an kontextspezifischer ITS-Forschung verdeutlicht.
Während frühere Arbeiten vielfältige Analysen und Lösungen für spezifische ITS-Probleme hervorgebracht haben, sind diese Lösungen nicht durchgängig kontextabhängig und vergleichbar, was zu dem Wunsch nach besser abgestimmten Lösungen in Forschung und Praxis führt.
Um dieses Informationsdefizit und den Mangel an Abstimmung zu beheben, habe ich eine ITE-Best-Practice-Ontologie erstellt.
Anhand dieser Ontologie habe ich einen Katalog bewährter Verfahren aus der vorhandenen Literatur zusammengestellt.
Anschließend habe ich Tooling-Lösungen vorgeschlagen und bewertet, die beschreiben, wie diese bewährten Verfahren in bestehende Umgebungen integriert werden können.

Die Erkenntnisse dieser Arbeit ermöglichen einen strukturierten Ansatz zur Verbesserung der ITE-Qualität.
Die Ontologie, der Katalog und die Algorithmen sind Beiträge für Forscher, die sich für das ITE-Verständnis und die -Verbesserung interessieren.
In der Praxis können der kontextbezogene Katalog und die Algorithmen genutzt werden, um Schlüsselbereiche für Verbesserungen zu identifizieren und organisatorische Prozesse zu automatisieren.

\tableofcontents

    \mainmatter

    \part{Foundation Staging}  \label{part:foundation}
    
\chapter{Introduction}  \label{ch:introduction}

\epigraph{The purpose of software engineering is to control complexity,\\not to create it.}{Dr. Pamela Zave}

\section{Problem Statement}  \label{sec:intro_problem}

\xgls{se} is the process of planning, developing, testing, and maintaining software~\cite{Sommerville_2011_Book}.
\xgls{se} organisations conduct various forms of \xgls{se}, ranging from more traditional models involving heavy processes and strict stakeholder hierarchy~\cite{Maalej_2013_MARKBook}, to more open models of light-weight processes and open stakeholder communication~\cite{Ernst_2012_empiRE}.
Despite the rich and diverse landscape of \xgls{se} processes, it is rare to find an \xgls{se} organisation today that does not use some form of \xgls{its} to support their \xgls{se} processes.
An \xgls{its} is a software tool that structures information in the form of independent ``issues'', which are individual units of work to be conducted within organisations.
For many teams, \xglspl{its} are the main place to collect and manage many types of \xgls{se} artefacts, including requirements~\cite{VanCan_2024_REFSQ}, development~\cite{Lüders_2022_RE}, and maintenance~\cite{Bettenburg_2008_FSE,Zimmermann_2010_TSE}.
This is the case, for example, with many \xgls{oss} projects hosted on GitHub such as Mastodon\footnote{\url{https://github.com/mastodon/mastodon}} and ReactJS.\footnote{\url{https://github.com/facebook/react}}
An \xgls{ite} is the \xgls{its}, and all surrounding contextual factors that interface with the \xgls{its}.
This involves the \xgls{se} processes and stakeholders that interface with the \xgls{its}.
The quality of an \xgls{ite} refers to the extent to which it supports the organisation, including the features offered by the central \xgls{its} and the organisationally defined processes around it.

The benefit of \xglspl{ite} for \xgls{se} organisations are manifold, including communication and collaboration~\cite{Bertram_2010_CSCW}, workflow management, and user support~\cite{Montgomery_2025_BookChapter}.
\xglspl{ite} have emerged as a central place for communication and collaboration, particularly among \xgls{oss} communities~\cite{Bertram_2010_CSCW,Mockus_2002_TOSEM}.
Git has also been tightly linked to other \xgls{se} processes using \xglspl{its}, for example, GitHub has automatic linking of issues and Git commits using a custom ID system.
\xglspl{its} also unmask the complex workflows within a \xgls{se} organisation, encouraging more involvement from a diverse set of stakeholders.
For example, software users can see the status of Feature Requests.
User support has also been enhanced by modern \xglspl{its} in that Bug Reports can be public and therefore shared among users.
Bug reports can also be linked directly to \xgls{se} artefacts such as user stories or release plans.
Overall, the benefits of \xglspl{its} have lead to widespread adoption in industry~\cite{6SenseJira_2024_Online,DatanyzeJira_2024_Online,EnlyftJira_2024_Online} and many empirical studies seeking to understand these tools~\cite{Bettenburg_2008_FSE,Zimmermann_2010_TSE,Heck_2013_IWPSE,Heck_2016_REJ,Halverson_2006_CSCW}.

While central to modern \xgls{se}, \xglspl{ite} are notoriously difficult to navigate.
They have captured the attention of \xgls{se} research for over three decades, producing thousands of empirical findings across different research communities such as
    ICSE\footnote{\url{https://conf.researchr.org/series/icse}},
    FSE\footnote{\url{https://conf.researchr.org/series/fse}},
    RE\footnote{\url{https://conf.researchr.org/series/RE}},
    EMSE\footnote{\url{https://link.springer.com/journal/10664}},
    IST\footnote{\url{https://www.sciencedirect.com/journal/information-and-software-technology}}, and
    JSS.\footnote{\url{https://www.sciencedirect.com/journal/journal-of-systems-and-software}}
The diversity of \xglspl{its} also speaks to their inherent complexity: new \xglspl{its} continue to emerge as each attempts to simplify and solve the problems of the previous ones.
While a few are well-known in \xgls{se}, such as Jira~\cite{Jira_2024_Online}, GitHub~\cite{GitHub_2024_Online}, GitLab~\cite{GitLab_2024_Online}, and Bugzilla~\cite{Bugzilla_2024_Online}, Wikipedia lists over 30 established \xglspl{its},\footnote{\url{https://en.wikipedia.org/wiki/Comparison_of_issue-tracking_systems}} in addition to the various other \xgls{its}-like tools that are not on the list such as asana~\cite{asana_2024_Online}, Smartsheet~\cite{Smartsheet_2024_Online}, and Monday~\cite{Monday_2024_Online}.
Jira is the most popular \xgls{its} in industry~\cite{6SenseJira_2024_Online,DatanyzeJira_2024_Online,EnlyftJira_2024_Online}.
In pop culture, Jira is repeatedly the target of jokes regarding its complexity and difficulty of use~\cite{JiraJoke1_2024_Online,JiraJoke2_2024_Online,JiraJoke3_2024_Online}.
Making fun of Jira has become such a normalised online behaviour, that there is even a website dedicated entirely to showcasing negative quotes and opinions about Jira~\cite{JiraJoke4_2024_Online}.
Research into \xglspl{its} has revealed some of this complexity (e.g., divergent expectations and difficulty with tooling~\cite{Bettenburg_2008_FSE,Zimmermann_2010_TSE,Fucci_2018_RESFQ}), but there is still much to be understood.
\textbf{Gap 1}: \textit{Our understanding of practitioner challenges with \xglspl{ite} is limited.}

\xglspl{its} manage many \xgls{se} artefacts and activities, including requirements, development, and maintenance.
It is known, to some extent, that \xglspl{its} contain artefacts beyond just Bug Reports; however, this has not been studied in-depth.
Studies have revealed the \xglspl{its} support ``just-in-time'' \xgls{re}~\cite{Ernst_2012_empiRE}, similar to \ltexdummy{requirements} processes conducted during ``Agile \xgls{se}''~\cite{Heck_2013_IWPSE}.
Studies have also investigated the user support aspect of \xglspl{its}~\cite{Montgomery_2017_RE}.
However, the line between requirements, development, maintenance, and user support is blurred within \xglspl{its}~\cite{VanCan_2024_REFSQ}, perhaps by design, or perhaps as a natural consequence of the needs of modern \xgls{se}.
\xglspl{its} are also a central tool for Agile \xgls{se}, where iteration and evolution are key.
Research investigating evolution within \xglspl{its} is rare~\cite{Heck_2013_IWPSE}, and there are still open questions regarding information and evolution within \xglspl{its}.
\textbf{Gap 2}: \textit{We are missing a holistic understanding of which artefacts and activities exist within \xglspl{its}, and what information exists and evolves.}

\xglspl{its} are highly customisable tools, offering personalisation through configuration and automation.
This customisation is a major benefit of \xglspl{its}, but it also leads to complexity.
How to use and configure parts of \xglspl{its} has also been extensively researched~\cite{Bettenburg_2008_FSE,Zimmermann_2010_TSE}.
However, these studies have focused on a specific \xgls{issue} type~\cite{Bettenburg_2008_FSE,Zimmermann_2010_TSE,Heck_2013_IWPSE,Heck_2014_JSEP} or process~\cite{Ernst_2012_empiRE,Thompson_2016_MSR,Merten_2016_REFSQ,Li_2018_APSEC,Tomova_2018_TSE,Stanik_2018_ICSME,Lueders_2019_RE} within \xglspl{its}, and not the entire \xgls{ite}.
Additionally, existing recommendations are often contextless, offered as ``fix all'' solutions aimed at a general environment.
\textbf{Gap 3}: \textit{Context-dependent \xgls{ite} recommendations for practitioners are understudied, including a structured way to document recommendations for future research, and communicate this to practitioners.}

\section{Objectives and Contributions}  \label{sec:intro_obj_cont}

The primary direction of this thesis is to \textit{investigate, describe, and improve the quality of \xglspl{ite} in both research and practice}.
This includes four \textbf{objectives}, listed here:

\newcommand*{\itemsepObjectives}{0mm}

{\setstretch{1}  
\begin{enumerate}[label=\textbf{O\arabic*},itemsep=\itemsepObjectives]
    \item Specify and derive concepts essential to understanding and working with \xglspl{ite}.
    \item Obtain an empirically grounded understanding of \xglspl{ite} in practice.
    \item Improve the state of \xglspl{ite} through theoretical and practical solutions.
    \item Recommend future directions for \xglspl{ite} research and application in practice.
\end{enumerate}}

These four objectives form the structure of this thesis:
Part~\ref{part:foundation}: \nameref*{part:foundation} (O1),
Part~\ref{part:problem}: \nameref*{part:problem} (O2),
Part~\ref{part:solution}: \nameref*{part:solution} (O3), and
Part~\ref{part:outlook}: \nameref*{part:outlook} (O4).
Within each part, there are several chapters, each with their own research questions that further explore and address the objectives.
Each research question is addressed using empirical methods that produce rigorous findings.
Across these objectives, research questions, and their associated findings, nine central \textbf{contributions} emerged:

\newcommand*{\contributionHeader}[2]{\vspace{2mm}\noindent{\nameref*{#1} (O#2)}\vspace{0mm}}
\newcommand*{\itemsepContributions}{0mm}
\newcommand*{\lightStatementContributions}[1]{\vspace{-3mm}\textcolor{lightgray}{\scriptsize(#1)}}

{  
    \setstretch{1}  

    \vspace{5mm}

    \contributionHeader{part:foundation}{1}
    \begin{enumerate}[label=\textbf{C\arabic*},itemsep=\itemsepContributions]
        \item ``\xglsfirst{ite}'' as a fundamental concept.
    \end{enumerate}

    \contributionHeader{part:problem}{2}
    \begin{enumerate}[resume,label=\textbf{C\arabic*},itemsep=\itemsepContributions]
        \item A grounded understanding of the problems practitioners face within \xglspl{ite}.  \label{cont:problem_practitioners}
        \item A dataset of 16 Jira repositories with 2.7 million issues and 30 million issues.
        \item A data-driven characterisation of \xgls{its} artefacts, activities, information, and evolution.
        \item The cross-study finding that ``context is key'' for \xgls{ite} problems and solutions.
    \end{enumerate}

    \contributionHeader{part:solution}{3}
    \begin{enumerate}[resume,label=\textbf{C\arabic*},itemsep=\itemsepContributions]
        \item An ontology for Best Practices for \xglspl{ite}.
        \item A catalogue of Best Practices for \xglspl{ite}.
        \item Algorithms to support the automatic detection of violations to the Best Practices.
    \end{enumerate}

    \contributionHeader{part:outlook}{4}
    \begin{enumerate}[resume,label=\textbf{C\arabic*},itemsep=\itemsepContributions]
        \item Tooling recommendations to support Best Practices for \xgls{ite} in industry.
    \end{enumerate}
}  

\section{Scope}

The scope of this thesis is the quality of \xglspl{ite}.
The scope does not include the quality of \xgls{se} processes in general, the quality of software, or the general quality of artefacts involved in the \xgls{se} lifecycle.
An \xgls{ite} involves factors including \xgls{se} processes; however, this thesis focuses only on constructs that are tightly coupled with \xglspl{its}.
For example, the study of customer involvement in Agile processes is outside the scope of this work, but the study of sprint lengths is inside the scope of the work because \xglspl{its} manage and are directly affected by sprint lengths.
However, this thesis is only concerned with sprint lengths (and similarly tightly coupled processes) to the extent that \xglspl{its} can impact and manage them.
In other words, this thesis is not concerned with what sprint lengths are best for Agile, but rather how an \xgls{ite} can support and guide the management of sprint lengths as a feature of \xglspl{its}.
This applies to all \xgls{se} processes discussed in this thesis that are tightly coupled with \xglspl{its} and \xglspl{ite}.

\section{Outline}

\newcommand*{\outlineRefPart}[1]{\large Part \hyperref[{#1}]{\ref*{#1}. \nameref*{#1}}}
\newcommand*{\outlineRefCh}[2]{Ch \hyperref[{#1}]{\ref*{#1}. #2}}

\begin{figure}[bh]
    \centering
    \makebox[0pt]{\includegraphics[width=1\textwidth]{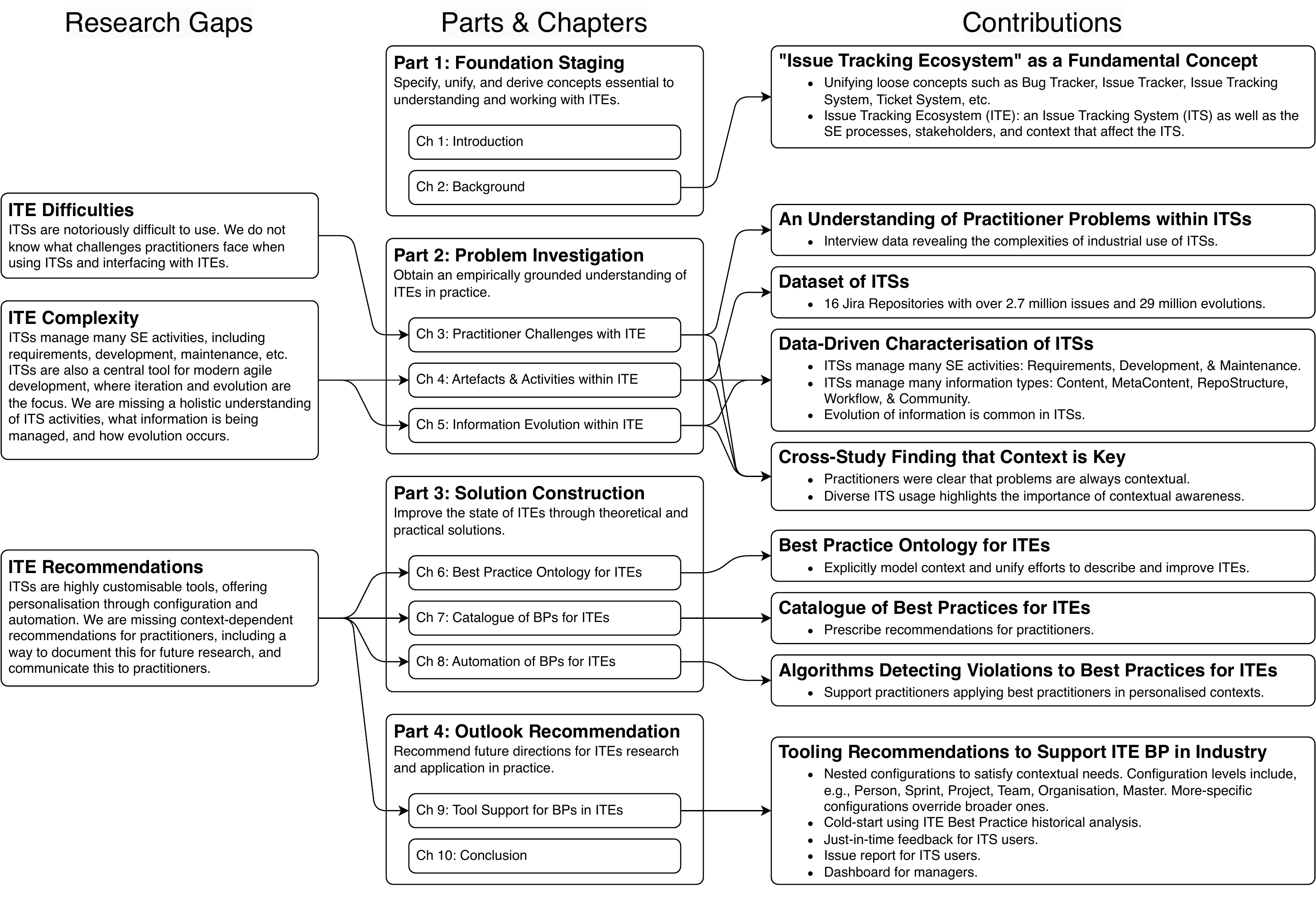}}
    \caption{Thesis overview: research gaps, parts, chapters, and contributions.}
    \label{fig:thesis_overview}
\end{figure}

The thesis is structured in four main parts: \nameref*{part:foundation}, \nameref*{part:problem}, \nameref*{part:solution}, and \nameref*{part:outlook}.
Figure~\ref{fig:thesis_overview} provides a visual overview of the thesis structure, from research gaps, through parts and chapters, and then finally the contributions.
Here is a textual overview of the parts, with a description for each chapter.

\begin{description}
    \item[\outlineRefPart{part:foundation}] \hphantom{.}
    \begin{description}
        \item[\outlineRefCh{ch:introduction}{Introduction}] Overview of the problem, objectives, contributions, and scope.
        \item[\outlineRefCh{ch:background}{Background}] Description of key concepts in this thesis.
    \end{description}
    \item[\outlineRefPart{part:problem}] \hphantom{.}
    \begin{description}
        \item[\outlineRefCh{ch:challenges}{Challenges}] Investigation of \xgls{ite} problems in industry through interviews with 26 \xgls{se} practitioners who utilise \xglspl{its} regularly.
        \item[\outlineRefCh{ch:activities}{Artefacts \& Activities}] Investigation of \xgls{se} artefacts and activities in \xglspl{its} through archival data analysis of 16 Jira \xglspl{its} consisting of 2.7 million issues.
        \item[\outlineRefCh{ch:evolution}{Information \& Evolution}] Investigation of information within \xglspl{its} and evolution of that information through archival analysis of 13 Jira \xglspl{its} consisting 1.3 million issues and 13 million evolutions.
    \end{description}
    \item[\outlineRefPart{part:solution}] \hphantom{.}
    \begin{description}
        \item[\outlineRefCh{ch:ontology}{Ontology}] Formalisation of \xgls{ite} Best Practices through the empirical construction of an ontology designed to structure \xgls{ite} quality research and offer an approachable format for practitioners.
        \item[\outlineRefCh{ch:catalogue}{Catalogue}] Consolidation of \xgls{ite} quality research into a catalogue of \xgls{ite} Best Practices to offer a starting point for both researchers and practitioners.
        \item[\outlineRefCh{ch:algorithms}{Algorithms}] Creation and replication of algorithms to automatically detection violations of \xgls{ite} Best Practices in the catalogue.
    \end{description}
    \item[\outlineRefPart{part:outlook}] \hphantom{.}
    \begin{description}
        \item[\outlineRefCh{ch:tooling}{Tooling}] Proposal of tooling features central to the creation of a recommender system for \xgls{ite} Best Practices in practice.
        \item[\outlineRefCh{ch:conclusion}{Conclusion}] Summarising the thesis.
    \end{description}
\end{description}

\vfill

\chapter{Background}  \label{ch:background}

\epigraph{I would rather have questions that can't be answered than answers that can't be questioned.}{Richard Feynman}

In this chapter, I outline the fundamental concepts used in this thesis.
The concepts in this thesis are all within the scope of \xgls{se} and concern the task of improving \xgls{se} processes.
In particular, this thesis describes \xglspl{its} (such as Jira~\cite{Jira_2024_Online}, GitHub~\cite{GitHub_2024_Online}, and GitLab~\cite{GitLab_2024_Online}).
While \xglspl{its} originally stored primarily maintenance items (such as Bug Reports), they have grown to support most \xgls{se} processes, including \xgls{re} and development.
The topic of maintenance in \xglspl{its} has received much research attention in the past two decades~\cite{Bettenburg_2008_FSE,Zimmermann_2010_TSE}, while the topic of requirements has not received nearly as much~\cite{Heck_2013_IWPSE,Heck_2016_REJ,VanCan_2024_REFSQ}.
Accordingly, this chapter describes \xgls{re} and how Agile practices have come to use and benefit from \xglspl{its}.
Finally, this chapter discusses empirical methods, and how they are applied within this thesis.

\statementPublication{\cite{Montgomery_2022_REJ,Montgomery_2025_BookChapter}}

\section{Requirements Engineering}

\xgls{re} is the process of gathering, documenting, verifying, and managing stakeholder needs, system constraints, and environmental factors for a system (often software) to be implemented.
\xgls{re} involves five fundamental phases: elicitation, analysis, specification, validation and verification, and management~\cite{Thayer_1997_Book,Maalej_2013_MARKBook,Pohl_2016_Book}.\footnote{The exact number of phases and the exact names used varies across sources, but they agree on the underlying concepts of these phases.}
\xgls{re} has a number of different forms and interpretations, including traditional, agile, and change-based~\cite{Maalej_2013_MARKBook,Mendez_2021_Lecture}.
Each form shapes the interpretations of the definitions, processes, artefacts, outcomes, and research involved.
This section outlines these three forms in sufficient detail to understand their similarities and differences, and discusses which forms are relevant to this thesis in which way.

There are five phases of \xgls{re}~\cite{Thayer_1997_Book,Maalej_2013_MARKBook}.
\textit{Elicitation} is the process of gathering requirements for a system.
Traditionally, this has been done through direct contact with important stakeholders, but has evolved into a much more complex and comprehensive ecosystem including data- or user-driven \xgls{re}~\cite{Pagano_2013_RE,Guzman_2014_RE,Maalej_2015_RE,Maalej_2015_IEEESoftware}.
\textit{Analysis} is the process of understanding the requirements and achieving consensus through modelling and discussions with various stakeholder groups.
\textit{Specification} is the process of documenting the requirements in some medium (textual or modelled).
\textit{Validation} is the process of checking if the system requirements---as documented---represent the system the stakeholders really want and need (e.g. ``are we building the right system'').
\textit{Verification} is the process of checking if the system itself is correct, via some well-defined metrics such as completeness and compliance (e.g. ``are we building the system right'')~\cite{Arora_2019_ESE}.
Finally, \textit{management} is the process of working with, updating, and maintaining the requirements throughout the duration of the project.

\subsection{Traditional RE}

\xgls{tre} is the first conceptualisation and formalisation of \xgls{re}, often called ``Waterfall \xgls{re}''~\cite{Petersen_2009_PROFES,Kassab_2018_IPCC,Laplante_2022_Book}.
The metaphor of ``Waterfall'' comes from the clear and well-defined phases that Traditional \xgls{re} is based on, whereby passing from one phase of \xgls{re} to the next is a one-way process (you cannot go back up the waterfall).
This form of \xgls{re} has the aim of creating upfront, well-structured, and complete documents representing the functional and non-functional requirements (and constraints) for a system~\cite{Eckhardt_2016_ICSE}.
These documents include a Problem Statement, a \xgls{srs}, and a \xgls{sdd}.
The Problem Statement outlines the gap between the current and desired system, often framed as a distinct set of ``problems'' to be addressed.
In some forms, a Problem Statement can contain much of what is included in a full \xgls{srs}, including functional and non-functional requirements, but in a shorter, non-complete, and high-level summary form.
The \xgls{srs} fully describes the requirements for a system, communicated through a description of the existing system and surrounding environment, a list of functional and non-functional requirements, use cases, and a glossary of domain-specific terms.
An \xgls{srs} can be hundreds of pages long.
The final document found in the Traditional \xgls{re} process is the \xgls{sdd}.
The \xgls{sdd} crosses the boundary from problem space into solution space, detailing aspects of the system to be built.
An \xgls{sdd} contains low-level descriptions of the system and object design, including class and architectural diagrams, as well as aspects of interface design.
In addition to the documentation created for the system, Traditional \xgls{re} can also involve pre-contract documents such as a Request for Quotes, Request for Proposals, and Business Proposals.
I outline these documents used in Traditional \xgls{re} to compare with those of the following \xgls{re} type.
I will not be discussing these documents further in this thesis.

\subsection{Agile RE}  \label{sec:background_agile_re}

\xgls{are} (also referred to as ``Agile \xgls{se}'') is the response to the major drawbacks found in Traditional \xgls{re}, primarily the long, gated, non-iterative phases.
The Agile Manifesto~\cite{AgileManifesto_2001_SD}, released in February 2001 by 17 prominent \xgls{se} researchers, consultants, and industry leaders, outlines a set of values for Agile \xgls{se}:
\begin{description}[itemsep=-2mm]
    \item[Individuals and interactions] over processes and tools.
    \item[Working software] over comprehensive documentation.
    \item[Customer collaboration] over contract negotiation.
    \item[Responding to change] over following a plan.
\end{description}

The modern conceptualisation of ``Agile'' involves adopting one of a few models, including Kanban, Scrum, Lean, Extreme Programming, and Feature-Driven Development~\cite{Alsaqqa_2020_iJIM}.
In principle, Agile \xgls{se} is simply the application of iterative feedback cycles to the primary phases of \xgls{se}, mainly \xgls{re}, software development, and system maintenance.
One common form of Agile \xgls{se} is Scrum.
The Scrum process involves small co-located teams working tightly together in small development ``sprints'' of 1--4 weeks.
A Scrum team is organised and managed by a combination of the roles Product Owner and Scrum Master.
The Scrum Master is a meta-role responsible for managing and aligning team practice with the Scrum framework.
The Product Owner is the most requirements-centric role, as they are concerned with the needs of their customers, users, and stakeholders, and guiding the team to create a better software product for them.

In Scrum, a popular form of requirements documentation is called a ``User Story''.
User Stories are one way to capture the needs and desires of software users.
Each User Story captures a small, self-contained unit of development work framed from the perspective of one actual user, conceptualised into a user group.
User Stories are traditionally written with the form ``As a [user persona], I want to [action], so that I [can accomplish this goal].''
Since User Stories are small and self-contained, there is no inherent way to build larger structures.
For this, agile teams use another artefact type called the ``Epic'', which is a high-level goal for software development.
Epics have their own description, and a group of User Stories contained within them.
Epics can also be used to form larger structures by grouping other Epics within them.
These Epics and User Stories are collected in an \xgls{its} in a list called the Product Backlog.
The Product Backlog doesn't have a limit to the number of Epics and User Stories contained within it, but it is the responsibility of the Product Owner to keep it clean and useful.
Product Backlogs are often stored within \xglspl{its} along with other \xgls{se} artefacts.

\subsection{Change-Based RE}  \label{sec:background_change_based_re}

\xgls{cre} is an emergent form of \xgls{re} focused on product maintenance and evolution, and is conducted primarily within \xglspl{its}~\cite{Mendez_2021_Lecture}.
While Traditional \xgls{re} and \xgls{are} are both methodologies designed for \textit{product creation}, Change-Based \xgls{re} has emerged as an important methodology for long-lived software systems where \textit{product maintenance and evolution} are the focus.
Change-Based \xgls{re} utilises \xglspl{its} as a central hub where maintenance tasks (such as Bug Reports) are collected and prioritised.
The current user base can submit Feature Requests to be considered for integration into the evolving product vision.
\ltexignore{\xgls{are} and Change-Based \xgls{re} are often used together by Agile teams who use an \xgls{its} to manage their \xgls{se} processes.}

The importance of Change-Based \xgls{re}, as an additional \xgls{re} methodology beyond Agile, is that Change-Based \xgls{re} is agnostic to how the original software product was formed.
Long-lived software systems can adopt a Change-Based \xgls{re} process as a sustainable way to keep a software product updated and relevant over time.
\xgls{oss} teams use Change-Based \xgls{re} through a publicly available \xgls{its} platform such as GitHub.
This allows them to keep the product alive through contributions from their active user base.
\ltexignore{This also allows them to maintain a useable product as software standards change and new hardware is released.}
Finally, it also allows \xgls{oss} products to stay relevant through community suggestions for new features and the community effort of developing desired features.
This applies to both Agile products and other products, as long as their maintainers desire this Change-Based \xgls{re} paradigm for their software product.

\subsection{Requirements Quality}

The quality of requirements refers to the individual characteristics of requirements that both lead to a successful and cost-effective system, and solve the user's needs~\cite{Davis_1993_CS,Montgomery_2022_REJ}.
Requirements quality has been the topic of research for many decades, including the seminal 1994 Standish Chaos Report describing the perceived importance of requirements on software project \ltexignore{success~\cite{Standish_1994_ChaosReport}}.
This report outlines project success factors such as ``clear statement of requirements'' and ``clear vision and objectives'', as well as project failure factors such as ``incomplete requirements''.
These factors are forms of quality for requirements, often called ``quality factors'' or ``quality attributes''.
In 2007, Kamata and Tamai empirically validated the claims of the Standish Chaos Report through an investigation of 32 industrial software projects focusing on the impact of \xgls{re} on quality, time, and cost calculations~\cite{Kamata_Tamai_2007_CS}.
Their research found that projects with an acceptable level of quality across all sections of their requirements were within the time and cost budgets.
These findings have also been confirmed more \ltexdummy{recently~\cite{Mendez_2017_EMSE,Wagner_2019_TOSEM}}.

The quality of requirements is closely related to the quality of \xglspl{ite}.
Requirements are often managed \textit{within} \xglspl{ite}, and are central to the processes therein.
In the case of requirements issues in \xglspl{its}, \textit{requirements quality} applies rather directly.
Other artefact types in \xglspl{its} (such as development or maintenance issues) are similar to requirements because all issues are actions that the maintainers should perform.
For the broader \xgls{ite} around the \xgls{its}, \ltexdummy{requirements quality} attributes such as ``completeness'', ``correctness'', and ``consistency'' can apply to many of the processes being conducted.
Requirements quality is not an all-encompassing concept, but there is considerable overlap between requirements quality factors and the quality of \xglspl{ite}.

\section{Issue Tracking Systems and Ecosystems}

An \xgls{its} is a tool used by \xgls{se} organisations to manage the development and maintenance of software.
Each \textit{issue} in an \xgls{its} represents a stand-alone unit of work to be performed as part of the project, or a problem to be solved.
The \textit{issue type} represents the main purpose of the issue and the work it comprises: for instance, a ``Bug Report'' to be fixed or a ``User Story''  to be implemented.
Within an \xgls{its}, issues are assigned to \textit{projects}, which are mutually exclusive bins that structure the work and teams within an organisation.
Each project usually has certain goals (e.g. the development of a certain tool), involves certain stakeholders, and follows some shared (explicit or tacit) practices and workflows. For example, ``Hadoop Common'' is an Apache project and ``JBoss Application Server'' is a RedHat project.
Each issue is composed of several issue \textit{fields}, where each field has a name and a value.
The fields represent the information relevant for the issue management: including the understanding, planning, and resolution of the issues~\cite{Zhang_2016_CJ}.
Common examples of fields include Summary, Description, Priority, and Comments.
When an issue is created, basic information such as its summary and description are stored.
Subsequently, the issue evolves until it gets resolved, meaning that the values of different fields are updated.
Each of these \textit{issue evolutions} has a timestamp, an author, the field name, as well as the value before and after.

\subsection{Benefits of Issue Tracking Systems}

\xglspl{its} are ubiquitous in the \xgls{se} community.
When learning about the workflow of a \xgls{se} organisation, it is inevitable that one must ask about which \xgls{its} they are using.
They could be using a simplified workflow with Trello~\cite{Trello_2024_Online}, a long-lived tool such as Trac~\cite{Trac_2024_Online}, or embracing the full complexity of Jira~\cite{Jira_2024_Online}.
Regardless, \xglspl{its} clearly play an important role in \xgls{se} organisations.
Looking closely at the features and fields offered by \xglspl{its}, there are more specific reasons \xglspl{its} are so popular and important.
One such reason is that \xgls{oss} communities use them as a \textit{central tool for communication and collaboration}~\cite{Bertram_2010_CSCW,Mockus_2002_TOSEM}.
The central tool for \xgls{oss} communities was once mailing lists, where each email thread represented a single discussion about a bug, feature, or wider topic.
Now, those discussions are held in the comments section of issues, further enhanced by the other features offered by issues trackers such as labelling and linking~\cite{Borg_2013_CSMR,Lueders_2019_RE,Lüders_2022_RE,Lüders_2022_MSR,Lüders_2022_RE}.
Another reason for their popularity is that modern \xglspl{its} are often used as \textit{an enhancement to traditional versioning systems such as Git}.
While Git captures most of what is necessary when it comes to versioning software development efforts, it rarely captures or links to the other \xgls{se} artefacts such as requirements or Bug Reports.
\xglspl{its} enable a rich enhancement through traceability between Git commits and the rationale behind them.

From an \xgls{re} perspective, \xglspl{its} have created a direct avenue for \textit{just-in-time requirements}~\cite{Ernst_2012_empiRE} from diverse stakeholder groups.
While there is merit in traditional \xgls{re} processes in certain contexts---which often have a top-down approach to requirements planning---\xglspl{its} enable just-in-time requirements where individual needs and specific requests are combined with community involvement (``Crowed \xgls{re}''~\cite{Groen_2017_IEEESoftware}).
For companies interested in understanding ``the crowd'', traditional \xgls{re} requires user studies, surveys, or interviews.
With properly managed and maintained \xglspl{its}, they have a constant stream of just-in-time requirements to shape their product direction.

\xglspl{its} have also enabled \textit{an open and transparent approach to managing \xgls{se} processes} through issue statuses and public conversations~\cite{Ruparelia_2010_SEN}.
The community involved in (or just interested in) the development of a particular product can see the workflow of requirements as they are created, discussed, assigned to sprints, worked on, tested, and delivered.
This transparency allows users to directly see the version a particular requirement is attached to, and know when and how the outcomes of that requirement will appear in their software.
Similarly, users invested in particular Bug Reports can be involved in many ways.
They can see if their problem is already reported, or if the similar Bug Reports have similar details.
They can also check the status and know if it is being addressed or ignored by the maintainers, and they can also see when the finalised fix for that bug is merged into which version release.
This kind of workflow management transparency is a powerful tool for managing an ever-shifting community of users.

Finally, one reason \xglspl{its} have gained popularity is the ability to \textit{conduct direct user support} in them, alongside related \xgls{se} activities.
User support differs from other maintenance tasks such as managing bug and security reports in that user support is directly concerned with the thoughts, feelings, and problems the customer is dealing with.
In other words, attention is given to this customer and their state of mind, and less attention is given to the software issue they are experiencing.
This is a central component of customer relationship management~\cite{Reinartz_2004_JMR}.
\xglspl{its} support this workflow, and allow direct linking to related Bug Reports, Feature Requests, user stories, and release schedules.
This creates forward traceability from user concerns to organisational action, and backwards traceability (rationale) for related organisational decisions.

Overall, the importance of \xglspl{its} on modern \xgls{se} is diverse and impactful across many aspects of the \xgls{se} lifecycle.
Many organisations leverage this power in different ways, and \xgls{se} researchers continue to strive to understand and improve \xglspl{its} as well as associated processes.

\subsection{Issue Tracking Ecosystems (ITEs)}

There are many terms to describe tools such as GitHub~\cite{GitHub_2024_Online}, Jira~\cite{Jira_2024_Online}, and RedMine~\cite{RedMine_2024_Online}, such as bug tracker~\cite{Bortis_2011_CHASE}, issue tracker, \xglsfirst{its}, issue repository, and ticket system.
Many of these terms exist for historical reasons.
For example, ``bug tracker'' is an older term from a time when these tools stored exclusively Bug Reports.
Even recent literature, however, uses terms interchangeably, such as \xgls{its} and ``backlog''~\cite{VanCan_2024_REFSQ}.
Seemingly, there is little difference between the underlying concepts these terms describe and how these terms are used.

These terms, despite containing words such as ``system'', solely refer to the software tool.
\xgls{se} research, however, also involves the study of people, their processes, their plans, desires, and interactions with various tools.
\xglspl{its} are tools designed specifically to support \xgls{se} processes such as development and maintenance, they are used by diverse stakeholders, and they are configured differently across companies, teams, and projects.
As such, it is important to consider these factors when researching \xglspl{its}, otherwise the results will likely suffer from threats to construct validity, and any recommendations for industry will be missing relevant context factors.

To address these shortcomings, I introduce a new term: \glsxtrlong{ite}.
\begin{quote}
    An \textbf{\xglsfirst{ite}} is a combination of an \xgls{its}, and all surrounding contextual factors that affect the \xgls{its}, including the \xgls{se} processes used within it, the stakeholders who interact with it, the company they work for, the team they work with, and the project they are working on.
\end{quote}

This concept has the key advantage that it explicitly includes the context (compared to \textit{alongside} or \textit{tangential}).
This acknowledgement has benefits.
The first benefit is simply spreading awareness that these context factors are always involved, and should be considered when studying \xglspl{its}.
It is often the case that research discussing \xglspl{its} ignores potentially relevant context factors, and one reason for this could be the tool-centric names such as ``issue tracker'' and ``bug tracker''.
By recognising the larger ecosystem encapsulating the \xgls{its}, it will be more obvious that researchers should consider these factors in their studies.
Second, this explicit modelling acknowledges the interface between \xglspl{its} and their surroundings, which draws attention to \xglspl{its} in other areas of research.
For example, the study of agile methods is largely one of process models and human interaction, while the \xglspl{its} play a major role in shaping how agile methods are applied and adopted in industry contexts.
Where there is an \xgls{its} present and involved, there is an \xgls{ite}.
Finally, by explicitly including the context factors in the definition, studying an \xgls{its} \textit{without} those context factors requires an explanation why they are \textit{not} relevant to the study.
This encourages a necessary explanation that is often missing from studies of \xgls{ite}.

Accordingly, I use the term \xglsfirst{its} when referring to the tool itself, and \xglsfirst{ite} when referring to the tool \textit{and} the surrounding context as described above.
I will not use other terms such as issue tracker and bug tracker, as these terms are all synonyms of the more universal term \xgls{its}.
In general, I will use \xgls{ite} where possible, and only use \xgls{its} when I want to highlight the tool itself.

\section{Quality, Smells, and Patterns}  \label{sec:background_qsbp}

\subsection{Quality in Software Engineering}

The concept of ``quality'' in \xgls{se} research is \ltexdummy{widespread~\cite{Parra_2019_ISSREW}}, including requirements quality~\cite{Li_2016_CAiSE,Femmer_2017_REW,Parra_2018_ICSECP,Ferrari_2019_ASE,Montgomery_2022_REJ,Winter_2020_REFSQ,Femmer_2019_IEEES}, code quality~\cite{Iftikhar_2024_IST}, test quality~\cite{Fischbach_2020_ESEM}, and maintenance quality~\cite{Bettenburg_2008_FSE,Zimmermann_2010_TSE}.
A central concept in quality is that of quality attributes~\cite{Frattini_2022_RE}, which are the specific forms of quality that apply within a given context.
For example, quality attributes for requirements include causality~\cite{Fischbach_2021_REFSQ,Fischbach_2020_RE}, conditionals~\cite{Fischbach_2021_PROFES}, completeness~\cite{Eckhardt_2016_RE}, and consistency~\cite{Montgomery_2022_REJ}, while code quality includes other attributes such as cohesion, coupling, and fault-proneness~\cite{AlDallal_2017_TSE}.
Each sub-area of quality research features rich and diverse perspectives on what quality attributes exist, why they are important, and how to improve them in automated and manual ways.

\fixoverfull{
    \xglspl{its} store many forms of \xgls{se} artefacts, including requirements~\cite{Heck_2013_IWPSE,VanCan_2024_REFSQ}, development items~\cite{Johnson_2003_CSE,Janak_2009_PhDThesis}, testing tasks~\cite{Fischbach_2020_ESEM}, and maintenance tasks~\cite{Bettenburg_2008_FSE,Zimmermann_2010_TSE}.
    The quality of maintenance items in \xglspl{its} has received much attention in the past few decades~\cite{Zhang_2015_SCIS}, and many notable findings have already been produced~\cite{Bettenburg_2008_FSE,Zimmermann_2010_TSE}.
    Software quality does not apply to development items in \xglspl{its}, since development items are a form of to-do list for the developers.
    Development items in \xglspl{its} are much closer to being requirements than they are to being software.
    The quality of requirements items in \xglspl{its} has received some attention, including Feature Requests~\cite{Heck_2013_IWPSE} and user stories~\cite{VanCan_2024_REFSQ}.
    However, research into the quality of requirements in \xglspl{its} is by far the least studied area of \xglspl{its}~\cite{Montgomery_2022_REJ}.
}

\subsection{Smells}

The concept of a ``smell'' in \xgls{se} nomenclature is that of a potential problem.
Fowler and Beck define \textit{code smells} as ``violations of coding design principles''~\cite{Fowler_2018_Book}.
The concept of a smell, however, is broader than just source code, and has been applied to requirements~\cite{Femmer_2017_JSS}, tests~\cite{Tufano_2016_ASE}, and \xgls{oss} community interactions~\cite{Scacchi_2002_IEEPS,Tamburri_2019_TSE}.
A key characteristic of a smell is that it \textit{might} be a problem, and it \textit{might not} be a problem.
The act of detecting it does not mean that a problem exists; rather, the situation must be investigated by someone to further understand the existence and extent to which it is a problem.
Common code smells include duplicate code, long parameter lists, dead (unused) code, and large classes.
While each of these things could be a problem, there are also good reasons for each of these situations to exist in a production code base, depending on the context.
Overall, ``smell'' is a well-known concept that has been applied to several \xgls{se} concepts.

\xgls{ite} smells have been proposed by various researchers, including Tamburri et al.~\cite{Tamburri_2016_IEEESoftware}, Telemaco et al.~\cite{Telemaco_2020_IEEEAccess}, Qamar et al.~\cite{Qamar_2022_IST}, Borg~\cite{Borg_PhDThesis_2015}, Tuna et al.~\cite{Tuna_2022_ICSESEIP}, L{\"u}ders~\cite{Lüders_2023_PhDThesis}, and Prediger~\cite{Prediger_2023_MSc}.
In accordance with existing literature on smells, they all define and utilise \xgls{ite} smells as a hint that something \textit{might be} wrong, but requires further investigation to confirm or deny.
Before \xglspl{ite} smells were formalised, \xgls{ite} \textit{problems} had already been discussed for decades, including work by Halverson et al.~\cite{Halverson_2006_CSCW}, Bettenburg et al.~\cite{Bettenburg_2008_FSE}, Zimmerman et al.~\cite{Zimmermann_2010_TSE}, and Heck and Zaidman~\cite{Heck_2013_IWPSE}.
Given the way in which these articles describe \xgls{ite} problems, it is clear that they are also referring to smells, rather than problems.

The key advantage to smells is that a minimal amount of knowledge is needed about the context to flag something as a smell.
Smells must be further assessed by an intelligent agent (a person or system) to understand if it is a problem in that context.
However, this also means that a lot of the important information and work to be done is not captured within the concept of a smell and the real problem of understanding and assessing the existence of an actual problem has not been addressed.
For this, a different concept is needed.

\subsection{Patterns and Antipatterns}

Patterns in the \xgls{se} domain are essential to the condensing, packaging, and communication of knowledge.
Design patterns communicate repeatable code solutions to common programming situations~\cite{Beck_1996_ICSE}.
Style guides package and describe agreed-upon stylistic constructs when programming within certain teams or organisations~\cite{Adhya_2015_TC}.
Architectural patterns guide software architects in the design of larger systems, drawing from knowledge gained from decades of software design~\cite{Kuchana_2004_Book}.
These are just a few examples of \xgls{se} patterns.
The documentation of patterns contains certain elements designed specifically to effectively communicate that knowledge.
The common elements of patterns include context, problem, and solution~\cite{Alexander_1997_Book}.
The \textit{context} positions the mind of the developer to understand when this pattern may be applicable.
The \textit{problem} outlines the abstract situation that is undesired, and therefore can be resolved by applying this pattern.
The \textit{solution} describes the way in which the problem can be addressed.
\xgls{ite} quality patterns have been proposed by Aranda and Venolia~\cite{Aranda_2009_ICSE} and Eloranta et al.~\cite{Eloranta_2016_IST}.
Similar to that of other \xgls{se} patterns, these articles outline attributes for each ``pattern'' that give them structure.
The patterns described by Aranda and Venolia~\cite{Aranda_2009_ICSE} are only characterised by a name and a description, and therefore are missing many essential descriptive elements traditionally used within patterns.
Eloranta et al., however, provide a list of \xgls{ite} quality antipatterns, each of which is described with: name, context, solution, consequences, exceptions, and company recommendations~\cite{Eloranta_2016_IST}.
To the best of my knowledge, this is the first time researchers have applied a more formal and complete pattern-like structure to \xgls{ite} quality.

\section{Empirical Methods}

In this section, I describe the empirical methods applied in this thesis.
Each method is explained in detail, touching on all aspects that are relevant for the studies presented in this thesis.
In this way, the methods only have to be fully explained once.
When presenting specific individual studies I conducted, I briefly summarise the high-level methodological details relevant for the study, but reference this chapter for the majority of the information and reasoning behind the methods themselves.
The descriptions of the individual studies, therefore, can focus on the fine-grained methodological details of the specific study being discussed.

I apply four methods for data collection and analysis: interviews, historical data analysis, \xgls{ta}, and ontology building.
I only describe two of the four methods in this section, based on their respective complexity and level of notoriety in the \xgls{se} research community.
Interviews and historical data analysis (often called ``repository mining'') are well-known methods, with highly cited guides on how to apply them in \xgls{se} research.
This is further emphasised by the \xgls{se} research community's heavy use of methodology articles outlining empirical guidelines for how to plan, execute, and document such studies.
\xgls{ta} and ontology-building, however, are less-applied methodological constructs within the \xgls{se} community that have little well-known methodological guidance from within the \xgls{se} community, and therefore result in them being misapplied, over simplified, and often creatively altered without justification~\cite{Ralph_2018_TSE}.
To clarify my use of \xgls{ta} and ontology building, I describe the methodological basis of my studies and which style of these methods I apply in this work.

\subsection{Thematic Analysis}  \label{sec:method_thematic_analysis}

\xglsfirst{ta} is a qualitative data analysis method for ``identifying, analysing, and reporting patterns (themes) within data [...] as it minimally organises and describes data in (rich) detail''~\cite{Braun_2006_QRP}.
The result of \xgls{ta} is a set of themes that describe the data at a high-level, where each theme is broken down into ``codes'' that further describe the data.
\xgls{ta} is a rigorous data analysis method that enables researchers to draw qualitative insights from textual datasets, both large and small.
In this thesis, I utilised both large historical datasets of \xglspl{its} and interview notes.
I used \xgls{ta} to extract qualitative insights from the data.
Large-scale quantitative analysis of datasets can reveal many important generalised findings, but there is often the risk of threats to construct validity if the dataset is not understood well enough.
To counteract this threat and gain different perspectives on the data, I applied \xgls{ta} in Chapters~\ref{ch:challenges}, \ref{ch:activities}, \ref{ch:evolution}, and \ref{ch:catalogue}.

\xgls{ta} is a process of iteratively reviewing data and forming categories of information until conceptual saturation is achieved.
As with all qualitative methods, the goal is not repeatability~\cite{Ralph_2018_TSE}.
The goal of \xgls{ta} is to guide and support the researcher such that the best quality extraction is conducted, \textit{given the reliance on the researcher as a knowledge expert in this task}.
The ``best quality extraction'' is one with high reliability.
\xgls{ta} does this in three ways: initial decision guidance when designing the analysis, well-defined phases that iterate over the data in specific ways, and a process of documenting the results that creates traceability from the findings back to individual data points.

\subsubsection{Pre-Analysis Decisions}

The first step of \xgls{ta} is to explicitly consider four decisions for the analysis: level of detail, style of approach, level of analysis, and epistemological lens~\cite{Braun_2006_QRP}.

\textbf{Level of Detail.}
Consider whether the goal is to achieve a rich overview of the data, or a detailed account of one aspect.
For a rich overview, summarising and describing the entire dataset is a priority, and ``some depth and complexity is necessarily lost''~\cite{Braun_2006_QRP}.
The focus is on capturing all meaningful aspects of the data, at the cost of time that can't be spent diving into more specific areas.
For a detailed account of one aspect, the analysis is limited to a more in-depth and nuanced investigation of one (or a few) aspects of the entire dataset.
The focus is on describing as much (meaningful) detailed as is possible for this subset of the data.

\textbf{Style of Approach.}
Consider whether the analysis will follow an inductive or theoretical approach.
In an inductive approach, the themes identified are ``strongly linked to the data themselves'', which means that the extracted information is data-driven and more representative of a researcher-agnostic extraction process~\cite{Braun_2006_QRP}.
In a theoretical approach, the analysis is driven by a theoretical or analytical interest, such as an existing model or framework.
The analysis is then much more researcher-driven, and tightly linked to this pre-existing theoretical grounding.

\textbf{Level of Analysis.}
Consider whether you want semantic or latent themes.
Semantic themes represent the surface or explicit meanings present in the data.
Latent themes go beyond the \ltexdummy{surface level} meanings, and ``identify or examine the underlying ideas, assumptions, and conceptualisations that are theorised as shaping or informing the semantic content of the data''~\cite{Braun_2006_QRP}.

\textbf{Epistemological Lens.}
Consider whether you are approaching the analysis with an essentialist or constructivist lens.
Essentialism (and realism) assumes a relatively straightforward relationship between meaning, experience, and language, which means a fairly straightforward extraction process.
In Constructionism, ``meaning and experience are socially produced [...], rather than inhering within individuals''~\cite{Braun_2006_QRP}.
This means the analysis is far more focused on the social constructs that could lead to the meaning and motivations within the data.

\subsubsection{Phases of Analysis}

\xgls{ta} has six phases:
    1) familiarising yourself with the data,
    2) generating initial codes,
    3) searching for themes,
    4) reviewing themes,
    5) defining and naming themes, and
    6) producing the report.
These phases are both flexible and iterative.
The goal is not to get through them as quick as possible, but rather to understand and utilise the true nature and purpose of each phase until their purpose has been achieved.
Understanding the fundamental concept of ``saturation'' is essential to orchestrating these phases successfully.

\textbf{Phase 1: Familiarising yourself with the data.}
It is essential to first read through all the data, at least once---and likely more than that, to gain an initial understanding of what to expect in the analysis.
Failing to do this step results in inconsistent analysis results, where the initial work is not as informed as the final work, given the familiarity that you bring to the final work that is not present at the beginning.
By fully reviewing all data first, including actively searching for meaning and patterns, you leave this phase with a good understanding of what to expect during your analysis.

\textbf{Phase 2: Generating initial codes.}
This phase involves reading through the data, and forming small and specific categories called ``codes''.
These codes are the basic element of the data that can be assessed meaningfully, regarding the studied phenomenon.
This phase is complete once all the data has been coded (or actively \textit{not} put into a code).
In the case of a detailed account of one specific part of the data, only data relevant to the one specific part of the data needs to be coded, and the rest will be marked as ``not relevant''.

\textbf{Phase 3: Searching for themes.}
This phase involves forming connected ideas across the codes, called ``themes''.
These themes connect the data at a higher, conceptual level, and represent a more analysis-like grouping of the data.
Whereas the codes are extracted using more straightforward mappings of meaning, the themes now involve interpreting the codes, and forming groupings that the researcher believes best fit the analysis goal.
This phase is complete once all codes are grouped under a theme (or sub-theme).
This phase is more about ideation and coverage than it is about conciseness and a well-formed set of themes.

\textbf{Phase 4: Reviewing themes.}
In this phase, the researchers iteratively review and refine the themes for overall consistency and representativeness across the dataset (level one).
Then, the evidence under each code is thoroughly reviewed and cleaned such that it is representative of the code and theme itself (level two).
This phase is one of the most critical---and yet most overlooked---phases in qualitative analyses (such as content analysis)~\cite{Braun_2006_QRP,Cruzes_2011_ESEM}.
While many researchers will naturally iterate over the themes as they create them, it is also important to scrutinise the work \textit{after the forming of the final themes} with a full analysis and cleaning process.
Level one of this phase, as described by Braun and Clarke~\cite{Braun_2006_QRP}, is about reviewing the all codes within each theme, and asking if they form a coherent pattern.
While certain codes might have fit best under a certain theme during phase 3, perhaps a comparison of all codes in that theme now reveals that that code is not a good fit for this theme overall.
Level one is also about reviewing all themes compared to each other, and evaluating the final group as representative of the overall findings (or not).
Level two of phase 4 is an in-depth review of each piece of evidence under each code, with the context of the themes in mind.
Every evidence piece is read, and re-considered as a candidate to remaining sorted under this code and theme.
The primary reason for this second analysis is the additional final context of the final codes and themes.
Evidence that may have made sense under a certain code at the beginning, might no longer make sense with the full context of all themes and codes combined.
For example, the code ``quick learner'' could easily contain the evidence piece ``she regularly picks up new knowledge on weekends''.
However, if the final theme map sorts the code ``quick learner'' under ``machine learning'', then the stated evidence piece may no longer fit, since it should also fit with the context of the relative theme.
This process is iterative, and is often supported by visual mapping tools such as mind-mapping, facet lists, or taxonomy-like structuring.

\textbf{Phase 5: Defining and naming themes.}
This phase is primarily about considering the presentation of the themes: how are they named, how will they be presented, do they tell a cohesive story, do they all fit together well, are they representative of the data and analysis you are now so familiar with?
While not the primary purpose of this phase, you can also continue to refine the themes themselves, if the process of considering the presentation leads to an obvious change that should be made.
For example, perhaps in the process of naming a theme, you discover that the theme is too broad, and therefore needs to broken down.
At the end of this phase you will have a set of presentation themes, each with a name, description, and a well-known set of codes under each.

\textbf{Phase 6: Producing the report.}
In this phase, the analysis and findings are described.
While writing about data analyses is nothing new to researchers, the importance of this phase for qualitative work should not be underestimated.
While quantitative results can be manually reviewed, re-analysed, and even reproduced, qualitative findings are limited to the interpretation presented by the researcher.
With good evidence tracing, qualitative findings can be manually reviewed by other researchers, but the rest relies on the report produced by the researcher.
It is critical that the report provides a ``concise, coherent, logical, non-repetitive and interesting account of the story the data [tells] within and across the themes''~\cite{Braun_2006_QRP}.
When discussing the themes, it is important that the ``write-up [provides] sufficient evidence of the themes within the data---i.e., enough data extracts to demonstrate the prevalence of the theme''~\cite{Braun_2006_QRP}.
In addition to the replication package, the researcher should provide ``vivid examples [...] which capture the essence of the point''~\cite{Braun_2006_QRP}.
In other words, the report should describe the findings in such a way that the reader understand the overall findings as the researcher sees them.

\subsection{Taxonomy and Ontology Development}  \label{sec:method_tax_ont}

In this thesis, I develop multiple taxonomies (and a single ontology) using empirical evidence, as well as through secondary studies.
While the methodological design, development, and reporting of each taxonomy is self-contained and reported in each respective chapter, there are a few key overarching concepts that need to be explained and clarified.
First, I will define my usage of the terms ``taxonomy'' and ``ontology'' in this thesis, given the diversity of ways in which the \xgls{se} research community uses related terms.
Second, I will explain the process I followed to construct the taxonomies that were formed from a methodology \textit{other than} \xgls{ta}, and how I constructed the ontology.
Third, I will discuss how taxonomies and ontologies relate to theory in \xgls{se} research.

\subsubsection{Terminology: Taxonomy vs Ontology}  \label{sec:methods_tax_vs_ont}

In 2004, Garshol wrote that ``the term taxonomy has been widely used and abused to the point that when something is referred to as a taxonomy it can be just about anything, though usually, it will mean some sort of abstract structure''~\cite{Garshol_2004_JIS}.
This is still the case today, where terms such as ``taxonomy'', ``ontology'', ``framework'', ``model'', and ``topology'' are often used interchangeably in \xgls{se} research~\cite{Nickerson_2013_EJIS,Ralph_2018_TSE}.
For this reason, I define here the terms ``taxonomy'' and ``ontology'' as I use them, within the context of this thesis.

Functionally, the goal of building both taxonomies and ontologies is to \textit{organise information} for use by some \textit{specific actor or system}~\cite{Gruber_1995_Online,Gruber_2009_Online}.
The primary means of information organisation for both of them is through a process called \textit{classification}, where objects are grouped relative to their \textit{subjects}~\cite{Garshol_2004_JIS}.
A ``subject'' is the specific classification chosen for a set of objects.
Importantly, the selected subjects must be relevant and useful to the actor or system in some \textit{meaningful} way.
For example, if we take a list of sorting algorithms and classify them for a lecturer who will teach them to students, a meaningful classification (the subject) could be average sorting-time complexity (e.g., $O(n)$, $O(n log(n))$, and $O(n^2)$).
Contrastingly, a non-meaningful classification (subject) could be the number of letters in each of their names (e.g., $4$, $5$, and $6$).
These named subjects then become the \textit{controlled vocabulary}~\cite{Gruber_1993_KA,Gruber_2009_Online,Garshol_2004_JIS} through which all future algorithms must be discussed (within the context of this classification).

Taxonomies are best known for their use of \textit{hierarchy} to organise \textit{terms} (objects and subjects).
Another lesser known form of taxonomies is that of a \textit{faceted taxonomy}, where terms are organised under \textit{facets} (which can also be objects and subjects).
Hierarchies generally form a tree structure (although terms in a taxonomy can have more than one parent~\cite{Ralph_2018_TSE}), whereas facets form simple single-dimension lists.
The relationship that a hierarchical taxonomy uses between terms is called the ``broader/narrower'' relationship.
This is because taxonomies contain \textit{conceptual} hierarchies, where information higher in the hierarchy is ``broader'', and lower is ``narrower''~\cite{Garshol_2004_JIS}.
The fundamental defining concept of a taxonomy is that it provides \textit{structure}, whether that be through hierarchy or facets.
As defined by Garshol, taxonomies are ``a subject-based classification that arranges the terms in the controlled vocabulary into a hierarchy \textit{without doing anything further}''~\cite{Garshol_2004_JIS}.
The final part of that definition, ``without doing anything further'', provides a clear and logical boundary between taxonomies and ontologies, since ontologies \textit{can} do other things to support---and go beyond---structuring~\cite{Gruber_1993_KA}.

Ontologies are best known for their use of properties (also known as dimensions) and relationships to describe terms.
An ontology is ``a model for describing the world that consists of a set of types, properties, and relationship types''~\cite{Garshol_2004_JIS}.
One possible relationship type is the ``broader/narrower'' relationship, thus allowing ontologies to contain hierarchical taxonomies---as they often do~\cite{Garshol_2004_JIS}.
The additional tools provided by ontologies, namely ``properties'' and other ``relationship types'', allows terms to be further \textit{specified}.
For example, we can reuse the sorting algorithms example, and define a subject that is ``sorting algorithm'', with properties such as ``average sorting speed'', ``fastest sorting speed'', and ``maximum memory requirement''.
Thus, ontologies provide more tools for describing terms than taxonomies.

In short, both taxonomies and ontologies \textit{classify}, but taxonomies also \textit{structure} while ontologies can both \textit{structure} and \textit{specify}.
Thus, ontologies are a conceptual extension of taxonomies, despite not every ontology containing a taxonomy (hierarchical or faceted).
This is useful to keep in mind, since recommendations directed at taxonomies or ontologies therefore apply universally across the two, as long as the concepts they are directed at are understood and applied correctly (those concepts being classification, structure, and specification).
In my thesis, I produce an ontology that is based on a central hierarchical taxonomy.

The literature on \ltexdummy{taxonomies and ontologies} uses these words rather interchangeably~\cite{Garshol_2004_JIS}, and I will be citing and summarising the work as they have stated it.
However, I will note when I disagree with their use of terminology (based on the foundational definitions above) and therefore apply their recommendations to a different term (but same underlying concept).

\subsubsection{Steps to Build and Evaluate Taxonomies and Ontologies}  \label{sec:methods_tax_build}

Taxonomies and ontologies are a common research output produced by \xgls{se} research, but the methods through which they are created lack structure and transparency~\cite{Nickerson_2013_EJIS,Kundisch_2021_BISE}.
In response to this situation, Nickerson et al.~\cite{Nickerson_2013_EJIS} (and later updated by Kundisch et al.~\cite{Kundisch_2021_BISE}) outline recommendations for taxonomy construction in the \xgls{is} domain (which apply rather easily to the constructs in this thesis).
While both articles refer to ``taxonomy'', they are referring to the concept described in this thesis as ``ontology'' (due to their description and inclusion of properties on the subjects).
However, in describing recommendations for ontologies, some recommendations also apply to the construction of taxonomies.
Accordingly, I refer to (and apply) their recommendations to my taxonomies and ontology \textit{where it makes sense}.

Nickerson et al. recommend seven primary ontology development phases, which I use to structure the following description of building an ontology~\cite{Nickerson_2013_EJIS}.
There are two separate paths through the ontology development phases, and I will only describe the path I took: empirical-to-conceptual.
Here are the seven phases I applied in this thesis.

\textbf{Determine Meta-Characteristic.}
Ontologies are designed to be used for a specific purpose~\cite{Gruber_1993_KA,Gruber_1995_IJHCS,Gruber_2009_Online,Nickerson_2013_EJIS}.
While they should represent the world as it is, and therefore have an objective description of the objects they seek to describe, there are different ways to frame those objects.
This means that there are many different, correct and useful ontologies for describing the same set of objects.
However, it is also possible to create unuseful framings, such as the previous example given regarding classification where the number of letters in the name is used to characterise sorting algorithms.
Nickerson et al. call this framing of a \textit{useful} concept the ``meta-characteristic'' of an ontology~\cite{Nickerson_2013_EJIS}.
It is recommended to pick this meta-characteristic at the beginning of the ontology creation process to avoid ``naïve empiricism'', a situation in which \ltexdummy{``a large number of related and unrelated characteristics are examined in the hope that a pattern will emerge''~\cite{Nickerson_2013_EJIS}}.
They also note, however, that the meta-characteristic often ``does not become clear until part way through the [ontology] development process''~\cite{Nickerson_2013_EJIS}.
When picking the meta-characteristic, Nickerson et al. recommend considering who the users are, and how they intend to use the ontology.
This lends itself to the functional perspective on ontologies that drives many of the following methodological recommendations~\cite{Nickerson_2013_EJIS}.

\textbf{Determine Ending Conditions.}
The process of creating a conceptual model from empirical evidence is inherently iterative, since any new observation in the data could affect all previous developed concepts~\cite{Nickerson_2013_EJIS}.
Accordingly, it is necessary to identify ending conditions to this iterative process.
In other words, when is the ontology creation process complete?
Another important consideration is that of quality.
Although the ontology creation process may be finished, objectively, how do we check the quality of the final ontology, and how might we further improve it?
Nickerson et al. define two sets of ending conditions: objective conditions and subjective conditions~\cite{Nickerson_2013_EJIS}.
The objective ending conditions answer the more concrete question of being finished or not, while the subjective ending conditions address the concept of quality.

There are no objective criteria that can outline---with certainty---when a qualitative process such as content analysis is complete, and this also applies to building ontologies from empirical data.
There are, however, tips on what things to look out for when trying to ``finish'' the process, as well as \ltexdummy{a number of} signs that the iteration is likely not done.
The recommendations by Nickerson et al.~\cite{Nickerson_2013_EJIS} are a mix of those two things.
Presented below, verbatim from their article~\cite{Nickerson_2013_EJIS}, are the recommended \textit{objective ending conditions} for the ontology creation process.
When building my ontology, I utilised these objective ending conditions at the end of each iteration, acting as a guide as to when to stop iterating.
\ltexignore{
\begin{itemize}[nosep]
    \item ``All objects or a representative sample of objects have been examined.''
    \item ``No object was merged with a similar object or split into multiple objects in the last iteration.''
    \item ``At least one object is classified under every characteristic of every dimension.''
    \item ``No new dimensions or characteristics were added in the last iteration.''
    \item ``No new dimensions or characteristics were merged or split in the last iteration.''
    \item ``Every dimension is unique and not repeated (i.e., there is no dimension duplication).''
    \item ``Every characteristic is unique within its dimension (i.e., there is no characteristic duplication within a dimension).''
    \item ``Each cell (combination of characteristics) is unique and is not repeated (i.e., there is no cell duplication).''
\end{itemize}}

Due to the context-rich nature of qualitative research---including forming models from empirical data---it is important to use \textit{subjective ending conditions} for iteration guidance.
Nickerson et al. propose a set of five subjective ending conditions for building taxonomies.
Furthermore, they are clear that these five are the ``minimal'' ones, and that the researcher ``may wish to add more subjective conditions to these based on the researcher's particular view''~\cite{Nickerson_2013_EJIS}.
Presented here, verbatim from their article~\cite{Nickerson_2013_EJIS} (but only showing a portion of the original description), are the recommended subjective ending conditions for the ontology creation process:
\begin{description}
    \item[Concise] ``An ontology should contain a limited number of dimensions and a limited number of characteristics in each dimension because an extensive classification scheme with many dimensions and many characteristics may exceed the cognitive load of the researcher and thus be difficult to comprehend and apply.''
    \item[Robust] ``A useful ontology should contain enough dimensions and characteristics to clearly differentiate the objects of interest. An ontology with few dimensions and characteristics may not be able to adequately differentiate among objects.''
    \item[Comprehensive] ``There are two interpretations of this condition. 1) A useful ontology can classify all known objects within the domain under considerations. 2) A useful ontology includes all dimensions of objects of interest.''
    \item[Extendible] ``A useful ontology should allow for inclusion of additional dimensions and new characteristics within a dimension when new types of objects appear. An ontology that is not extendible may soon become obsolete.''
    \item[Explanatory] ``A useful ontology contains dimensions and characteristics that do not describe every possible detail of the objects but, rather, provide useful explanations of the nature of the objects under study or of future objects to help us understand the objects. An ontology that simply describes objects may be of interest initially but will have little value in understanding the objects being classified.''
\end{description}
When building my ontology, I refer to these subjective ending conditions list at the end of each iteration, acting as a guide as to when to stop iterating.

\textbf{Pick the Approach.}
As with any qualitative conceptual modelling, there are two main approaches: inductive and deductive.
Nickerson et al. describe these two approaches as ``empirical-to-conceptual'' and ``conceptual-to-empirical''~\cite{Nickerson_2013_EJIS}.
Moving forward, I will use the more accepted terms ``inductive'' (replacing ``empirical-to-conceptual'') and ``deductive'' (replacing ``conceptual-to-empirical'')\footnote{In my opinion, the terms used by Nickerson et al. are misleading because they imply that it is possible to work towards empirical data using conceptual work.
The terms ``inductive'' and ``deductive'' describe two techniques that produce the same result (some model), but via two approaches (bottom-up vs top-down).}.
It is recommended to use the inductive approach if you have ``significant data available'' and little knowledge or understanding of the domain, and to use the deductive approach if you have ``little data available'' and significant understanding of the domain.
In the case of having both significant data available and significant understanding of the domain, it is recommended to pick the approach that fits best~\cite{Nickerson_2013_EJIS}.

\textbf{Identify Subset of Objects.}
The researcher needs to describe the population of objects to be studied, and then pick the subset they will analyse as part of the ontology development process.
There are many reasons to create an ontology, and therefore there are many potential priorities when picking a population and selecting the subset to be analysed.
If the purpose of the ontology is to become theory, then the population should be the actual population of objects that exist under study, and the sample should either be the entire population, or a representative sample.
If the purpose of the ontology is to describe some sub-aspect of the population, or act as a tool for particular actors within a system, then the population should be selected accordingly (less need for a representative sample).
In such cases, the subset of studied objects ``are likely to be the ones with which the researcher is most familiar or that are most easily accessible''~\cite{Nickerson_2013_EJIS}.
Nickerson describes the potential subset samples as ``a random sample, a systematic sample, a convenience sample, or some other type of sample''~\cite{Nickerson_2013_EJIS}.
In summary, the population and subset need to be selected according to the purpose of the ontology.

\textbf{Identify Common Characteristics and Group Objects.}
The researcher then identifies common characteristics across the objects.
Importantly, these characteristics should be related to, born from, or perspectives on the meta-characteristic.
Without the meta-characteristic as the lens of the analysis, it is too easy to create characteristics that are unhelpful to the purpose of the ontology, and thus irrelevant to the goal of creating a useful ontology.
The characteristics should also separate the objects in meaningful ways~\cite{Nickerson_2013_EJIS}.
If a characteristic is defined such that all objects have the same value, then the characteristic is not separating the objects.
Since the process is iterative, it is okay to create characteristics that are later removed.
What might seem to be a good differentiator at first, may have very little discriminatory power over all the objects, and therefore needs to be removed.
At the end of this phase, there should be many characteristics that group the objects in different ways.

\textbf{Group Characteristics into Dimensions to Create the Ontology.}
In this final phase, the researcher now groups the characteristics into dimensions.
It is possible to group these characteristics via statistical methods, but a common approach is to manually group them via visual graphing~\cite{Nickerson_2013_EJIS}.
These dimensions then become the ``properties'' of the ontology.

\textbf{Review Ending Conditions.}
Ontology development is an iterative process, and as such there needs to both cycles of improvement, and conditions describing when to break this cycle.
As described above under ``Determine Ending Conditions'', Nickerson et al. define a number of objective and subjective ending conditions~\cite{Nickerson_2013_EJIS}.
Once these conditions have been met, it is then assumed that the ontology is complete and ready to be communicated.

\subsubsection{Theory Building in SE Research}  \label{sec:methods_tax_theory}

Theory enables scientific fields to collect and cumulate knowledge that generalises across many (if not all) known contexts~\cite{Sjøberg_2008_Book}.
The advantage of theory is that generalised claims can be formulated, tested, strengthened, and refuted, thus providing a structure for the iterative nature of science itself.
While the building blocks of theory are claims, propositions, hypotheses, and empirical evidence (among other constructs)~\cite{Gregor_2006_MIS,Sjøberg_2008_Book}, failing to form theory from these building blocks leaves the process stuck in its infancy.
Broadly speaking, theory is ``a system of ideas for explaining some phenomenon''~\cite{Ralph_2018_TSE,Sjøberg_2008_Book,Gregor_2006_MIS}.
There are many views on what theory is~\cite{Sjøberg_2008_Book}, as well as different perspectives depending on which epistemological lens is being applied~\cite{Kitchenham_2013_IST,Ralph_2018_TSE}.
There are also different conceptualisations and classifications of theory~\cite{Gregor_2006_MIS,Sjøberg_2008_Book}.
In this thesis, I am working with the concept of taxonomic theory, the methodological construct most prevalent in my work.
I will be focusing on the classification summarised by Paul Ralph in his methodological guidelines for taxonomies in \xgls{se} research~\cite{Ralph_2018_TSE}.
While Ralph refers to ``taxonomy'', I interpret the work as applying to ontologies, based on the work's description and inclusion of properties and relationships.

There are three types of theory: variance theories, process theories, and ontologies.
Variance theory seeks to explain the world in terms of how one variable (dependent variable) reacts in response to changes in another variable (independent variable).
Process theory seeks to explain, understand, and even predict how an entity changes and develops over time~\cite{Ralph_2018_TSE}.
Ontological theory, on the other hand, is a type of theory that seeks to explain and understand entities and systems as they are.
Process theories and ontologies are closely related, as three of the four process theory types \textit{include ontologies}~\cite{Ralph_2018_TSE}; however, not all ontologies are part of a process theory.
While variance theories are the prominent theory type in \xgls{se} research~\cite{Ralph_2018_TSE}, this thesis focuses on and describes ontological theory.

Ontologies (and taxonomies) are rather common in \xgls{se} research, but they are often not stated as ``theory''.
There are many potential reasons for this, including the belief that ontologies are not theory~\cite{Gregor_2006_MIS,Ralph_2018_TSE}.
Two additional likely reasons are the lack of understanding as to \textit{why ontologies are theory}, and why some other simple categorisations are not theory.
As described above, there are many dimensions and perspectives to theory, such that there is no consensus across scientific communities as to the meaning of theory~\cite{Ralph_2018_TSE}.
However, the broad definition of theory presented above (``a system of ideas for explaining some phenomenon''), requires only that ontologies \textit{explain} some real-world \textit{phenomenon}.
Ontologies are commonly built from secondary studies of a phenomenon, as well as from singular empirical studies of a phenomenon (a weaker form of ontology building)~\cite{Ralph_2018_TSE}.
In this way, ontologies are concerned with real-world \textit{phenomena}.
While it is possible to create an arbitrary classification of some phenomenon based solely on conjecture or a rigourless process, these are not ontologies~\cite{Ralph_2018_TSE}.
Ontologies are designed to, among other things, ``identify, describe and understand the entities and events in a domain''~\cite{Ralph_2018_TSE}, which are all forms of \textit{explaining}.
Thus, ontologies are a valid form of theory, when built correctly and described adequately.

The output of \xgls{se} research is often prescriptive, while theory should be explanatory.
These goals should not overlap because they can lead to \xgls{se} theory including elements of prescription, and can lead to validity issues in the constructed theory~\cite{Ralph_2018_TSE}.
The overlap between \xgls{sdm}s and process theory is a common place for misconception~\cite{Ralph_2018_TSE}, and relevant for this thesis.
When producing a theory that explains how a certain process in \xgls{se} works, it is tempting to also add how it \textit{should} work if there is an identified deficiency.
It is also tempting to take recommendations on how things should be done, grounded in empirical investigations, and frame them as theory.
In my thesis, I produce a catalogue of recommendations for \xgls{se} processes.
These recommendations are grounded in investigations of existing \xgls{se} contexts, including case studies, interviews, observations, and surveys.
However, these recommendations are intended to be prescriptive, and therefore are not described as process theory.

\subsubsection{Propositions in Software Engineering Research}  \label{sec:background_propositions}

Propositions are predictions about the world that may be deduced logically from theory~\cite{Shanks_2002_AJIS,Sjøberg_2008_Book,Runeson_2012_Wiley}.
Sjøberg et al. describe the process of theory building in \xgls{se} as a five-step process, where the first step is to define the constructs of the theory, and the second is to define the propositions of the theory.
Propositions are the stepping stone into a fully formed theory, as they represent relationships between the constructs defined in the first step.
The combination of the propositions (with some explanations, scope, and empirical evidence) form the singular construct that is the theory in question~\cite{Sjøberg_2008_Book}.
For the purpose of this thesis, however, I would like to focus on propositions themselves.

The first step of theory building is defining the constructs~\cite{Sjøberg_2008_Book}.
This is fundamental to both the theory and the propositions, as propositions can only describe relationships between the constructs.
One evaluation metric of a good theory is \textit{parsimony}, which is ``the extent to which unnecessary constructs and propositions are excluded''~\cite{Sjøberg_2008_Book}.
Therefore, it is important to minimise the constructs collected and the propositions constructed.
Sjøberg et al.~\cite{Sjøberg_2008_Book} discuss five ways to make a contribution from the perspective of constructs within a theory.
For the purpose of this thesis, only the first way applies: ``defining new constructs as the basis for building a new theory about some phenomena [...] they might conceive phenomena that have been the focus of prior theories, but in a different way''.

The second step of theory building is defining the propositions~\cite{Sjøberg_2008_Book}.
These propositions relate the constructs together in different ways.
This could be quantitatively, using statistical statements, or more broadly as generalisations to be tested and further refined.
For example, a more broad proposition would be to say ``the use of a \glsxtrshort{uml}-based development method positively affects communication'' (example from Sjøberg et al.~\cite{Sjøberg_2008_Book}).
A single theory is normally composed of many propositions, but the relative importance of each proposition differs.
The importance of the proposition relates to its usage within the theory.
Some propositions are more central, and combine with others to form the core of the theory, while others are more tangential, but still part of the whole framing that is the theory overall.
I define five propositions in Chapter~\ref{ch:ontology}, as a means of beginning the journey to a future theory.

\section{Summary}

\xglspl{its} are complex tools that play a fundamental role in \xgls{se} organisations, including during the requirements, development, and maintenance phases of \xgls{se}.
\xglspl{ite} are complex systems that involve both an \xgls{its} (one or more), and all surrounding contextual factors that affect the \xgls{its}.
\ltexignore{\xglspl{ite} have quality aspects, and those are related to concepts such as smells and Best Practices.}
The concept of ``quality'' for \xgls{ite} has been framed in different ways, including as patterns, antipatterns, and smells.
With the support of different empirical methods such as interviews, historical data analysis, \xgls{ta}, and ontology building, I will investigate the reported difficulties that practitioners have with \xglspl{ite} in the next chapter.

    \part{Problem Investigation}  \label{part:problem}
    
\chapter{Practitioner Challenges with Issue Tracking Ecosystems}  \label{ch:challenges}

\epigraph{Organisation is not a goal in itself, it is a tool. Don't get caught up in the illusion of productivity and get distracted from the actual task.}{Unknown}

\begin{figure}[hb]
    \centering
    \includegraphics[width=0.5\textwidth]{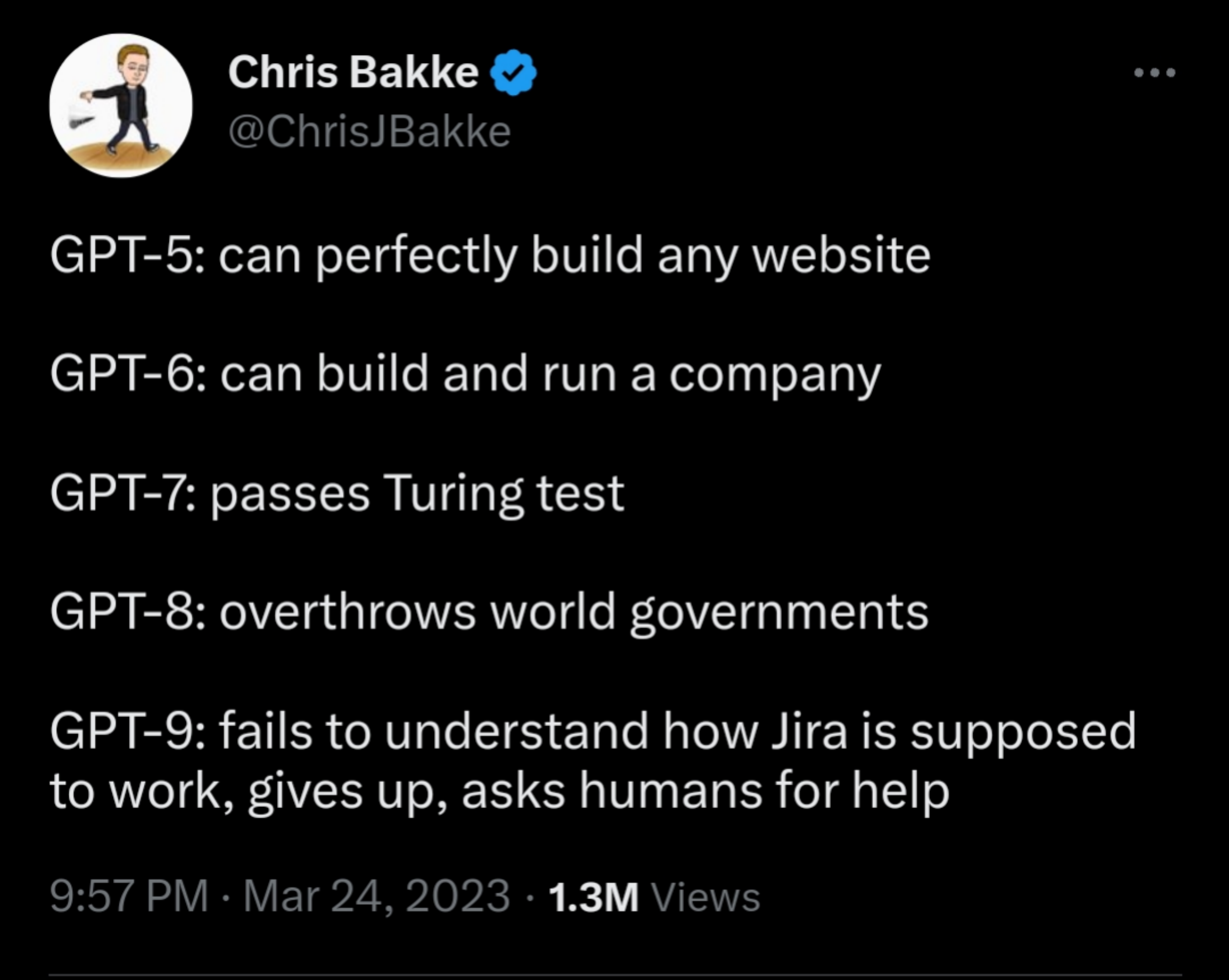}
    \caption[Twitter user Chris Bakke jokes about the complexity of Jira.]{Twitter user Chris Bakke jokes about Jira being so complex that an AI---that has already taken over the world---has to ask humans for help.}
    \label{fig:jira_gpt}
\end{figure}

In this chapter, I investigate the claim that \xglspl{ite} are difficult to interact with by conducting an interview study with practitioners who regularly interface with \xglspl{ite}.
\xglspl{ite} are commonly reported in pop-culture as difficult to interact with (see Fig.~\ref{fig:jira_gpt}), but the studied details of such difficulties are lacking.
Researching the thoughts and opinions of practitioners is an important part of empirically grounding our \ltexdummy{research assumptions~\cite{Franch_2017_RE,Franch_2017_RE,Franch_2020_TSE,Franch_2022_TSE,Vogelsang_2019_ST}}.
To understand the difficulties faced by practitioners when using \xglspl{ite}, I conducted an interview study with 26 industry practitioners.
This empirical investigation was conducted in collaboration with my colleagues \ltexdummy{\nameCL}, \ltexdummy{\nameCR}, and \ltexdummy{{\nameWM}}.
I asked practitioners about the problems they face when using \xglspl{ite}.
The results show three key areas where \xgls{its} difficulties arise: \xgls{ite} information problems, \xgls{ite} workflow problems, and \xgls{ite} organisational problems.
In this chapter, I discuss the details regarding these results, including the contextual complexity of interpreting the results.
Overall, this investigation highlights key problems faced by practitioners and draws attention to the context-specific nature of \xglspl{ite}.

\section{Research Methodology} \label{sec:challenges_method}

My primary objective with this chapter is to understand what problems practitioners face when using \xglspl{ite}.
For this, I have a singular research question:
\begin{description}
    \item[RQ] What \textit{problems} do software practitioners usually encounter when using \xglspl{ite}?
\end{description}

To investigate this research question, I conducted an interview study with 26 practitioners.
The interview involved mostly open questions with the goal of following the interviewee along their thought-processes regarding problems in \xglspl{ite}.
Detailed interview notes were recorded by two interviewers, which were later transferred to a digital format for analysis.
We then analysed the interview notes using content analysis.

\subsection{Interview Participants}

Given the objective to gather a broad and rich understanding of \xgls{ite} problems and contextual factors in practice, we sought a diverse set of participants along five primary dimensions: participant role, years of experience, types of \xglspl{its}, company size, and industries.
A representative sample was not sought, as we did not aim to generalise the qualitative observations.
We contacted \xgls{se} companies in our network and asked for recommendations for experienced employees who would fit the sampling scheme.
All participants worked at \xgls{se} companies and had at least one year of experience using \xglspl{its}.
In total, we interviewed 26 practitioners working in Germany, Canada, and Poland, listed in Table~\ref{tab:Participants}.
The participants had a range of work experience from 1.5--25 years (median 7).
They held various roles, such as Developer, Manager, and Product Owner.
Consistent across them all was the regular task of developing software, albeit less for the participants with more managerial roles.

\begingroup

    \newcommand\its[1]{\scriptsize#1}           
    \newcommand\oit[1]{\color{gray}\tiny(#1)}   

    \setlength{\tabcolsep}{2pt}         
    \renewcommand{\arraystretch}{1.0}   

    \begin{table}[ht]
        \small      
        \centering  

        \caption{Overview of study participants.}
        \label{tab:Participants}

        \begin{tabular}{@{} r lll lll lll @{}}
            \toprule
            \textbf{\#ID} & \textbf{Role} & \textbf{YoE} & \textbf{G} & \textbf{ITS~\oit{other known ITSs}} & \textbf{ITS~Size} & \textbf{Industry} & \textbf{CS} \\
            \midrule
            \textbf{P01} & Product Owner & 7   & m & \its{Jira}                                  & 10k-100k & Automotive  & L \\
            \textbf{P02} & Developer     & 4   & m & \its{Trac}                                  & 10k      & Engineering & M \\
            \textbf{P03} & Developer     & 10  & m & \its{Trac}                                  & 10k      & Engineering & M \\
            \textbf{P04} & Manager       & 4   & m & \its{Jira}                                  & 10k-100k & Maritime    & M \\
            \textbf{P05} & Manager       & 6.5 & m & \its{Jira} \oit{Asana}                      & 100k-1m  & Database    & L \\
            \textbf{P06} & Developer     & 6   & m & \its{Jira} \oit{Trello}                     & 100k-1m  & Energy      & M \\
            \textbf{P07} & Developer     & 24  & m & \its{Jira} \oit{Watson}                     & 1m+      & Media       & L \\
            \textbf{P08} & Product Owner & 5   & m & \its{Jira}                                  & 100k-1m  & Automotive  & L \\
            \textbf{P09} & Developer     & 10  & m & \its{Jira} \oit{Mantis, Watson, Trac}       & 1m+      & Media       & L \\
            \textbf{P10} & Manager       & 5.5 & m & \its{Trac}                                  & 10k      & Engineering & M \\
            \textbf{P11} & Developer     & 8   & f & \its{Jira} \oit{Custom}                     & 10k-100k & Consulting  & M \\
            \textbf{P12} & Manager       & 11  & m & \its{Jira} \oit{Brainstorm, Watson}         & 1m+      & Media       & L \\
            \textbf{P13} & Developer     & 3.5 & m & \its{Jira} \oit{GitHub, Bugzilla}           & 1k-10k   & Engineering & L \\
            \textbf{P14} & Product Owner & 6   & m & \its{GitLab} \oit{GitHub, Jira}             & 100-1k   & Engineering & S \\
            \textbf{P15} & Developer     & 25  & m & \its{RedMine} \oit{GitLab}                  & 1k-10k   & Research    & L \\
            \textbf{P16} & Developer     & 1.5 & f & \its{GitLab} \oit{GitHub}                   & 100-1k   & Engineering & S \\
            \textbf{P17} & Developer     & 15  & m & \its{GitHub, GitLab} \oit{Jira, Savanna}    & 100-1k   & Research    & L \\
            \textbf{P18} & Developer     & 3.5 & m & \its{ServiceNow} \oit{Jira, SpiraTest}      & 1k-10k   & Technology  & L \\
            \textbf{P19} & Developer     & 16  & m & \its{GitLab} \oit{RedMine, Jira, GitHub}    & 10k-100k & Research    & L \\
            \textbf{P20} & Developer     & 9   & m & \its{GitLab, GitHub} \oit{Jira, BaseCamp}   & 1k-10k   & Software    & M \\
            \textbf{P21} & Developer     & 15  & m & \its{Mantis} \oit{Jira, Trello, GitHub}     & 100-1k   & Medical     & L \\
            \textbf{P22} & Developer     & 4   & m & \its{Jira} \oit{Miro, Trello, GitHub}       & 10-100k  & Consulting  & L \\
            \textbf{P23} & Developer     & 23  & m & \its{Custom, GitHub}                        & 10-100k  & Research    & L \\
            \textbf{P24} & Manager       & 15  & m & \its{Jira, Azure}                           & 10-100k  & Consulting  & L \\
            \textbf{P25} & Manager       & 17  & m & \its{Jira}                                  & 1k-10k   & Medical     & L \\
            \textbf{P26} & Product Owner & 3.5 & f & \its{Jira, Azure}                           & 1k-10k   & Engineering & L \\
            \bottomrule
            \multicolumn{8}{@{} p{14cm}}{\footnotesize{
            Study participants by \textbf{Role}, \textbf{Y}ears \textbf{o}f \textbf{E}xperience, \textbf{G}ender, currently used \textbf{I}ssue \textbf{T}racking \textbf{S}ystem, \textbf{I}ssue \textbf{T}racking \textbf{S}ystem \textbf{Size}, \textbf{Industry} and \textbf{C}ompany~\textbf{S}ize~\cite{EntSize_2024_Online}.
            }} \\
        \end{tabular}
    \end{table}
\endgroup

All participants utilised an \xgls{its} in their \xgls{ite}, and had a range of experience using different \xglspl{its} throughout their careers.
The majority (14/26) primarily used Jira (in alignment with market share~\cite{6SenseJira_2024_Online,EnlyftJira_2024_Online,DatanyzeJira_2024_Online}), with six more participants having experience with Jira.
Others used GitHub~\cite{GitHub_2024_Online}, GitLab~\cite{GitLab_2024_Online}, Trac~\cite{Trac_2024_Online}, Azure~\cite{Azure_2024_Online}, Bugzilla~\cite{Bugzilla_2024_Online}, Trello~\cite{Trello_2024_Online}, asana~\cite{asana_2024_Online}, Mantis~\cite{Mantis_2024_Online}, SpiraTest~\cite{SpiraTest_2024_Online}, RedMine~\cite{RedMine_2024_Online}, BaseCamp~\cite{BaseCamp_2024_Online}, Miro~\cite{Miro_2024_Online}, and Savanna~\cite{Savanna_2024_Online}.
The size of their \xglspl{its} ranged from just a few hundred up to over a million issues.
The company sizes ranged from tens to thousands of employees, covering different industries including automotive, medical, consulting, and energy.

\subsection{Interview Procedure}

We used interviews as the main method for several reasons.
First, data analysis and simulation studies are common on \xgls{its} data, particularly based on open-source data~\cite{Feller_2000_ICIS,Scacchi_2006_SPIP,Scacchi_2007_FSE,Scacchi_2007_AC,Scacchi_2010_FoSER}.
While useful for characterising recorded phenomena, these studies do not reflect the experiences, priorities, and reasoning of actual \xgls{se} practitioners.
Only focusing on data analysis and lab studies can create a gap between research and practice.
Interviews, however, are well-suited to collect perceptions and reasoning of human subjects.
In contrast to surveys (which usually include closed questions, focus on representativeness and quantification, and usually assume preliminary hypothesis and sets of variables to measure), interviews usually focus on open questions and enable follow-up clarifications.
This is particularly important for us to gather the \textit{context} when a certain problem matters.
Interviews also enable asking \textit{``why'' follow-ups}, which is crucial for understanding practitioners' perceptions.

When designing the interviews, we aimed at maximising realism, diversity, and reliability.
Realism implies that our observations should reflect real \xgls{ite} interactions in real project environments.
We thus decided to gather insights only from practitioners who worked for years in industry, and excluding students and researchers.
By maximising diversity, we tried to cover diverse perspectives, including diverse roles, companies, industries, and \xglspl{its}.
The aim was not to seek for representative results, rather, we aimed to draw a comprehensive picture of \xgls{ite} problems.
To maximise the reliability of results, we took measures to reduce researchers' interference and potential bias as far as possible---as the goal was to summarise the perceptions of practitioners, which might or might not affirm current findings in research.

We conducted 1-hour semi-structured interviews over video calls, guided by an interview protocol.
To encourage an open dialogue regarding their true thoughts on the quality of their company's \xgls{ite} practices, we refrained from recording the interviews.
To minimise observer bias, \textit{two interviewers} were present in each interview session, who both could ask follow-up or clarification questions.
We used slides to guide the interview sections.
The two interviewers met within {\mytilde}24 hours of the interview for 15--60 minutes to discuss the results and align the notes.
15 interviews were conducted in English and 11 in German.
The language was decided based on the preference of the participant.
All German interviews were conducted by native German speakers who are also fluent in English, who later translated their notes into English.

Each interview consisted of three main parts: problems related to using \xglspl{ite}, perceptions of smells related to \xgls{ite} Best Practices, and opinions on tooling to manage these problems and smells.
Only the first of these three parts is discussed in this chapter, whereas part 2 is discussed in Chapter~\ref{ch:catalogue} and part 3 is discussed in Chapter~\ref{ch:tooling}.
We started with a short welcome session ({\mytilde}10 minutes) where we introduced ourselves and the study and asked about their role, experience, and work context.
We only introduced key concepts when they were necessary to avoid biasing their answers to earlier questions.
Relevant to this chapter, we asked the interviewees about challenges they encountered during their work with \xglspl{ite}.

\subsection{Analysis}

Following the interviews, we digitised the 3,675 recorded meeting ``notes'' for analysis.\footnote{Each ``note'' corresponds to a complete thought, usually a sentence.}
We gave each note a unique ID based on the participant, interviewer, and interview question, allowing for evidence tracing to support the analyses.
We conducted a \xgls{ta} of the entire set of notes to produce the qualitative findings for the research question.

\textbf{Thematic Analysis of Interview Notes.}
The findings of the interviews were extracted using \ltexdummy{\xgls{ta}~\cite{Braun_2021_Book}}.\footnote{\ltexdummy{\nameCL}, \ltexdummy{\nameCR}, and I conducted phases 1--6, while \ltexdummy{\nameWM} was involved in phases 3--6.}
We independently read through all the digitised notes, to familiarise ourselves with the data (phase 1).
Then, we independently extracted preliminary codes (phase 2).
We then met to compare, discuss, and group the codes into a single agreed-upon set (phase 3).
During these shared meetings, we also began phase 4, where we created an initial set of themes based on our shared understanding of the codes.
All four researchers met to discuss and refine the themes and codes based on our shared knowledge.
We then named and defined the themes (phase 5), and selected representative examples (phase 6).

\textbf{Replication Package.}
I created a replication package as part of this thesis~\cite{Montgomery_2025_PhDThesisReplicationPackage}.
This package includes all research artefacts discussed in this chapter, except for the digitised meeting notes.
Statements and knowledge about \xglspl{ite} can be quite critical (both for the business and the individuals), and so to protect the privacy of the interviewees, the digitised notes are not archived.
I do, however, share the remaining artefacts which have the IDs embedded in them, allowing for interpretation of the observations in context: you can trace analysis to the participant table and connect to their anonymous context factors.

\section{Results: ITE Common Problems}

In this section, I describe the recurrent problems the participants observed in their daily work with \xglspl{ite}.
These results are from the \xgls{ta} applied to the participant statements.

\subsection{ITE Information Problems}
Participants described difficulties retrieving information from the \xgls{ite} and maintaining a general understanding of the overall \xgls{ite} state.

\textbf{P1. Missing Issue Information.}
Eighteen participants mentioned problems related to the accuracy and completeness of information in their \xglspl{its}.
This includes information that is incorrect, irrelevant, or missing, as well as vague definitions and insufficient details in issue descriptions (confirming results by Zimmerman et al.~\cite{Zimmermann_2010_TSE}).
These problems were more often mentioned by developers (12/16) than managers (2/6).
P07 said that the ``main problem is the people who did not enter enough information''.
There is simply too much to be filled in across too many diverse and irrelevant workflows, so users revert to skipping the field or filling in incorrect information.
Without the correct information, resolving issues takes a longer time due to required clarifications, or may not happen at all.
One context factor for this appears to be when \xglspl{ite} are used by many untrained or external people.

\textbf{P2. Issue Overload.}
Seventeen participants mentioned the problem of too many issues (both open and closed) within their \xglspl{its}.
P18 described ``[getting] lost navigating around Jira''.
The consequences of this problem include a lack of awareness of existing issues, leading to duplicates and overlapping issues, missed links, and issues being forgotten.
As context factors, participants stated older and larger \xglspl{its} (as they have likely accumulated more issues).
Issue overload appears to affect managers more than developers.

\textbf{P3. Zombie Issues.}
Twelve participants mentioned the problem of inactive and abandoned issues within their \xglspl{its}, which some referred to as ``Zombie Issues''.
These participants (primarily developers), described the problem as a lack of attention towards issues that were still open, but were waiting for some action to be taken.
Specific causes for this problem include when the reporter and assignee are the same person (accountable only to themselves), and low-priority issues (minimal pressure to act).
P07 said some issues get ignored, and that ``it happens to all developers'' due to a ``focus on other topics''.
Zombie Issues may lead to a clogged \xgls{its}, and duplicate issues (due to unawareness of existing issues).
The context factors for this problem include older and larger \xglspl{its} (where the stressor is more issues), as well as multiple disjoint teams working loosely together (which leads to uncertainty about respective responsibilities).

\textbf{P4. Ineffective Search.}
Ten participants mentioned the problem of ineffective search within their \xglspl{its}.
They described difficulties searching for specific issues, as well as classes of issues such as open Epics associated with a certain project, due to limitations of the search features in the \xgls{its}.
Even with sophisticated filtering, the exact phrases required to locate the issues were not easily obtainable.
This would result in the search returning no issues, or too many issues related to non-important aspects of the search term (confirming findings from Heck and Zaidman~\cite{Heck_2013_IWPSE}).
P09 mentioned that ``there are many synonyms, which makes it harder to find something''.
Difficulties in locating relevant existing issues lead to a less connected \xgls{its}, as well as duplicate issues.
Ineffective Search is particularly severe in larger \xglspl{its}.

\textbf{P5. Lack of Comprehensive Overview.}

Ten participants mentioned the lack of a comprehensive overview.
They described difficulties gaining a holistic understanding of their \xglspl{its}, as well as their own tasks and responsibilities.
They expressed a desire for a better grasp of the big picture and the dependencies between issues.
P08 stated that \ltexdummy{``it is hard to keep an overview of the huge backlog, and keep the projects separate from each other''}.
Participants described the cause as missing features in their \xgls{its}.
While some \xglspl{its} facilitate custom dashboards, these must be created by the users, who are themselves often not sure what to visualise.
The consequence of a lack of an overview is the repeating need to manually find things, as well as the need to remember (or write down) exactly what to keep track of.
The desire for such a dashboard was mentioned by half of the managers (3/6) and product owners (2/4), but less than a third of developers (5/16).

\subsection{ITE Workflow Problems}
Most participants described a complex and nuanced trade-off between the bloat that exists for more complicated workflows, and the need for more complete workflow coverage.
This includes the problems: Workflow Bloat, Missing Issue Information, Information Islands, Divergent Needs, Lack of a Fitting Workflow, and Unclear Workflow.

\textbf{P6. Workflow Bloat.}
Nineteen participants mentioned the problem of complex workflows within their \xgls{ite}, which some referred to as ``Workflow Bloat''.
This includes too many issue fields, required steps, and over-engineered workflows with numerous edge cases and excessive maintenance overhead.
P13 mentioned that there are ``too many link types, a whole drop-down menu''.
Another challenge is the amount of information that must be entered while creating an issue, which can be tedious and time-consuming.
P17 reported a ``fine granularity of issue fields with a very granular workflow'', which sometimes leads to ``fitting yourself into the workflow, instead of just doing the work''.
This can be particularly problematic when the required information is not known at issue creation time, or when the \xgls{its} is configured with overly strict rules.
P02 and P20 mentioned examples where one small bug fix needed several hours of documentation.
The consequences of Workflow Bloat include confusion regarding what needs to be done and uncertainty regarding the importance of the required information.
P18 mentioned that in Jira ``there is so much information presented that is often not used''.
This can lead to users ignoring important fields, or entering the wrong information.
This can also result in fields having no meaning, e.g., one participant mentioned that too many incoming Bug Reports were marked as the highest priority to get attention~\cite{Saha_2015_MSR}, but are actually not a high priority.
One context factor for Workflow Bloat is older \xglspl{its}, given the possibility for outdated fields and processes.
Another is when many teams with different internal workflows all use the same \xgls{its} configuration.
A more powerful \xgls{its} can also be a context factor, given the possibility for more complex workflows.
As P05 mentioned, ``Jira tries to mimic a perfect world where everything is known''.

\textbf{P7. Lack of Workflow Enforcement.}
Thirteen participants mentioned the problem of a Lack of Workflow Enforcement in their \xglspl{its}.
This problem includes a lack of enforcement of required fields, accepted norms of \xgls{its} usage, and workflow steps that are assumed to be consistent.
About half of the participants (more managers than developers) mentioned that \xglspl{its} can allow ``too much freedom'', resulting in erroneous user input, such as misusing specific properties.
P06 mentioned that the ``environment property is a large text field where sometimes the description is erroneously entered''.
Consequences of a Lack of Workflow Enforcement include missing or incorrectly located information, and skipping statuses, e.g. ``testing'' or ``quality assurance'' of implementations.
However, care must be taken when addressing Lack of Workflow Enforcement through automated restrictions, as they may lead to additional Workflow Bloat, especially in the presence of an Unclear Workflow.
Moreover, if there are exceptions to the enforced rule, or the rule affects teams with Divergent Needs, a Lack of a Fitting Workflow may arise from preventing certain behaviour.

\textbf{P8. Lack of Workflow Support.}
Six participants mentioned the problem of a Lack of Workflow Support in their \xgls{its}.
Examples of desired automation include setting an issue from ``in progress'' to ``in review'' once a fixing commit was made, or closing a task once all sub-tasks were closed.
The consequence of this lack of automation is more manual work needed, which may then be done incorrectly or not at all.
If automation support is implemented, workflow automation can help ease the problem of Workflow Bloat by taking away manual steps and therefore making the process less tedious.

\textbf{P9. Lack of a Fitting Workflow.}
Five participants mentioned the problem of a Lack of a Fitting Workflow in their \xglspl{its}.
They described limitations in issue fields and workflow options.
P26 said, ``there are so many exceptions in real life; the process is valid for 90\% of use cases, but the 10\% of cases do not fit, so you have to cheat the process to make it work for these 10\%''.
A consequence of this problem is information being stored in the wrong location or lost altogether, as there is no appropriate place for it.
This lack of information can lead to uncertainty.
Lightweight \xglspl{its} might not offer advanced customisation options and could potentially be more at risk of experiencing this problem.
Working to address a Lack of a Fitting Workflow has the potential to incur Workflow Bloat through newly added fields or options.
Conversely, Workflow Bloat itself may lead to Lack of a Fitting Workflow, if strictly defined workflows make it difficult to represent edge cases in the \xgls{its}.
Users need to find a balance between Workflow Bloat and Lack of a Fitting Workflow.

\textbf{P10. Unclear Workflow.}
Five participants mentioned the problem of Unclear Workflow in their \xgls{its}.
This includes unclear definitions, vague meanings, and differing definitions of issue fields, e.g., what do ``done'' and ``fixed'' mean for a bug versus an epic?
P05 mentioned, ``Jira tries to establish a naming standard (e.g., blocked and resolved), but what does ``resolved'' actually mean?''.
This problem may lead to users spending more time than intended on the issue documentation, or alternatively skipping steps because of decision fatigue.

\subsection{ITE Organisational Problems}
Some participants described problems involving the collaboration across organisational contexts and individual personnel.

\textbf{P11. Lack of Mindset \& Discipline.}
Fourteen participants mentioned the problem of a Lack of Mindset \& Discipline within their \xgls{its}.
These participants explained that users must adopt the right mindset and discipline to keep the \xglspl{its} up to date, requiring motivation and integration into daily work.
P15 and P22 both mentioned that good discipline is needed.
P22 elaborated, saying that ``\xglspl{its} must be used properly to be successful, this has nothing to do with the \xgls{its} itself''.
This problem also includes inappropriate communication, waiting for people, or difficulty choosing an estimate or due date.
Managers are often responsible for organising issues and keeping them up-to-date, which can lead to developers creating issues with lower quality.
It is problematic if other users do not see the value in correctly creating and updating issues if the processes are too complex or tedious, or if the overhead is too much.
As Meyer et al.~\cite{Meyer_2014_FSE} observed, developers feel productive when they close many or big tasks, whereas creating or organising issues feels unproductive.
If the \xgls{its} is not kept up to date, this problem can lead to other issues, such as Zombie Issues or Missing Issue Information.

\textbf{P12. Scoping Issues is Hard.}
Eleven participants mentioned that Scoping Issues is Hard in their \xglspl{its}.
They described difficulties breaking apart and appropriately scoping large issues.
The participants described that \xgls{its} users may struggle with uncertainty and ambiguity, resulting in
issues being scoped incorrectly or lacking clarity in their scope.
Managers were affected by this more than developers.
Users might be uncertain when an issue is done and close issues too early, leading to issues needing to be reopened.
As P14 pointed out: ``what is the definition of `done' for this work?''.

\textbf{P13. Divergent Tracking Needs.}
Ten participants mentioned the problem of divergent needs in their \xgls{ite}, particularly across teams and projects.
This includes variances in issue fields and available properties, but also entire workflows supported by the \xgls{its}.
As described by P17, there is a ``heterogeneous usage of Jira, even within the same project, which means no one is following the same workflow''.
The consequence of Divergent Needs is that \xglspl{its} become too diverse with the number of available options and start exhibiting Workflow Bloat, thereby creating confusion as to the correct options for a given workflow.
This can lead to communication issues and a lack of understanding about the processes being used by different teams, including their own.
One context factor for this problem is larger companies with many projects and teams, all using the same \xgls{its} instance.

\textbf{P14. Information Islands.}
Ten participants mentioned the problem of multiple disjoint systems and missing integration in their \xgls{ite}, leading users to duplicate, split, or omit information across systems.
Such disjoint systems include Git, communication and documentation tools, and even multiple \xglspl{its} being used in parallel.
A common reported problem was the use of multiple \xglspl{its} within the same organisation, with slightly different use cases (e.g. one public and one private \xgls{its}), which lead to many duplicate and missing issues across the systems.
As P20 mentioned, there are ``multiple sources of truth, including Slack and Jira''.
P19 similarly mentioned, ``Many things [are] not stored in the ticket: inside other communication tools, such as MatterMost''.
One cause of this problem is the use of many similar systems, without useful integration.
The consequences of Information Islands include duplicate and missing issues, as well as increased effort required to create, maintain and retrieve the information across systems.

\textbf{P15. Difficult Customisation.}
Eleven participants mentioned the problem of customising an \xgls{its} to the workflow and needs for specific projects and teams.
Some \xglspl{its}, such as Jira, can be too complex and have too many features, leading to ``technology overload''~\cite{KarrWisniewski_2010_CHB} and confusion about how to use certain features correctly.
P17 said, ``being able to configure Jira properly is a difficult task that requires advanced knowledge of development workflows and software development practices''.
P21 said that it is ``difficult to use these tools out-of-the-box'' as it ``requires a lot of customisation''.
The more familiar a Jira administrator or user is with the features, the fewer problems are likely to occur.

\textbf{P16. Issue Linking is Cumbersome.}
This includes issues where linking requires multiple clicks or where the link type cannot be easily changed.
This was only mentioned by a few participants.
They confirmed that creating links in some \xglspl{its}, such as Jira, can be cumbersome, as it is only possible to add and delete links, not change their type or the involved issues.
P12 said, ``Link types can't be changed. If I set a link type wrongly, I need to set a second link [...]''.
Incorrect and missing links make it harder to gain a holistic overview (Lack of Comprehensive Overview) and can hinder efficient navigation and organisation within the \xglspl{its}.

\section{Related Work}

\textbf{Quality of Issue Reports.}
The quality of issue data (P1) is a critical prerequisite for effective issue resolution (P8--P10).
Bettenburg et al.~\cite{Bettenburg_2008_FSE} and Zimmerman et al.~\cite{Zimmermann_2010_TSE} first studied gaps between information contained in Bug Reports and information needed by developers to fix the bugs (P1, P7, P8, P10, P11, P13).
Since then, plenty of follow-up works investigated particular aspects of Bug Report quality.
Huo et al.~\cite{Huo_2014_ICSME} compared Bug Reports written by developers vs users and found a significant difference (P10, P11) that impacts prediction models.
Chaparro et al.~\cite{Chaparro_2019_ESECFSE} focused on the quality of steps-to-reproduce in the reports, while Davies and Roper~\cite{Davies_2014_ESEM} focused on observed behaviour and expected results (P10, P11).
These works led to improvements in \xglspl{its} for creating good Bug Reports: such as assisting reporters to include relevant and essential information (P1, P8, P10), incentivising good quality Bug Reports (P8), and merging additional information into existing issues~\cite{Breu_2010_CSCW,Just_2008_IEEESymposium}.
My work is complementary.
Instead of focusing on the reporter perspective, I studied the \xgls{its} process as it interferes with the work of developers, managers, and product owners.
I also studied various issue types beyond Bug Reports, such as requirements issues and development tasks.

Only a few studies investigated the quality of issues beyond Bug Reports.
Heck and Zaidman~\cite{Heck_2016_REJ} defined three levels of Feature Request completeness: basic, required, and optional.
They found that all Feature Requests fulfilled the basic completeness, but only 54\% fulfilled ``required''.
Seiler and Paech~\cite{Seiler_2017_REFSQ} ran interviews on the problems with Feature Requests in \xglspl{its} and found that unclear feature descriptions, insufficient traceability, and fragmentation of feature knowledge are common in practice.
My results confirm their findings and add more context (e.g., when details or links matter in practice and why).
Moreover, I cover other issue types such as user stories, epics, and tasks---which are frequently used in modern \xglspl{its}~\cite{Montgomery_2022_MSR}.

\textbf{Mining Issue Repositories.}
Recent \xgls{its} datasets include millions of issues with thousands of comments and \ltexdummy{links~\cite{Montgomery_2022_MSR}}.
Several studies have highlighted that large data volumes in \xglspl{its} can be overwhelming for manual handling~\cite{Anvik_2005_OOPSLA,Regnell_2008_REFSQ,Fucci_2018_ESEM,Baysal_2013_ICSE} (P2, P4, P5).
Thus, routine tasks such as prioritisation and planning become challenging and time-consuming~\cite{Heck_2013_IWPSE,Ernst_2012_empiRE, Baysal_2013_ICSE_2} (P8--P10).

There are numerous \xgls{its} mining studies, often to retrieve relevant issues or predict particular fields.
Common research goals include issue classification, issue assignment, issue prioritisation, or duplicate detection~\cite{Cavalcanti_2014_JSEP,Zou_2020_TSE} (P4).
Classification aims to identify or correct the issue type (bug, Feature Request, or enhancement) based on issue properties using \xgls{nlp} and machine learning~\cite{Merten_2016_RE,Perez_2021_ICPC}.
Issue assignment aims at predicting an assignee for an issue, e.g., by analysing the change history, affected components, code ownership~\cite{Xia_2013_WCRE}, issue text and code similarity~\cite{Stanik_2018_ICSME}, or previous contributions~\cite{Rocha_2016_SANER}.
Issue prioritisation aims at predicting the severity, priority, or ranking of issues, based on issue properties~\cite{Menzies_2008_ICSM,Li_2022_ESEM, Tian_2012_WCRE, Lamkanfi_2011_CSMR,Tian_2013_ICSM, Li_2022_ESEM,Izadi_2022_EMSE}.
I used a different research method and focus on a different perspective.
Instead of data mining, I conducted an interview study for an-depth understanding of how practitioners use issues and issue properties, as well as problems they encountered.

\section{Discussion}

\textbf{Context is key.}
Our results highlight that there is no ``one size fits all approach'' for intelligent information retrieval and automation for \xglspl{ite}.
It all depends.
Which information is expected in certain issues and which not?
Which automations are considered to hinder stakeholder's work and which not?
What assistance is considered helpful?
How should the issue and its attributes best evolve?
Should certain workflows be strictly enforced?
Should certain discussions and collaboration take place in certain tools?
The answers to these questions depend on the organisational context, much of which is embedded directly in the \xgls{ite}.
What may sound like a simple finding is particularly crucial for future \xgls{se} research.
Automated approaches, e.g. to retrieve related or duplicate issues, predict the priority, or assign the issues, have been so far rather universal.
Our results suggest that researchers should consider these confounding context factors more carefully, as they might strongly impact the evaluation of intelligent \xgls{its} solutions.
Our results include some context factors, such as the issue type, the role of the stakeholder, the \ltexdummy{age and size} of \xglspl{its}, and the practices followed by a project or a team.
How universal the impact of context factors on certain \xglspl{ite} or types of projects remains unclear.
I hope that our work inspires follow-up studies to formalise, quantify, and measure the impact of these context factors.
The main takeaway is that context is key, and needs to be considered when confronting these problems and attempting to improve the quality of \xglspl{ite}.

\textbf{Conflicting perspectives.}
The participants shared conflicting perspectives on many of their problems with \xglspl{ite}, which comes down to the context-sensitive nature of these problems.
The clearest example is the difference between the \ltexdummy{problems Workflow Bloat and Lack of a Fitting Workflow}: the more workflow support that is added for one set of stakeholders, the more workflow bloat that is added for a different set of stakeholders.
It is tempting to imagine an ideal world where a tool supports all relevant workflows, but does not have anything extra that would constitute workflow bloat.
However, this assumes that such a tool could exist, that the correct configuration is known, and that there are no conflicting configuration settings.
From the interview results, it is our understanding that the problem is not about the tool itself, but the context of the stakeholders using the tool.
Multiple diverse groups of stakeholders (within the same organisation) all trying to use the same \xgls{its} instance will ultimately find themselves lacking some workflow process, and also experiencing workflow bloat.
One such solution to this could be to use multiple \xgls{its} instances, but this can create siloed workflows where cross-cutting operations like search may not function correctly.
Another solution may lie in automation support, as hinted by some interviewees.

\textbf{Data heterogeneity.}
The heterogeneity of \xglspl{its} and issue data in practice also sheds light on evaluating machine learning models trained on issue data.
This heterogeneity influences the issue data, its structure, and semantics.
It seems thus crucial that models (e.g., to predict duplicate issues or who should fix a defect) should be evaluated across different \xglspl{ite} and different contexts---not only on popular benchmarking datasets such as GitHub or Bugzilla.
In this line, L{\"u}ders et al.~\cite{Lüders_2022_MSR} recently showed that duplicate prediction models developed on Bugzilla data are much less accurate with Jira data: in this case due to the greater variety of issue links in Jira.
I argue that researchers should either precisely scope the \xgls{ite} context targeted by their ML models or extend the evaluation to heterogeneous \xgls{its} contexts.

\textbf{Implications for practitioners.}
Our results also have implications for \textit{SE practitioners} concerning 1) the archiving of issue data, 2) the configuration of the trackers, and 3) the training of their users.
First, managers and administrators should carefully consider limiting the size and diversity of issue data, particularly in large long-lasting projects.
Archiving old issue data could reduce the overload and stress when searching for relevant issue data.
Second, configuring \xglspl{its} seems like a highly crucial task that should be revisited regularly, possibly consulting different stakeholders.
If, for instance, old issue types or link types are not used any more, they should be cleaned from the trackers.
Our results suggest that restricting users too much can make them frustrated, but if the degree of freedom is too high with little validation, many mistakes can occur.
Too many issue properties and value options can also become counterproductive.
For example, Herraiz et al.~\cite{Herraiz_2008_MSR} argued for simplifying the issue report form in Eclipse.
The degree of detail and freedom depends on the specific needs and processes of the team or organisation.
Moreover, most \xglspl{its} offer powerful workflow configuration features, for instance Jira for automation.\footnote{\url{https://www.atlassian.com/software/jira/guides/expand-jira/automation-use-cases}}
Our results suggest that some teams seem unaware of (or unwilling to) use these features.
Third, training and educating software practitioners about \xgls{its} processes seems crucial.
Overall, the use of \xglspl{its} seems ad-hoc without common principles, rules, or (explicit) guidelines.
This can be tackled in part by generating awareness and regular training, for example, on how to create and evolve issues, what are good and bad practices, and what are powerful options to configure and analyse \xglspl{its}.

\section{Summary}

I identified common \xgls{ite} challenges by interviewing 26 experienced practitioners from diverse domains.
Identified challenges include the efficient retrieval of relevant information from the \xgls{ite}, maintaining and navigating workflows within the \xgls{ite}, and managing the \xgls{ite} usage on an organisational level.
These challenges become increasingly severe as \xglspl{its} age and various stakeholders with various habits, roles, and practices get involved.
Default search features and universal issue property recommenders only address these challenges in part.
These results motivate researching intelligent \xgls{ite} approaches based on mining issue data and considering \textit{context factors} such as issue types, practitioners roles, previous interaction with the issues, and diversity of the \xgls{ite}.
The results also highlight how software teams are striking the balance between automation and flexibility in \xglspl{ite}.
While some problems are directly related to the tooling, many problems are more related to the \textit{context-specific} way in which people work.
The results highlight some needed areas of improvement in \xglspl{ite}, which may come in the form of improved tooling, but will more likely come in the form of automation support and well-defined processes.
These results lead to questions regarding the complexity and diversity of \xglspl{ite}.
What kinds of information are stored inside these systems?
How often do the issues evolve?
I investigate and answer these questions in the following chapter.
These findings also raise questions about how to address these problems.
Automation seems like a necessary next step, but what recommendations are appropriate, and under which circumstances?
I seek to address this issue in Part~\ref{part:solution}, beginning in Chapter~\ref{ch:ontology}.

\chapter{Artefacts and Activities within Issue Tracking Ecosystems}  \label{ch:activities}

\epigraph{The secret of getting started is breaking your complex, overwhelming tasks into small manageable tasks, and then starting on the first one.}{Mark Twain}

In this chapter, I investigate the types and prevalence of different artefacts and activities in \xglspl{its}.
\xglspl{its} have existed for many decades, and began as a tool to track bugs in software as ``Bug Reports''~\cite{Bettenburg_2008_FSE}.
As \xgls{se} shifted towards Agile practices, involving less documentation and more iterative improvement, there was a need for more lightweight documentation and even better support for bug fixing.
The \xgls{its}, a tool commonplace in most \xgls{se} organisations, slowly grew to accommodate those needs.
Artefacts like Epics, User Stories, and Support Tickets are now represented as ``issues'' within \xglspl{its}.
However, the full extent of artefacts and activities represented in \xglspl{its} is not well known.
In this chapter, I first compile a dataset of 16 publicly available Jira \xgls{its} repositories.
The dataset has {\varNumUsedIssueTypes} unique issue types across the 16 \xglspl{its}.
I conducted a \xgls{ta} to investigate, understand, and categorise the available issue ``types'' across the organisations' \xglspl{its}.
From this analysis, I found 12 artefact types (thematic codes), which I categorised into 3 software activities (thematic themes).
In this chapter, I produced a thematic mapping of the artefacts and activities represented in 16 different Jira \xglspl{its}.
I also collected and publicly released a dataset of 16 public Jira \xgls{its} repositories, available on Zenodo.\footnote{\url{https://doi.org/10.5281/zenodo.5882881}}
The results reveal a diverse ecosystem of artefacts and activities within \xglspl{its}.
The next chapter investigates the information within issues, as well as the \ltexdummy{evolution frequency}, time, and ownership of different artefacts and activities in \xglspl{its}.

\statementPublication{\cite{Montgomery_2022_MSR}}

\section{Research Methodology}  \label{sec:activities_method}

My primary objective in this chapter is to explore and understand the different artefacts and activities documented within \xglspl{its}.
I define three research questions that guide this work.

\begin{description}
    \item[\textbf{RQ1}] What artefacts and activities are documented within \xglspl{its}?
    \item[\textbf{RQ2}] How prevalent are the artefacts and activities within \xglspl{its}?
    \item[\textbf{RQ3}] Which artefacts and activities co-occur within \xglspl{its}?
\end{description}

To investigate the artefacts and activities within \xglspl{its}, I first collected and curated a dataset of 16 public Jira repositories.
This dataset serves as the foundation for much of the research presented in this thesis.
I discuss the details of the collection and curation of this data in the next section (Section~\ref{sec:activities_dataset}).
Jira is a tool composed of ``issues'', where each issue is of a certain ``type''.
This type can be, for example, ``Bug'' or ``Feature Request''~\cite{Hesse_2016_IST}, to name a few popular and well-studied types in \xgls{se} research.
The issue type indicates the general organisational category the issue belongs to and the workflow processes it must undergo from ``open'' to ``close'', thereby clearly separating its use from issues of other types.
By investigating the utilised issue types in a given set of \xglspl{its}, I can uncover the artefacts and activities contained within the \xglspl{its}.

The difficulty of this task, however, is that Jira is a highly customisable tool, where each organisation defines which of the default options they will use, and which things they will customise.
Therefore, issue type usage is not standardised, despite the likelihood of similar underlying organisational processes.
A preliminary analysis of the dataset (presented below) revealed 352 unique issue types across the \xglspl{its} (see Table~\ref{table:jira_datasource}).
Given this large number of unique issue types, it is unclear which issue types are used by each \xgls{its}, how to group them, and then how to unify them across the \xglspl{its} for more general analyses.

To address this investigation and unification problem, I conducted a \xgls{ta}~\cite{Braun_2006_QRP,Cruzes_2011_ESEM} of all issue types across all 16 of the \xglspl{its} to provide a unified mapping between the issue types.
I followed the recommendations by Braun and Clarke~\cite{Braun_2006_QRP} and Cruzes and Dyba~\cite{Cruzes_2011_ESEM}.
Before beginning the analysis, I first decided on three major factors of the analysis: the type of analysis, the style of approach, and the level of analysis~\cite{Braun_2006_QRP}.
For the type of analysis, I chose a Rich Overview.
The goal of this analysis was to get an overall understanding of the artefacts and activities contained within \xglspl{its}, which requires all facets of the data to be considered.
For the style of approach, I selected a Theoretical approach.
I was interested in mapping the \xgls{its} data as it related to contemporary \xgls{se} artefacts and activities (theoretical), rather than finding emerging ideas that might be imbedded in the data (inductive).
Despite the choice for a theoretical approach, reviewing and understanding the data was still an integral part of the analysis, just with the top-down theoretical lens of existing \xgls{se} constructs.
For the level of analysis, I selected Latent Themes.
I was interested in understanding more than just what the issue type name, summary, and description had to say, I was interested in why these issue types existed, and what they were being used for.
This type of analysis, which requires me to interpret beyond the presented semantic information, required a deeper layer of interpretation known as Latent analysis.
The final choice, then, was a Theoretical \xgls{ta} providing a Rich Overview that produced Latent Themes.

The primary analysis was on the names of the issue types.
The issue Summaries and Descriptions were also analysed to cross-reference the expected artefact type based on the name of the issue.
I followed a five phase \xgls{ta} process, as recommended by Braun and Clarke~\cite{Braun_2006_QRP} and Cruzes and Dyba~\cite{Cruzes_2011_ESEM}.
First, I familiarised myself with the data by looking at the issue type descriptions in the repos and 50 example issues for each type.
Second, I merged issue types with similar names and meaning, and identified initial phrases or codes describing the types.
Third, I iterated through the codes and example issues and identified an initial set of themes.
Fourth, I reviewed and reduced the themes during two peer-discussion sessions with researchers familiar with \xglspl{its}.
I finalised the themes by forming definitions for each.

\section{Issue Tracking Dataset}  \label{sec:activities_dataset}

Here I describe the dataset I collected for this research, which later served as the foundation for many of the empirical investigations I performed.

\subsection{Why Jira}

Over the last two decades, \xgls{se} research has intensively studied issues and \xglspl{its}~\cite{Hassan_2008_FSM,Banerjee_2015_BIGDSE}, often based on Bugzilla~\cite{Bugzilla_2024_Online,Lazar_2014_MSR} and GitHub~\cite{GitHub_2024_Online}.
The primary research focus has been on specific issue types, such as \textit{Bug Reports}, to either improve the quality therein~\cite{Bettenburg_2008_FSE,Zimmermann_2010_TSE} or to predict properties such as severity~\cite{Lamkanfi_2010_MSR, Lamkanfi_2011_CSMR}, assignee~\cite{Jeong_2009_FSE}, and duplicate reports~\cite{Wang_2008_ICSE,Deshmukh_2017_ICSME,He_2020_ICPC}.
Another prominent but rather understudied \xgls{its} is Jira.
Jira~\cite{Jira_2024_Online} is an agile planning platform that offers features such as scrum boards, kanban boards, and roadmap management.
Jira's ticket-centric design mimics that of GitHub\footnote{\url{https://docs.github.com/en/issues}}, GitLab\footnote{\url{https://docs.gitlab.com/ee/user/project/issues/}}, and Bugzilla.
The benefits of Jira over other \xglspl{its} include a history of issue changes, complex issue linking networks, and a diverse set of custom field configurations across organisations.
According to 6Sense~\cite{6SenseJira_2024_Online}, Datanyze~\cite{DatanyzeJira_2024_Online}, and Enlyft~\cite{EnlyftJira_2024_Online}, Jira is by far the most popular tool in the \xgls{its} and agile project management markets.\footnote{To the best of my knowledge, there is no empirical evidence on the magnitude of Jira usage in industry.}
Yet, it is under-represented in \xgls{se} research: Google Scholar \textit{article title search} (1980--2023) for ``GitHub'' returns 5,840 results, ``Jira'' returns 747.\footnote{GitHub: \url{https://scholar.google.com/scholar?q=allintitle\%3A+Jira&hl=en&as_sdt=0\%2C5&as_ylo=1980&as_yhi=2023} \& Jira: \url{https://scholar.google.com/scholar?hl=en&as_sdt=0\%2C5&as_ylo=1980&as_yhi=2023&q=allintitle\%3A+GitHub&btnG=}}
I believe the lack of wide-spread research on Jira is due to the lack of available Jira data to study.
While Ortu et al. released a dataset of four Jira repos in 2015~\cite{Ortu_2015_PROMISE}, our dataset is larger and more diverse.

Overall, Jira is a more popular and complex \xgls{its}, which opens up many important avenues for research.
Given the feature-rich nature of Jira, other \xglspl{its} tend to offer a subset of the features available in Jira.
Findings that apply to Jira datasets, can likely be applied to other \xglspl{its}.
Additionally, Jira keeps track of historical changes to issues, which is a useful tool for researchers.
This enables a special type of evolutionary investigation that other \xglspl{its} either do not offer, or only offer in partial ways.
For these reasons, I chose Jira as my primary data source and tool for investigating \xglspl{its}.

\subsection{The Jira Dataset}

\begin{table}[ht]
    \small                      
    \setlength\tabcolsep{1pt}   
    \centering                  
    \captionsetup{skip=0pt}     

    \caption{Dataset consisting of 16 public Jira repositories.}
    \label{table:jira_datasource}

    \begin{tabular}{lrrrrrrrrrr}
        \multicolumn{11}{c}{\footnotesize{\makecell{
            Column names: \textbf{D}ocumented \textbf{I}ssue \textbf{T}ypes; \textbf{U}sed \textbf{I}ssue \textbf{T}ypes; \\
            \textbf{D}ocumented \textbf{L}ink \textbf{T}ypes; \textbf{U}sed \textbf{L}ink \textbf{T}ypes; \\
            \textbf{Ch}anges per \textbf{I}ssue; \textbf{Co}mments per \textbf{I}ssue; \textbf{U}nique \textbf{P}rojects}}} \\
        \toprule
        Jira repo     & Born & Issues    & DIT & UIT & Links   & DLT & ULT & Ch/I & Co/I & UP   \\
        \midrule
        Apache        & 2000 & 1,014,926 & 48  & 49  & 264,108 & 20  & 20  & 10   & 5    & 657  \\
        Hyperledger   & 2016 & 28,146    & 9   & 9   & 16,846  & 6   & 6   & 12   & 2    & 36   \\
        IntelDAOS     & 2016 & 9,474     & 4   & 11  & 2,667   & 12  & 12  & 15   & 3    & 7    \\
        JFrog         & 2006 & 15,535    & 30  & 22  & 3,303   & 17  & 10  & 9    & 1    & 35   \\
        Jira          & 2002 & 274,545   & 52  & 38  & 110,507 & 19  & 18  & 21   & 3    & 123  \\
        JiraEcosystem & 2004 & 41,866    & 122 & 40  & 12,439  & 19  & 18  & 15   & 2    & 153  \\
        MariaDB       & 2009 & 31,229    & 11  & 10  & 14,950  &  -  & 6   & 12   &  -   & 24   \\
        Mindville     & 2015 & 2,134     & 2   & 2   & 46      &  -  & 4   & 3    &  -   & 10   \\
        Mojang        & 2012 & 420,819   & 1   & 3   & 215,821 & 6   & 5   & 8    & 2    & 16   \\
        MongoDB       & 2009 & 137,172   & 31  & 37  & 92,368  & 15  & 13  & 17   & 3    & 95   \\
        Qt            & 2005 & 148,579   & 19  & 14  & 41,426  & 10  & 10  & 12   & 3    & 38   \\
        RedHat        & 2001 & 353,000   & 74  & 55  & 163,085 & 23  & 19  & 13   & 2    & 472  \\
        Sakai         & 2004 & 50,550    & 43  & 15  & 20,292  & 7   & 7   & 10   & 4    & 55   \\
        SecondLife    & 2007 & 1,867     & 10  & 12  & 674     & 9   & 5   & 30   & 8    & 3    \\
        Sonatype      & 2008 & 87,284    & 16  & 21  & 4,975   & 10  & 9   & 7    & 4    & 17   \\
        Spring        & 2003 & 69,156    & 13  & 14  & 14,716  & 7   & 7   & 8    & 3    & 81   \\
        \midrule
        Sum           &      & 2,686,282 & 485 & 352 & 978,223 & 180 & 169 &      &      & 1,822 \\
        Median        &      & 59,853    & 18  & 14  & 15,898  & 10  & 10  & 12   & 3    & 37   \\
        Std Dev       &      & 251,973   & 31  & 16  & 81,744  & 6   & 5   & 6    & 2    & 178  \\
        \bottomrule
    \end{tabular}
\end{table}

I collected data from 16 public Jira repositories (repos) containing 1,822 projects and 2.7 million issues.
The dataset includes historical records of 32 million changes, 9 million comments, as well as 1 million issue links that connect the issues in multiple ways.
Table~\ref{table:jira_datasource} shows an overview of the data.
The table lists the following attributes for each Jira: the year it first was used, the number of issues, number of documented and used issue types, number of documented and used link types, number of changes per issue, number of comments per issue, and number of unique projects.\footnote{Documented issue types may be removed over time, but their use is still recorded in the data, leading to a lower number of documented issue types than used issue types.}
Apache is the largest repo with 1 million issues and 10.5 million changes.
Mindville and SecondLife are the smallest \xglspl{jr} with \mytilde2,000 issues each.
The number of unique projects is particularly important because each \xgls{jr} contains documentation for multiple projects, for example, 657 projects for Apache.
While the dataset has 16 different repositories, it represents 1,822 actual projects that can be studied.
The data and code is publicly available with an open-source licence~\cite{Montgomery_2025_PhDThesisReplicationPackage}.

Other public datasets, such as the Bugzilla dataset provided by Lazar et al.~\cite{Lazar_2014_MSR}, mainly include Bug Reports and other maintenance issues.
Recent user stories datasets are rather small, and allow for neither a comparison between different issue types nor for a study of the issue evolutions~\cite{Dalpiaz_2018_DS,Pena_2020_DS}.
For this research, I only used 13 of the 16 available Jira repos because two of the repos (MariaDB and Mindville) contain no comments (an important issue field for the comparative analysis), and Mojang only contains bugs.

Each \xgls{jr} has a number of documented issue types, but only a subset are actually used in the projects.
As shown in Table~\ref{table:jira_datasource}, 485 types are defined while only 352 are used.
Notably, the median number of used issue types is 14, which is significantly higher than expected from research discussing \xglspl{its} (Bug Reports~\cite{Bettenburg_2008_FSE,Zimmermann_2010_TSE}, Feature Requests~\cite{Fitzgerald_2011_RE,Merten_2016_RE}, Tasks~\cite{Ernst_2012_empiRE}, or technical debt~\cite{Bavota_2016_MSR,Xavier_2020_MSR}).
The projects also vary in terms of the number of documented (186) and used (169) link types.
The median number of used link types is 10, which is considerably more than the mainly well-researched ``duplicate'', or ``depends/relates'' types.

\begin{figure}[t]
    \centering
    \includegraphics[width=\textwidth]{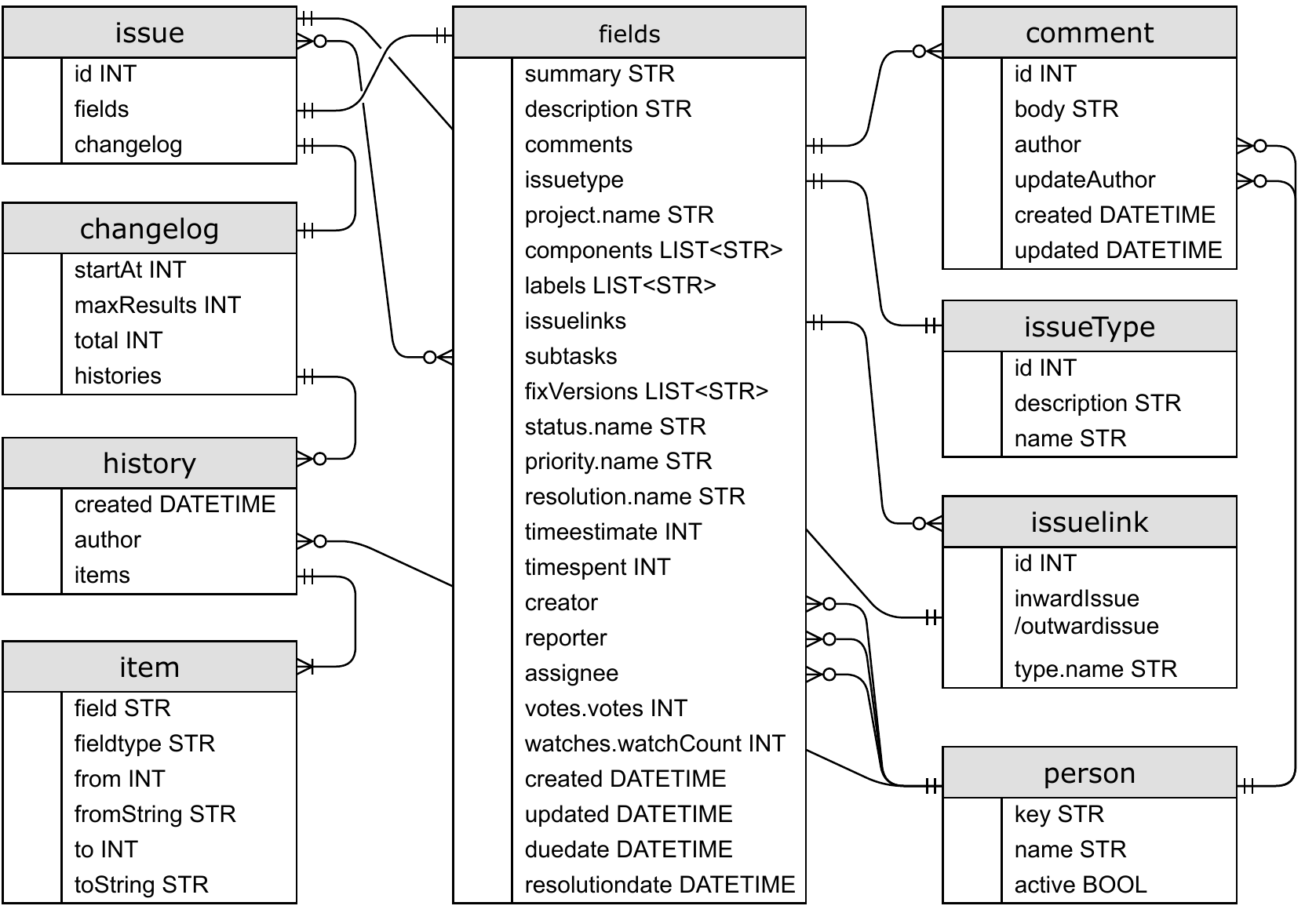}
    \caption{Jira MongoDB database scheme.}
    \label{fig:jira_database_erd}
\end{figure}

\begin{table}[ht]
    \small                          
    \centering                      
    \setlength\tabcolsep{2.5pt}     
    \captionsetup{skip=0pt}         

    \caption{Issue fields.}
    \label{table:jira_issue_fields}

    \begin{tabularx}{\textwidth}{l X c}
        \toprule
            \textbf{Issue Field} & \textbf{Description}  \\
        \midrule
            summary             & A brief one-line summary of the issue.  \\
            description         & A detailed description of the issue.  \\
            comments            & Community discussion on each issue.  \\
            issuetype           & The issue purpose within the organisation.  \\
            project             & The parent project to which the issue belongs. \\
            components          & Project component(s) to which this issue relates.   \\
            labels              & Labels to which this issue relates.  \\
            issuelinks          & A list of links to related issues.  \\
            subtasks            & Sub-issues to this issue; can only be one level deep.  \\
            fixVersions         & Project version in which the issue was (or will be) fixed.  \\
            status              & The stage the issue is currently at in its lifecycle.  \\
            priority            & The issue importance in relation to other issues.  \\
            resolution          & A record of the issue's resolution, once resolved or closed.  \\
            timeestimate        & Estimated amount of time required to resolve the issue.  \\
            timespent           & Amount of time spent working on this issue.  \\
            creator             & The person who created the issue. \\
            reporter            & The person who found/reported the issue. Defaults to the creator unless otherwise assigned. \\
            assignee            &  The person responsible to resolve the issue. \\
            votes               & Number of people who want this issue addressed. \\
            watches             & Number of people watching this issue.  \\
            created             & Time and date this issue was entered into Jira.  \\
            updated             & Time and date this issue was last edited.  \\
            resolutionedate     & Time and date this issue was resolved.  \\
            duedate             & Time and date this issue is scheduled to be completed.  \\
        \bottomrule
    \end{tabularx}
\end{table}

The Jira data is stored in MongoDB.
Figure~\ref{fig:jira_database_erd} describes the data as an \xgls{erd}.
It is not possible to create a full and reliable \xgls{erd} for a document database such as MongoDB, unless a strict structure is enforced on all documents, and that is not the case for Jira across different repositories.
Figure~\ref{fig:jira_database_erd}, therefore, is a simplification of the real data structure.
I have diagrammed the key objects in the data, simplified structures where the complexity is self-explanatory, and extrapolated the data types to give an idea of what to expect.
The primary documents to expect in the dataset are ``issues'', each unit of which represents a single issue within Jira.
The two primary documents nested within each issue are ``fields'' and ``changelog''.
The fields document contains the current attributes of the issue, and is described in Table~\ref{table:jira_issue_fields}.
The issue field descriptions are primarily from the Jira official documentation.\footnote{\url{https://confluence.atlassian.com/adminjiraserver/issue-fields-and-statuses-938847116.html}}
The ``changelog'' stores the changes that have occurred to this issue.
When an issue is changed, a new ``history'' is saved, where each changed attribute is stored as an ``item''.
``Issuelinks'' connect issues to each other and are stored on both linked issues.
``Comments'' are made by the community and form a discussion around the issue.

The dataset is available on Zenodo as a MongoDB dump file.
The raw data includes a README file, the scripts used to download the data, the scripts used to produce the tables and figures in this paper, and a licence (CC BY 4.0\footnote{\url{https://creativecommons.org/licenses/by/4.0/}}).
In principle, using this dataset is as easy as importing the MongoDB dump file and modifying the included JupyterLab notebook to explore the data.
The data can also be queried using standard MongoDB queries.
Figure~\ref{fig:jira_database_erd} can be used to guide the initial exploration.

The data was downloaded using a Python script utilising the Jira API exposed by public Jira \xglspl{its}.
I obtained the list of repos through a manual search for public \xglspl{jr} on the internet.
This search involved reviewing GoogleScholar results for ``Jira'' and Google search results discussing a public \xgls{jr}.
With the public Jira URLs, the data was downloaded using the Jira REST API V2\footnote{\url{https://developer.atlassian.com/cloud/jira/platform/rest/v2}} and a Python script within a JupyterLab notebook.
In total, \mytilde50 GB of data was downloaded and stored in MongoDB.
The initial data download was performed in May 2021 and the data was updated in January 2022.\footnote{The \xgls{ta} was performed using the May 2021 data.}
Unfortunately, the MariaDB and Mindville \xglspl{jr} are no longer publicly available, so their data was not updated.

\section{Results}

\subsection{Artefacts and Activities in Issue Tracking Ecosystems}  \label{sec:artefacts_activities_findings}

\begin{table}[t]
    \footnotesize               
    \centering                  
    \def\arraystretch{1.1}      
    \setlength\tabcolsep{2pt}   

    \caption{Homogenized issue types in the Jira dataset.}
    \label{table:issue_types}

    \begin{tabular}{p{50mm}rp{50mm}}
        \toprule
            Issue types unified and grouped into themes (in bold) & Count & Examples of original  names of the issue types\\
        \midrule
        \textbf{Requirements} & \textbf{386,536} &  \\
            \qquad Epic & 4,290 & Epic, Roadmap item, Initiative \\
            \qquad Story & 20,441 & User Story, Requirement, Story \\
            \qquad New Feature & 49,793 & Feature, New Feature \\
            \qquad Feature Request & 20,175 & Brainstorming, Feature Request \\
            \qquad Improvement Suggestion & 291,837 & Suggestion, Improvement, Wish \\
        \textbf{Development} & \textbf{245,457} &  \\
            \qquad Task & 139,003 & Task, Dev Task, Technical task \\
            \qquad Sub-Task & 93,763 & Sub-Task, Dev Sub-task \\
            \qquad Quality Assurance & 9,872 & Test, \xgls{qa} Task, Performance \\
            \qquad Documentation & 2,819 & Doc API, Docs Task, Doc UI \\
        \textbf{Maintenance} & \textbf{695,710} &  \\
            \qquad Bug Report & 684,740 & Bug, Incident, Defect, Issue \\
            \qquad Legacy Upgrade & 7,338 & Dependency upgrade, Backport  \\
            \qquad Continuous Integration & 3,632 & Release, Build Failure, Tracker \\
        \midrule
            Total Combined & 1,327,703 & \\
        \bottomrule
    \end{tabular}
\end{table}

The \xgls{ta} produced a mapping between each issue type in each Jira and a unified set of themes and codes.
Table~\ref{table:issue_types} shows the complete results.
Every issue type falls into exactly one theme and code.
This mapping (stored as a JSON) allows one to programmatically ask questions such as ``give me the `Requirements' issues for Apache''.
In the following, I describe the five themes in more detail.

\textbf{Requirements} activities are documented in \xglspl{its} through lightweight representations~\cite{Ernst_2012_empiRE} such as epics, user stories, and Feature Requests.
Both top-down and bottom-up requirements are captured in \xglspl{its}.
I found top-down epics and stories, as well as bottom-up \ltexdummy{Feature Requests~\cite{Giuliano_2018_CASCON}}.
This shows the breadth of requirements knowledge maintained within \xglspl{its}.

\textbf{Development} issue types track development activities such as what needs to be done and who is doing it.
Example issue types include Task, Technical Task, Dev Task, and Sub-Task.

\textbf{Maintenance} issue types correspond to maintenance activities documented within \xglspl{its} through bottom-up (Bug, Defect, Incident) and top-down (Technical Debt, Documentation, \ltexdummy{\xgls{qa} Task}) work.
This includes quality assurance, legacy upgrades, and continuous integration.

\textbf{User Support} issues directly assist users of software systems regarding how to use the product.
While Bug Reports focus on \textit{issues with the code or product}, issues submitted to the User Support theme are focused on \textit{the user and their use or interpretation of the code or product}.
Example issue types include Support Request, Problem Ticket, IT Help, and Question.

\textbf{Other} issue types are not representative across the Jira repos, or could not be categorised at all.
Example issue types include New Project, GitHub Integration, Fug, and Spike.

\subsection{Prevalence of Artefacts and Activities}

Requirements, Development, and Maintenance issues are all common, as detailed in Table~\ref{table:issue_types}.
Of the 1,327,703 issues in the dataset, 386,536 (29\%) are Requirements, 245,457 (19\%) are Development, and 695,710 (52\%) are Maintenance issues.
While Table~\ref{table:issue_types} displays the counts across all repos, it lacks a more nuanced perspective on the prevalence and usage of these issue types.
For this, I investigate how the issues are used \textit{across the projects}, since each project represents a distinct organisational unit within each \xgls{jr}.

\begin{figure}[th]
    \centering
    \includegraphics[width=0.8\textwidth]{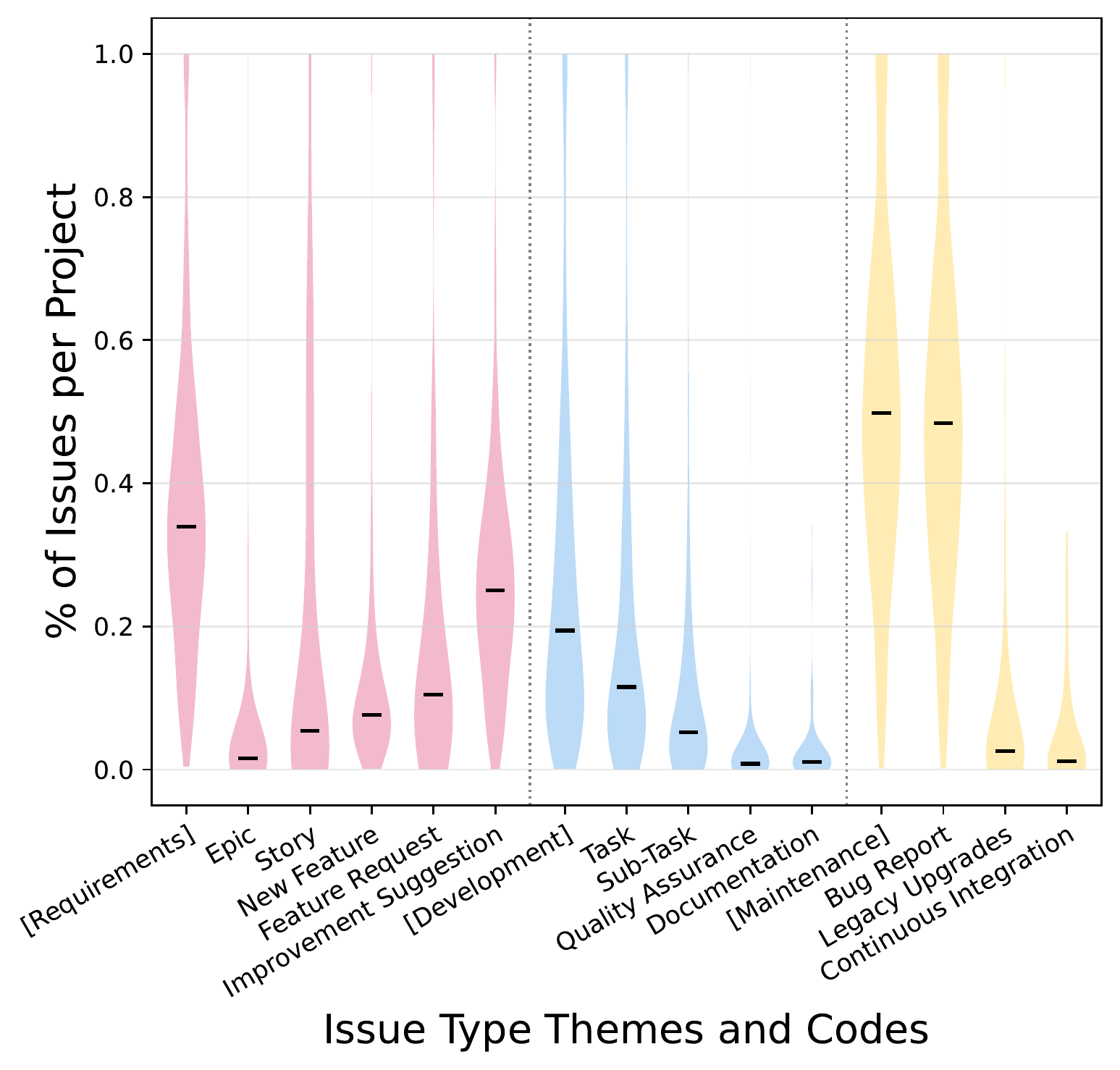}
    \caption{Distribution of issue types per project.}
    \label{fig:iss_dist}
\end{figure}

Figure~\ref{fig:iss_dist} shows the ratios of each issue type per project.
Maintenance issues are the most popular across the projects, followed by Requirements and Development.
For half of the studied projects, at least a third of the issues are Requirements (median 34\%).
This suggests that different projects seem to use requirements issues differently.
For 25\% of the studied projects, requirements issues seem even predominant, accounting for 47--81\% of all issues (upper quartile to the upper cap).\footnote{The replication package~\cite{Montgomery_2025_PhDThesisReplicationPackage} includes all these numbers, and the box plots.}
Overall, Improvement Suggestions are the most popular Requirement type across the projects, followed by Feature Requests.
However, as Figure~\ref{fig:iss_dist} shows, a significant ratio of studied projects are heavily Story-driven.
In 25\% of the projects, at least 40\% of issues represent User Stories.
Unsurprisingly, Epics (usually used to organise user stories) are rare in most projects, with a median prevalence of 2\%.
Development has a median prevalence of 19\%, and largely consists of Tasks (12\%) and Sub-Tasks (5\%).
Maintenance has a median prevalence of 50\%, composed almost entirely of Bug Reports with a median prevalence of 48\%.

\subsection{Co-Occurrence of Artefacts and Activities}

Jira is an \xgls{its} that has a high degree of customisability, which allows different organisations and projects to choose which artefacts and activities they use.
While the above analysis describes the existence and prevalence of different activities and artefacts, it does not describe how they are used together.
There are 1,477 projects within the 13 Jiras used for this analysis, so it is not reasonable to review each one in detail to get an understanding of their usage patterns.
Instead, I apply co-occurrence analysis to investigate which artefacts and activities are used together.
The result is a ranked list of usage pattern sets which give an idea of which artefacts and activities are being used together within these \xglspl{its}.

Traditionally, co-occurrence analysis is applied in contexts where items are grouped in various ways, such as text analysis and purchase patterns analysis.
The primary goal is to surface which items commonly occur with other items, in small sets such as two, three, or four.
This kind of analysis can be further complicated by considering time (which items come first), as well as magnitude (how many items at a time, or cost of the items).
Instead of considering small sets such as two, three, or four, I decided to analyse the complete usage sets of artefacts and activities.
That means, for each project, I consider all artefacts and activities being used and compare those to other projects with identical or similar usage sets.
This enables the analysis of the full usage across all artefacts and activities.

The ``usage patterns of artefacts and activities'', defined by which issue types are present in each project, is not immediately clear from the raw data.
For example, what constitutes ``usage'' of an issue type?
Clearly, the existence of issues with a certain issue type counts as usage, but how many times must a project use an issue type before it should be counted as part of the usage pattern?
If a project opens a single Continuous Integration ticket to experiment, or two Sub-Tasks to see if it fits their workflow, or a User Story thinking it is a Task, does this count as part of the usage pattern?
In this work, I define ``usage'' as using an issue type at least 5 times, or at least 10\% of all issues (for projects with less than 50 issues).
My assertion is that after 5 it is very unlikely to be a mistake or a small test.
This reduces the total number of co-occurrence usage patterns from 215 to 186, so it does have an impact on the results, but not a considerable one.
The unfiltered results (215 sets) are available in the replication package~\cite{Montgomery_2025_PhDThesisReplicationPackage}.

\begin{table}[th]
    \small                          
    \centering                      
    \captionsetup{skip=0pt}         
    \newcommand*{\colspace}{4.1mm}  

    \caption{Top 30 issue type co-occurrences}
    \label{table:issuetype_cooccurrence}

    \begin{tabular}{ r | r r r | wc{\colspace} wc{\colspace} wc{\colspace} wc{\colspace} wc{\colspace} | wc{\colspace} wc{\colspace} wc{\colspace} wc{\colspace} | wc{\colspace} wc{\colspace} wc{\colspace}}
        \toprule
            & \multicolumn{3}{c|}{Project Distribution} & \multicolumn{5}{c|}{Requirements} & \multicolumn{4}{c|}{Development} & \multicolumn{3}{c}{Maintenance} \\
            Set & Count & \multicolumn{1}{c}{\%} & Sum \%
                                    & Ep & St & NF & FR & IS & Ta & SB & QA & Do & BR & LU & CI \\
        \midrule
            1  & 139 & 9.41 &  9.41 &    &    & NF &    & IS & Ta &    &    &    & BR &    &    \\
            2  & 130 & 8.80 & 18.21 &    &    &    &    & IS &    &    &    &    & BR &    &    \\
            3  & 118 & 7.99 & 26.20 &    &    & NF &    & IS & Ta & SB &    &    & BR &    &    \\
            4  & 112 & 7.58 & 33.78 &    &    & NF &    & IS & Ta & SB & QA &    & BR &    &    \\
            5  &  86 & 5.82 & 39.61 &    &    &    &    &    &    &    &    &    & BR &    &    \\
            6  &  71 & 4.81 & 44.41 &    &    & NF &    & IS &    &    &    &    & BR &    &    \\
            7  &  55 & 3.72 & 48.14 &    &    &    &    & IS & Ta &    &    &    & BR &    &    \\
            8  &  44 & 2.98 & 51.12 &    &    &    &    &    & Ta &    &    &    &    &    &    \\
            9  &  38 & 2.57 & 53.69 &    &    & NF &    & IS & Ta &    &    &    & BR & LU &    \\
            10 &  36 & 2.44 & 56.13 &    &    &    &    &    & Ta &    &    &    & BR &    &    \\
            11 &  29 & 1.96 & 58.09 &    &    &    &    & IS & Ta & SB &    &    & BR &    &    \\
            12 &  21 & 1.42 & 59.51 &    &    &    & FR &    & Ta &    &    &    & BR &    &    \\
            13 &  20 & 1.35 & 60.87 &    &    &    &    & IS &    & SB &    &    & BR &    &    \\
            14 &  18 & 1.22 & 62.09 &    &    & NF &    & IS & Ta &    & QA &    & BR &    &    \\
            15 &  17 & 1.15 & 63.24 &    & St &    &    &    & Ta & SB &    &    & BR &    &    \\
            16 &  17 & 1.15 & 64.39 &    &    &    & FR & IS & Ta &    &    &    & BR &    &    \\
            17 &  16 & 1.08 & 65.47 &    &    &    &    & IS &    &    &    &    &    &    &    \\
            18 &  16 & 1.08 & 66.55 &    &    &    & FR & IS & Ta & SB &    &    & BR &    &    \\
            19 &  15 & 1.02 & 67.57 & Ep & St &    &    &    & Ta & SB &    &    & BR &    &    \\
            20 &  12 & 0.81 & 68.38 & Ep &    & NF &    & IS & Ta &    &    &    & BR &    &    \\
            21 &  12 & 0.81 & 69.19 &    &    &    & FR &    & Ta & SB &    &    & BR &    &    \\
            22 &  11 & 0.74 & 69.94 &    &    & NF &    & IS &    & SB &    &    & BR &    &    \\
            23 &  11 & 0.74 & 70.68 &    &    & NF &    &    &    &    &    &    &    &    &    \\
            24 &  10 & 0.68 & 71.36 &    &    & NF &    &    &    &    &    &    & BR &    &    \\
            25 &   9 & 0.61 & 71.97 &    &    & NF &    &    & Ta &    &    &    & BR &    &    \\
            26 &   9 & 0.61 & 72.58 & Ep &    & NF &    & IS & Ta & SB &    &    & BR &    &    \\
            27 &   9 & 0.61 & 73.19 &    & St &    &    &    &    &    &    &    &    &    &    \\
            28 &   9 & 0.61 & 73.80 &    & St &    &    &    &    &    &    &    & BR &    &    \\
            29 &   8 & 0.54 & 74.34 &    &    & NF &    & IS & Ta & SB & QA &    & BR & LU &    \\
            30 &   8 & 0.54 & 74.88 &    & St &    &    & IS & Ta &    &    &    & BR &    &    \\
        \bottomrule\addlinespace[1mm]
        \multicolumn{16}{l}{\scriptsize{\makecell[l]{
            \textbf{Requirements}: (Ep) Epic (St) Story (NF) New Feature (FR) Feature Request (IS) Improvement Suggestion.\\
            \textbf{Development}: (Ta) Task (SB) Sub-Task (QA) Quality Assurance (Do) Documentation.\\
            \textbf{Maintenance}: (BR) Bug Report (LU) Legacy Upgrades (CI) Continuous Integration.}}} \\
    \end{tabular}
\end{table}

Table~\ref{table:issuetype_cooccurrence} shows the top 30 results of the issue type usage analysis across the entire dataset of 13 Jira repos.
On the right-hand side of the Table, there are the three \xgls{its} activities and the 12 types of artefacts grouped under them.
On the left-hand side, there are three project distribution values: the count of projects (1,477 projects in total), the \% of projects (count/1,477), and the cumulative sum of the \% of projects up to that point.
The far-left value is the set number used for referencing the individual co-occurrences.
To walk through an example, in set 5 of Table~\ref{table:issuetype_cooccurrence}, only ``BR'' (Bug Report) is present, and all other columns are left blank.
This represents the rather common usage pattern of \xglspl{its}, which is to use it \textit{solely} as a ``Bug Tracker''.
86 projects in this dataset follow this usage pattern, which represents 5.82\% of all projects (total of 1,477).
If you add up the \% of all rows above and including set 5, you get 39.61\%, which represents the total percentage of projects that follow a usage pattern in set 5 or above.
In other words, the top 5 usage patterns are used by \mytilde40\% of projects.

There are 186 unique usage patterns across the 1,477 projects, where 75\% of all projects use only 30 different patterns (Table~\ref{table:issuetype_cooccurrence}).
The most common pattern, followed by 139 projects (\mytilde10\% of all projects), is to utilise New Features, Improvement Suggestions, Tasks, and Bug Reports.
This represents a sort of bottom-up \xgls{re}, simple task breakdown, and documenting bugs.
The second most common pattern, followed by 130 projects (\mytilde9\%), is to only utilise Improvement Suggestions and Bug Reports.
The third most common pattern, followed by 118 projects (\mytilde8\%), is the same as the most common but with the addition of Sub-Tasks.
Across these three most common usage patterns, it already applies to \mytilde25\% of all projects in this dataset.
This shows a strong conformance to particular usage patterns.
Additionally, there are 186 unique usage patterns with many projects with completely custom usage patterns only utilised by one or two projects.
This shows a diversity of usage patterns.

These results highlight the context-sensitive nature of \xglspl{its}: there are many unique ways they are configured and utilised, even within a single \xgls{its} such as Jira.
Recommendations for how to use \xglspl{its} need to be aware of these different usage patterns, rather than assume that \xglspl{its} are a universal tool with universal usage patterns.
Even in the top 30 usage patterns presented in Table~\ref{table:issuetype_cooccurrence}, there are projects that only use Requirements (rows 17, 23, and 27), only use Development (row 8), and only use Maintenance (row 5).
Therefore, researchers---and practitioners---should consider the importance of understanding their data and check their assumptions before applying analyses or solutions designed to support \xgls{its} usage.

\FloatBarrier

\section{Related Work}

\xglspl{its} have been studied from particular perspectives such as Bug Reports or Feature Requests, while a full understanding of this ecosystem remains rather limited.
Researchers started studying requirements engineering in agile settings~\cite{Lucassen_2016_REFSQ, Berntzen_2023_EMSE, Madampe_2022_TSE}.
However, these studies have focused on the perceptions of practitioners using interviews, surveys, and field observations.
My work is complementary as it uses data analysis methods and focuses on different requirements types in addition to user stories.
Recently, work by van Can and Dalpiaz~\cite{VanCan_2024_REFSQ,VanCan_2025_IST} investigated the existence of requirements in \xglspl{its}.
They conducted a manual content analysis of 1,636 issues, manually labelling them as different types of requirements.
Their analysis was focused on solely requirements, and was a small set of 1,636 issues.
They did, however, go into much greater detail with these issues, looking at the text in each description to understand the true nature of the artefact they were labelling.
As such, their analysis is complementary to mine, as their findings speak to the nature of assumed issue types and actual issue descriptions.
My analysis assumes the validity of the issue types, to make broader analyses and findings across more data.
Together, more generalised claims can be made about the true nature of artefacts in large \xglspl{its}.

\section{Discussion}

Our results show that, while Bug Reports and Feature Requests are indeed prominent in \xglspl{its} across projects, there are other prominent and interesting issue types that have a notable impact on how \xglspl{its} are used.
Epics and Stories, for example, show prominence across projects.
This implies a varied usage of \xglspl{its}, depending on which issue types are being utilised.
When approaching an organisation's \xgls{its}, one should first understand how they are utilising it, before proposing various recommender systems.
This includes systems such as machine learning techniques to categorise Bug Reports vs Feature Requests, or quality control of ``requirements'', both of which require an understanding of the extent their \xgls{its} utilises Stories vs Feature Requests.
There is also a high prevalence of a diverse set of Requirements issue types (representing many styles and phases of Requirements Engineering), while Development and Maintenance issue types are quite homogenous.
Development issues are mostly Task and Sub-Tasks, representing work to be done, and Maintenance issues are almost entirely Bug Reports.
This highlights a rather underrepresented aspect of \xgls{its} analysis and support in both research and practice: \xglspl{its} as a primary tool for Requirements Engineering processes and documentation.
Future work should further dissect these diverse issue types and seek to understand patterns of \xgls{its} usage, using the findings of prevalence and evolution as a foundation.

\section{Summary}

While \xgls{se} literature has repeatedly studied Bug Reports and Feature Requests, I investigated the prevalence of various issue types representing requirements, development, and maintenance.
I performed this analysis on 1.3 million issues from 16 distinct organisations across 1,477 \xgls{oss} projects.
The results highlight that Bug Reports and Feature Requests are indeed pervasive, but so are Stories, Improvement Suggestions, and Development Tasks.
Requirements account for 34\% of all issues in a project (median across all projects), Development accounts for 19\% of all issues, and Maintenance is 50\%.
The co-occurrence analysis revealed that 51\% of projects use only eight different sets of artefacts and activities.
The most popular configuration is to use New Features, Improvement Suggestions, Tasks, and Bug Reports.
The second most popular configuration uses only Improvements Suggestions and Bug Reports.
While the majority use only eight configurations, there are 186 usage sets across the 1,477 projects.

These results show the diversity and complexity that exists in \xglspl{its} on an artefact and activity level.
However, there are other levels of information and complexity to be investigated.
For example, what information exists \textit{within} these artefacts?
How often---and in what way---does this information evolve?
This additional layer of complexity was already discussed by our participants in Chapter~\ref{ch:challenges}.
They discussed problems with missing information (P1), workflow information (P6--P10), scoping issue information (P12), information islands (P14), and cumbersome issue linking (P16).
I investigate these problems in further detail in the following chapter.
Additionally, the findings from this chapter further highlight the diversity of information within \xglspl{its}.
This diversity means additional emphasis on understanding and prescribing solutions that are context-aware, and therefore more likely to address the underlying problems.
I propose a context-aware solution in Part~\ref{part:solution}, beginning in Chapter~\ref{ch:ontology}.

\chapter{Information Evolution within Issue Tracking Ecosystems}  \label{ch:evolution}

\epigraph{The only constant in life is change.}{Heraclitus}

In this chapter, I investigate the types of information found in \xglspl{its}, as well as the extent to which these information types evolve across the dimensions of frequency, time, and ownership.
\xglspl{its} have become a central place for \xgls{se} processes to be conducted and documented, leading to ecosystems as diverse and complex as the \xgls{se} process itself.
While certain dimensions of \xglspl{its} have been the attention of intense study~\cite{Bettenburg_2008_FSE,Zimmermann_2010_TSE}, the entirety of information within \xglspl{its} is not yet well understood.
\xglspl{its} focus on communication~\cite{Bertram_2010_CSCW} and iteration~\cite{Heck_2013_IWPSE,Shi_2013_RE}, with per-issue comments~\cite{Arya_2019_ICSE} and community contributions.
The iterative community effort is a key feature that involves the evolution of issues.
Most \xglspl{its} keep track of the changes made to their issues, enabling both quantitative and qualitative investigation.
Despite the rich data source that is \xgls{its} evolution data, it has yet to be fully explored.

To investigate the information within \xglspl{ite} and how that information evolves, I conducted an exploratory investigation of 13 Jira \xglspl{its} using the empirical method: historical data analysis.
I applied \xgls{ta} to over 20,000 issue field samples to uncover and organise the information within \xglspl{its}.
I then use descriptive statistics applied to various cross-sections of the data, also in combination with the artefact types from Chapter~\ref{ch:activities}, to quantify and describe the evolution occurring across the information types.
The results of the \xgls{ta} revealed five issue information themes: Content, MetaContent, RepoStructure, Workflow, and Community.
There are 20 codes under those themes that represent information available in issues, such as Summary and Description for the Content theme.

As a result of these investigations, we now know the distributions of major artefact types in \xglspl{ite}, the involvement of different stakeholder groups, and the distribution of evolutions across many dimensions of \xglspl{ite}.
For each of the findings, future researchers can go into great detail within the replication package~\cite{Montgomery_2025_PhDThesisReplicationPackage}, which contains the full data and scripts, as well as other pre-compiled figures and findings that didn't make it into this thesis.
We now have a grounded understanding of the usage of \xglspl{ite}, with more questions to be further explored by future researchers interested in particular facets of this data and context.
The findings of this chapter, in combination with the industrial perspective on \xgls{ite} challenges (Chapter~\ref{ch:challenges}) and our findings regarding requirements in \xglspl{ite} (Chapter~\ref{ch:activities}), present a unique opportunity to address these challenges with a theory-based solution.
The next chapter will propose such a theory, with a focus on addressing these challenges and supporting future investigation of \xgls{ite} problems and complexities.

\section{Research Methodology}

My primary objective with this chapter is to explore and understand evolving issue information used within \xgls{ite}.
To this end, I define three research questions:

\begin{description}
    \item[\textbf{RQ1}] What issue information is evolving?
    \item[\textbf{RQ2}] How does issue information evolve?
    \begin{description}
        \item[\textbf{RQ2A}] How frequently does issue information evolve?
        \item[\textbf{RQ2B}] When does issue information evolve?
        \item[\textbf{RQ2C}] Who is evolving issue information?
    \end{description}
    \item[\textbf{RQ3}] What textual changes are occurring when Requirements evolve?
\end{description}

I define ``issue information'' as the different fields that exist in issues, within a given \xgls{its}.
For example, a standard issue in any \xgls{its} has the issue fields Summary and Description.
Additional common issue fields include the Creator, Assignee, Labels, Priority, and Status.
To investigate the issue information contained within the Jira dataset, I performed a \xgls{ta} of all available issue fields across the repositories.
This process is described in detail in Section~\ref{sec:evolution_thematic_analysis}.

To investigate how issue information evolves, I analysed the evolution history of issues.
I was interested in three primary perspectives: frequency, time, and ownership.
I performed a number of cross-analyses to understand how evolution occurs across subdivisions of the data, such as the issue information types, as well as the issue artefacts and activities from Chapter~\ref{ch:activities}.

\subsection{ITS Dataset}  \label{sec:evolution_dataset}

\begin{table}[ht]
    \small                      
    \centering                  

    \caption{Research data: 13 public Jira repositories.}
    \label{table:datasource}

    \begin{tabular}{@{}lrrrrrr@{}}
    \toprule
    Jira          & Born & Projects & Issues & Evolutions & Authors  \\
    \midrule
    Apache        & 2000 & 892   & 724,226   & 7,663,616  & 96,368  \\
    Hyperledger   & 2016 & 36    & 19,544    & 176,354    & 1,740   \\
    IntelDAOS     & 2015 & 6     & 6,134     & 73,554     & 112     \\
    JFrog         & 2006 & 32    & 8,001     & 62,097     & 2,493   \\
    Jira          & 2002 & 143   & 134,297   & 1,249,338  & 69,329  \\
    JiraEcosystem & 2004 & 152   & 28,289    & 256,298    & 3,851   \\
    MongoDB       & 2009 & 96    & 62,370    & 607,296    & 11,698  \\
    Qt            & 2003 & 44    & 79,480    & 779,291    & 17,614  \\
    RedHat        & 2001 & 504   & 180,890   & 1,609,278  & 16,171  \\
    Sakai         & 2004 & 50    & 24,376    & 217,310    & 1,330   \\
    SecondLife    & 2009 & 16    & 451       & 6,729	  & 193     \\
    Sonatype      & 2008 & 28    & 4,893     & 52,787     & 1,681   \\
    Spring        & 2003 & 95    & 54,752    & 491,818    & 14,549  \\
    \midrule
    Sum           &      & 2,094 & 1,327,703 & 13,245,766 & 237,129 \\
    Median        &      & 50    & 28,289    & 256,298    & 3,851   \\
    Std Dev       &      & 245   & 187,035   & 1,976,001  & 28,712  \\
    \bottomrule
    \end{tabular}
\end{table}

To address these research questions, I started with the same dataset collected and described in Chapter~\ref{ch:activities}.
The full dataset contains 16 public Jira repos containing 1,822 projects and 2.7 million issues.
For this chapter, I only used 13 of the 16 available Jira repos because two of the original repos (MariaDB and Mindville) contain no comments (an important issue field for the comparative analysis), and Mojang only contains bugs.

I cleaned the data by removing issues that are not yet resolved, as well as unexplainable and unusable data (e.g., issue created date is null).
Jira also stores the fields entered at the issue creation time in the evolution data.
When an issue is created, {\mytilde}12 fields are filled in on average (mean), often with the default value of null.
I exclude these ``creational'' evolutions from my analysis to focus on issue evolution, not issue creation.
As a result, I removed 16,518,405 creational data points from the dataset, leaving 13,245,766 evolutions of issues fields.

Table~\ref{table:datasource} shows an overview of the research data for this investigation.
For each repo, it lists the creation year, the number of unique projects, issues, evolutions, and evolution authors.
The research data consists of 13 Jira repos, 2,094 projects, 1.3 million issues, 13 million post-creational evolutions, and 237k authors.
The count of 2,094 projects is higher than the 1,822 projects reported in Chapter~\ref{ch:activities} due to the difference in how the projects are counted: Chapter~\ref{ch:activities} reports unique \textit{final} projects, whereas here I am reporting all unique projects ever assigned, which is found in the evolution history.
Chapter~\ref{ch:activities} does not analyse the evolution of issues, and therefore does not consider these intermediate values.
The data and scripts are available in my replication package~\cite{Montgomery_2025_PhDThesisReplicationPackage}.
The package includes additional information in the form of box plots, medians and quartiles, and the results of various additional investigations.

\subsection{Thematic Analysis}  \label{sec:evolution_thematic_analysis}

My goal was to produce a set of generalised issue fields across the \xglspl{jr}, and group them into information themes for further analysis.
Unifying the issue fields and forming the information themes was a non-trivial process due to the 2,562 unique issue fields in the dataset.
Despite the large number of unique issue field names, there is still a standard set of fields that most \xgls{jr} utilise, and many of these unique issue field names are synonyms.
Therefore, I sought to unify these fields into a smaller set of summarised fields.
To address these issues, I conducted a \xgls{ta}~\cite{Braun_2006_QRP,Cruzes_2011_ESEM} of issue fields across the 13 Jira repos.
I defined two analysis criteria for the investigation: 1) each field must exist in all the Jira repos, and 2) each field must evolve.
The first criterion is to obtain a \textit{generalised} set of issue fields: they must exist in all studied repositories.
The second criterion is to obtain \textit{evolving} issue fields: they must all evolve (not all fields evolve with the dataset).

\xgls{ta} is characterised by initial decisions: the type of analysis, the style of approach, and the level of analysis~\cite{Braun_2006_QRP}.
For the type of analysis, I chose a Rich Overview.
The goal was to get an overall understanding of the issue information contained within \xglspl{its}, which requires all facets of the data to be considered.
For the style of approach, I selected an Inductive approach.
I was interested in unifying and grouping the issue information given the emergent ideas I would find by looking closely at each of the issue fields.
For the level of analysis, I selected Semantic Themes.
I was interested in understanding the surface information presented when viewing examples of these fields in the data, without interpreting beyond what was described.
In summary, I conducted an Inductive \xgls{ta}, providing a Rich Overview that produced Semantic themes.
\xgls{ta} involves five main phases, presented here.


\textbf{In phase 1}, I familiarised myself with the data by looking through issue field samples.
For each issue field, I looked in three locations: 1) the final state of the fields in issues, 2) the history of the fields in issues, and 3) the documentation provided by each \xgls{jr} that explains the field.
I used the first two locations to construct a full history of the field throughout the lifetime of the issue.\footnote{The history alone is not sufficient, since fields that are set but never changed are not listed in the history.}
The third location helped disambiguate field names and usage across \xglspl{jr}.
I utilised this familiarisation process to already begin systematically searching for issue fields that satisfied the two analysis criteria.
I found that there were a fair number of issue fields that shared the same names, and even more issue fields that were close in name, but not identical.
Inspecting examples of these issue fields, it became clear that some organisations had simply chosen to rename the fields, even though they were using the issue fields for the original purpose.

\textbf{In phase 2}, I generated an initial set of codes.
Within the 13 \xglspl{jr}, there are 2,562 unique field names across the three studied locations (union), and only 22 unique field names that exist in all three of these locations (intersection).
When adding the restriction that each field name must exist in all 13 \xglspl{jr}, there are then only 10 unique field names across the three locations.\footnote{Exact details of these fields available in the replication package~\cite{Montgomery_2025_PhDThesisReplicationPackage}.}
My goal in applying the analysis criteria, however, was to go beyond exact field names and look for different or modified names given the non-standard nature of \xglspl{jr}.
Here is an example of why this is necessary: issue ``links'' are an important field within Jira, but they are called ``issuelinks'' in the issues and ``Link'' in the histories.
Without this analysis, issue links would have been missed.
I used a script to randomly select issues to investigate, and recorded this process in an output file.
I iteratively inspected (and documented) 20,374 issue field samples across the three distinct issue field locations.
20 issue fields satisfied the analysis criteria, and formed the final set of codes for the \xgls{ta}.

\textbf{In phase 3}, I re-focused the analysis on the broader level of themes across the 20 codes from phase 2.
Using the 20,374 field samples, I investigated the way in which these fields were being used within each \xgls{jr}.
I built the initial set of themes based on how the issue fields were being used to support issues within the organisational purpose of Jira.
For example, some issue fields such as IssueType and Parent were clearly being used to provide structure to the issues within Jira; fields such as Summary and Description were the core content of the issues; and fields such as Status and Priority were being used to support the workflow of issues.
The result of this phase was an initial set of themes with all codes mapped underneath.

\textbf{In phase 4}, I challenged the themes by reviewing them with four other researchers in the field, iterating over them, and checking additional data points.
I had some movement within the themes, e.g., Comments was as a controversial field because it contains content-like information, but ultimately can be used to communicate anything and therefore fit much better under Community.
The result of this phase was a final set of themes.

Finally, \textbf{in phase 5}, I defined each theme and verified that each issue field matched that definition, forming a final set of five themes that contain 20 codes with sufficient mutual exclusion.
I discuss each theme and code in detail in the following subsections.
Much of our understanding of the fields comes from the \xgls{ta}, while some information was gained through official Jira documentation.\footnote{Jira docs on issue fields: \url{https://confluence.atlassian.com/adminjiraserver/issue-fields-and-statuses-938847116.html} and at \url{https://confluence.atlassian.com/jirasoftwareserver/advanced-searching-fields-reference-939938743.html}}

\subsection{Evolution Analysis}  \label{sec:evolution_evolution_analysis}

To quantify the extent to which evolution is occurring within \xglspl{its}, I compare how the issue information and the \xgls{se} artefacts and activities collected in Chapter~\ref{ch:activities} evolve across a generalised set of \xglspl{its}.
This process is exploratory in nature, and thus involves observing, understanding, and interpreting real-world data.
While interviews and content analysis provide much value in contextualising detailed and rich information, they do not allow for an overview over large amounts of data, particularly when looking to understand cross-organisation behaviour.
For this, I chose the ``Sample Study'' strategy as described by Stol and Fitzgerald~\cite{Stol_2018_TOSEM}.
This strategy involves either the collection of data through mass questionnaires, or finding large datasets through data mining.
I selected data mining and performed a historical data analysis of the large dataset.
This enabled me to investigate issue evolution across many contexts, and provide generalised evidence towards the exploratory objective.
With these techniques, I conducted an empirical study of 13 Jira \xglspl{its}, across 1.3 million issues with 13 million evolutions.
I investigated the frequency of the evolutions, the timing (relative to the \ltexdummy{issue creation}), and the ownership (who changed the field).
For the description field, I also performed a manual analysis of a random sample to understand what was changing syntactically and semantically.

Before performing the analyses, I first created an evolution dataset from the raw Jira data.
By default, Jira data is stored in a JSON format as fully detailed with an \xgls{erd} in Figure~\ref{fig:jira_database_erd}.
While the JSON format is easy to interface with, and has advantages for certain types of selective queries, it is not an ideal format for queries that need to look across the entire dataset.
For this, I created a single Pandas DataFrame\footnote{\url{https://pandas.pydata.org/docs/reference/api/pandas.DataFrame.html}} that stores the data in an SQL-like format that is quick to perform full dataset queries.
The process of weaving together the evolution history was not a trivial process, given the rather unorthodox way in which Jira stores the history of each issue.
Additionally, different fields are stored in different ways within the history.
The resulting DataFrame and scripts to produce the DataFrame are available in my thesis replication package~\cite{Montgomery_2025_PhDThesisReplicationPackage}.
The evolution dataset has 30 million evolutions from the 1.3 million issues.
Of the 30 million evolutions, 17 million evolutions (56\%) are the ``creational'' evolutions: records of fields that were set when the issue was created.
This leaves 13 million evolutions (44\%) that are post-creational evolutions: records of fields that were changed after the issue was created.
For the entire analysis of RQ2 and RQ3, ``evolution'' refers to ``post-creational evolution''.

I use box plots to present the results for RQ2A and RQ2B.
Box plots summarise a single dimension of data (list of values), visualising the median, the upper- and lower-quartiles (edges of the box), and the upper- and lower-whiskers (the lines drawn at each end of each line).
The median (middle value in a sorted dataset) is resistant to outliers, which is important for visualising the results.\footnote{\xglspl{its} have inexplicably strange outliers, a great subject for future research.}
Each box represents the \xgls{iqr}, where the top and the bottom of the box represents the median of the upper and lower halves of the data (split on the original median).
In other words, the \xgls{iqr} (box) represents the middle 50\% of the data: the most expected values.
Finally, the upper- and lower-whiskers represent the highest and lowest values within the dataset, to a maximum of 1.5x the \xgls{iqr}.
This is the popular choice\footnote{\url{https://en.wikipedia.org/wiki/Box_plot}} for visualising the whiskers, and is the default option of the Python visualisation library matplotlib.\footnote{\url{https://matplotlib.org/stable/api/_as_gen/matplotlib.pyplot.boxplot.html}}
I do not visualise outliers beyond the whiskers.\footnote{To see all data points, see the replication package~\cite{Montgomery_2025_PhDThesisReplicationPackage}, and turn on outliers.}

I split the data into Requirements, Maintenance, and Development issues.
Each box plot figure visualises sets of three box plots, where the orange box (the first box) represents the Requirements issues, followed by the Maintenance issues, and then the Development issues.
Each box plot axis is labelled with their respective data scale, which varies from frequencies to distributions, and normal scale to \ltexdummy{(sym-)log} scale.\footnote{The ``Symlog'' scale (symmetrical log) allows values between 0 and 1, and is used in the analysis so that values close to 0 can be interpreted and visualised.}
All scripts used to format the data and produce the figures are available in the replication package~\cite{Montgomery_2025_PhDThesisReplicationPackage}.

\subsection{Issue Content Analysis}

To address RQ3, I conducted an open-coding content analysis to investigate requirements evolution patterns in the content of issues.
I performed an exploratory open-coding of 156 issues (78 each) with one or more evolutions to the description (for a total of 315 evolutions) across all \xglspl{jr}.
I stratified the random selection across the 13 \xglspl{jr} and the three issue types (Requirements, Maintenance, and Development), selecting four issues from each, thereby arriving at $156 = 13 \times 3 \times 4$.
The goal of this process was to gain a preliminary understanding of content changes occurring across all \xglspl{jr} and issue types and to identify change recurring reasons, which I call content evolution patterns.
Given the goal of a ``preliminary understanding'', I did not aim for representativeness or concept saturation, but rather diversity.
That being said, I did find by the end of the open-coding process that the patterns were saturated, although I did not design or track this systematically.
The aim was to explore the existence of patterns rather than quantifying them.
Once the task was complete, I met to with {\nameWM} to discuss, combine, and agree on the patterns found.
Overall, we found a similar set of patterns but with different frequency and with slightly different names.
Discussing the findings based on examples led to a shared and uniform pattern list.
The full pattern extraction data, including labelled samples, is included in the replication package~\cite{Montgomery_2025_PhDThesisReplicationPackage}.

\section{Results: Information Types}  \label{sec:evolution_info_types}

The results of the \xgls{ta} applied to the issue fields reveals a rich ecosystem of information that covers many aspects of the \xgls{se} lifecycle.
Here, I present the themes and codes that emerged from this analysis.
Table~\ref{table:issue_information} shows the themes, codes, and issue field names that emerged from this analysis.
There are five issue information themes: Issue Content, Issue MetaContent, Issue RepoStructure, Issue Workflow, and Issue Community.

\begin{table}[t]
    \small      
    \centering  

    \caption{Information themes and codes, and issue field names.}
    \label{table:issue_information}

    \begin{tabular}{p{0.25\textwidth}p{0.65\textwidth}}
        \toprule
            \textbf{Themes} and Codes & Associated Field Names \\
        \midrule
        \textbf{Issue Content} & \\
            \qquad Summary          & summary \\
            \qquad Description      & description \\
        \textbf{Issue MetaContent}  & \\
            \qquad Labels           & labels, Labels \\
            \qquad Environment      & environment, Environment, Environment: other information \\
            \qquad VersionsAffected & versions, versions.name, Version \\
            \qquad VersionsFixed    & fixVersions, fixVersions.name, Fix Version \\
        \textbf{Issue RepoStructure} & \\
            \qquad IssueType        & issuetype, issuetype.name, Issue Type, type, issueType \\
            \qquad Project          & project, project.name, Project, Project Name \\
            \qquad Components       & components, components-name, Component \\
            \qquad Parent           & parent, Parent, Parent Issue, IssueParentAssociation, Parent Link \\
            \qquad IssueLinks       & issuelinks, Link \\
        \textbf{Issue Workflow}     & \\
            \qquad CreatedDate      & created  \\
            \qquad ResolvedDate     & resolutiondate \\
            \qquad Status           & status, status.name, Status, Current Status \\
            \qquad Priority         & priority, priority-name, Priority \\
            \qquad Resolution       & resolution, resolution.name, Resolution \\
        \textbf{Issue Community}    & \\
            \qquad Creator          & creator, creator displayName \\
            \qquad Reporter         & reporter, reporter displayName \\
            \qquad Assignee         & assignee, assignee displayName \\
            \qquad Comments         & comment, comments, Comment \\
        \bottomrule
    \end{tabular}
\end{table}

\subsection{Issue Content}

The Issue Content theme covers issue fields that form the core information of an issue.
The two codes in this theme are the \textit{Summary} and the \textit{Description}.

The \textbf{Summary} field acts as a title for the issue, providing a brief overview of the information contained in the issue.
For example: ``Allow toggling every single projector's ability to cast a shadow separately'' is a summary of STORM-2147 from SecondLife\footnote{\url{https://jira.secondlife.com/browse/STORM-2147}}.
Arguably, the Summary is the most important field for representing the issue.
Important tasks in \xglspl{its} include searching for issues, prioritising issues, assigning issues to people, and managing the backlog.
These tasks will likely be done by only reading the Summary field.
As such, a well-written Summary should touch on all important high-level aspects of the issue.
There was only a single issue field name for Summary found during the analysis, which is ``summary''.
This perfect alignment across the 13 \xglspl{jr} and in all three locations (issue final state, issue history, and issue documentation) is likely due to the importance of the field and the forced structure by Jira itself.

The \textbf{Description} field contains the full-text details for the issue.
Arguably, the Description is the second most important field in the issue (after the Summary).
When the Summary is insufficient, or the reader wants a complete understanding of the issue, they will go to the Description.
Issue Descriptions come in many forms, from simple small paragraphs to multisection deep dives.
Additionally, the form of information in Descriptions is not always natural language prose, but can also include headers, code blocks, log outputs, steps-to-reproduce, and more.
Accordingly, the processes of understanding, unpacking, interpreting, and presenting the results are all subject to increased complexity.
In principle, all necessary information is listed in the Description, but in reality, some information is listed elsewhere in the issue, and some information is not listed at all.
For example, it is common for important Description-like information to only be included in the Comments field.
Same as Summary above, the Description only had a single field name across all \xgls{jr} and locations: ``description''.

\subsection{Issue MetaContent}

The Issue MetaContent theme covers fields that enhance the Content fields.
Importantly, these are distinguished from the Issue Content fields in that MetaContent information is shared across many issues (e.g., shared labels and environments), whereas Content information is unique.
The four codes in this theme are the \textit{Labels}, \textit{Environment}, \textit{VersionsAffected}, and \textit{VersionsFixed}.

The \textbf{Labels} field contains tag-like information shared across issues.
An example of Labels can be seen in the Bug Report MCPE-31887\footnote{\url{https://bugs.mojang.com/browse/MCPE-31887}} which has multiple labels referring to the various Minecraft items that the Bug Report relates to: acacia, birch, button, oak, pick, pick-block, pickblock.
These labels enable two primary functions within the \xgls{its}: meta-description and categorisation.
They add meta-description in the form of at-a-glance descriptors that further add context.
By looking at the labels, a reader can gain a much quicker understanding of the issue.
Labels also add a deeper form of categorisation within an \xgls{its}, since these labels can be used to query the issues.
For example, a developer working on the above Bug Report may want to view open Bug Reports with the ``pick-block'' label.
This would allow them to identify related issues that may add valuable context to this Bug Report, or perhaps there is an opportunity to fix multiple Bug Reports at once if they share this underlying context.

The final three thematic codes for the Issue MetaContent theme are all label-like, except they have specific types of information within them.
The \textbf{Environment} field focuses on hardware- or software-related information.
The \textbf{VersionsAffected} field contains the software version(s) affected by the content described in the issue.
The \textbf{VersionsFixed} field contains the software version(s) fixed by the content described in the issue.

\subsection{Issue RepoStructure}

The Issue RepoStructure theme covers fields that provide structure to \xglspl{jr}.
The four codes in this theme are the \textit{IssueType}, \textit{Project}, \textit{Parent}, and \textit{IssueLinks}.

The \textbf{IssueType} field describes the role this issue plays within the \xgls{its}.
For example, the issue MCPE-31887\footnote{\url{https://bugs.mojang.com/browse/MCPE-31887}} has the IssueType ``BugReport'', while the issue JRASERVER-7784\footnote{\url{https://jira.atlassian.com/browse/JRASERVER-77884}} has the IssueType ``Suggestion''.
These two types greatly distinguish how these issues are handled within these organisations.
Bug Reports are issues that likely need to be addressed or else the software will not continue to function as it should.
Suggestions are a form of button-up requirements that \textit{could} be implemented, if the developers believe it enhances the software in a way that benefits the stakeholders enough.
Other examples of IssueTypes includes Epic, Story, Feature Request, Work Item, and Quality Assurance.
See Chapter~\ref{ch:activities} for a comprehensive breakdown of IssueTypes, and a thematic mapping into artefacts and activities.

The \textbf{Project} field is the primary top-down organisational construct used to provide structure to \xglspl{jr}.
Every \xgls{jr} is composed of a set of projects, and every issue is organised into exactly one project.
The idea of a Project is that it is mutually exclusive within the \xgls{jr}.
The concept of a Project, however, is different across \xglspl{jr}.
Some \xglspl{jr} have true project-like values that represent different internal products or large initiatives inside companies.
Other \xglspl{jr} use the Project field to distinguish between teams such as \xgls{qa} or Testing.
Regardless of the type and granularity of information stored in the Project field, it is always the case that an issue can only have exactly one value assigned to the Project field.
The \textbf{Components} field also acts as a sort of top-down organisational construct within \xgls{jr}; however, Components are not mutually exclusive like Projects, and a single issue can have multiple components assigned to it.
Components are also used in diverse ways by organisations.
The \textbf{Parent} field contains a link to the parent issue that this issue is organised under.
The use of the Parent field naturally forms tree-like structures within the data, since every issue can only have one parent, but a parent issue can have multiple child issues.
The concept of a parent is generic, and therefore can be used for many reasons.

The \textbf{IssueLinks} field is a list of links to other issues that are related in some way.
See Montgomery et al.~\cite{Montgomery_2022_MSR} for a comprehensive breakdown of issue link types.
Many \xglspl{its} offer the ability to link issues together via specialised ``link'' types.
These links relate issues to each other, allowing stakeholders to understand and capture the relationships between different issues, making it easier to find information, and structuring overall project knowledge~\cite{Sandusky_2004_MSR}.
For example, a reported issue might be part of a major bug, a bug might contribute to a specific requirement, or two issues might refer to the same topic.
Issue links can also be described as \textit{horizontal traceability} as they relate artefacts (issue reports) on the same level~\cite{Heck_2014_JSEP,Gotel_2012_BOOK}.
Issue links (and their types if used in the respective tool) are listed either on the corresponding issue page (e.g. JIRA, Bugzilla) or in the comments sections (e.g. GitHub Issues).
There are different link types, such as Duplicate, Relate, Block, and Subtask, which serve different purposes and can be used to indicate various relationships between issues.

\subsection{Issue Workflow}
The Workflow theme covers fields that support issues in their lifecycle from creation to resolution.
The five codes are the \textit{Status}, \textit{Priority}, \textit{Resolution}, \textit{CreatedDate}, and \textit{ResolvedDate}.

The \textbf{Status} field marks the current stage of the lifecycle an issue is in.
Organisations, projects, and teams define workflows for different types of issues, where each workflow consists of stages an issue must go through from ``Opened'' to ``Closed''.
The Status field stores the current stage of the issue.
The collective set of workflow stages forms a set of nodes in a potential workflow diagram.
To get the actual workflow diagram for a given organisation, the history of issues needs to be investigated to understand the order the Statuses are normally used.
The \textbf{Priority} field records the importance of an issue in relation to other issues.
The \textbf{Resolution} field marks the reason for closing an issue.
The \textbf{CreatedDate} and \textbf{ResolvedDate} mark the beginning and end of an issue lifecycle.
However, the CreatedDate cannot change, and any changes to the ResolvedDate are not tracked in the evolutions. I kept these two fields in the analysis because they enable time-related analysis (particularly RQ2B).
Individual use cases need to investigate the value of the minimal natural language content stored in these fields.
The only field name found during the analysis is: ``created''.
The \textbf{ResolvedDate} field records the date the issue was resolved.

\subsection{Issue Community}
The Community theme covers fields that describe the community of stakeholders and their involvement in the issue.
The four codes are the \textit{Creator}, \textit{Reporter}, \textit{Assignee}, and \textit{Comments}.

Issue Community covers fields that describe the community of stakeholders and their involvement in the evolution of an issue.
The \textbf{Creator} field lists the person who opened the issue in Jira.
The \textbf{Reporter} field contains the person who initially brought up the issue (which is before issue creation).
The \textbf{Assignee} field lists the person responsible for the issue.
The \textbf{Comments} field lists all the comments made on this issue.
For example, ``That could be fun to implement, although I won't have time for it for a while'' from a comment on QTBUG-76315,\footnote{\url{https://bugreports.qt.io/browse/QTBUG-76315}} where a developer conveys interest but also time constraints which led to this issue not being done, despite being 4 years old.
Another example would be ``I think you are right and [because] three months have passed, I can fix it for you. Thank you for the issue and the suggested code changes.'' on ZOOKEEPER-4703\footnote{\url{https://issues.apache.org/jira/browse/ZOOKEEPER-4703}}
In most \xglspl{its}, the comments are displayed in ascending chronological order after the Description.
Each comment tends to be just a few sentences long, but it can also be multiple paragraphs.
The content of Comments includes questions, personal updates, requests for information, and additional information.
The comments may contain relevant information not included in the Description.
For example, a comment might contain the log output of the error reported in the description.
This can happen as \xgls{its} maintainers comment and request additional information, and the original issue reporter replies with the requested information instead of updating the original Description.
The worst form of this is when conflicting information is listed in the comments, or when information specifically designed for other fields is listed in the comments (such as the Status field).
The consequence of this spreading of information is that all future readers must consume all comments to understand the issue.
A complete interpretation requires analyses or scripts that align the information listed in all three of the Summary, Description, and Comments fields.

\section{Results: Information Evolution}

\subsection{Frequency of Issue Evolutions}

\begin{figure}[th]
    \centering
    \includegraphics[width=.5\textwidth]{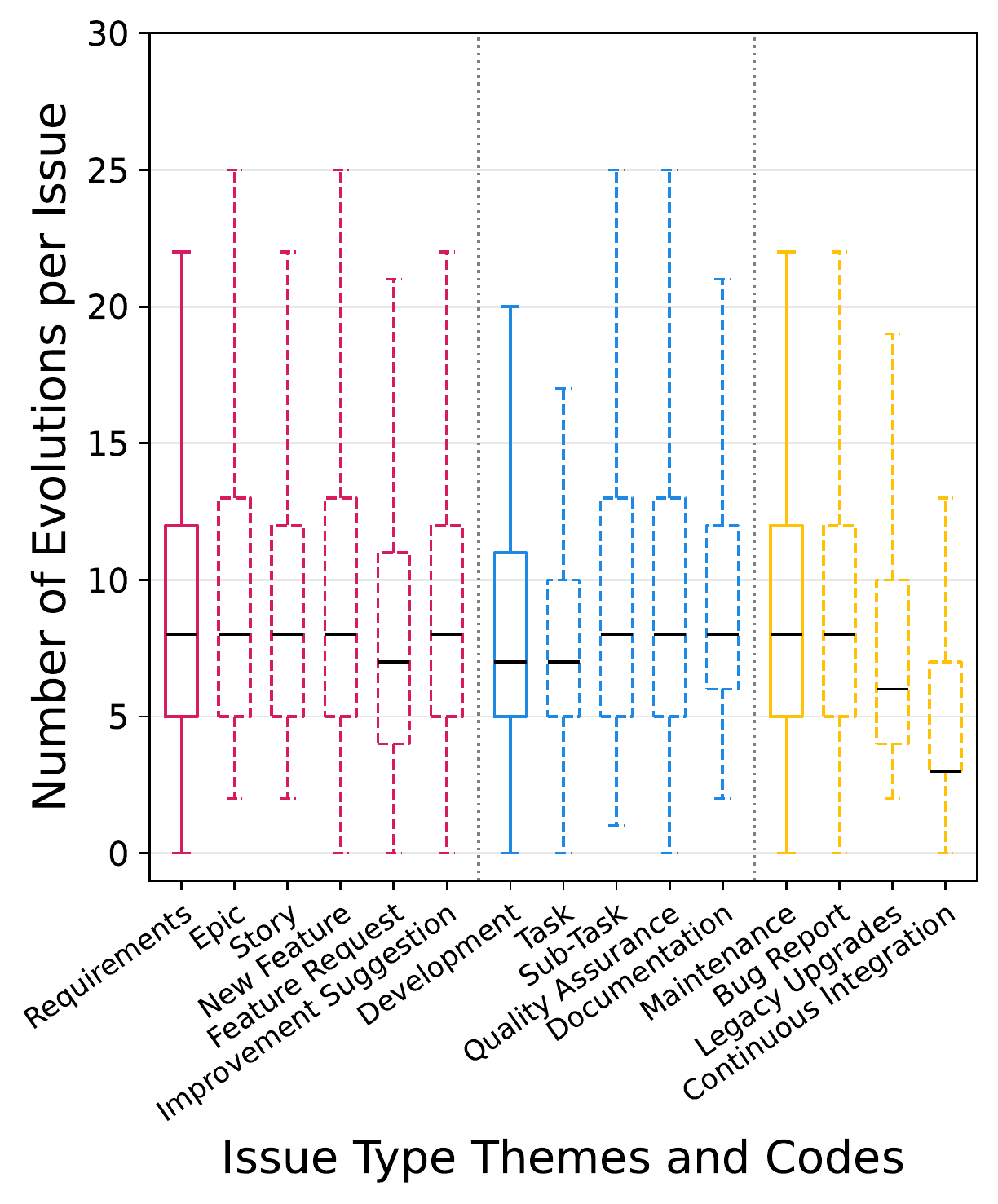}
    \caption{Number of evolutions per issue type.}
    \label{fig:iss_evo_freq}
\end{figure}

\begin{figure}[th]
    \centering
    \includegraphics[width=\textwidth]{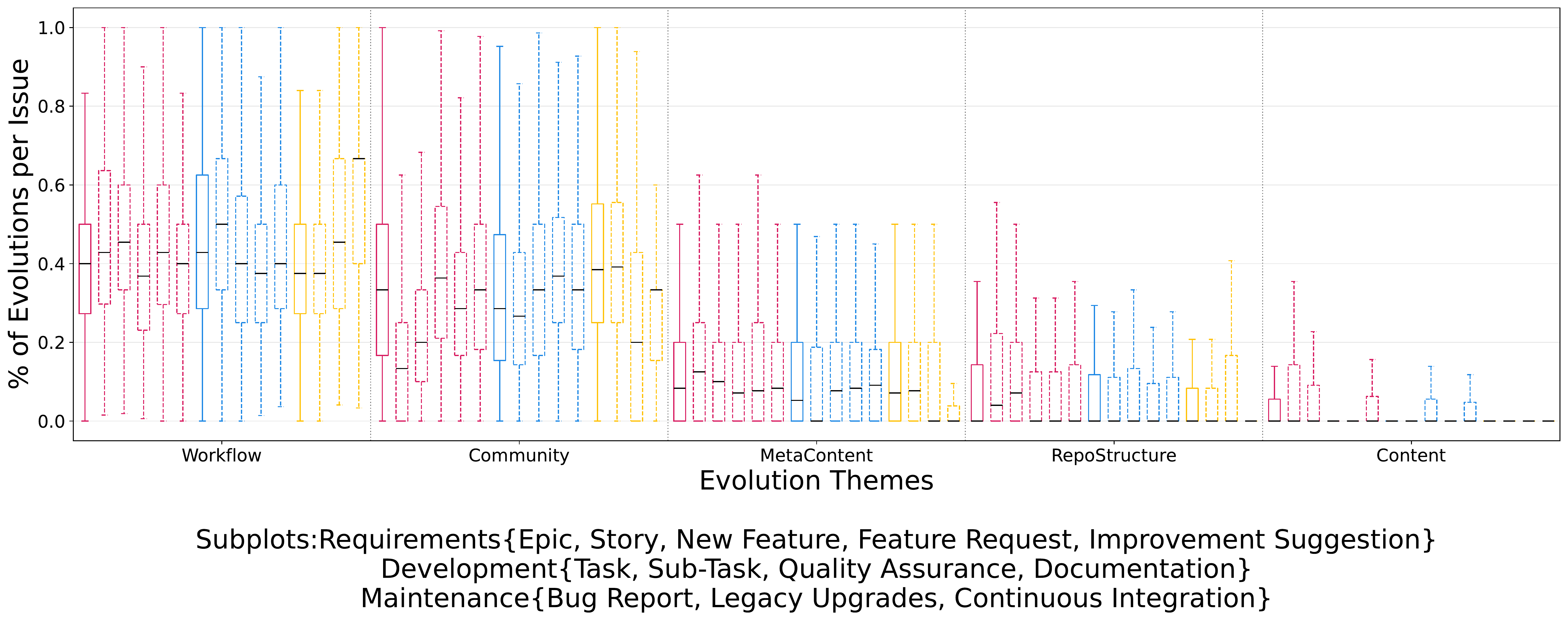}
    \caption{Distribution of evolutions across issue information themes, per issue type.}
    \label{fig:iss_evo_dist_evo_themes}
\end{figure}

To answer RQ2A, I counted the number of evolutions to each issue, and broke that down into issue types and information types.
I present the results of this analysis in Figures~\ref{fig:iss_evo_freq} and \ref{fig:iss_evo_dist_evo_themes}.

Requirements evolve a median of {\mytilde}8 times in their lifetime, with an \xgls{iqr} of 5--12 evolutions.
This is the same for Maintenance issues, while Development issues evolve slightly less.
Epics and New Features have about 5 more evolutions than the rest of the requirements types (up to 25 in total), which is the same for Task and Sub-Task for Development.
I visualise the number of evolutions per issue type in Figure~\ref{fig:iss_evo_freq}.
Note that all Epics and Stories evolve at least two times while all other requirements types have some issues that did not evolve at all (lower cap goes to 0).
All issues in the dataset are resolved, so an issue with zero post-creational evolutions was created in an already-resolved state.
This suggests that Epics and Stories go through a more structured process, requiring at least some interaction before resolution.

Looking at the distribution of evolutions across the information themes (Fig.~\ref{fig:iss_evo_dist_evo_themes}), we see that \textit{Workflow} evolutions occur the most, at {\mytilde}40\% (median) of all evolutions on Requirements issues, while Content evolutions occur the least, with a median of 0\% and an upper-quartile of 6\%.
Both Development and Maintenance share similar distributions, although there are some differences---particularly in Community evolutions.
For \textit{Community} evolutions, Epics have the least percentage of evolutions at 13\%, similarly low as Stories at 20\%.
This highlights that Epics and Stories have fewer discussions or changes of ownership (or both) than other types of issues.
This is in contrast to \textit{Content} evolutions, where Epics have the highest evolution percentage, followed second by Stories.
In other words, after creation, issue Content barely changes for all issue types, except sometimes for Epics and Stories.

\subsection{Issue Evolution Time}

\begin{figure}[th]
    \centering
    \includegraphics[width=.6\textwidth]{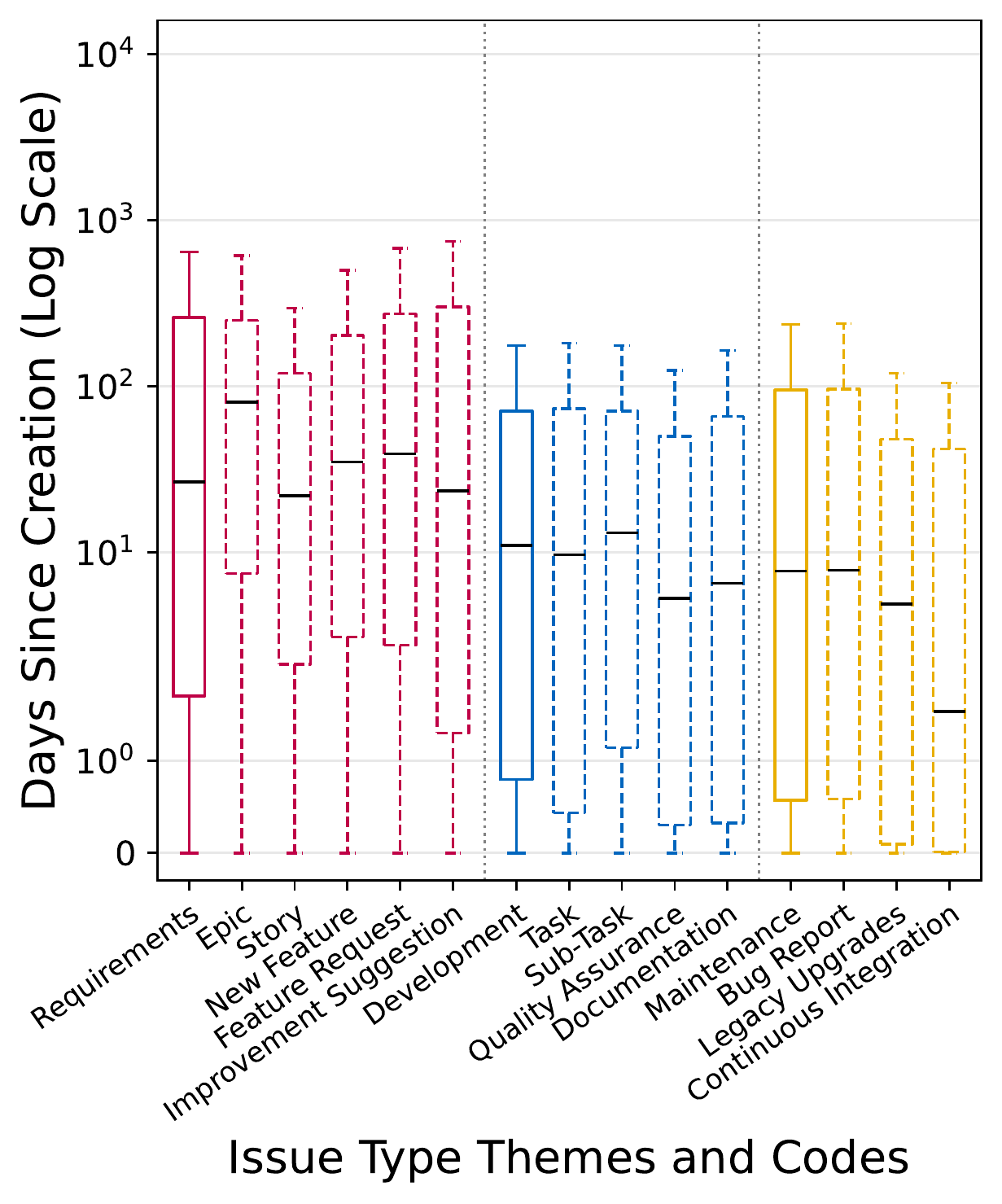}
    \caption{Evolution time after issue creation, per issue type.}
    \label{fig:iss_evo_time_iss_type}
\end{figure}

On average, Requirements continue to evolve 27 days after their creation (median), compared to Development at 11 days and Maintenance at 8 days.
Epics continue to evolve the longest of all Requirements types at 81 days, while Stories evolve the shortest at just 22 days.
Although Stories evolve the shortest of all Requirements types, this is still double that of Development and almost triple that of Maintenance.
I visualise these results in Figure~\ref{fig:iss_evo_time_iss_type} by plotting the time-of-change offset from issue creation (in days), for each evolution in each of their respective subgroups.\footnote{Note the y-axis is in log scale, which makes these values appear visually closer than they are. Technically, in Symmetrical Log, to plot values between 0 and 1: \url{https://en.wikipedia.org/wiki/Logarithmic_scale\#Extensions}.}
The upper quartiles show an even larger difference with Requirements at 260 days, compared to Development at 72 days and Maintenance at 94 days.
In summary, Requirements issues evolve 2--3x longer than the other issues.
Even though all issue types tend to evolve the same number of times ({\mytilde}8 times), Requirements evolutions need 2--3 times a longer period after creation, with Epics being even more exaggerated at 7--10x longer than non-Requirements.

\begin{figure}[th]
    \centering
    \includegraphics[width=\textwidth]{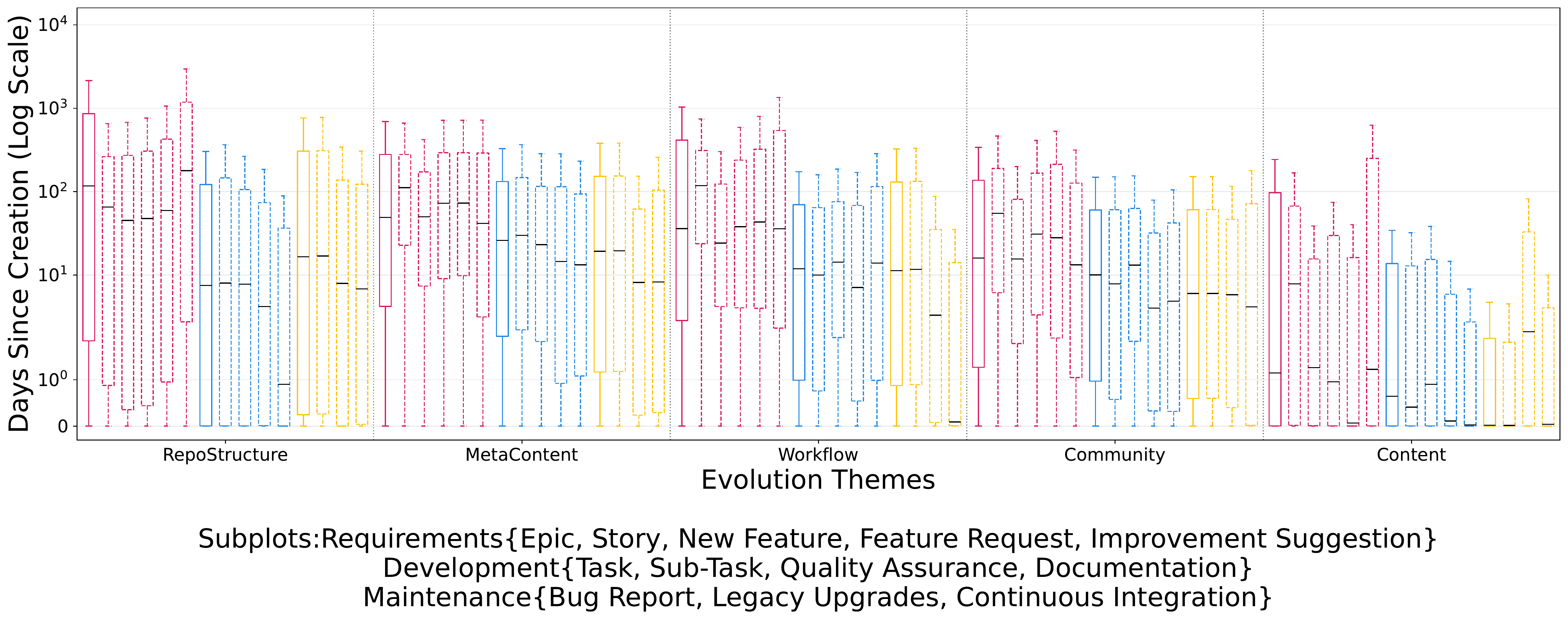}
    \caption{Evolution time after issue creation, per evolution theme and issue type.}
    \label{fig:iss_evo_time_evo_themes}
\end{figure}

When breaking down Requirements into information themes (Fig.~\ref{fig:iss_evo_time_evo_themes}), I see that RepoStructure continues to evolve 118 days later, compared to MetaContent at 49 days, Workflow at 36 days, Community at 16 days, and Content at just 1 day.
In other words, while structural information such as project, component, and links continued to be organised within the \xglspl{its} for months (sometimes years), 50\% of the Content evolutions occurred immediately after creation.
Looking at RepoStructure, Requirements continue to evolve much longer (118 days) than Development (8 days) and Maintenance (17 days).
This shows a higher uncertainty to Requirements organisation than other issue types.
One potential explanation is that some issues originally filed as bugs might emerge as new requirements while working on them.
Another explanation is that requirements issues might simply be more complex.

Looking into the five Requirements types in dotted red in Figure~\ref{fig:iss_evo_time_evo_themes}, Epic Content evolutions continue to evolve much longer at a median of 8 days, compared to the Requirements overall (1 day).
This suggests either a more careful editing and update of Epics or new details that emerge after opening the Epic.
Epic Workflow evolutions also occur much later at 118 days compared to Requirements overall at 36 days (and even later compared to Development at 12 days and Maintenance at 11 days).
This shows the need for more time when it comes to working with Epics compared to all other issue types.

\begin{figure}[th]
    \centering
    \includegraphics[width=\textwidth]{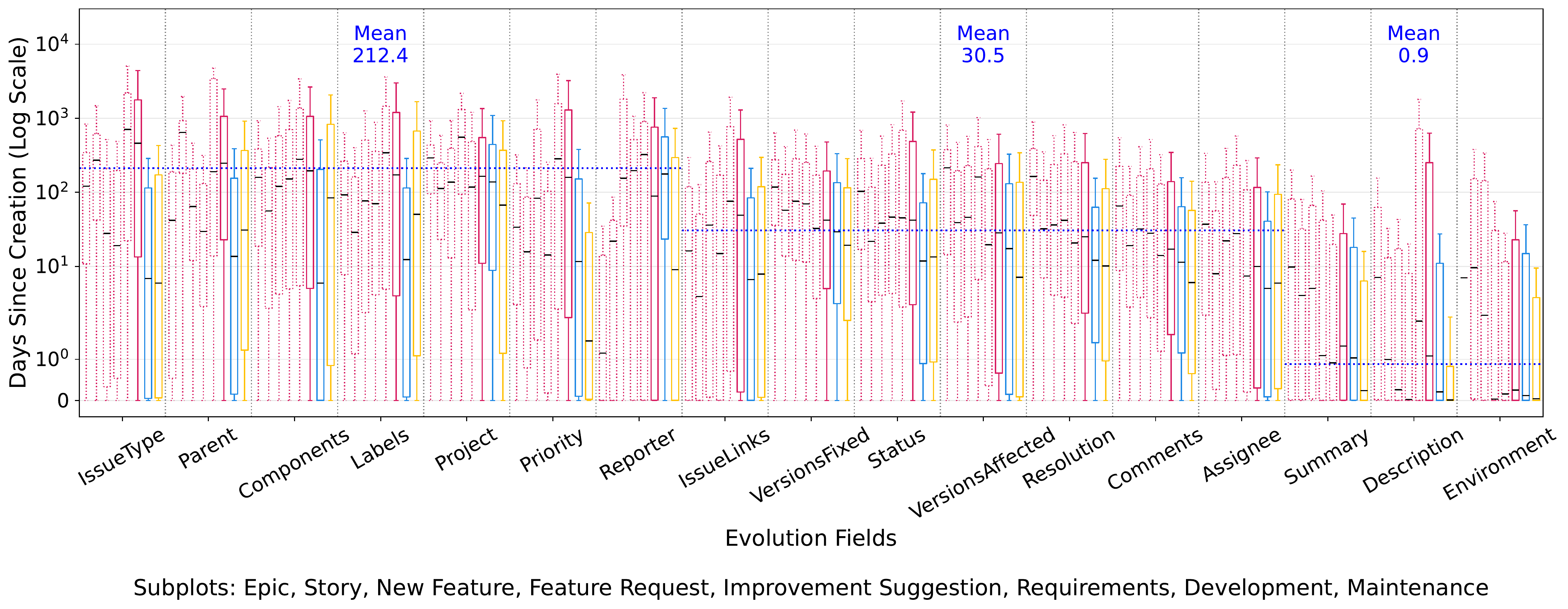}
    \caption{Evolution time after issue creation across issue fields, per issue type.}
    \label{fig:iss_evo_time_iss_fields}
\end{figure}

When analysing the individual information fields, I see large differences between them: some fields evolve for months (medians) and even years (upper-quartiles), while other fields evolve for just a few hours.
I applied the Jump Method\footnote{Sort the data and split on largest gaps N times: \url{https://en.wikipedia.org/wiki/Determining_the_number_of_clusters_in_a_data_set}} to cluster the Requirements medians, which revealed three distinct magnitudes of evolution time: months, weeks, and days.\footnote{At this level of analysis, there are technically five groups, but two non-presented groups contain just one field each, at both ends of the figure. Given the nature of largest-gap analysis applied to log-scale data, I deemed it useful to combine those two outer groups with their representative larger groups. It makes no difference to the analysis of the inner three groups regarding the grouping.}
I present these findings in Figure~\ref{fig:iss_evo_time_iss_fields} where the three distinct magnitudes are visualised as horizontal dotted blue lines representing the mean of the Requirements medians in each magnitude group.
The first magnitude, consisting of seven fields, evolves for an average of 7 months (212 days).
The second magnitude also contains seven fields and evolves for an average of 4 weeks (30 days).
The final magnitude contains just three fields and evolves for an average of 22 hours.
There are many notable differences between Requirements and the other types.
IssueType is the most extreme, with Requirements IssueType evolutions occurring a median of 460 days after creation, while Development and Maintenance are only 7 and 6 days, respectively.
Other notable differences include Parent, Labels, Priority, Reporter, Summary, and Description.

Looking into the five Requirements types in dotted red, most issue fields show a large difference between the Requirements types.
Most notably, Epics and Stories are rather consistently different from the other Requirements types (and the Requirements median).
For example, looking at IssueType and Parent, Epics and Stories are one magnitude higher than most other Requirements types.
For \ltexdummy{Reporter}, Epics and Stories evolve for a median of 1 and 27 days, respectively, compared to the Requirement's median of 89 days.
Regarding Resolution, Epics have a median evolution time of 164 days, compared to the Requirement's median of 25 days.
Finally, for Description, Epics evolve a median of 7 days, compared to the Requirement's average of 1 day.
These are just some examples of the extreme differences visualised in Figure~\ref{fig:iss_evo_time_iss_fields}.

\subsection{Issue Evolution Stakeholder}

\subsubsection{Method and Visualisation}

I compared the Jira-defined ownership roles against who is evolving issues, and split those observations along two primary dimensions: activities (from Chapter~\ref{ch:activities}) and information types.
I present the results of this analysis in Table~\ref{table:ownership}.
There are three ownership roles defined in Jira: Creator, Reporter, and Assignee.
The three ownership roles are not mutually exclusive.
For example, Gina can ``report'' a Feature Request and enter it into the system by ``creating'' the issue, and then she ``assigns'' herself to be responsible for it (Gina is now all three roles at once)~\cite{Anvik_2006_ICSE}.
It is also possible to interact with an issue even if you are not one of those roles, which therefore defines a fourth mutually exclusive role: Non-Owner.
When an issue is evolved, I check and record the evolution author against the current state of the four roles.
Table~\ref{table:ownership} visualises the distribution of evolutions performed by the respective ownership groups, described across activities and information types.
The first two data rows represent the distribution of evolutions done by either an Owner (a Creator, Reporter, or Assignee), or a Non-Owner (none of Creator, Reporter, or Assignee).
Each column of these two rows add up to 100\%, as all evolutions are performed by either an Owner, or a Non-Owner.
The next and final seven rows represent subsets of Owners, where the row names denote which subsets the evolver was in: uppercase means ``within set'' and lowercase means ``not in set''.
The final seven data rows are all subsets of the Owner roles: all combinations of \textbf{C}reator, \textbf{R}eporter, and \textbf{A}ssignee.

\newcommand{\abhi}[1]{\textbf{#1}}
\newcommand{\ph}[1]{\makebox[\widthof{Xiii}][c]{#1}}
\newcommand{\rotHead}[1]{\rotatebox{45}{#1}}

\begin{table}[th]
    \small                      
    \centering                  
    \def\arraystretch{1.2}      
    \setlength\tabcolsep{0.5pt} 

    \caption{Ownership Distribution Heatmap across Evolution Types and Issue Types}
    \label{table:ownership}

    \centerline{                
    \begin{tabular}{r|c||ccc||ccccc||ccccccccccccccc|}
        \toprule
        Scope & \multicolumn{1}{c||}{\textbf{S1}} & \multicolumn{3}{c||}{\textbf{S2}} & \multicolumn{5}{c||}{\textbf{S3}} & \multicolumn{15}{c|}{\textbf{S4}} \\
        \midrule
        {EvoType} & {All} & \multicolumn{3}{c||}{All} & \ph{\rotHead{Con}} & \ph{\rotHead{Met}} & \ph{\rotHead{Rep}} & \ph{\rotHead{Wor}} & \ph{\rotHead{Com}} & \multicolumn{3}{c|}{\rotHead{Con}} & \multicolumn{3}{c|}{\rotHead{Met}} & \multicolumn{3}{c|}{\rotHead{Rep}} & \multicolumn{3}{c|}{\rotHead{Wor}} & \multicolumn{3}{c|}{\rotHead{Com}} \\
        {IssType} & \ph{All} & \ph{R} & \ph{M} & \ph{D} & \multicolumn{5}{c||}{All} & \ph{R} & \ph{M} & \multicolumn{1}{c|}{\ph{D}} & \ph{R} & \ph{M} & \multicolumn{1}{c|}{\ph{D}} & \ph{R} & \ph{M} & \multicolumn{1}{c|}{\ph{D}} & \ph{R} & \ph{M} & \multicolumn{1}{c|}{\ph{D}} & \ph{R} & \ph{M} & \ph{D} \\
        \midrule
        Owner & {\setlength\tabcolsep{10pt}}{\cellcolor[HTML]{8E8E8E}}\color[HTML]{F1F1F1}48 & {\cellcolor[HTML]{FD9446}}\color[HTML]{000000}43 & {\cellcolor[HTML]{909090}}\color[HTML]{F1F1F1}47 & {\cellcolor[HTML]{6F6F6F}}\color[HTML]{F1F1F1}58 & {\cellcolor[HTML]{252525}}\color[HTML]{F1F1F1}79 & {\cellcolor[HTML]{AAAAAA}}\color[HTML]{F1F1F1}39 & {\cellcolor[HTML]{9C9C9C}}\color[HTML]{F1F1F1}43 & {\cellcolor[HTML]{757575}}\color[HTML]{F1F1F1}56 & {\cellcolor[HTML]{A4A4A4}}\color[HTML]{F1F1F1}41 & {\cellcolor[HTML]{D14501}}\color[HTML]{F1F1F1}69 & {\cellcolor[HTML]{1C1C1C}}\color[HTML]{F1F1F1}81 & {\cellcolor[HTML]{0A0A0A}}\color[HTML]{F1F1F1}87 & {\cellcolor[HTML]{FDA159}}\color[HTML]{000000}38 & {\cellcolor[HTML]{B2B2B2}}\color[HTML]{000000}37 & {\cellcolor[HTML]{888888}}\color[HTML]{F1F1F1}49 & {\cellcolor[HTML]{FDA55F}}\color[HTML]{000000}37 & {\cellcolor[HTML]{A3A3A3}}\color[HTML]{F1F1F1}41 & {\cellcolor[HTML]{676767}}\color[HTML]{F1F1F1}60 & {\cellcolor[HTML]{F77B28}}\color[HTML]{F1F1F1}50 & {\cellcolor[HTML]{7B7B7B}}\color[HTML]{F1F1F1}54 & {\cellcolor[HTML]{444444}}\color[HTML]{F1F1F1}71 & {\cellcolor[HTML]{FDA762}}\color[HTML]{000000}36 & {\cellcolor[HTML]{9E9E9E}}\color[HTML]{F1F1F1}43 & {\cellcolor[HTML]{999999}}\color[HTML]{F1F1F1}44\\
        Non-Owner & {\cellcolor[HTML]{7E7E7E}}\color[HTML]{F1F1F1}52 & {\cellcolor[HTML]{EE6511}}\color[HTML]{F1F1F1}57 & {\cellcolor[HTML]{7D7D7D}}\color[HTML]{F1F1F1}53 & {\cellcolor[HTML]{9F9F9F}}\color[HTML]{F1F1F1}42 & {\cellcolor[HTML]{DCDCDC}}\color[HTML]{000000}21 & {\cellcolor[HTML]{666666}}\color[HTML]{F1F1F1}61 & {\cellcolor[HTML]{727272}}\color[HTML]{F1F1F1}57 & {\cellcolor[HTML]{999999}}\color[HTML]{F1F1F1}44 & {\cellcolor[HTML]{6B6B6B}}\color[HTML]{F1F1F1}59 & {\cellcolor[HTML]{FDB77A}}\color[HTML]{000000}31 & {\cellcolor[HTML]{E1E1E1}}\color[HTML]{000000}19 & {\cellcolor[HTML]{ECECEC}}\color[HTML]{000000}13 & {\cellcolor[HTML]{E5590A}}\color[HTML]{F1F1F1}62 & {\cellcolor[HTML]{5F5F5F}}\color[HTML]{F1F1F1}63 & {\cellcolor[HTML]{858585}}\color[HTML]{F1F1F1}51 & {\cellcolor[HTML]{E25508}}\color[HTML]{F1F1F1}63 & {\cellcolor[HTML]{6C6C6C}}\color[HTML]{F1F1F1}59 & {\cellcolor[HTML]{A9A9A9}}\color[HTML]{F1F1F1}40 & {\cellcolor[HTML]{F87F2C}}\color[HTML]{F1F1F1}50 & {\cellcolor[HTML]{929292}}\color[HTML]{F1F1F1}46 & {\cellcolor[HTML]{C8C8C8}}\color[HTML]{000000}29 & {\cellcolor[HTML]{E15307}}\color[HTML]{F1F1F1}64 & {\cellcolor[HTML]{6F6F6F}}\color[HTML]{F1F1F1}57 & {\cellcolor[HTML]{757575}}\color[HTML]{F1F1F1}56\\
        \midrule
        $CRa$ & {\cellcolor[HTML]{B3B3B3}}\color[HTML]{000000}37 & {\cellcolor[HTML]{FDA863}}\color[HTML]{000000}36 & {\cellcolor[HTML]{A9A9A9}}\color[HTML]{F1F1F1}40 & {\cellcolor[HTML]{C7C7C7}}\color[HTML]{000000}30 & {\cellcolor[HTML]{5E5E5E}}\color[HTML]{F1F1F1}64 & {\cellcolor[HTML]{BFBFBF}}\color[HTML]{000000}33 & {\cellcolor[HTML]{9F9F9F}}\color[HTML]{F1F1F1}42 & {\cellcolor[HTML]{D9D9D9}}\color[HTML]{000000}23 & {\cellcolor[HTML]{929292}}\color[HTML]{F1F1F1}46 & {\cellcolor[HTML]{EB610F}}\color[HTML]{F1F1F1}59 & {\cellcolor[HTML]{3C3C3C}}\color[HTML]{F1F1F1}73 & {\cellcolor[HTML]{999999}}\color[HTML]{F1F1F1}44 & {\cellcolor[HTML]{FDAD69}}\color[HTML]{000000}34 & {\cellcolor[HTML]{C2C2C2}}\color[HTML]{000000}32 & {\cellcolor[HTML]{BABABA}}\color[HTML]{000000}35 & {\cellcolor[HTML]{FD9344}}\color[HTML]{000000}43 & {\cellcolor[HTML]{A3A3A3}}\color[HTML]{F1F1F1}41 & {\cellcolor[HTML]{9D9D9D}}\color[HTML]{F1F1F1}43 & {\cellcolor[HTML]{FDD1A3}}\color[HTML]{000000}22 & {\cellcolor[HTML]{D4D4D4}}\color[HTML]{000000}25 & {\cellcolor[HTML]{E0E0E0}}\color[HTML]{000000}19 & {\cellcolor[HTML]{FD8F3E}}\color[HTML]{F1F1F1}44 & {\cellcolor[HTML]{868686}}\color[HTML]{F1F1F1}50 & {\cellcolor[HTML]{B4B4B4}}\color[HTML]{000000}36\\
        $CRA$ & {\cellcolor[HTML]{BABABA}}\color[HTML]{000000}34 & {\cellcolor[HTML]{FD9E54}}\color[HTML]{000000}39 & {\cellcolor[HTML]{CECECE}}\color[HTML]{000000}27 & {\cellcolor[HTML]{8C8C8C}}\color[HTML]{F1F1F1}48 & {\cellcolor[HTML]{CACACA}}\color[HTML]{000000}29 & {\cellcolor[HTML]{B8B8B8}}\color[HTML]{000000}36 & {\cellcolor[HTML]{B4B4B4}}\color[HTML]{000000}36 & {\cellcolor[HTML]{9C9C9C}}\color[HTML]{F1F1F1}43 & {\cellcolor[HTML]{D2D2D2}}\color[HTML]{000000}25 & {\cellcolor[HTML]{FDB06E}}\color[HTML]{000000}33 & {\cellcolor[HTML]{DEDEDE}}\color[HTML]{000000}20 & {\cellcolor[HTML]{919191}}\color[HTML]{F1F1F1}47 & {\cellcolor[HTML]{FDA057}}\color[HTML]{000000}38 & {\cellcolor[HTML]{C9C9C9}}\color[HTML]{000000}29 & {\cellcolor[HTML]{8A8A8A}}\color[HTML]{F1F1F1}49 & {\cellcolor[HTML]{FDA159}}\color[HTML]{000000}38 & {\cellcolor[HTML]{C9C9C9}}\color[HTML]{000000}29 & {\cellcolor[HTML]{949494}}\color[HTML]{F1F1F1}46 & {\cellcolor[HTML]{F98230}}\color[HTML]{F1F1F1}48 & {\cellcolor[HTML]{B8B8B8}}\color[HTML]{000000}35 & {\cellcolor[HTML]{757575}}\color[HTML]{F1F1F1}56 & {\cellcolor[HTML]{FDB77A}}\color[HTML]{000000}31 & {\cellcolor[HTML]{E0E0E0}}\color[HTML]{000000}19 & {\cellcolor[HTML]{ADADAD}}\color[HTML]{000000}38\\
        $crA$ & {\cellcolor[HTML]{CCCCCC}}\color[HTML]{000000}28 & {\cellcolor[HTML]{FDCB9B}}\color[HTML]{000000}24 & {\cellcolor[HTML]{C1C1C1}}\color[HTML]{000000}32 & {\cellcolor[HTML]{DEDEDE}}\color[HTML]{000000}20 & {\cellcolor[HTML]{F6F6F6}}\color[HTML]{000000}7 & {\cellcolor[HTML]{C5C5C5}}\color[HTML]{000000}30 & {\cellcolor[HTML]{DFDFDF}}\color[HTML]{000000}20 & {\cellcolor[HTML]{BFBFBF}}\color[HTML]{000000}33 & {\cellcolor[HTML]{CFCFCF}}\color[HTML]{000000}26 & {\cellcolor[HTML]{FEECDA}}\color[HTML]{000000}7 & {\cellcolor[HTML]{F7F7F7}}\color[HTML]{000000}6 & {\cellcolor[HTML]{F5F5F5}}\color[HTML]{000000}8 & {\cellcolor[HTML]{FDC590}}\color[HTML]{000000}26 & {\cellcolor[HTML]{AEAEAE}}\color[HTML]{000000}38 & {\cellcolor[HTML]{E8E8E8}}\color[HTML]{000000}15 & {\cellcolor[HTML]{FEDCB9}}\color[HTML]{000000}17 & {\cellcolor[HTML]{CACACA}}\color[HTML]{000000}28 & {\cellcolor[HTML]{F2F2F2}}\color[HTML]{000000}10 & {\cellcolor[HTML]{FDBF86}}\color[HTML]{000000}28 & {\cellcolor[HTML]{ABABAB}}\color[HTML]{000000}39 & {\cellcolor[HTML]{D4D4D4}}\color[HTML]{000000}24 & {\cellcolor[HTML]{FDCD9C}}\color[HTML]{000000}24 & {\cellcolor[HTML]{C9C9C9}}\color[HTML]{000000}29 & {\cellcolor[HTML]{DADADA}}\color[HTML]{000000}22\\
        $cRa$ & {\cellcolor[HTML]{FFFFFF}}\color[HTML]{000000}1 & {\cellcolor[HTML]{FFF5EA}}\color[HTML]{000000}0 & {\cellcolor[HTML]{FFFFFF}}\color[HTML]{000000}1 & {\cellcolor[HTML]{FFFFFF}}\color[HTML]{000000}1 & {\cellcolor[HTML]{FFFFFF}}\color[HTML]{000000}0 & {\cellcolor[HTML]{FFFFFF}}\color[HTML]{000000}0 & {\cellcolor[HTML]{FFFFFF}}\color[HTML]{000000}0 & {\cellcolor[HTML]{FFFFFF}}\color[HTML]{000000}0 & {\cellcolor[HTML]{FEFEFE}}\color[HTML]{000000}1 & {\cellcolor[HTML]{FFF5EB}}\color[HTML]{000000}0 & {\cellcolor[HTML]{FFFFFF}}\color[HTML]{000000}0 & {\cellcolor[HTML]{FFFFFF}}\color[HTML]{000000}0 & {\cellcolor[HTML]{FFF5EB}}\color[HTML]{000000}0 & {\cellcolor[HTML]{FFFFFF}}\color[HTML]{000000}0 & {\cellcolor[HTML]{FFFFFF}}\color[HTML]{000000}1 & {\cellcolor[HTML]{FFF5EB}}\color[HTML]{000000}0 & {\cellcolor[HTML]{FFFFFF}}\color[HTML]{000000}0 & {\cellcolor[HTML]{FFFFFF}}\color[HTML]{000000}0 & {\cellcolor[HTML]{FFF5EB}}\color[HTML]{000000}0 & {\cellcolor[HTML]{FFFFFF}}\color[HTML]{000000}0 & {\cellcolor[HTML]{FFFFFF}}\color[HTML]{000000}0 & {\cellcolor[HTML]{FFF4E9}}\color[HTML]{000000}1 & {\cellcolor[HTML]{FEFEFE}}\color[HTML]{000000}1 & {\cellcolor[HTML]{FDFDFD}}\color[HTML]{000000}2\\
        $Cra$ & {\cellcolor[HTML]{FFFFFF}}\color[HTML]{000000}0 & {\cellcolor[HTML]{FFF5EA}}\color[HTML]{000000}0 & {\cellcolor[HTML]{FFFFFF}}\color[HTML]{000000}0 & {\cellcolor[HTML]{FFFFFF}}\color[HTML]{000000}0 & {\cellcolor[HTML]{FFFFFF}}\color[HTML]{000000}0 & {\cellcolor[HTML]{FFFFFF}}\color[HTML]{000000}0 & {\cellcolor[HTML]{FEFEFE}}\color[HTML]{000000}1 & {\cellcolor[HTML]{FFFFFF}}\color[HTML]{000000}0 & {\cellcolor[HTML]{FFFFFF}}\color[HTML]{000000}0 & {\cellcolor[HTML]{FFF5EA}}\color[HTML]{000000}0 & {\cellcolor[HTML]{FFFFFF}}\color[HTML]{000000}0 & {\cellcolor[HTML]{FFFFFF}}\color[HTML]{000000}0 & {\cellcolor[HTML]{FFF5EA}}\color[HTML]{000000}0 & {\cellcolor[HTML]{FFFFFF}}\color[HTML]{000000}0 & {\cellcolor[HTML]{FFFFFF}}\color[HTML]{000000}0 & {\cellcolor[HTML]{FFF4E8}}\color[HTML]{000000}1 & {\cellcolor[HTML]{FFFFFF}}\color[HTML]{000000}1 & {\cellcolor[HTML]{FFFFFF}}\color[HTML]{000000}0 & {\cellcolor[HTML]{FFF5EB}}\color[HTML]{000000}0 & {\cellcolor[HTML]{FFFFFF}}\color[HTML]{000000}0 & {\cellcolor[HTML]{FFFFFF}}\color[HTML]{000000}0 & {\cellcolor[HTML]{FFF5EB}}\color[HTML]{000000}0 & {\cellcolor[HTML]{FFFFFF}}\color[HTML]{000000}0 & {\cellcolor[HTML]{FEFEFE}}\color[HTML]{000000}1\\
        $cRA$ & {\cellcolor[HTML]{FFFFFF}}\color[HTML]{000000}0 & {\cellcolor[HTML]{FFF5EB}}\color[HTML]{000000}0 & {\cellcolor[HTML]{FFFFFF}}\color[HTML]{000000}0 & {\cellcolor[HTML]{FFFFFF}}\color[HTML]{000000}0 & {\cellcolor[HTML]{FFFFFF}}\color[HTML]{000000}0 & {\cellcolor[HTML]{FFFFFF}}\color[HTML]{000000}0 & {\cellcolor[HTML]{FFFFFF}}\color[HTML]{000000}0 & {\cellcolor[HTML]{FFFFFF}}\color[HTML]{000000}0 & {\cellcolor[HTML]{FFFFFF}}\color[HTML]{000000}0 & {\cellcolor[HTML]{FFF5EB}}\color[HTML]{000000}0 & {\cellcolor[HTML]{FFFFFF}}\color[HTML]{000000}0 & {\cellcolor[HTML]{FFFFFF}}\color[HTML]{000000}0 & {\cellcolor[HTML]{FFF5EB}}\color[HTML]{000000}0 & {\cellcolor[HTML]{FFFFFF}}\color[HTML]{000000}0 & {\cellcolor[HTML]{FFFFFF}}\color[HTML]{000000}0 & {\cellcolor[HTML]{FFF5EB}}\color[HTML]{000000}0 & {\cellcolor[HTML]{FFFFFF}}\color[HTML]{000000}0 & {\cellcolor[HTML]{FFFFFF}}\color[HTML]{000000}0 & {\cellcolor[HTML]{FFF5EB}}\color[HTML]{000000}0 & {\cellcolor[HTML]{FFFFFF}}\color[HTML]{000000}0 & {\cellcolor[HTML]{FFFFFF}}\color[HTML]{000000}0 & {\cellcolor[HTML]{FFF5EB}}\color[HTML]{000000}0 & {\cellcolor[HTML]{FFFFFF}}\color[HTML]{000000}0 & {\cellcolor[HTML]{FFFFFF}}\color[HTML]{000000}0\\
        $CrA$ & {\cellcolor[HTML]{FFFFFF}}\color[HTML]{000000}0 & {\cellcolor[HTML]{FFF5EB}}\color[HTML]{000000}0 & {\cellcolor[HTML]{FFFFFF}}\color[HTML]{000000}0 & {\cellcolor[HTML]{FFFFFF}}\color[HTML]{000000}0 & {\cellcolor[HTML]{FFFFFF}}\color[HTML]{000000}0 & {\cellcolor[HTML]{FFFFFF}}\color[HTML]{000000}0 & {\cellcolor[HTML]{FFFFFF}}\color[HTML]{000000}0 & {\cellcolor[HTML]{FFFFFF}}\color[HTML]{000000}0 & {\cellcolor[HTML]{FFFFFF}}\color[HTML]{000000}0 & {\cellcolor[HTML]{FFF5EB}}\color[HTML]{000000}0 & {\cellcolor[HTML]{FFFFFF}}\color[HTML]{000000}0 & {\cellcolor[HTML]{FFFFFF}}\color[HTML]{000000}0 & {\cellcolor[HTML]{FFF5EB}}\color[HTML]{000000}0 & {\cellcolor[HTML]{FFFFFF}}\color[HTML]{000000}0 & {\cellcolor[HTML]{FFFFFF}}\color[HTML]{000000}0 & {\cellcolor[HTML]{FFF5EB}}\color[HTML]{000000}0 & {\cellcolor[HTML]{FFFFFF}}\color[HTML]{000000}0 & {\cellcolor[HTML]{FFFFFF}}\color[HTML]{000000}0 & {\cellcolor[HTML]{FFF5EB}}\color[HTML]{000000}0 & {\cellcolor[HTML]{FFFFFF}}\color[HTML]{000000}0 & {\cellcolor[HTML]{FFFFFF}}\color[HTML]{000000}0 & {\cellcolor[HTML]{FFF5EB}}\color[HTML]{000000}0 & {\cellcolor[HTML]{FFFFFF}}\color[HTML]{000000}0 & {\cellcolor[HTML]{FFFFFF}}\color[HTML]{000000}0\\
        \bottomrule\addlinespace[1mm]
        \multicolumn{25}{c}{\scriptsize{\makecell{
            Col Names: \abhi{R}equirements, \abhi{M}aintenance, \abhi{D}evelopment, \abhi{Con}tent, \abhi{Met}aContent, \abhi{Rep}oStructure, \abhi{Wor}kflow, \abhi{Com}munity.\\
            Row Names: \abhi{Evo}lution\abhi{Type}, \abhi{Iss}ue\abhi{Type}, \abhi{C}reator, \abhi{R}eporter, and \abhi{A}ssignee. Uppercase: within set, lowercase: not in set.\\
            Cell Colours: All cells use the same heatmap intensity scale, and \abhi{R}equirements data is orange.
        }}} \\
    \end{tabular}}
\end{table}

The Owner role subsets:
\begin{itemize}[nosep]
    \item \textbf{Data-Filled Sets}
    \begin{itemize}[nosep]
        \item $CRA$: The evolver is all three: Creator, Reporter, and Assignee.
        \item $CRa$: The evolver is Creator and Reporter, but not Assignee.
        \item $crA$: The evolver is only the Assignee.
    \end{itemize}
    \item \textbf{Mostly Empty Sets}
    \begin{itemize}[nosep]
        \item $Cra$: The evolver is only the Creator.
        \item $CrA$: The evolver is Creater and Assignee, but not Reporter.
        \item $cRa$: The evolver is only the Reporter.
        \item $cRA$: The evolver is Reporter and Assignee, but not Creator.
    \end{itemize}
\end{itemize}

Each column of these seven rows also adds up to 100\%, but 100\% of the Owner evolutions (not all evolutions).
The final four rows have almost no data in them.
Looking at these subsets, they represent situations where the Creator is never the same person as the Reporter.
Given the low percentage of data (less than 1\%), this is quite a rare occurrence.
Every cell is additionally shaded with a different intensity of orange or grey, where the intensity is equal to the value of the number shown.
Requirements have been visualised in orange.
Table~\ref{table:ownership} visualises the data in four column scopes: (\textbf{S1}) all data without dimensional splits, (\textbf{S2}) split along the activities, (\textbf{S3}) split along evolution type, and (\textbf{S4}) split across evolution and activity.

\subsubsection{Results}

The evolution distribution between Owners and Others is almost equal at 48\% and 52\%, respectively.
The bottom three rows show a similar result, with all three sub-roles being fairly similar, around 1/3rd of the evolutions each.
This is why I dive deeper, going into detail across the information themes and issue types: averaged across enough data, equal distributions tend to arise.
All results are presented in Table~\ref{table:ownership}.

Requirements issues are more collaborative among Non-Owners.
Specifically, Requirements Content evolutions are authored by Non-Owners 31\% of the time, compared to Development at 13\% and Maintenance at 19\%.
This shows a consistent type of community involvement among requirements in \xglspl{its} that is less so in the traditional issue types Development and Maintenance.
Improvement Suggestions are consistently the issue type with the most Non-Owner evolutions \textit{across all information themes}.
This suggests that this issue type is popular among Non-Owners.
This contrasts with Development issues, where Owners are the majority contributors across the information themes.
Specifically, Development Content evolutions are made by Owners 87\% of the time, split 49/51 for Meta Content, 60\% for RepoStructure, 70\% for Workflow, and slightly less at 43\% for Community.

Looking at the bottom three data-filled rows of Table~\ref{table:ownership}, Epics consistently stand out as being evolved by the Creator who is also the Reporter, but \textit{not} the Assignee (across information themes).
This is particularly the case for Workflow and Community information themes.
This is likely some kind of product owner or manager who is working on (evolving) the issue, despite having assigned it to someone else.

\textbf{Inspecting the first scope (S1)}, the evolution distribution between Owners and Others is almost equal at 48\% and 52\%, respectively.
The Owner subsets share an almost equal split as well at 34\%, 37\%, and 28\%.
These sets ($Cra$, $CrA$, $cRa$, and $cRA$) represent evolutions made by someone who is either the Creator, or the Reporter, but not both (logical XOR).
This means that within the \xglspl{jr}, considering evolutions made by a Creator or Reporter, 99.34\% of the time they also hold the other role---in other words, they are ``Creator-Reporters''.
$CRA$ represents evolutions made by someone holding all three roles, which accounts for 34\% of all Owner evolutions.
$CRA$~and~$crA$ represent all Assignee evolutions (Creator-Reporters or not), which accounts for 62\% of all Owner evolutions, whereas $CRa$ represents non-Assignee evolutions (who are Creator-Reporters), which only accounts for 37\% of the Owner evolutions.
In other words, Assignees make 62\% of the Owner evolutions.
$CRa$~and~$CRA$ represent all Creator-Reporter evolutions (Assignees or not), which accounts for 71\% of all Owner evolutions, whereas $crA$ represents non-Creator-Reporters, which only accounts for 28\% of the Owner evolutions.
In other words, Creator-Reporters make 71\% of the Owner evolutions.

\textbf{Inspecting S2}, Requirements evolutions have the highest contribution rate from Others at 57\%, compared to Maintenance at 53\% and Development at 42\%.
Looking at the Owner evolutions sets of S2 compared to S1, there are small differences.
Evolutions made by $CRA$ are lower in Maintenance issues (27\%), and higher in Development issues (45\%).

\textbf{Inspecting S3}, Content evolutions are largely made by Owners (79\%), with a majority made by non-Assignees (64\%), and very few made by non-Creator-Reporters (7\%).
MetaContent, RepoStructure, and Community evolutions are made by Others \mytilde60\%, and Workflow evolutions are leaning towards Owners at 56\%.

\textbf{Inspecting S4}, we see Requirements Content evolutions are largely made by Owners at 69\%, and even more so for Maintenance at 81\% and Development at 87\%.
Similar to all Content evolutions, Requirements Content evolutions made by Owners are made more by non-Assignees (59\%) and very rarely by non-Creator-Reporters (7\%).
This pattern is more pronounced in Maintenance Content evolutions, with non-Assignees accounting for 73\% of Owner evolutions, and non-Creator-Reporters accounting for only 6\%.
Looking at Requirements across the other evolution types in S4, Workflow evolutions are evenly split at 50/50, while MetaContent, RepoStructure, and Community have a 2/3rds distribution of Others making the evolutions.
For Development Workflow evolutions, Owners account for 71\% of evolutions.
Other notable Requirements Owner evolution distributions include $CRA$ Workflow evolutions at 48\%, and $CRa$ Community evolutions at 44\%.

\FloatBarrier

\subsection{Patterns of Content Evolutions}

\renewcommand{\ph}[1]{\makebox[\widthof{XXXX}][c]{#1}}
\renewcommand{\rotHead}[1]{\rotatebox{40}{#1}}
\definecolor{c_red}{HTML}{D81B60}
\definecolor{c_blue}{HTML}{1E88E5}
\definecolor{c_yellow}{HTML}{FFC107}
\def\hmr#1{\cellcolor{c_red!\fpeval{100*#1/50}}#1}  
\def\hmb#1{\cellcolor{c_blue!\fpeval{120*#1/50}}#1}  
\def\hmy#1{\cellcolor{c_yellow!\fpeval{120*#1/50}}#1}  

\begin{table}[t]
    \footnotesize                   
    \setlength\tabcolsep{2.5pt}     
    \def\arraystretch{1.3}          
    \centering                      

    \caption{Occurrences of observed content evolution types.}
    \label{table:content_types}

    \begin{tabular}{@{}rccccc@{}}
        & \ph{\rotHead{Requirements}} & \ph{\rotHead{Development}} & \ph{\rotHead{Maintenance}} & Total & \%      \\
        \toprule
        Issues     & 52           & 52          & 52          & 156   &         \\
        Evolutions & 105          & 115          & 95         & 315   & 100\%   \\
        \midrule
        T1         & \hmr{41}           & \hmb{31}          & \hmy{22}          & 94    & 29.8 \% \\
        T2         & \hmr{23}           & \hmb{35}          & \hmy{24}          & 82    & 26.0 \% \\
        T3         & \hmr{14}           & \hmb{26}          & \hmy{25}          & 65    & 20.6 \% \\
        T4         & \hmr{8}            & \hmb{10}          & \hmy{9}           & 27    & 8.6 \%  \\
        T5         & \hmr{5}            & \hmb{7}           & \hmy{3}           & 15    & 4.8 \%  \\
        T6         & \hmr{7}            & \hmb{2}           & \hmy{5}           & 14    & 4.4 \%  \\
        T7         & \hmr{6}            & \hmb{1}           & \hmy{5}           & 12    & 3.8 \%  \\
        T8         & \hmr{1}            & \hmb{3}           & \hmy{2}           & 6     & 1.9 \%  \\
        \bottomrule
    \end{tabular}
\end{table}

Regarding the manual analysis of textual evolutions to the Summary and Description fields, I found eight content evolution types.
Table~\ref{table:content_types} shows the distribution of the eight types across Requirements, Development, and Maintenance, sorted by occurrences in Requirements, and magnitude highlighted with a heatmap.

\textbf{T1: Rewording and Refinement.}
The most common type, accounting for 29.8\% of content evolutions analysed, was rewording and refinement of existing information in issue descriptions.
Identifiers (such as version numbers) and links are updated and words are replaced by more precise ones.
In a few cases, entire descriptions were rewritten, although they appeared to be reworded for clarity and not semantic changes.
It was common for additional descriptive details to be added to issue descriptions, refining the formulation to narrow the focus of the statements.
Identifiers such as project IDs, product names, and version numbers were added to make the statements more precise without adding new information.

\textbf{T2: Adding New (Contextual) Information.}
A common type was to add entirely new information to the issue, usually to clarify the context.
Most often, this involved appending additional sentences or paragraphs to the end of the description, which suggests that this information was discovered or requested after starting the work on the issue.
Examples include adding new supporting information such as stack traces, error logs, code snippets, screenshots, and rationale for the issue or its criticality.
Additionally, I observed examples of quality assurance details being added such as acceptance criteria, testing details, and steps to reproduce.

\textbf{T3: Style Evolution.}
I observed multiple changes targeting the visual cleanliness of the descriptions.
These types include removing newlines and other symbols such as dashes and slashes, adding Jira formatting such as ``code'' and ``noformat'' blocks as well as formatting hyperlinks, and converting prose sentences into bullet or numbered lists.

\textbf{T4: Adding Missing Description.}
I observed that sometimes the description is kept empty during the issue creation and is added in a follow-up evolution.
These descriptions were added anywhere from minutes to days later.
In the sample, I observed that it was mostly the original issue creator who came back and added the missing description.

\textbf{T5: Removing Small Details.}
Descriptive details were occasionally removed.
Examples include removing Jira project information and identifiers such as release versions.

\textbf{T6: Adding Meta Content.}
I observed multiple changes where meta content is added to the description, usually as one or two sentences at the top of the description (e.g. starting with ``note that ...'').
This meta-content describes the issue in the larger context of the \xgls{jr} or organisation.
For instance, product owners mention that this issue is a duplicate, summarise related updates, or mention that the issue is addressed in a different project or \xgls{its}.

\textbf{T7: Correcting Spelling, Grammar, and Punctuation.}
Language errors occur when describing issues.
I observed that these are often fixed in follow-up edits.
These content evolutions included addressing spelling mistakes, missing spaces between words, and punctuation.

\textbf{T8: Adding External Links.}
I observed that multiple changes to the descriptions included adding links to external resources for further reading and context.
Examples include reference reading, links to source code, and additional project details found on mailing lists.

\section{Related Work}

Li et al.~\cite{Li_2012_EASE} performed a preliminary literature review on evolution in requirements, summarising findings across 125 primary studies.
They identified four phases of requirements evolution.
The final phase, evolution tracking, involved two activities: issue management and evolution measurement.
Only 7\% of the studies were concerned with evolution measurement.
The majority of the studies focused on other aspects such as Management Process (41\%), Model Evolution (36\%), and Impact Analysis (10\%)~\cite{Li_2012_EASE}.
These articles were predominantly concerned with traditional \xgls{re} practices rather than requirements in \xglspl{its}.

Heck and Zaidman~\cite{Heck_2013_IWPSE} performed an exploratory investigation into the evolution of Feature Requests in \xglspl{its}.
Their work specifically focused on Feature Request duplicates and the reasons that lead to them.
While their work highlighted the need to understand issue evolution in \xglspl{its}, our study goes beyond just Feature Requests.
I quantitatively explore the actual issue evolution in \xglspl{its} across all notable fields, comparing requirements, maintenance, and development issues and identifying several differences between Feature Requests and other types of requirements.

Jayatilleke and Lai~\cite{Jayatilleke_2018_IST} conducted a systematic literature review of requirements change management and report on the causes of change, the processes to manage it, the techniques developed to work with and avoid it, and decision-making for change management.
Their results describe traditional requirements artefacts and traditional change management processes.
However, the authors refer to the benefits of agile methods in change management.
They describe the benefit of ``minimal documentation using user stories which do not require long and complex specification documents'', thus ``the possibility of dramatic and constant changes is reduced''.
Our work showed that requirement issues did, in fact, evolve: about 8 times on average and up to 25 times.
As I observe rather few changes to the content, I expect that requirements might evolve outside the \xgls{its}, leading to the creation of a new issue whenever necessary.

Bug reports are a common focus for \xgls{its} research due to their importance to the software maintenance and evolution processes in open-source \xglspl{ite}.
Bettenburg et al. and Zimmerman et al. studied the quality of Bug Reports and, in particular, the needs of developers using those Bug Reports~\cite{Bettenburg_2008_FSE,Zimmermann_2010_TSE}.
The primary finding from their work was the extreme mismatch between the needs of developers, and the information provided by reports.
In particular, they note that incomplete information was the most common problem faced by \ltexdummy{developers~\cite{Zimmermann_2010_TSE}}.
This prompted a wave of research focused on improving Bug Reports.
Herzig et al. investigated issues labelled as Bug Reports and found that roughly 1/3rd of Bug Reports are misclassified, leading to the conclusion that data needs to better classified~\cite{Herzig_2013_ICSE}.
Research focusing on automatically classifying issues as Bug Reports then became popular in an attempt to help \xglspl{its} and clarify data~\cite{Terdchanakul_2017_ICSME,Zhou_2016_JSE}.
This research, however, is solely focused on Bug Reports.

Multiple studies have focused on the broader needs for and usefulness of \xglspl{its} to support software organisations.
This includes researchers asking what information stakeholders need to know~\cite{Maalej_2014_EmpiRE} and how tool support can be tightened ~\cite{Fucci_2018_ESEM,Saito_2017_RE,Xuan_2012_ICSE}.
Maalej et al. found that information needs were dependent on organisational role and project phase~\cite{Maalej_2014_EmpiRE}.
Fucci et al. discovered issues of information overload, tool limitations, and needs to understand dependencies between issues and requirements reuse~\cite{Fucci_2018_ESEM}.
The works of Bavota and Russo~\cite{Bavota_2016_MSR} and of Xavier et al.~\cite{Xavier_2020_MSR} show how \xglspl{its} are being repurposed to manage self-admitted technical debt by using special issue types.
Finally, several studies have been published about Feature Request detection in \xglspl{its}~\cite{Fitzgerald_2011_RE,Merten_2016_RE,Seiler_2017_REFSQ} as well as detection in other issue-like sources such as app stores~\cite{Maalej_2016_RE,Johann_2017_RE}.
Our work confirms the importance of \xglspl{its} for \xgls{re} practice, with both benefits and drawbacks to their use focusing on \textit{some} issue types such as Bug Reports, Feature Requests, and technical debt.
However, previously published work has covered neither the full breadth of \xglspl{its} nor requirements evolution specifically.

Finally, another related area is customer relationship management, which reflects the process of understanding users, customers, and stakeholders in general, and how they impact the evolution of requirements.
Customer relationship management involves the use of artefacts, tools, and workflows to successfully initiate, maintain, and sometimes terminate customer relationships~\cite{Reinartz_2004_JMR}.
\xglspl{its} can be used as a customer relationship management tool, managing customer feature suggestions, customer incident reports, and general support tickets~\cite{Kabbedijk_2009_RE,Merten_2016_RE}.
Previous work suggested that support tickets can be used to predict and prevent the loss of customers~\cite{Montgomery_2017_RE, Montgomery_2018_REJ, Montgomery_2017_RE_a}, furthering the value of documenting customer relationship management practices.
I found that 52\% of \ltexdummy{requirements} evolutions are performed by stakeholders outside the ownership roles, revealing the importance of these stakeholders in the process.
A deeper analysis of stakeholder interactions is required to expand on the findings presented in this article in relation to customer relationship management.

\section{Discussion}

\textbf{Process-Centred vs.~Community-Centred Requirements.}
The results reveal that requirements are prevalent in \xglspl{ite}, but different projects use various types of requirements differently.
While overall, Improvement Suggestions and Feature Requests are the most popular types in the entire dataset, comparing the single projects shows that Stories (combined with Epics) are overwhelmingly used by a {\mytilde}quarter of all analysed projects.
This reveals at least two different \textit{styles of requirements usage} in (\xgls{oss}) \xglspl{its}: a process-centred requirements usage (such as Epics and Stories which are central to agile processes), and a community-centred requirements usage (such as Feature Requests and Improvement Suggestions which are bottom-up issues by users and other stakeholders who want or should have a say on the software).
Two additional results confirm this observation.
First, community evolutions occur less for Epics and Stories than for the other types of requirements, also compared to Development and Maintenance issues.
Second, when requirements Content (summary and description) evolves, it is only for Epics and Stories (Fig.~\ref{fig:iss_evo_dist_evo_themes})---implying that these two types are likely more carefully maintained by the development team, while the others rather follow the goal of ``accepting or rejecting'' them, a typical elicitation and analysis goal.

Researchers and tool vendors should carefully distinguish between these different types of requirements when studying and supporting \xgls{re} within \xglspl{ite}.
It is also interesting to understand how the requirement usages are interlinked, whether story-driven projects represent a ``clean'' version of requirements focusing on implementation and validation while community-driven projects focus more on elicitation, negotiation, and analysis of requirements.

\textbf{Operative vs. Corrective Evolutions.}
The manual analysis of the large number of fields used to manage issues across 13 well-known communities sheds light on what information evolves after the creation of a requirement issue in an \xgls{its}.
I reduced more than 2,000 fields in the dataset into 20 fields with distinct goals and semantics.
These 20 fields represent common ground for various heterogeneous projects using the highly customisable \xgls{its} Jira.
Each of the fields by themselves can be the focus of a future comparative study, to understand, for example, how Resolution information or Comments emerge in various contexts and issue types.
While some fields like Priority, Status, or Assignee have attracted notable research attention over the past decade, others, such as Links, have not.
The results suggest indeed that these fields bear differences, e.g. in how long they ``live'' or by whom they get updated.
I also grouped the issue fields into five themes, which represent different meanings likely with different impacts when changed: Content, MetaContent, RepoStructure, Workflow, and Community.
This also served as a needed focus for our analysis, rather than 20 fields.
Our quantitative results also reveal multiple differences between the information themes.
Most notably, I found differences between Content and Workflow, although also between Content and others.
For instance, Content usually only evolves within a few hours from the issue creation, while Workflow and Community usually evolve within weeks and up to months for structural information (RepoStructure) such as Parent issue or IssueType itself.
Additionally, the fields have clear differences between the different types of requirements, as discussed above.

Generally, Workflow field evolutions reflect the \textit{operative} nature of \xglspl{its}: to track status and manage workflows.
Content field evolutions, on the other hand, rather reflect \textit{corrective} actions: including the rewording and refinements of issues as well as the addition of new contextual information, needed for understanding and resolving the issue.
Controlling for this difference is crucial to studying and assisting issue evolution in \xglspl{ite}.
So far, most research on requirements' evolution focuses on the corrective nature.
The goal is to reduce the ambiguity of requirements, remove inconsistency, and ``develop'' the requirements from brief goals or needs into detailed use cases, user stories or models that can be validated, implemented, and tested.
On the other hand, traditional \xgls{its} research, originating from bug tracking and maintenance, rather focuses on the operative evolution: on how to reduce duplicates, assist issue allocation, or reduce the resolution time as much as possible.
Future work should bridge these two perspectives, as both are important, and further support and understanding are needed for both of them.

\section{Summary}

In this chapter, I investigated the information types and evolution behaviour within \xglspl{its}.
I applied both quantitative and qualitative techniques to a large dataset of public Jira repos~\cite{Montgomery_2022_MSR} consisting of 1.3 million issues, 13 million evolutions, and 2,094 projects.
I performed this analysis on 1.3 million issues from 13 distinct organisations across 2,094 \xgls{oss} projects.
\ltexignore{The Inductive \xgls{ta} revealed five information type themes in the data: Content, MetaContent, RepoStructure, Workflow, and Community.}
There are also 20 codes under those five themes that further break down and specify the information types.
The historical data analysis revealed both similarities in evolution behaviour and many differences.
Across all activities, issues evolve a median of {\mytilde}8 times in their lifetime.
Across the information themes, however, Workflow accounts for {\mytilde}40\% of all evolutions, whereas Content accounts for almost none.
Evolution time is not that different with requirements at 27 days after their creation (median), compared to development at 11 and maintenance at 8.
When looking at information themes, however, RepoStructure was the longest evolving evolution theme (10--100 days, medians), in contrast to Content evolutions which occurred, on average, within just the first 22 hours (median).
These differences in time evolution time became even more exaggerated when looking at the individual fields where some fields evolve for months (and sometimes years), whereas others consistently evolve for just a few days or weeks.
The results also show that Owners and Non-Owners both evolve issue information roughly the same (48\% and 52\%, respectively).
Finally, I observed eight types of changes occurring to the issue Content, including rewording and refinement, adding new contextual information, and style edits.
I also observed that information is mostly added over time to the issue description, with very little removal or semantic rewording.

This chapter revealed the types of information and evolution that occur within \xglspl{its}.
These findings further highlight the diversity of information and processes that occur within \xglspl{its}.
With the collective ``problem investigation'' findings from Part~\ref{part:problem}, we now know much more about \xglspl{ite}.
We know the problems practitioners face when using \xglspl{ite}.
We have learned specific insights regarding the inner complexities of \xglspl{its} including the artefacts, activities, information, and evolution within \xglspl{its}.
Overall, we are aware of the types of problems practitioners face, and the types of environments within which they operate.
In the following chapter, I transition into Part~\ref{part:solution}: \nameref*{part:solution}.
In the first chapter, I propose a solution to these problems that considers the diversity, depth, and complexity of the studied environments.
The following chapters of Part~\ref{part:solution} elaborate on and further contribute to this primary solution.

    \part{Solution Construction}  \label{part:solution}
    
\chapter{Best Practice Ontology for Issue Tracking Ecosystems}  \label{ch:ontology}

\epigraph{Taxonomy is described sometimes as a science and sometimes as an art, but really it's a battleground.}{Bill Bryson}

\xgls{its} problems, complexities, and proposed solutions are all multidimensional and context-dependent (see Chapters~\ref{ch:challenges},~\ref{ch:activities},~\&~\ref{ch:evolution}), yet the description of the proposed solutions hardly ever captures these dimensions or context factors.
This may lead to missed information during investigations, duplicate work by researchers and practitioners seeking to address these problems, and solutions that are too vague and broad to be adequately applied.
We know from the previous chapters that \xglspl{ite} are: complex, rich environments with different artefacts, activities, information, and evolution, and practitioners struggle to use them effectively.
We also know from research that there are many proposed solutions to individual problems in isolation from the rest of the \xgls{ite}.
\ltexdummy{We} are missing a holistic theory that attempts to capture these complexities for structuring existing knowledge of \xglspl{its}, and for guiding future research in comparison to what exists and how to structure their findings and \xgls{its} solutions.

In this chapter, I construct an ontology that captures the multidimensional and context-dependent nature of solutions to problems in \xglspl{ite}: the Best Practice Ontology for \xglspl{ite}.
I built this ontology using existing research on quality factors for \xglspl{ite}, existing theory involving \xgls{ite} ``smells'' and ``Best Practices'', and with inspiration from existing constructs in the area of quality factors for \xglspl{ite}.
The ontology has five sections and 16 dimensions, each of which is explained in detail.
This chapter introduces and explains the ontology, and the next chapter lists a catalogue of Best Practices as collected and analysed as part of the ontology-building process.
As a result of this work, future research now has a structured ontology to guide their analyses.
Additionally, the ontology itself (as well as the follow-up catalogue) is now open to be challenged, falsified, and improved with future evidence and competing ontological structures.
Following this chapter, I present a catalogue of \xgls{ite} Best Practices, followed by listing a number of detection and repair methods for \xgls{ite} Best Practices.

\section{Research Methodology}

My primary objective with this chapter is to introduce an ontological model designed to describe \xgls{ite} Best Practices.
For this qualitative model-building activity, I defined no research questions.
For the development of the ontology, I followed the recommendations of Nickerson et al.~\cite{Nickerson_2013_EJIS} (later updated by Kundisch et al.~\cite{Kundisch_2021_BISE}).
I have described these recommendations in full in Section~\ref{sec:methods_tax_build}, and will only be describing here, in this section, the choices made within the framing of those recommendations.
Nickerson et al. recommend seven primary taxonomy development phases, with two potential paths that can be taken depending on the ``approach'' chosen~\cite{Nickerson_2013_EJIS}.
I selected the inductive approach (as described in Section~\ref{sec:ontology_approach}), and so the seven phases I followed are: determine meta-characteristic, determine ending conditions, pick the approach, identify subset of objects, identify common characteristics and group objects, group characteristics into dimensions to create taxonomy, and review ending conditions~\cite{Nickerson_2013_EJIS}.
Finally, I tested the external robustness of the final ontology by applying it to structure Best Practices that \textit{could benefit \xglspl{ite}}, but do not yet have the right framing to do so.

\subsection{Determine Meta-Characteristic}  \label{sec:ontology_meta_characteristic}

The purpose of the meta-characteristic is to consider and define the important perspectives that your taxonomy/ontology will be a part of.
This helps the construction of the dimensions such that the output is useful to the right user groups~\cite{Nickerson_2013_EJIS}.
The primary reason for constructing this ontology is to support practitioners in improving the quality of their \xgls{ite}, by giving them guidance in the form of information, recommended processes, and algorithms.
This highlights ``\xglspl{its}'' as the main \textit{artefact} of interest, ``practitioners'' as the main \textit{user group} of interest, and ``guiding'' and  ``documenting'' as the main \textit{uses} of interest.
A secondary reason for constructing this ontology is to support researchers in investigating and documenting how practitioners interact with \xglspl{ite}, with a focus on prescribing processes that will help improve the quality of their \xgls{ite}.
This highlights ``researchers'' as an additional user group of interest, and ``investigating'', ``documenting'', and ``prescribing'' as additional uses of interest.

The artefacts of importance for the meta-characteristic are the \textit{\xgls{its} overall}, and the \textit{issues} within them.
\xglspl{its} are a tool used to manage \xgls{se} processes, and they are also used as documentation.
\xglspl{its} are complex tools that support many use cases across various \xgls{se} processes, for many stakeholder groups.
As such, it is important that the ontology can handle the different use cases possible within \xglspl{its}, as well as the various processes and stakeholder groups.
If the ontology is not specific enough, it won't be clear which use cases to do with which processes are affected, and which stakeholders are involved.
The issues within an \xgls{its} are unique and atomic artefacts, each representing a unit of work that needs to be completed, each with properties (dimensions) that characterise it.
The ontology should be aware of this complex artefact, and also utilise the structure where possible to simplify instructions and definitions.

The two primary users of this ontology will be \xgls{se} practitioners and \xgls{se} researchers.
For the practitioners, I have identified three different user groups: \textit{managers} (e.g., team leads, and product owners), \xgls{its} \textit{maintainers}, and \textit{developers}.
The manager user group is defined as someone who needs, wants, or could benefit from an overview of the quality of an \xgls{its}.
The manager also adds issues to the \xgls{its} in the form of top-down plans and instructions for developers, and uses the \xgls{its} to keep track of the progress of the issues their developers are working on.
The \xgls{its} maintainer user group is defined as someone who configures the \xgls{its}.
The developer user group is defined as someone who utilises the \xgls{its} as a tool for knowing what needs to be done, tracking and completing tasks, and communicating with other stakeholders regarding the issues.
Depending on the company context, either the manager or maintainer will also document their \xgls{ite} Best Practices for (re-)use by colleagues.
For the researchers, I have identified two user groups: \textit{consumers} and \textit{contributors}.
It is also worth mentioning that ``developer'' is being used here to describe a broad set of roles in traditional \xgls{se}, such as programmer, tester, documenter, and compliance checker.
All these roles share the same user group of ``Developer'' since they ``develop'' something in response to what is described in \xgls{its} issues.
The consumers user group is defined as someone who is interested in the state-of-the-art research on \xgls{ite} Best Practices.
The contributors user group is defined as someone who is interested in modifying or creating Best Practices based on empirical research.
These five user groups outline the relevant stakeholders to be considered for the meta-characteristic.

From the five user groups above, I have identified four primary uses for the ontology: guiding, investigating, documenting, and prescribing.
The ontology should support managers, maintainers, and developers by guiding them through what an \xgls{ite} Best Practice is, why it needs to exist in their specific context, how to implement it, and what algorithmic support they can expect from the system.
The ontology should support managers and maintainers by helping them document their existing \xgls{ite} Best Practices for (re-)use by their colleagues.
The ontology should support researchers in investigating existing \xgls{ite} Best Practices phenomena, and subsequently modifying existing ontological items, or adding new ones.
The ontology should support researchers in prescribing future \xgls{ite} Best Practice behaviour, given the empirical evidence from investigating \xgls{ite} usage highlights a strong recommendation for practitioners.

In summary, the \textit{meta-characteristic} is ``the information required to guide practitioners in using and documenting \xgls{ite} Best Practices within their specific organisational context, and the information required to support researchers in investigating and documenting \xgls{ite} Best Practices, including the prescription of future \xgls{ite} Best Practices.''

\subsection{Determine Ending Conditions}

I apply the recommended ending conditions by Nickerson et al.~\cite{Nickerson_2013_EJIS}, both objective and subjective.
There are eight recommended objective ending conditions and five recommended subjective ending conditions.
See Section~\ref{sec:methods_tax_build} for a full list of these ending conditions.

\subsection{Pick the Approach}  \label{sec:ontology_approach}

I chose the inductive approach as the primary means of creating the ontology, but I also integrate deductive thinking into the overall design.
The concept of Best Practices for \xglspl{ite} is not new (although not yet framed this way).
The concept of ``smells'' and ``recommendations'' for \xglspl{ite} are certainly not \ltexdummy{new~\cite{Bettenburg_2008_FSE,Zimmermann_2010_TSE,Qamar_SEAA_2021,Qamar_2022_IST}}.
Considering this existing research, I decided that the best approach to building this ontology was to review existing practices for discussing \xgls{ite} quality, and combine that knowledge into a single ontological model.
However, as I have been studying this topic for my entire PhD, I also have a significant amount of knowledge and understanding of both \xglspl{ite} and the state-of-the-art research.
Thus, I decided to integrate deductive passes over the ontology during the overall inductive approach.
My decision when to incorporate these deductive passes was not systematic or based on clear criteria, but as an additional component not required by the inductive approach, I viewed these passes as strictly beneficial with no clear downside.
I would compare my deductive passes to that of Phase 4 of \xgls{ta} (as proposed by Braun and Clarke~\cite{Braun_2006_QRP}), whereby as the researcher you need to step back, and ask yourself if the model that is forming is cohesive and representative of your goal.
Without such a perspective, it is possible for your inductive analysis to produce something empirically grounded, but missing higher-level meaning and purpose~\cite{Braun_2006_QRP}.

\subsection{\ltexignore{Identify Subset of Objects}}

Ontology development can be either inductive or deductive (see Section~\ref{sec:methods_tax_build}).
I chose the inductive approach because there is an abundance of empirical data on \xglspl{ite} and the quality thereof.
I defined which types of objects are important for the taxonomy, and which subset of the population to sample.
For the objects of interest, I identified two types of objects that are important for the taxonomy: existing classification systems for quality aspects of \xgls{ite} and the quality aspects themselves.
The \textbf{first} type represents the quintessential objects that this ontology is designed to unify and improve.
These are existing efforts to structure quality aspects of \xglspl{ite} as units (e.g., communication and analysis).
My ontology is designed to capture, unify, and structure this area of research.
The \textbf{second} type represents the unstructured form of the objects that this ontology seeks to organise.
These objects can (presumably) be structured to fit into the taxonomy easily, and act as \textit{internal robustness checks} during the ontology creation process (thereby impacting the formation of the ontology).
Each paragraph below describes these types in more detail, and describes the subset of the population that was sampled.

The first object type contained articles structuring and specifying \xgls{ite} quality aspects, and served as the primary ontology development dataset.
The goal for this type was to review existing models for structuring and reporting best-practice-like concepts, which could then inform the creation of the ontology.
To find these objects, I searched for articles that discussed some named concept synonymous with ``\xgls{ite} Best Practices'' such as ``recommendations'', ``smells'', ``patterns'', and ``antipatterns''.
While there is an abundance of literature involving quality aspects for \xglspl{ite}, it is rare that these articles also attempt to structure the objects beyond giving each a name and a description.
Therefore, I based the data collection for this rare object type on key literature and snowballing.
I started with key literature on the topic---well-known and recent literature on \xgls{its} Best Practices, and then I snowballed forward and backwards searching for similar research.
This approach is acknowledged by Nickerson et al. who describe the collected objects for ontology development ``likely to be the ones with which the researcher is most familiar or that are most easily accessible [...] a convenience sample''~\cite{Nickerson_2013_EJIS}.
The goal for collecting this type of object, however, was not completeness.
While collecting as many articles as I could find was the goal, I did not follow a strict secondary-study protocol~\cite{Petersen_2008_EASE,Petersen_2015_IST,Kitchenham_2007_TR} that would have increased the confidence and validity of a completeness claim.
Despite the lack of knowledge regarding completeness, the first object-type dataset was collected using acknowledged rigorous methods, and is sufficient for the purpose of forming an ontology~\cite{Nickerson_2013_EJIS}.

The second object type contained quality aspects for \xglspl{ite}, regardless of the structure used to discuss them.
The goal for this type was to iteratively check the robustness of the ontology \textit{during development}.
In other words, to check if the ontological properties could handle the important parts of each object, and that the properties were differentiating the objects well enough~\cite{Nickerson_2013_EJIS}.
This object type was intended to only include ``internal'' objects, which are objects that are already within the context of quality aspects for \xglspl{ite}.
Given the shared context, these objects should be easily structured and specified by the ontology.
To find these objects, I started with the articles from the first object type, which each already had such quality aspects.\footnote{I did not find any articles for the first object type that only discussed structure, without also noting examples of \xgls{ite} quality aspects from previous articles or ones from new empirical work conducted in the article.}
I collected additional articles that described \xgls{ite} quality aspects without mentioning a structure beyond a name and a description.
The data collection process for these articles was also based on key literature.
The study of \xglspl{its}, and the broader environment and context of \xglspl{ite}, is a well-studied area that has been popular for the past 20+ \ltexdummy{years~\cite{Bettenburg_2008_FSE,Zimmermann_2010_TSE}}.
Similar to the first object type, the goal here was \ltexdummy{not completeness}.
Rather, the goal with this object type was to check the internal robustness of the ontology.

\subsection{Identify Common Characteristics and Group Objects}  \label{sec:ontolog_characteristics}

Some common characteristics arose from the articles discussing quality factors for \xgls{ite} Best Practices.
First, it was common for the articles to discuss something negative that is to be avoided.
This is commonly referred to as a ``smell'' across different areas, including \xgls{its} quality and code quality (among others).
However, even without a mentioned construct, \textit{avoiding negative outcomes} appeared to be the main driver behind many articles describing quality aspects for \xglspl{ite}.
Second, it was common for the articles to discuss some kind of \textit{algorithmic contribution}.
It seems most \xgls{se} researchers are interested in engineering solutions to the quality issues that exist in \xglspl{ite}.
Combined, these two concepts (avoiding negative outcomes and algorithmic detection) were by far the most common characteristics to be discussed.

Less common, but still present in some articles, is the concept of recommending some kind of process to avoid these negative outcomes.
This could be as simple as recommending the implementation of the algorithmic detection, or as complicated as recommending a process change within the company, or a configuration change within the \xgls{its} itself.
The classic ``What Makes a Good Bug Report'' by Zimmerman et al.~\cite{Zimmermann_2010_TSE} recommends both process changes (such as ``provide feedback on Bug Reports'') and \xgls{its} configuration changes (such as ``provide tool support to collect information'').
Finally, present in some articles, but exceedingly rare, was the characteristic of context.
Given the meta-characteristic described above in Section~\ref{sec:ontology_meta_characteristic}, and my findings from Chapters~\ref{ch:challenges}, \ref{ch:activities}, and \ref{ch:evolution}, I was looking for \textit{contextual factors} to focus the discussion on a particular set of organisational factors.
Most articles do not discuss context at all, which implies that their \xgls{its} quality factor discussion applies to all \xglspl{its} in all organisations.
Those rare articles that did discuss context, were largely reported in the methodology section when describing the organisation, \xgls{ite}, or dataset they were working with.
There was minimal discussion of context and how it applies to the findings and recommendations when it comes to quality factors in \xglspl{ite}.
Despite the rarity of this characteristic, some articles did address context, including Eloranta et al.~\cite{Eloranta_2016_IST}, who explicitly model context as a dimension in their catalogue of scrum antipatterns.
Overall, I noticed the concepts of avoiding negative outcomes, recommendations to be better, and contextual factors.
These characteristics are well aligned with the meta-characteristic, supporting the activities of guiding and prescribing.

\subsection{Group Characteristics into Dimensions to Create Taxonomy}

The results of this phase in the iterative process is the ontology itself.
Every iteration led to the creation and refinement of dimensions based on the meta-characteristic described in Section~\ref{sec:ontology_meta_characteristic} and the characteristics that began to form in Section~\ref{sec:ontolog_characteristics}.
For full details of the formed dimensions of the final ontology, see Section~\ref{sec:ontology_ontology} below.

\subsection{Review Ending Conditions}

Creating this ontology was an iterative process that took many cycles, each time reviewing the ending conditions listed in Section~\ref{sec:methods_tax_build}.
As the subset of reviewed objects is not the full population, it was also possible to find more objects and continue the cycles.
However, for the purpose of creating this ontology, the initial sample of objects was sufficient for creating a thoroughly grounded ontology.

\section{The Ontology}  \label{sec:ontology_ontology}

The ontology has five sections and 16 dimensions.
The five sections are Meta, Summary, Recommendation, Context, and Violation.
I structure and summarise these sections and dimensions in Table~\ref{tab:ontology_structure}.
In the following subsections, I describe each ontology section and dimension, with a focus on linking them to the meta-characteristics when possible.

\afterpage{%
    \ActivateWarningFilters[largefigure]  
    \clearpage
    \vspace*{\fill}  
    
\newcommand*{\ontHeaderMain}[1]{\multicolumn{2}{l}{\large #1}}
\newcommand*{\ontHeaderSecondary}[1]{\small{\textbf{#1:} }}
\newcommand*{\ontContent}[1]{\small{#1}}

\begin{table}[!ht]
    \centering                              
    \setlength{\LTleft}{-20cm plus -1fill}
    \setlength{\LTright}{\LTleft}

    \begin{longtable}{p{.01\textwidth}p{.95\textwidth}}

        \caption{Best Practice Ontology for ITEs.}  
        \label{tab:ontology_structure} \\

        \toprule
            \ontHeaderMain{Meta*} \\
                & \ontHeaderSecondary{Name}\ontContent{The name of the Best Practice.} \\
                & \ontHeaderSecondary{Source}\ontContent{Where the Best Practice came from; often a research article.} \\
        \midrule
            \ontHeaderMain{Summary} \\
                & \ontHeaderSecondary{Objective}\ontContent{The purpose of the Best Practice; the goal of implementing it. If possible, don't describe what needs to be done, but rather what the (positive) outcome will be.} \\
                & \ontHeaderSecondary{Motivation}\ontContent{Explanation of the concepts involved, including the background, importance, and potential benefits.} \\
        \midrule
            \ontHeaderMain{Recommendation} \\
                & \ontHeaderSecondary{Process}\ontContent{The recommended steps to take to follow the Best Practice.} \\
                & \ontHeaderSecondary{ITS}\ontContent{Specific instructions or configurations for the \xgls{its} involved.} \\
                \midrule
                \ontHeaderMain{Context} \\
                & \ontHeaderSecondary{Stakeholder Benefits}\ontContent{Outline who benefits from this Best Practice being implemented. Separate each stakeholder group who is affected differently. Be clear by first naming the stakeholder group, followed by the benefits.} \\
                & \ontHeaderSecondary{Stakeholder Costs}\ontContent{Same as ``Stakeholder Benefits'', except, outline the costs associated with different stakeholder groups.} \\
                & \ontHeaderSecondary{ITS Scope}\ontContent{The scope of the \xgls{its} required to understand the conformance to this Best Practice. Examples include a single issue, pairs of issues, and the entire \xgls{its}.} \\
                & \ontHeaderSecondary{Issue Types}\ontContent{The specific issue types this Best Practice applies to. Examples include Bug Report, User Story, Epic, and Work Item.} \\
                & \ontHeaderSecondary{Inclusion Factors}\ontContent{Specific context factors that \textit{should} be fulfilled for this Best Practice to be a good recommendation for a specific individual, project, team, or organisation.} \\
                & \ontHeaderSecondary{Exclusion Factors}\ontContent{Specific context factors that \textit{should not} be fulfilled for this Best Practice to be a good recommendation for a specific individual, project, team, or organisation. These factors can be explicit exceptions to the Inclusion Factors above, or they can be key indicators that this Best Practice is not well-suited to a given context.} \\
        \midrule
            \ontHeaderMain{Violation} \\
                & \ontHeaderSecondary{Smells}\ontContent{The outcomes that suggest this Best Practice is being violated. These are not necessarily \textit{negative} outcomes, but good first indicators that something is wrong.} \\
                & \ontHeaderSecondary{Consequences}\ontContent{Potential negative outcomes that could happen as a result of not following this Best Practice.} \\
                & \ontHeaderSecondary{Causes}\ontContent{Potential reasons why the Best Practice was not followed, leading to a violation.} \\
                & \ontHeaderSecondary{Algorithmic Detection}\ontContent{The pseudocode that automatically detects violations of the Best Practice in the associated \xgls{its}. If possible, actual code is also appreciated.} \\
        \bottomrule \\[-10pt]
            \multicolumn{2}{c}{\parbox{0.95\textwidth}{%
                \footnotesize * In the catalogue presented in Chapter~\ref{ch:catalogue}, the Meta information (name and source) are listed as a title above the table, and are not included in the table itself.
            }}
    \end{longtable}
\end{table}

    \vspace*{\fill}  
    \clearpage
    \DeactivateWarningFilters[largefigure]  
}

Ontologies can have a mix of properties (dimensions) and relationships.
As described in Section~\ref{sec:method_tax_ont}, the most important focus when building an ontology is the specific purpose for which it is built.
In the ontology presented in this chapter, the primary purpose is to structure and convey specific types of information, which an ontology does well through properties.
The only relationships conveyed through this ontology are the ``broader/narrower'' relationships naturally described by the taxonomy-like structuring that exists in the grouping of the objects.

\subsection{Meta}

The Meta section describes the ontological structure of the Best Practice itself.
The dimensions within the Meta section are the \textit{name} and \textit{source} of the Best Practice.
The name is a succinct and distinct natural-language way to refer to this Best Practice.
The Best Practice name should be both catchy and intuitive, such that the whole Best Practice is well summarised within the name itself.
The source is the place where this Best Practice comes from.
This is often a research article, but it can also be a blog post or an organisational statement.
The source dimension structures and encourages backwards traceability to where the Best Practice began.
Importantly, this dimension is not limited to a single value, and should instead list all sources that have contributed to the Best Practice as it is currently formed.

\subsection{Summary}

The Summary section describes the high-level purpose of the Best Practice.
The dimensions within the Summary section are the \textit{objective} and \textit{motivation} of the Best Practice.
The objective is a succinct description of the ultimate goal of applying this Best Practice.
For example, ``Achieve good Bug Report quality'' is a common objective found in nine of the Best Practices in the catalogue.\footnote{The abundance of Best Practices with this objective can be explained by the \xgls{se} research community's focus on Bug Report quality for the past 20 \ltexdummy{years~\cite{Halverson_2006_CSCW,Bettenburg_2008_FSE,Zimmermann_2010_TSE}}.}
The motivation is a longer explanation of the concepts involved with this Best Practice, including the background, gaps, problems, solutions, where it came from, and potential benefits.
The purpose of this field is not to be complete in reporting on the Best Practice, but rather to justify (to ``motivate'') the need for such a Best Practice.
While the need for a Best Practice \textit{might} be clear from the objective, it is not always obvious why such an objective is necessary or desirable.
The motivation is designed to be a dimension open to a little interpretation that has flexibility built in to how it is written for a given Best Practice.

\subsection{Recommendation}

The Recommendation section describes the actions that should be taken to follow the Best Practices.
This section is catering directly to the ``guiding'' and ``prescribing'' aspects of the meta-characteristic for practitioners.
The dimensions within the Recommendation section are the \textit{process} and the \textit{\xgls{its} recommendations} of the Best Practice.
The \textit{process} is a simple description of the steps an individual, team, or organisation would take to follow this Best Practice.
For example, \refBPTable{LinkDuplicates} has a recommended process of ``always include a link to the duplicate bug when referencing it in another bug.''
It should be short and to the point, and exclude the complexities of implementing it in teams and organisations.
The \textit{\xgls{its} recommendation} includes specific instructions or configurations for the \xgls{its} involved.
For example, the ``Link Duplicates'' Best Practice mentioned above has the \xgls{its} recommendation of ``have a separate `duplicated by' like type and use that when referencing duplicate bugs''.
The reason the \xgls{its} recommendation is separate from the process recommendation is due to the artefact of the meta-characteristic: the \xgls{its}.
The \xgls{its} is a central, dynamic, and configurable tool at the heart of organisations implementing these \xgls{ite} Best Practices, and should be leveraged.
Recommendations for the \xgls{its} are then central to the conceptualisation of \xgls{ite} Best Practices, and in most cases there should be an accompanying recommendation for the \xgls{its}.

\subsection{Context}  \label{sec:ontology_context}

The Context section describes important factors that affect the application of the Best Practices.
With just the Summary, Recommendation, and Violation sections, it is possible to understand and apply the Best Practice principles.
However, these outcomes would be the result of many hidden factors at play, which happen to be in place for the Best Practice to be successful.
The Context section holds the answer to the classic response to the application of Best Practices and smells in organisations: ``it depends''.
In Chapter~\ref{ch:challenges}, I found that many problems discussed by the practitioners were founded on an accepted uncertainty with which to interpret their statements.
These problems, and the Best Practices that may help mitigate them, are reliant on hidden context factors that can be surfaced.
Using the meta-characteristic, and the iterative cycles of inductive and deductive reasoning from the ontology creation process, I surfaced six dimensions that form this Context section: Stakeholder Benefits, Stakeholder Costs, \xgls{its} Scope, Issue Types, Inclusion Factors, and Exclusion Factors.

The \textit{Stakeholder Benefits} dimension describes who benefits from this Best Practice being implemented.
As a starting place for the question ``would this Best Practice be good for us?'' knowing who would benefit from such a practice is key.
If no stakeholders in your organisation would benefit, then the Best Practice should not be applied.
Conversely, multiple benefiting stakeholder groups implies a strong reason to implement the Best Practice.
This dimension should be described from the perspective of individual stakeholder groups.
Separate each stakeholder group who is affected differently by this Best Practice.
Be clear when describing the benefits by first explicitly listing the stakeholder group, followed by the benefits.
For example, \refBPTable{GoodBugReport} describes the stakeholder benefits as: ``\textit{Developers}: Get the information they need. \textit{Reporters}: Get their reports resolved faster.''
Benefits for one group of stakeholders often come at the cost of other stakeholders.
This is the reason for the second Context dimension: Stakeholder Costs.

The \textit{Stakeholder Costs} dimension describes who pays a cost for the implementation of the Best Practice.
This could be a cost associated with setting up the Best Practice, enforcing the Best Practice, or interacting with it daily.
For example, \refBPTable{GoodBugReport} describes the stakeholder costs as: ``\textit{Reporters}: Have to put more effort into submitting the required information.''
This Best Practice is quite pervasive in the \xgls{oss} community in places like GitHub, where the submission of a Bug Report usually starts with a template that you must fill out.
The template requests information such as software version, computer version, expected behaviour, actual behaviour, screenshots of GUI issues, and sometimes even a minimum working example of a development project that experience the bug.
This requires a lot of time to gather, which is a cost for the reporter that needs to be acknowledged.
Given the pervasiveness of the ``Good Bug Report'' Best Practice, this cost is assumed to be reasonable.
Same as the Stakeholder Benefits, the Stakeholder Costs should separate each stakeholder group who is affected differently by this Best Practice.
The result of the first two Context dimensions then is two lists of stakeholders and how they benefit from or pay for the Best Practice.

The \textit{\xgls{its} Scope} dimension outlines the scope of the \xgls{its} required to understand the conformance to the Best Practice.
Given the central nature of the \xgls{its} to \xgls{ite} Best Practices---as outlined by the meta-characteristic highlighting the \xgls{its} as the primary artefact of interest, it is important to understand how much of the \xgls{its} is involved with the Best Practice.
For example, the \refBPTable{GoodBugReport} Best Practice only requires a single issue to check conformance, while \refBPTable{LinkDuplicates} requires the entire \xgls{its} to check conformance.
This dimension, then, acts as a proxy for complexity of the Best Practice, including things such as algorithmic detection complexity, understanding its impact, and applying it in practice.
As a starting place for organisations looking to improve their \xgls{ite} through Best Practices, it is recommended to start with Best Practices where the scope is a single issue.
Once organisations become comfortable with \xgls{ite} Best Practices, they can look to implement more complex ones with greater \xgls{its} scopes.

The \textit{Issue Types} dimension outlines the specific issue types this Best Practice applies to.
As fully described in the meta-characteristic, \xglspl{its} are complex systems that manage \ltexdummy{many different} processes within an organisation.
In Chapter~\ref{ch:activities}, I found that those activities spanned the \xgls{se} lifecycle, including requirements, development, and maintenance.
These different activities and processes are largely delineated by the ``issue type'' assigned to every issue.
In this way, issues with different types represent very different artefacts and activities within an \xgls{its}, and therefore should be treated differently and quite distinctly by the Best Practices.
While a Best Practice can apply to all issue types, such as \refBPTable{RespectfulCommunication}, many Best Practices only apply to specific issue types, such Bug Reports.
Accordingly, the Issue Types dimension of the Context section captures this information, and can support organisations deciding which Best Practices to use.
If an organisation only manages Bug Reports in their \xgls{its}, then only Best Practices that apply to Bug Reports should be considered.

The \textit{Inclusion Factors} dimension describes specific context factors that \textit{should} be fulfilled for this Best Practice to be a good recommendation for a specific individual, project, team, or organisation.
The first four Context dimensions listed above are specific factors that should be considered for the inclusion of Best Practices within a given context; however, given the diversity of \xgls{ite} Best Practices and the organisations that implement them, it is appropriate to have a general catch-all dimension where additional inclusion factors can be described.
For example, many of the Best Practices including \refBPTable{OrderedProductBacklog} and \refBPTable{StoryPointsOverHours} only apply to organisations that apply agile practices.
Accordingly, these Best Practices have some form of ``Team uses Agile'' or ``Team uses Scrum'' listed within the Inclusion Factor dimension.
If an organisation does not apply Agile, then these Best Practices have no meaning to them and should not be applied.
Many of the Best Practices listed in the catalogue in Chapter~\ref{ch:catalogue} have ``None'' as a value for this dimension.
However, the value of ``none'' is likely due to a lack of empirical evidence and understanding of the Best Practices, rather than because there are no inclusion factors.

The \textit{Exclusion Factors} dimension describes the specific context factors that \textit{should not} be fulfilled for this Best Practice to be a good recommendation for a specific individual, project, team, or organisation.
These factors can be explicit exceptions to the Inclusion Factors, or stand-alone exceptions.
For readability and understandability, it is preferred to frame an arbitrary context factor in the positive and put it in the Inclusion Factors, rather than framed in the negative and included in the Exclusion Factors.
The reason for both inclusion and exclusion factors is the need to negate portions of Inclusion Factors.
For example, \refBPTable{RecommendedSprintLength} has the Inclusion Factor \ltexignore{``Team uses Agile (Scrum)''}, and the Exclusion Factor ``If the development must be synchronized with external work that has slower pace (e.g. hardware development), longer Sprints may be justified''.
So both the inclusion and the exclusion factor are needed to describe these additional context factors that affect the application of this Best Practice.
Both the Inclusion Factors and the Exclusion Factors are a key component of the meta-characteristic when it comes to \textit{prescribing} Best Practices to practitioners.
Without such dimensions, the ontology is missing a critical piece of its core purpose.
This can also be extended to imply that Best Practices with no listed inclusion or exclusion factors are also missing a critical piece of their core purpose.

\subsection{Violation}  \label{sec:ontology_violation}

The Violation section describes aspects of the negative consequences that come from not following this Best Practice.
The concept of violating the Best Practice is important.
While the positive outcomes of the objective described in the Summary section should be motivating enough, including the negative consequences forms a full picture of the potential effects of the Best Practice on the organisation.
Additionally, as described above in Section~\ref{sec:ontolog_characteristics}, a major motivator for the Violation section is the current ``negative to be avoided'' rhetoric in the research community regarding \xgls{its} quality factors.
In other words, it is popular to propose ``smells'' that are to be avoided, instead of ``practices'' that are to be followed.
To capture these aspects of the Best Practices, the Violation section has four dimensions: \textit{smells}, \textit{consequences}, \textit{causes}, and \textit{algorithmic detection}.

The \textit{smells} dimension represents the classic representation of \xgls{its} quality factors as ``smells'', and lists the outcomes that suggest the Best Practice is being violated.
The concept of ``smells'' in \xgls{se} nomenclature is well described, understood, and utilised, as fully detailed in Section~\ref{sec:background_qsbp}.
With the placement of smells inside this ontology, it should be understood that the concept of a smell is a subset of what an \xgls{ite} Best Practice describes and offers.
To create and describe an \xgls{ite} Best Practice, much more information is needed, and much more thought must be put into the process.
For \xgls{its} quality factor recommendations (including those aimed at the broader \xgls{ite}), it is possible and beneficial to frame them as an \xgls{ite} Best Practice instead of a Smell.
This is not based solely on the requirement of more information and thought, it also has to do with what is lacking in the concept of a Smell itself: context.
I elaborate on this in Section~\ref{sec:ontology_propositions}, where I introduce a set of Propositions.

The \textit{consequences} dimension lists the potential negative outcomes that could happen as a result of not following the Best Practice.
For example, the consequences of violating the ``Good Bug Report''~\ref{tab:bp_good_bug_report} Best Practice includes ``developers are slowed down by poorly written Bug Reports''.
For \textit{researchers}, this is a prime area for \textit{investigating}: do these consequences normally occur as a result of violating this Best Practice, and are they negative?\footnote{The classic example being the assumptive statement that ``passive voice leads to ambiguity in requirements'', which, although is a potential consequence of a Best Practice violation, is highly debated whether it actually does lead to ambiguity, and whether that ambiguity is negative or not~\cite{Frattini_2024_EMSE}.}

The \textit{causes} dimension highlights the potential reasons why the Best Practice was not followed, which subsequently lead to the violation.
This is an area of interest to both practitioners and researchers, as the answer to ``why'' questions are both interesting and useful.
If an organisation is trying to follow a Best Practice, and they are following the recommended process, but not seeing the positive outcomes described in the objective, they need to know where to look for problems.
The ``causes'' dimension is such a place where potential reasons for violations should be listed.
For example, using the ``Good Bug Report''~\ref{tab:bp_good_bug_report} Best Practice again, the causes include ``reporters don't know what information to provide, \ltexignore{reporters don't want to put in the time to provide all required elements,} and reporters don't know hot to get all the required elements''.
Investigating these causes in an organisation will likely lead to necessary changes required to see the positive outcomes from this Best Practice.

Finally, the \textit{algorithmic detection} dimension lists pseudocode that automatically detects violations of the Best Practice in the associated \xgls{its}.
Actual code is also appreciated, but rare and perhaps less useful given that pseudocode is language-agnostic.
The difference between smells and algorithmic detection is the certainty with which they operate.
A smell is often described as a high-level conceptual aspect to look out for, while algorithmic detection should describe specifics that can actually be coded.
Even if a smell is specific in what it describes (which is not uncommon), the smell is still describing a situation in which the Best Practice \textit{might} be violated.
The algorithmic detection, on the other hand, should only reveal situations in which the Best Practice has indeed been violated.

\notoc\subsection{The Ontology Modelled with Thesis Constructs}

\afterpage{%
    \ActivateWarningFilters[largefigure]  
    \clearpage
    \vspace*{\fill}
    \begin{figure}[ht]
        \centering
        \makebox[0pt]{\includegraphics[width=1\textwidth]{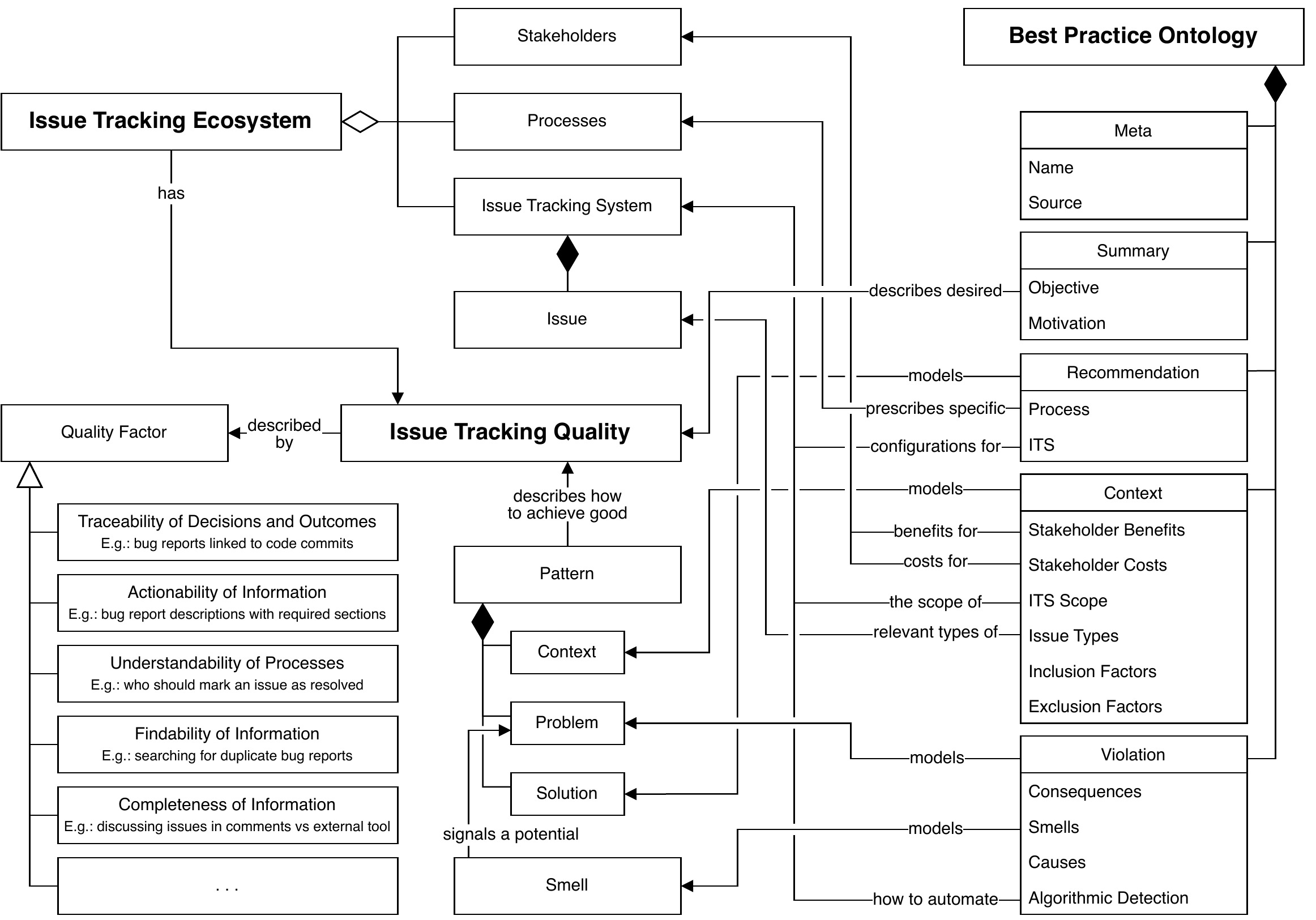}}
        \caption{Conceptual model of ontology and thesis constructs.}
        \label{fig:thesis_constructs}
    \end{figure}
    \vspace*{\fill}
    \clearpage
    \DeactivateWarningFilters[largefigure]  
}

I created the ontology to unify important and related constructs, and address particular shortcomings discussed in previous chapters.
Figure~\ref{fig:thesis_constructs} models these constructs alongside the ontology using a \glsxtrshort{uml} class diagram.
This model visualises how the ontology is related to these constructs.
Figure~\ref{fig:thesis_constructs} begins by modelling the three primary constructs: \xglsfirstplural{ite}, Issue Tracking Quality, and the Best Practice Ontology.

\textbf{\xglspl{ite}} are aggregates of \textit{stakeholders}, \textit{processes}, and an \textit{\xgls{its}}.
The Stakeholders include the \xgls{se} organisation, developers, managers, and clients.
The Processes include all workflows that govern how the stakeholders interact with the \xgls{its}.
The \xgls{its} is composed of \textit{issues}, each of which has an \textit{issue type}.
Together, these form the basic conceptualisation of the primary systems under study.
The full extent of \xglspl{its} constructs is discussed in Chapter~\ref{ch:background}.

\textbf{Issue tracking quality} is described by different \textit{quality factors}, for example, \textit{Traceability of Decisions and Outcomes} and the \textit{Findability of Information}.
All the listed quality factors can be found in the catalogue of Best Practices for \xglspl{ite} in the next chapter, including many more quality factors not listed in Figure~\ref{fig:thesis_constructs}.
Issue Tracking Quality can be obtained by following well-defined \textit{Patterns}, which themselves include a description of \textit{context}, \textit{problem}, and \textit{solution}.
The Context section of a pattern describes the expected factors that need to be present before the problem is likely to occur.
The Problem section describes an undesirable situation, such as a specific type of poor quality output.
The Solution section describes how to fix or improve the situation once the problem has been found.
\textit{Smells} signal a potential problem, but are lacking an awareness of context, which adds uncertainty to the detection.
Together, these concepts form the basic conceptualisation of Issue Tracking Quality in this thesis.

The \textbf{Best Practice Ontology} itself is described above in Section~\ref{sec:ontology_ontology}.
This ontology relates to \xglspl{ite} and Issue Tracking Quality in multiple ways.
The \textit{Objective} describes the desired issue tracking quality.
The \textit{Recommendation} section models the solution part of a pattern.
This includes the process recommendations for \xgls{ite} processes (to adhere to the Best Practice), and the \xgls{its} configuration recommendations.
The Context section models the context part of a pattern.
This includes benefits and costs for \xgls{ite} stakeholders, the \xgls{its} scope this Best Practice applies to, and the relevant issue types.
Finally, the Violation section models the problem part of a pattern.
This includes the Smells which models the concept of a smell, and the Algorithmic Detection which describes how to automate the detection of violations within the \xgls{its}.

Overall, the Best Practice Ontology for \xglspl{ite} is highly coupled with the core concepts presented and investigated in this thesis.
It draws inspiration from---and models---existing constructs such as Patterns and Smells.
It captures and describes core concepts such as the stakeholders and processes of \xglspl{ite}.
Most importantly, the ontology models these concepts in discrete categories, which offers a unified way to research, discuss, and apply these Best Practices.

\section{Propositions Towards Future Theory}  \label{sec:ontology_propositions}

In this section, I discuss the Best Practice Ontology for \xgls{ite} in relation to ontological theory, and what claims I seek to make with such a construct.
In Section~\ref{sec:methods_tax_theory}, I discuss taxonomic theory (which I refer to here as ontological theory), and in Section~\ref{sec:background_propositions}, I introduced the concept of propositions in theories.
I combine all these concepts together, to form a set of propositions towards future theory in the area of \xgls{ite} Best Practices.

The line between an ontology, and ontological theory, is not well described in the literature~\cite{Ralph_2018_TSE}.
In Section~\ref{sec:methods_tax_theory}, I discuss taxonomic theory, including the importance of grounding taxonomic and ontological models in rigorous secondary studies.
Ralph states that not all taxonomies are taxonomic theory~\cite{Ralph_2018_TSE}, but himself does not draw a clear line---as perhaps one is not possible, and requires consideration in each individual case.
As such, I will make my own attempt at drawing the line for this specific ontology.
I believe that this ontology lacks the rigorous data collection necessary to claim this ontology is also a theory, from the perspective of the data used to develop it.
There are other conditions for which an ontological theory should be evaluated, but the data from which it was formed is a central place of scrutiny~\cite{Ralph_2018_TSE}.
Another place of scrutiny, as discussed by Sjøberg et al.~\cite{Sjøberg_2008_Book} and Runeson et al.~\cite{Runeson_2012_Wiley}, is ``empirical support'': ``the degree to which a theory is supported by empirical studies that confirm its validity''.
While the formation of the ontology has been conducted using known rigorous secondary-study methods, there have been no follow-up studies on the theory itself to provide any empirical data on its utility or effectiveness.
For this reason, I refrain from calling this ontology a theory.
Instead, I frame five propositions, working towards a future theory.

I do not claim that these propositions \ltexdummy{form} a theory, nor do I claim that these propositions are all a subset of a single theory.
Rather, my claim is that these propositions each hold on their own, and some may be combined into a larger theory one day.
With additional work, empirical results, and reflection, these propositions can be the start of an ontological theory of \xgls{ite} Best Practices.
More importantly, however, these propositions should be challenged: refuted, strengthened, or replaced.
In the process of doing these things, our knowledge about quality factors for \xglspl{ite} will grow and mature.

\clearpage  
\begin{mybox}{Proposition 1}
    The Context dimensions enable falsifiability.
\end{mybox}

It is easy to claim, frame, and create a ``Smell'' because all you need is a single context in which it \textit{might be perceived} to be a problem, and the Smell holds.
This has many benefits, including the quick definition and recommendation of Smells to practitioners.
However, from a scientific perspective, Smells lack falsifiability, and therefore lack the ability to be challenged and improved over time.
It is possible to continually add potential contexts in which people believe a smell could apply.
However, Smells are---by definition---suppose to be indicators of potential problems, to be investigated by the stakeholders within their own given context, who then act as decision makers regarding how problematic the situation is, and how to act on that information.
That is both the benefit and the downside of Smells.
\xgls{ite} Best Practices, on the other hand, explicitly model context across multiple dimensions, with the strong statement that a violation \textit{is a problem}, and it holds under the specific context factors described.
This has downsides as well, including the difficulty involved with identifying and empirically validating these contextual factors.
However, that is precisely the purpose of empirical falsifiability in science: evidence needs to be gathered that supports or refutes the claims (propositions), such that over time the claims are strengthened or weakened.
I believe it would be a mistake to apply this perspective to Smells by extending their definition or application, and instead I believe \xgls{ite} Best Practices serve this purpose well.

\begin{mybox}{Proposition 2}
    The Context dimensions positively affect Best Practice adoption.
\end{mybox}

One of the primary uses for the Best Practice Ontology is to structure Best Practices that will be used in industry.
The Context dimensions explicitly model factors that need to be considered when implementing the Best Practice (or whether to implement them at all).
While the openness of Smells is a benefit that it doesn't restrict too much, it also does not provide guidance regarding where and when to implement them.
\xgls{ite} Best Practices, on the other hand, give plenty of implementation guidance in the form of Stakeholders, \xgls{its} issue types, \xgls{its} scope, and assorted inclusion and exclusion factors.
A well-formed \xgls{ite} Best Practice should sufficiently answer the practitioner question: ``Should we implement this in our context?''.
Therefore, I posit that the Context dimensions positively affects Best Practice adoption.

\begin{mybox}{Proposition 3}
    The positive framing of the Summary and Recommendation sections positively affects Best Practice adoption.
\end{mybox}

Both the Summary and the Recommendation sections are designed to explain what the Best Practice does and how to follow the practices, from a positive perspective.
In other words, they recommend what the goal is, and how to achieve it.
This is in contrast to Smells, which describe what problem could exist, and how to avoid it.
For most trivial recommendations, they can be equally worded in the positive or the negative.
For example, the \refBPTable{GoodBugReport} Best Practice could be just as easily described as ``avoid long paragraphs without structure, and don't miss important information such as software version number'', and called a ``Messy Bug Report'' Smell.
Over time, however, an organisation selecting many things \textit{to avoid}, does not form the same goal-oriented mindset as selecting many things \textit{to aim for}.
While framing the negative things to avoid helps those trying to avoid them, it doesn't help guide people.
As described above, it will take more effort to understand where you want to go as an organisation, and therefore which Best Practices to select and curate; however, the result is a much clearer understanding of organisational goals.
Additionally, achieving goals is akin to positive reinforcement, which ``enhances employee performance by encouraging desired [behaviours] and eliminating negative ones''~\cite{Wei_2014_AJIBM}.
Avoiding a negative outcome is akin to not breaking rules, or abiding the law, which is generally expected of everyone and not something to be rewarded.
Therefore, I posit that the positive framing of the Summary and Recommendation sections positively affects Best Practice adoption.

In the \xgls{ite} Best Practice catalogue, there are some Best Practices that are worded in the negative, and some that are even worded in both forms.
For example, \refBPTable{ActiveBugReports} compared to \refBPTable{AvoidZombieBugs}, which are the same underlying Best Practice, just worded in the positive and in the negative.
Admittedly, this is not the purpose of \xgls{ite} Best Practices; however, this initial catalogue is designed to showcase the structuring ability of the ontology, not showcase perfect Best Practices.
With so many recommendations worded negatively, I decided to include a few negatively framed Best Practices that were rather difficult to frame positively.
Additionally, I included the above example with both a positive and negative framing because I wanted to showcase an interesting case that deserves further analysis.
With that in mind, I hope these Best Practices are quickly challenged, updated, and added to, as per the primary function of the ontological structure.

\begin{mybox}{Proposition 4}
    The Stakeholder Benefits dimension positively affects Best Practice compliance and satisfaction doing so.
\end{mybox}

\xgls{se} is ultimately about building tools for users, which hopefully support the users and make their life easier.
Research has shown that developer's awareness of user involvement and feedback positively affects their satisfaction~\cite{AmoakoGyampah_1993_IM,Bano_2017_EMSE,Zowghi_2018_AffectRE}.
Research has also shown that employees gain job satisfaction when they prioritise collective interests~\cite{Clercq_2019_PR}, such as helping each other within the organisation.
With these perspectives in mind, it can be deduced that knowing who is positively impacted by a Best Practice will increase both the likelihood that the employee follows the Best Practice, and that they are satisfied doing so.
This is in contrast to Smells, where following the recommendation will benefit ``management'', in some broad, disconnected sense.
Therefore, I posit that the Stakeholder Benefits dimension positively affects Best Practice compliance and satisfaction doing so.

\begin{mybox}{Proposition 5}
    The Stakeholder Costs dimension increases empathy when considering the implementation of a Best Practice.
\end{mybox}

\xgls{ite} Best Practices explicitly model who will be negatively impacted by the implementation of the Best Practices.
Research has shown that knowing who is impacted can enhance empathic responses due to the ``neural activation shared between self- and other-related experiences''~\cite{Singer_2009_ANYAS,Bernhardt_2012_ARN}.
In this way, by revealing whom to account for when considering the implementation of a Best Practice, the Stakeholder Costs dimension will likely increase empathy and therefore understanding of the tradeoffs involved.
Without such a mechanism, it can be tempting to put too many restrictions or Smell detections, with the hope of improving the quality of \xgls{ite}, without knowing---or caring---who is negatively impacted.
Therefore, I posit that the Stakeholder Costs dimension increases empathy when considering the implementation of a Best Practice.

\section{Summary}

In this chapter, I formed the Best Practice Ontology for \xglspl{ite} using an approach that utilises both inductive and deductive cycles.
The resulting ontology can be used to structure existing research on quality factors for \xglspl{its}, and support future research and usage of Best Practices for \xglspl{ite}.
The need for this ontology was investigated in previous chapters.
The Context section in particular is the direct result of findings from Part~\ref{part:problem}, where the results consistently and repeatedly revealed the prevalence of diversity and complexity in these systems.
This includes the diversity and complexity of the problems presented by practitioners using \xglspl{its}, the \ltexdummy{artefacts and activities} in \xglspl{its}, and the information and evolution within \xglspl{its}.
When we showed the results of these Best Practices to practitioners, they agreed with the existence of most of the smells, and further confirmed the context-sensitive nature of when these smells occur and when they are problematic.
With the ontology now formed, the next step is to apply this ontology to existing literature on quality factors for \xglspl{ite}.
While the ontology is a useful tool for researchers, practitioners will get much more value out of the formed Best Practices themselves.
In the next chapter, I created a catalogue of Best Practices for \xglspl{ite}.

\chapter{Catalogue of Best Practices for Issue Tracking Ecosystems}  \label{ch:catalogue}

\epigraph{Research is seeing what everybody else has seen, and thinking what nobody else has thought.}{Albert Szent-Györgyi}

In this chapter I collect, structure, and present a catalogue of 40 \xglsfirst{ite} Best Practices.
In the previous chapter (Chapter~\ref{ch:ontology}), I proposed the Best Practices Ontology for \xgls{ite}, but not any actualised Best Practices.
Previous research has contributed many solutions and insights to improve the state of \xgls{its} and their ecosystems, but they do not structure them in such a way that my ontology does.
To highlight one dimension of value my ontology brings, and to contribute a starting place for the actualised ontology, I present here a catalogue of \xgls{ite} Best Practices.
I leverage existing research on \xgls{its} ``smells'' and ``Best Practices''.
I investigate this area to collect relevant findings that are lacking a comparable structure.
The results of this catalogue building process are 40 \xgls{ite} Best Practices.
The findings of this analysis contribute a concrete list of \xgls{ite} Best Practices, which the community can now utilise, build on, challenge, and update over time.
Following this chapter, I present {\numItebpAlgs} algorithms designed to automatically detect and repair violations of the \xgls{ite} Best Practices presented in this chapter.
While specific Best Practices within this catalogue already recommend algorithmic approaches that should be followed to detect violations, these recommendations are often pseudocode or just plain-text descriptions.

\section{Research Methodology}

My primary objectives with this chapter are to \textit{form} and \textit{validate} a catalogue of \xgls{ite} Best Practices that conform to the ontology I introduced in Chapter~\ref{ch:ontology}.
For the artefact-based catalogue formation process, I do not propose any research questions.
The goal is to produce catalogue items, and by the nature of artefact-based research, it will be successful.
For the validation of the catalogue, I formulate a single research question:

\begin{description}
    \item[RQ1] How do practitioners perceive issue \textit{smells}: do they occur and are they problematic?
\end{description}

I explain the two methodology stages in greater detail in this section.
For the formation of the catalogue, I describe the dataset used to create the catalogue, and I outline four methodological considerations for the final catalogue.
For the validation of the catalogue, I explain the interviews I conducted, including which constructs I asked them about.
Notably, I asked them about \textit{smells} and not \textit{violations} or \textit{Best Practices}.

\subsection{Forming the Catalogue: Inductive Extraction}

Methodologically, I chose to construct the catalogue using an inductive approach, utilising existing research in the domain of \xgls{ite} (compared to a deductive approach, constructing the catalogue from high-level principles, common knowledge, or personal experience).

\subsubsection{Catalogue Dataset}  \label{sec:catalogue_dataset}

I used two primary datasets to extract the Best Practices: 1) the set of articles from Chapter~\ref{ch:ontology} that also contain itemised Best Practices, and 2) articles that do not offer any ontology-like structural elements but still list itemised Best Practices.

The first set of articles from which I extracted Best Practices is the same set from Chapter~\ref{ch:ontology}, the formation of the ontology.
These articles were selected for the ontology creation because of the structural elements they impose on Best Practices; furthermore, they also contained Best Practices themselves.
I list them here, in chronological order of publication date:

\begin{description}[itemsep=-2mm]
    \item[2013] Heck and Zaidman: \ltexignore{``An Analysis of Requirements Evolution in Open Source Projects: Recommendations for Issue Trackers''~\cite{Heck_2013_IWPSE}}.
    \item[2016] Eloranta et al.: \ltexignore{``Exploring ScrumBut---An empirical study of Scrum anti-patterns''}~\cite{Eloranta_2016_IST} (extended from Eloranta et al.~\cite{Eloranta_2013_APSEC}, with structural inspiration from Brown et al.~\cite{Brown_1998_Book} and Sutherland et al.~\cite{Sutherland_2020_ScrumGuide}).
    \item[2016] Tamburri et al.: ``The Architect's Role in Community Shepherding''~\cite{Tamburri_2016_IEEESoftware}.
    \item[2020] Telemaco et al.: ``A Catalogue of Agile Smells for Agility Assessment''~\cite{Telemaco_2020_IEEEAccess} (extended from Telemaco et al.~\cite{Telemaco_2019_CIbSE}).
    \item[2022] Qamar et al.: ``Taxonomy of bug tracking process smells: Perceptions of practitioners and an empirical analysis''~\cite{Qamar_2022_IST} (extended from Qamar et al.~\cite{Qamar_SEAA_2021}).
\end{description}

The second set of articles from which I extracted Best Practices is articles that mention some kind of quality attribute for \xglspl{its}.
I list them here, in chronological order of publication date:

\begin{description}[itemsep=-2mm]
    \item[2006] Halverson et al.: ``Designing Task Visualizations to Support the Coordination of Work in Software Development''~\cite{Halverson_2006_CSCW}.
    \item[2009] Aranda and Venolia: ``The secret life of bugs: Going past the errors and omissions in software repositories''~\cite{Aranda_2009_ICSE}.
    \item[2010] Zimmerman et al.: ``What Makes a Good Bug Report?''~\cite{Zimmermann_2010_TSE} (extended from Bettenburg et al.~\cite{Bettenburg_2008_FSE}).
    \item[2023] Prediger: ``Visualising Data and Best Practices in Jira Issue Repositories''~\cite{Prediger_2023_MSc}.
    \item[2023] L{\"u}ders: ``Mining and Understanding Issue Links Towards a Better Issue Management''~\cite{Lüders_2023_PhDThesis}.
\end{description}

\subsubsection{Methodological Considerations}

\textbf{Scope of \xgls{ite} Best Practices.}
With the inclusion of certain articles above, the scope of what is considered an ``\xgls{ite}'' Best Practice is called into question.
\xglspl{its} are complex tools that support and manage many \xgls{se} processes, and therefore there is a significant overlap of concepts that involve \xglspl{its}.
What differentiates an \xgls{ite} Best Practice from a Best Practice that only applies to Scrum (for example), is that there is some meaningful aspect of the Best Practice that involves an \xgls{its}.
For example, the Scrum Best Practice ``Product Owner is not Customer''~\cite{Eloranta_2016_IST} is not an \xgls{ite} Best Practice because this is a role-based process recommendation that doesn't involve \xglspl{its} in any meaningful way.
The Scrum\index{Scrum} Best Practice ``Ordered Product Backlog'', however, \textit{is} an \xgls{ite} Best Practice because multiple aspects of the process, recommendation, and benefits all involve an \xgls{its} in some way.
Another way of looking at this overlap issue is to consider the fundamental purpose of an \xgls{its}: to support \xgls{se} processes.
All \xgls{ite} Best Practices will support some \xgls{se} processes.
Therefore, all \xgls{ite} Best Practices are also Best Practices for some \xgls{se} process, such as Scrum.
However, just because \xgls{ite} Best Practices are also \xgls{se} Best Practices, the opposite is not always the case (as described and illustrated above).

\textbf{Extrapolating Beyond Explicit Statements}
The process of structuring these Best Practices involved extrapolating beyond the explicit statements made in the articles.
Quotations have been used in the catalogue items where exact information has been copied.
Other, non-quoted information is created using deductive thinking, applying the collective knowledge of this thesis to act as a guide for completing each catalogue item.
Best Practices with less quoted information required more extrapolation, and require closer attention by future research.

\textbf{Empirical Evidence in this Initial Catalogue.}
It should be emphasised that the listed \xgls{ite} Best Practices are merely the structuring of statements from other researchers, this thesis provides no further evidence that these objects hold under the conditions stated.
While the purpose of the Best Practice Ontology for \xglspl{ite} is to increase the value of Best Practices through structure and guidance, it can also make the information appear more trustworthy or rigorous.
This thesis does not provide any additional evidence that should make these Best Practices more trusted.
However, the explicit structure should make it easier for researchers to identify which Best Practices require more evidence, and in what way (which dimensions, for example).
Additionally, it should be emphasised that most of the research that derived the catalogue items does not provide direct empirical evidence of the benefits of the Best Practices, or the harmfulness of the violations.
Most of the research presents analytical arguments based on \xglspl{spm} being presented as \xgls{se} theory.
While there is a lot of value in this approach, it does not provide direct empirical evidence that the Best Practices, embedded in the stated context, will provide the stated benefits.
This is also true of research into \xgls{ite} smells, which either extrapolate from \xglspl{spm}, or gather opinions and insights from practitioners through interviews.
Neither provide causal evidence of the impacts of these smells, violations, and Best Practices.
Despite these considerations, this catalogue provides considerable value to \xgls{se} researchers and practitioners.
The structuring of these Best Practices exposes these considerations, and offers a path forward for structuring future evidence.

\textbf{The Use of ``None'' and ``Unknown'' in the Catalogue.}
In the catalogue, the values for the dimensions will occasionally be either ``None'' or ``Unknown''.
``None'' represents that there is currently no known value for this dimension.
This is quite common for both Inclusion and Exclusion Factors.
In the initial formation of the Best Practices in this catalogue, however, ``None'' should be interpreted with care.
It is not clear whether the information is simply not listed in the article, does not exist in the opinion of the original authors, or perhaps does not yet have any empirical evidence to warrant listing such information.
I have done my best to extract meaningful insights out of the articles to fill in the Best Practices, but at times much of the information was missing.
In cases where it was really not clear what should go in a particular dimension, I then used the value of ``Unknown''.
In particular, ``Unknown'' was listed in dimensions where there should be a value, based on the Best Practice and the article describing it, but I was unable to find or deduce the information from the article.
For example, there should always be at least one Stakeholder Benefit (otherwise the Best Practice has no reason to exist), but I could not deduce a meaningful stakeholder benefit for \refBPTable{AssigneeBugResolution}.
I decided to leave this Best Practice in the catalogue, despite this missing information, for the exact purpose of highlighting cases like this.
It is not clear what information should go here (it is ``Unknown''), which is evidence that perhaps this Best Practice is not yet well-formed enough to be formulated and disseminated as a ``Best Practice''.

\subsection{Validating the Catalogue: Interviews with Practitioners}

In the same set of interviews reported in Chapter~\ref{ch:challenges}, I asked the interviewees about their perception of a set of \xgls{ite} smells.
To summarise the important methodological components, I interviewed 26 practitioners working in Germany, Canada, and Poland (full details listed in Table~\ref{tab:Participants}).
They had a range of work experience from 1.5--25 years (median 7 years).
They held various roles such as Developer, Manager, Product Owner, Solution Architect, and Consultant.
All participants used an \xgls{its} in their current position and had a range of experience using different \xglspl{its} throughout their careers.
The majority (14/26) primarily used Jira, with six more participants having experience with Jira.
Others used GitHub~\cite{GitHub_2024_Online}, GitLab~\cite{GitLab_2024_Online}, Trac~\cite{Trac_2024_Online}, Azure~\cite{Azure_2024_Online}, Bugzilla~\cite{Bugzilla_2024_Online}, Trello~\cite{Trello_2024_Online}, asana~\cite{asana_2024_Online}, Mantis~\cite{Mantis_2024_Online}, SpiraTest~\cite{SpiraTest_2024_Online}, RedMine~\cite{RedMine_2024_Online}, BaseCamp~\cite{BaseCamp_2024_Online}, and Miro~\cite{Miro_2024_Online}.
The size of their \xglspl{its} ranged from just a few hundred up to over a million issues.
The company sizes ranged from tens to thousands of employees, covering different industries including automotive, medical, consulting, and energy.

In the interviews, I chose to ask the practitioners directly about smells (instead of violations or Best Practices) due to the approachability of the concept.
While feedback on the entire concept of \xgls{ite} Best Practices is desirable, I deemed the concept too complex and broad to be explained and critiqued in a single interview setting.
Smells, on the other hand, are well-described, understood, and already utilised within industrial settings (whether for \xglspl{its} or tangential reasons such as code quality, as discussed in Section~\ref{sec:background_qsbp}).
The Best Practice Ontology for \xglspl{ite} has five sections, one of which is ``Violation'', with the first dimension being ``Smells''.
The Violation section represents the opposite behaviour or system state than is expected from the Best Practice.
While the framing of a Best Practice is intentionally positive, and not negative (see Proposition 3 in Section~\ref{sec:ontology_propositions}), validating the existence and problematicness of the smells (violations) provides strong evidence towards the Best Practice overall.

The goal for interviewing the participants about the \xgls{ite} Best Practice smells was to get their perception on occurrence and problematicness.
I showed them a list of pre-collected \xgls{ite} smells, and asked them ``have you observed this smell'' and ``do you think it is problematic''.
The list of smells used is based on a subset of the \xgls{ite} Best Practices in the catalogue.
I explain below in more detail which smells were selected, and why.\footnote{The list of smells is easier to understand, in context, once introduced to the entire \xgls{ite} Best Practice catalogue.}
Two data analysis methods were applied to the interview notes: \xgls{ta}, and a closed-labelling process.

We first conducted a \xgls{ta} of the notes to produce the qualitative findings.
To gain an overview of the responses towards the \xgls{ite} smells, we then conducted a closed-coding analysis of the RQ2 interview notes, in collaboration with another researcher.
This was possible due to the trivial nature of the questions for each smell: ``have you observed this smell'' and ``do you think it is problematic''.\footnote{These are summarised; see our interview protocol for original questions.}
We did this by coding their responses into one of four mutually exclusive categories: yes, no, depends, or no answer.
We first coded the responses separately, and then met to resolve the disagreements.
Additionally, we applied the ``depends'' label when the participant explicitly stated that there are some situations in which the smell is problematic, and others where it is not (although the reason was not always stated).
The question of ``occurrence'' does not warrant the ``depends'' code, since the participant either experienced the smell or not.
We coded ``no answer'' when the participant had no comment, or their response was not conclusive enough to be coded into one of the other three categories.\footnote{For example, one participant said, ``it is interesting'' in response to one of the smells, and nothing further.}

\section{The Catalogue of Best Practices}

The catalogue consists of 40 \xgls{ite} Best Practices.
In this chapter, I will describe the catalogue overall, the groups of Best Practices, and I will go into detail regarding a few key Best Practices.
Given the space consumed by each Best Practice, the full catalogue is listed in the Appendix~\ref{ch:appendix_catalogue}, instead of directly in this chapter.
In both this chapter and in the appendix, there is a table of contents for the \xgls{ite} Best Practices, to accommodate a succinct overview and quick lookup.
The table of contents, Table~\ref{tab:catalogue_toc_catalogue} (and Table~\ref{tab:catalogue_toc_appendix} in the appendix), presents the Best Practices in three groups: Issue Properties, Issue Linking, and Issue Processes.
Issue Properties Best Practices involve issue fields.
Issue Linking Best Practices involve the linking of issues to other issues.
Lastly, Issue Processes Best Practices involve \xgls{its} processes, including Scrum and generic Agile principles such as the backlog and sprints.

\bpToC{catalogue}

The full catalogue of 40 \xgls{ite} Best Practices is listed in Appendix~\ref{ch:appendix_catalogue}, but here I will go into detail explaining the following Best Practices: Good Bug Report, Bug-to-Commit Linking, Consistent Properties, and Active Bug Reports (Avoid Zombie Bugs).
All five of these Best Practices are from the Issue Properties group.
This is an intentional decision, as L\"uders~\cite{Lüders_2023_PhDThesis} and Prediger~\cite{Prediger_2023_MSc} already discussed the Issue Linking and Issue Processes Best Practices in great detail.
Neither of their works structure those Best Practices in the ontological form as my work, but they do focus on them as a key contribution of their work.
Accordingly, I have structured those Best Practices into their respective groups of the \xgls{ite} Best Practices catalogue, but I do not go into any further detail.

\subsection{Formalising a Well-Researched Concept: Good Bug Report}

\afterpage{%
    \ActivateWarningFilters[largefigure]  
    \clearpage
    \vspace*{\fill}
    \bpTable[example_]{GoodBugReport}
    \vspace*{\fill}
    \clearpage
    \DeactivateWarningFilters[largefigure]  
}

I formed the ``Good Bug Report'' Best Practice from the seminal work by Zimmerman et al.~\cite{Zimmermann_2010_TSE} and Bettenburg et al.~\cite{Bettenburg_2008_FSE}.
This is one of the earliest examples of strong empirical evidence being collected on a quality attribute of an \xgls{its}.
Their primary objective was to understand what makes a good Bug Report from the perspective of developers.
Their conference article by Bettenburg et al.~\cite{Bettenburg_2008_FSE} and their follow-up journal article by Zimmerman et al.~\cite{Zimmermann_2010_TSE} provide rich detail regarding their objectives and findings.

I chose this Best Practice as a starting place in this chapter because it is a well-known work, with solid empirical findings, that was relatively easy to extract into the \xgls{ite} Best Practice format.
This is an example of how the ontology can be used to structure existing strong research findings into a format for communication and comparison.
Thus, I formed the ``Good Bug Report'' Best Practice in Table~\ref{tab:example_bp_good_bug_report} (and in Table~\ref{tab:bp_good_bug_report} in the Appendix).

The interesting thing about this \xgls{ite} Best Practice is what is not yet included in the table: a detailed analysis of all follow-up works on Bug Report quality, to transform the Best Practice in its current form into a much more rigorously detailed Best Practice.
I created this Best Practice from the work of Zimmerman and Bettenburg alone, but there have been many follow-up articles regarding what makes a good Bug Report.
Those additional works add evidence towards or against certain claims and recommendations presented in the Best Practice.
Given the extent of the research in this area, there is enough evidence to make conclusive claims and prescriptions for practitioners.
The Best Practice Ontology for \xglspl{ite} can offer support in structuring and delivering this information to practitioners.

\subsection{Offering a Clear and Simple Solution: Bug-to-Commit Linking}

\afterpage{%
    \ActivateWarningFilters[largefigure]  
    \clearpage
    \vspace*{\fill}
    \bpTable[example_]{BugToCommitLinking}
    \vspace*{\fill}
    \clearpage
    \DeactivateWarningFilters[largefigure]  
}

I formed the ``Bug-to-Commit Linking'' Best Practice from the work of Qamar et al.~\cite{Qamar_SEAA_2021,Qamar_2022_IST}.
This is a popular concept in practice, that has received moderate attention in research ~\cite{Bachmann_2010_FSE,Bissyande_2013_CSMR,Qamar_SEAA_2021,Qamar_2022_IST}.
The primary objective with this Best Practice is to foster traceability between bugs and their resolving commits.
Table~\ref{tab:example_bp_bug_to_commit_linking} showcases this Best Practice.

This Best Practice is a good example of a clear and simple ``\xgls{its} Recommendation'': ``Have a dedicated `Commit' field that is mandatory before the Bug Report can be marked as `Resolved'.''
The previous Best Practice (``Good Bug Report''), can be automatically checked for violations using \xgls{nlp} and conformance templates; however, the certainty with which it can be detected, the preferences and ambiguities related to natural language descriptions, and the capabilities of the stakeholders managing the \xgls{its} all play a major role in how well the violations can be detected.
For Bug-to-Commit Linking, violation detection is 100\% accurate and easy to implement.
If an organisation decides they want to adopt this Best Practice, there is no barrier for them to implement its automation.

\subsection{Establishing a Best Practice Using the Ontology: Consistent Properties}

\afterpage{%
    \ActivateWarningFilters[largefigure]  
    \clearpage
    \vspace*{\fill}
    \bpTable[example_]{ConsistentProperties}
    \vspace*{\fill}
    \clearpage
    \DeactivateWarningFilters[largefigure]  
}

I formed the ``Consistent Properties'' Best Practice from the work of L\"uders~\cite{Lüders_2023_PhDThesis}.
The objective of this Best Practice is to maintain up-to-date, consistent properties within issues, attempting to combat the habit of \xgls{its} stakeholders who add properties information to the description or comments section, without updating them in their correct place.
For example, when someone adds a comment that this issue is a duplicate of another, but they do not add the ``duplicate'' link connecting the two issues.
This leads to issue properties that are inconsistent with the up-to-date information available on this issue.
Table~\ref{tab:example_bp_consistent_properties} showcases this Best Practice.

While L\"uders mentions this Best Practice in a list of smells, she goes into no further detail.
This \xgls{ite} Best Practice is an example of a initial idea from someone, that is first established within the structure of the Best Practice Ontology for \xglspl{ite}.
I used the \xgls{ite} Best Practice Ontology, and my tacit knowledge of \xglspl{ite}, to deductively fill in the entire Best Practice.
The result is just enough information to communicate the Best Practice for its intended purpose.
Now that the \xgls{ite} Best Practice has been formed, it is open to scrutiny, and hopefully eventually empirical evidence to strengthen or refute the claims made.

\subsection{Surfacing and Amending Duplicate Work}

\afterpage{%
    \ActivateWarningFilters[largefigure]  
    \clearpage
    \vspace*{\fill}
    \bpTable[example_]{ActiveBugReports}
    \vspace*{\fill}
    \clearpage
    \DeactivateWarningFilters[largefigure]  
}
\afterpage{%
    \ActivateWarningFilters[largefigure]  
    \clearpage
    \vspace*{\fill}
    \bpTable[example_]{AvoidZombieBugs}
    \vspace*{\fill}
    \clearpage
    \DeactivateWarningFilters[largefigure]  
}

I formed the ``Active Bug Reports'' Best Practice from the work of Qamar et al.~\cite{Qamar_2022_IST,Qamar_SEAA_2021} and Aranda and Venolia~\cite{Aranda_2009_ICSE}, and I formed the ``Avoid Zombie Bugs'' Best Practices from the work of Halverson et al.~\cite{Halverson_2006_CSCW}.
The objective for both of these Best Practices is to cultivate meaningful open Bug Reports through regular activity or archiving.
They share the same objective, and indeed they are essentially the same Best Practice, except one is framed from the positive (``Active''), and one is framed from the negative (``Avoid'').

I chose these two Best Practices to discuss here precisely because they are essentially the same Best Practice.
When I first encountered these two Best Practices, I considered combining them; however, the task of synthesising ideas such as Best Practices is a complex and tedious one.
The purpose of this catalogue is not to claim that these Best Practices are the reduced form of all possible Best Practices, and so I did not perform this task.
The work of Qamar et al.~\cite{Qamar_2022_IST,Qamar_SEAA_2021} from 2022 references the work of Halverson et al.~\cite{Halverson_2006_CSCW} from 2006, but only in the related work section, and not in the formative section outlining the Best Practice.
There is no clear scientific connection between the two specific Best Practices, and so they live in isolation from each other.
I did not notice their connection until I formed them into \xgls{ite} Best Practices, despite having a strong familiarity with both works.
With an ontological structure such as the \xgls{ite} Best Practice ontology, hopefully more such examples can be found and resolved.

\FloatBarrier  

\section{Practitioner Feedback on the ITE Best Practice Smells}

\subsection{Selected ITE Smells}  \label{sec:issue_tracking_smells}

I selected 31 smells in collaboration with two other researchers who were also researching similar topics.
My focus was on Issue Property smells, while the focus of the other researchers was on Issue Link smells and Issue Process smells.
Table~\ref{tab:catalogue_smell_list} lists the 31 smells, as well as the associated \xgls{ite} Best Practices.
The mapping from smell to \xgls{ite} Best Practice is not one-to-one.
A smell can have multiple Best Practices if the Best Practices are more granular than the smell, and vice versa.
A smell can also have no \xgls{ite} Best Practice if the smell was created or discovered independently of the Best Practice catalogue.
This is the case for many of the Issue Link and Issue Process smells, since they were created independently of my research.
For the Issue Property category, the only smell without a Best Practice is ``too many issue types'', which was a smell we added just for the interviews, to compare with the findings from the ``too many link types'' smell.
As most smells come from the \xgls{ite} Best Practices, the origin of the smells can be traced to the articles described in Section~\ref{sec:catalogue_dataset}.
In particular, we asked about the occurrence of these smells, as well as their opinions on how problematic each smell is.

\afterpage{%
    \ActivateWarningFilters[largefigure]  
    \clearpage
    \vspace*{\fill}
    
\newcommand*{\tabRef}[1]{%
    {\bp{#1}{UID} [\makebox[\widthof{A.10}][l]{\ref{tab:bp_\bp{#1}{REF}}}] \bp{#1}{NAM}}%
}

\begin{table}[ht]
    \footnotesize                           
    \centering                              
    \renewcommand{\arraystretch}{0.8}       

    \caption{List of smells discussed in the interviews.}
    \label{tab:catalogue_smell_list}

    \begin{tabularx}{\textwidth}{@{} l l >{\setlength{\baselineskip}{0.7\baselineskip}} X l @{}}
        \toprule
            & \textbf{ID} & \textbf{Smell} & \textbf{ITE Best Practice} (ID [Table] Name) \\
        \midrule
            \parbox[t]{1mm}{\multirow{18}{*}{\rotatebox[origin=c]{90}{\textbf{Issue Property Smells}}}}
            & S1.1  & No or short (non-informative) description &
                    \tabRef{SufficientDescription} \\
            & S1.2  & Description too Long &
                    \tabRef{SuccinctDescription} \\
            & S1.3  & Issues are missing properties (assignee, priority, severity, \dots) &
            \begin{tabular}[t]{@{} l @{}}
                    \tabRef{SetBugReportAssignee} \\
                    \tabRef{SetBugReportPriority} \\
                    \tabRef{SetBugReportSeverity} \\
                    \tabRef{SetBugReportEnvironment} \\
            \end{tabular} \\
            & S1.4  & Issues are assigned to a team &
                    \tabRef{AssignBugsToIndividuals} \\
            & S1.5  & Often switching properties (status or assignee, \dots) &
            \begin{tabular}[t]{@{} l @{}}
                    \tabRef{AvoidAssigneePingPong} \\
                    \tabRef{AvoidStatusPingPong} \\
            \end{tabular} \\
            & S1.6  & Too many issue types & \\
            & S1.7  & No link to commit &
                    \tabRef{BugToCommitLinking} \\
            & S1.8  & Non-assignee resolved issue &
                    \tabRef{AssigneeBugResolution} \\
            & S1.9  & Ignored issue/delayed for too long &
            \begin{tabular}[t]{@{} l @{}}
                    \tabRef{ActiveBugReports} \\
                    \tabRef{AvoidZombieBugs} \\
            \end{tabular} \\
            & S1.10 & No comments or too many comments &
                    \tabRef{BugReportDiscussion} \\
            & S1.11 & Toxic discussions &
                    \tabRef{RespectfulCommunication} \\
            & S1.12 & Properties discussed in comments but not updated in issue &
                    \tabRef{ConsistentProperties} \\
        \midrule
            \parbox[t]{1mm}{\multirow{12}{*}{\rotatebox[origin=c]{90}{\textbf{Issue Link Smells}}}}
            & S2.1  & Issue without any links & \\
            & S2.2  & Too many link types / Link types with overlapping meaning &
                    \tabRef{MinimalLinkTypes} \\
            & S2.3  & Multiple link with differing types to the same issue &
                    \tabRef{SingularRelationships} \\
            & S2.4  & Known link mentioned in comments but not documented &
                    \tabRef{RecordLinks} \\
            & S2.5  & Mismatch between link types and properties                         & \\
            & S2.6  & Mismatch between linked issues regarding their status, due date, priorities, or estimates & \\
            & S2.7  & Epic without sub-issues or sub-issues without main-issue &
                    \tabRef{ConnectedHierarchies} \\
            & S2.8  & Circular dependencies between issues &
                    \tabRef{RealisticDependencies} \\
        \midrule
            \parbox[t]{1mm}{\multirow{14}{*}{\rotatebox[origin=c]{90}{\textbf{Issue Process Smells}}}}
            & S3.1  & Unplanned work added during sprint                                 &
                    \tabRef{AvoidUnplannedWork} \\
            & S3.2  & Too many complex issues are assigned to the same sprint            & \\
            & S3.3  & Issues are missing an estimate                                     &
                    \tabRef{EstimateAllItems} \\
            & S3.4  & Estimate-scales differ between issues/ unclear estimate-scales     & \\
            & S3.5  & Sprint does not end at scheduled time                              &
                    \tabRef{ConsistentSprintLength} \\
            & S3.6  & Sprints have different duration                                    &
                    \tabRef{ConsistentSprintLength} \\
            & S3.7  & Sprint length differs from recommended length                      &
                    \tabRef{RecommendedSprintLength} \\
            & S3.8  & Sprint has to be (repeatedly) delayed                              & \\
            & S3.9  & No acceptance criteria or too many                                 &
            \begin{tabular}[t]{@{} l @{}}
                    \tabRef{UseAcceptanceCriteria} \\
                    \tabRef{LimitAcceptanceCriteria} \\
            \end{tabular} \\
            & S3.10 & Acceptance criteria are not checked during testing                 & \\
            & S3.11 & Acceptance criteria are changed during sprint                      & \\
        \bottomrule
    \end{tabularx}
\end{table}

    \vspace*{\fill}
    \clearpage
    \DeactivateWarningFilters[largefigure]  
}

\subsection{Results}

I describe here the results of showing each participant our list of smells and getting their feedback on 1) have they experienced these smells, and 2) how problematic are they.
I visualise the closed-coded smell responses in Figure~\ref{fig:smell_responses}.
Each cell in the figure represents a response from one of the participants.
The cells are across each of the 31 smells and 26 participants.
The initial coding of responses showed 74\% agreement ($\kappa = 0.61$), with a final agreement of 100\%.
I added four participant context factors from Table~\ref{tab:Participants} to the left side of Figure~\ref{fig:smell_responses} to make it easier to cross-reference them.
The bottom and right sides of the figure include the counts of responses for each smell and participant.

All the smells were noted as occurring by at least some participants.
Most of the smells were also noted as problematic by the participants, but 98 of the 290 responses were either ``no'' or ``it depends''.
Given the lack of contradictory evidence in research, these statements are the most interesting.
For this reason, I will focus the analysis on these responses.

\subsubsection{Issue Property Smells}

Participants largely agreed that Issue Property smells occur, but 28 of their responses said they are unproblematic, and 31 responses noted a related context factor.
Here I discuss those smells and the stated context factors in decreasing order of disagreement.

\textit{Issues are assigned to a team (S1.4)} received the most disagreement of the Property (8/9 participants that answered).
Most of the reasons for disagreement mentioned that issues were intentionally assigned to teams as part of their workflow.
P01 and P26 mentioned that issues are assigned to a team first, then to an individual.
P01 and P24 said that issues are assigned to teams to ``engage'' the teams, with the main assignment left to them to ``decide autonomously''.
P12 said that team assignments can be used for downstream activities such as ``in review'', whereby the team responsible for reviewing the issue will be assigned.
Finally, P14 mentioned that this kind of workflow dynamic ``depends on the team''.

\textit{Non-assignee resolved issue (S1.8)} received many disagreements (7/10 participants).
Many participants mentioned that non-assignees resolving the issue is an explicit part of their process.
For example, P01 mentioned that he is closing tickets as part of his role, regardless of the assignee.
P06 said that issues are often resolved as part of a group meeting.
P26 mentioned that it only happens when the original assignee cannot work on it, so the assignee is changed, but the original assignee could still change the stated to ``resolved''.

\afterpage{%
    \ActivateWarningFilters[largefigure]  
    \clearpage
    \begin{figure}[ht]
        \begin{adjustbox}{addcode={\begin{minipage}{\width}}{%
            \caption[Participant responses on smell occurence and problematicness.]{Participant responses on whether the smells occur (left), and whether they think they are problematic or not (right).\\Green: Agree, Red: Disagree, Blue: It Depends, Blank: No Response.%
            \label{fig:smell_responses}}
            \end{minipage}},rotate=90,center}
            \includegraphics[width=1.6\textwidth]{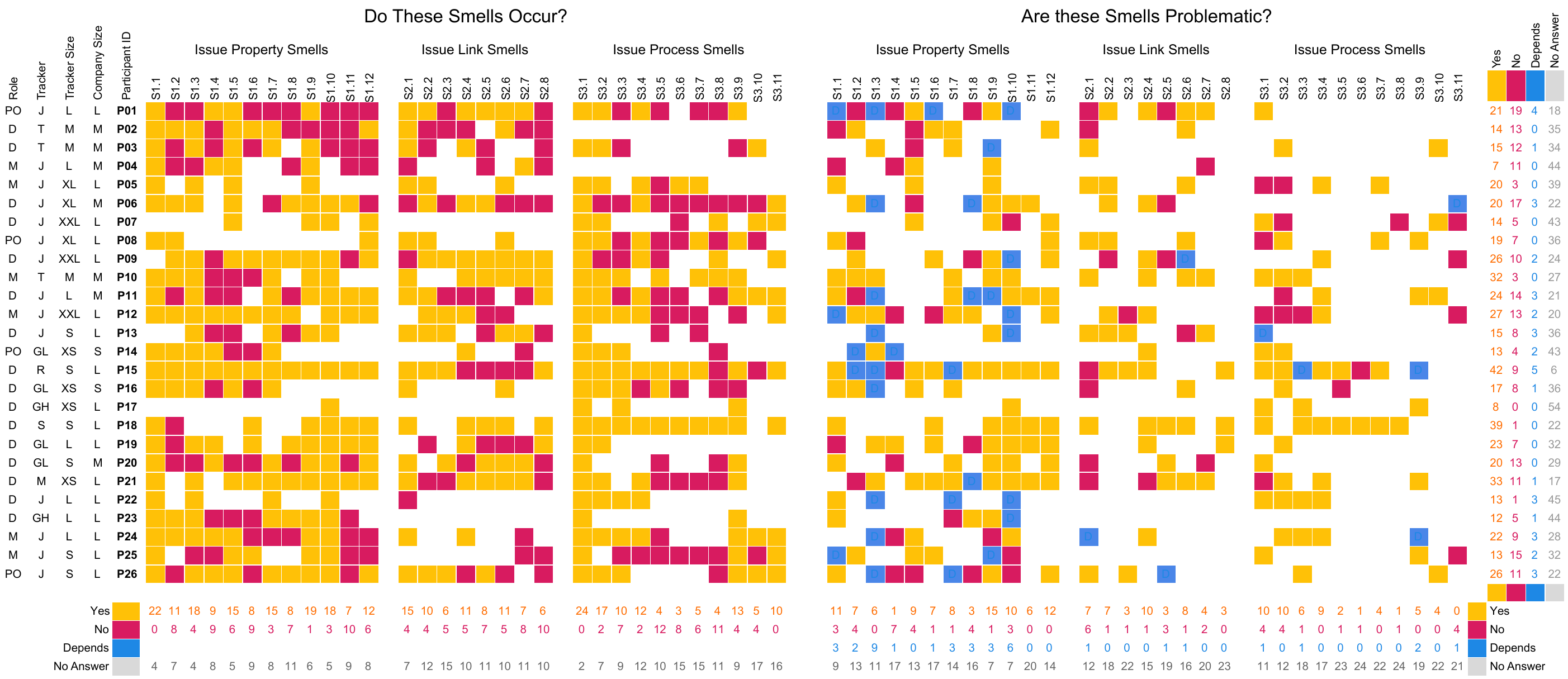}%
        \end{adjustbox}
    \end{figure}
    \clearpage
    \DeactivateWarningFilters[largefigure]  
}

\textit{Issues are missing properties (assignee, environment, priority, severity, ...) (S1.3)} also received many disagreements (9/15).
Three participants mentioned that fields may be blank at the beginning, but should later be filled.
P15 and P16 both said that fields such as the assignee and weight of the issue are assigned later, such as on the first day of the sprint.
P01 mentioned something similar, remarking that ``no assignee while the ticket is in progress would be considered bad''.
Participants also mentioned that it depends on the field, since fields such as Severity are important and cause for concern if missing (P26).
P16 said it depends on the field, and often the field is implied (known tacitly, given the context).
P22 mentioned that missing properties are ``not so important in small team settings, more in large settings''.
Interestingly, P06 and P24 mentioned that some fields are enforced by the \xgls{its}.
This means that they are so problematic if left empty, that the companies have put a technological barrier in place to prevent them from being blank.

\textit{No comments or too many comments (S1.10)} had 9/19 disagreements.
The primary disagreement about this smell is the number of comments considered bad.
P01 said he doesn't consider too many comments bad, but he also thinks that ``over 100'' would be a problem.
P07 and P26 both commented that ``no comments is not a smell'', with P26 elaborating ``because the daily [standup] should be used for that, [which then leads to] a description change''.
Others mentioned the number of problematic comments depends on the issue (P12), with long comments important to their role (P12), and ``unimportant stuff is the most problematic'' (P13).
P09 doesn't think it is a smell, but he also admitted that he ``doesn't want to read them all''.

\textit{Description too long (S1.2)} was another smell with split opinions (6/13).
Participants mentioned that long descriptions offer more information, which they value.
For example, P01 said long descriptions offer a ``better impression'', P08 values longer and more complex descriptions, and P11 said, ``the more information the better''.
Other participants said that it depends on the issue type.
P14 said it is ``context-dependent'' since long descriptions are not good for small Bug Reports, but epics often require longer descriptions.
P15 mentioned that long descriptions can be a problem, but noted that ``complex issues'' may require long descriptions.

\textit{\ltexignore{No} or short (non-informative) description (S1.1)} had 6/17 in disagreement.
The main reason participants said a short (or empty) description was not problematic was when the information could be found elsewhere.
P01 said that no description is usually bad, ``but there are [issue] types where a title is sufficient''.
P12 said that there is sometimes a comment that mentions external information, such as a Pull Request.
P04 mentioned that it is often sufficient if the issue has a Parent issue.
This hierarchical structure of issues can lead to well-defined Epics, with children User Stories that just contain a title and no description.
Participants P25 and P04 mentioned that ``too short'' is difficult to define.
P25 said, \ltexignore{``we like shorter descriptions, but single words can also be hard to interpret or remember''}.
One participant (P01) said that ``it depends on the context'' without further qualifier.
Finally, P12 also said that short descriptions \ltexignore{``are often not an issue, if the issue is created by me and assigned to me''}.

\textit{No link to commit (S1.7)} had 4/12 in disagreement.
P15 said that it depends on the issue type, since issues such as support requests don't have to be linked to anything.
P26 mentioned that it only matters if this issue and its link need to be tracked.
Participants mentioned that they had set up automation in their \xglspl{its} that automatically add a commit link when the issue is mentioned in the Git commit.
This is also the case for trackers such as GitHub that are already linked with the Git system.
These automations highlight the importance of linking.

\textit{Often switching properties (status, assignee, ...) (S1.5)} had 4/13 in disagreement.
P04 said that the assignee switching back-and-forth might be used by people ``to keep their statistics clean'' (fewer issues assigned to them).
It was also, however, used as an intended process similar to peer-review.
For example, P02 said it is ``part of their process to pass the issue back and forth between assignee and reviewer [as] their quality assurance''.

\textit{Ignored issue/delayed for too long (S1.9)} only had 4/19 in disagreement.
P03 mentioned that it is only a problem if the release is affected by it.
P11 said, \ltexdummy{``old backlog issues are often not important''}.
From an information perspective, P24 said the ``backlog is cared for'' because these ignored issues ``contain knowledge'' and therefore are important.
P24 added that some issues ``need some more time to ripen''.
P25 mentioned that how problematic these ignored issues are depends on the discipline of the team and the habits of interacting with the \xgls{its}.

\textit{Too many issue types (S1.6)} had 2/9 in disagreement, but no explicit contextual factors were mentioned.
The smells \textit{Toxic discussions (S1.11)} and \textit{Properties discussed in comments by not updated in issue (S1.12)} were stated to be problematic by all participants who gave a response (6 and 12, respectively).

\subsubsection{Issue Link Smells}

Participants largely agreed that Link Smells occur, but 15 of their responses disagreed that these smells are problematic, and 3 responses noted a dependent context factor.

\textit{Mismatch between link types and properties (S2.5)} had the most disagreement with 4/7 participants, but very few mentioned context factors.
P01 mentioned that if it happens, it tends to be with duplicate reports (wrong issue type of ``duplicate'' used instead of something else).

\textit{Issue without any links (S2.1)} had many disagreements with 7/14 participants.
P24 mentioned that if the issue is too small, then there is no need for links.
P02 said that certain types of issues, like support requests, often do not have or need links.

\textit{Mismatch between linked issues regarding their status, due date, priorities, or estimates (S2.6)} had 2/10 participants in disagreement.
P09 said that can happen, but only the status fields being out of sync matters, not the resolution.
Both P09 and P13 mentioned that this problem is largely solved by external communication regarding the issue.

\subsubsection{Issue Process Smells}

Participants largely agreed that Issue Process Smells occur, but 16 of their responses disagreed that these smells are problematic, and 5 responses noted a dependent context factor.

\textit{Acceptance criteria are changed during sprint (S3.11)} had full disagreement with all 5 participants that responded.
The consensus among the participants was that it is more important to update the acceptance criteria, than to leave it outdated when you know it should be changed.
As P25 stated, ``it is important to update the acceptance criteria if the change is needed''.

\textit{Unplanned work added during sprint (S3.1)} had 5/15 participants in disagreement.
The five participants who disagreed all mentioned that unplanned work was a normal part of sprint operations.
P08 said that it ``happens every sprint'' and is \ltexignore{``normal operations''}.
P12 emphasised that it is their ``default'' and represents standard procedure.
P21 added that this is an important topic, and it is not a problem ``as long as you get your work done''.

\textit{No acceptance criteria or too many (S3.9)} had 2/7 participants in disagreement.
P25 said that it ``depends on who is asking for the feature'', and noted that if any problem exists, it is more about no acceptance criteria than too many.
P24 said that it is not a problem since ``no acceptance criteria happens, and they are created after the fact''.

\textit{Too many complex issues are assigned to the same sprint (S3.2)} had 4/14 participants in disagreement.
P05 and P07 both said the smell is unproblematic because you can delay issues to the next sprint without repercussions.
P11 said complex issue count is not enough, since it is not a problem as long as the complex issues are distributed between people correctly.

\textit{Issues are missing an estimate (S3.3)} had 2/8 participants in disagreement.
P12 said that they don't do estimates because of their very complex code base, which always leads to wrong estimates anyway.
P15 mentioned that it depends on the sector you are working in, since not all companies have to agree on estimates ahead of time.

The following smells had disagreements, but no mentioned context factors: \textit{Sprint has to be (repeatedly) delayed (S3.8)}, \textit{Sprints have different duration (S3.6)} had 1/2 participants in disagreement, and \ltexignore{\textit{Sprint does not end at scheduled time (S3.5)}}.
The last three smells had no disagreements, with all participants who answered agreeing that the smells were problematic.
Those smells are \textit{Acceptance criteria are not checked during testing (S3.10)} (4 participants), \textit{Sprint length differs from recommended length (S3.7)} (4), and \textit{Estimate-scales differ between issues/ unclear estimate-scales (S3.4)} (9).

\section{Related Work}

\textbf{ITE Smells.}
This new area is the closest to our work, with preliminary studies that presented the starting point for our RQ2.
Aranda and Venolia~\cite{Aranda_2009_ICSE} conducted a survey and a mining study to extract coordination patterns such as ``forgotten'' and ``close-reopen'' bugs.
Eloranta et al.~\cite{Eloranta_2016_IST} conducted semi-structured interviews to identify 14 agile ``antipatterns''.
Recently, Qamar et al.~\cite{Qamar_SEAA_2021} proposed a taxonomy of 12 bug tracking process smells.
They surveyed 30 developers about these smells and mined their occurrences in 8 open-source projects~\cite{Qamar_2022_IST}.
Survey respondents also commented on actions they took to avoid the smells.
They observed a considerable amount of the smells in all projects, and the majority of surveyed practitioners agreed with the smells.
Our work covers 19 additional issue smells, including issue linking and process smells.
We ran hour-long interviews to uncover when and in which context the smells actually may lead to problems and why.
Finally, our goal is to understand the overall challenges in \xglspl{its}, including the smells and their management.
Our results discuss possible confounding factors that may explain such correlations.
Finally, Tuna et al.~\cite{Tuna_2022_ICSESEIP} studied the 12 smells by Qamar et al.~at JetBrains, surveying 24 developers at this company.
Similar to our results, they also found the perception of smell severity to vary across smell types, and that smell detection tools are considered useful for only six of the smells.
We studied a larger catalogue of smells by interviewing 26 practitioners in 19 different companies.

Halverson et al.~\cite{Halverson_2006_CSCW} observed collaboration antipatterns in \xglspl{its}.
Tamburri et al.~\cite{Tamburri_2014_JISA} defined and evaluated community smells~\cite{Catolino_ICSE_2021} that might lead to unforeseen project cost due to a `suboptimal' community.
They used the term ``community smell'' as organisational and social circumstances which cause mistrust, delays, and uninformed or miscommunicated architectural decision-making.
Palomba et al. observed that community smells contribute to the intensity of code smells~\cite{Palomba_2021_TSE}, which motivates our research on \xgls{ite} smells.

There are some studies, mostly surveys, evaluating the perception of \xgls{se} research among practitioners~\cite{Carver_2016_ESEM,Lo_2015_ESECFSE,Zou_2020_TSE}.
Zou et al.~\cite{Zou_2020_TSE} conducted follow-up interviews with 25 survey respondents provided further insight into their perspectives.
Zou et al.~\cite{Zou_2020_TSE} surveyed 327 practitioners to understand better how practitioners view these techniques.
The survey asked participants to rate the importance of various categories of automated Bug Report management techniques and provide their rationale.

\section{Discussion}

The findings highlight and confirm an important characteristic of \xgls{ite} smells: their relevancy to a given set of stakeholders is \textit{highly context-dependent}.
Research on \xgls{ite} smells up until this point has stated binary opinions on their applicability, usually related to a single easy-to-interpret contextual variable, such as ``uses agile''.
For example, if a company ``uses agile'', these smells apply to their company.
A similarly extreme---and incorrect---position to take would be to say that there are no shared context factors across interpretations of the smells, and every smell depends on the individual who interprets it.
\xgls{ite} Best Practices \textit{apply across consistent context factors}, but much more research needs to be conducted to formally understand and label these context factors on a per-best-practice basis.
For example, given the context factors ``company uses acceptance criteria'' and ``acceptance criteria apply to issue types x, y, z'', it is expected that the smell ``no acceptance criteria or too many'' applies.
While individual developers may disagree on a per-issue basis, having that smell auto-detected within their \xgls{its} is likely going to raise the quality of the \xgls{ite} process over time.
Currently, our understanding and definitions of \xgls{ite} smells are missing empirically grounded context factors.

Good and bad \xgls{ite} practices strongly vary among the studied practitioners and their organisations.
The results highlight that the perceived relevance and severity of 31 smells from the literature are context-dependent.
What is considered risky for some was a deliberate, intended procedure for others.
For instance, practitioners generally agreed that issues should not be ignored for too long, descriptions should be informative, and knowledge should not remain hidden in comments.
However, smells like ``issue ping-pong''~\cite{Halverson_2006_CSCW}, lack of comments on an issue, and issues being assigned to a team~\cite{Qamar_SEAA_2021} were an intentional part of their workflows.
This has several implications.
First, teams and even single users should be able to configure specific smell detection and management approaches:  not only what should be detected as smells but also at what threshold and for whom.
Such configuration can be challenging for administrators, particularly at the setup time of the \xgls{its}.
My work provides guidance for (a) what may matter in which context and (b) how to involve stakeholders in internal smell configuration study.

One interesting research direction is to ``learn'' the smells and their severity by analysing the interactions of stakeholders within an \xgls{ite}.
Such learned smells can assist \xgls{its} users externalise and assess their practices and potential smells.
Moreover, it remains unclear whether the perceptions of practitioners always reflects the objective practice.
Therefore, analytical approaches that visualise the overall state of \xglspl{ite} and the evolution of certain issues can be informative no matter whether a particular observation is an actual smell or not.

Interviewees discussed many broad problems they are experiencing while using \xglspl{ite}, but very few specific problems with granular solutions.
This highlights the need for more granular approaches to improving quality in \xglspl{ite}---such as well-defined context-dependent smells.
When asked about problems with their \xglspl{its}, large, unbounded, and vague problems surfaced, such as information overload, and workflow bloat.
These are real and serious problems that affect the daily lives of \xgls{se} practitioners, but they cannot be addressed directly.
\xgls{ite} smells, on the other hand, are designed to be specific units of potentially problematic situations, that can be automatically identified and explained to the user.
For example, Zombie Issues are a real problem that is easy to explain to users and can also be formulated as a concrete smell, such that a script can automatically notify the correct involved stakeholders when it is detected at sufficient levels; however, if left unattended (not identified automatically or manually), this leads to a general sense of anxiety and uncertainty with the \xgls{its}, since there are hundreds (if not thousands) of ageing issues that clog up the system.
This then further leads to additional problems such as searching difficulties (due to too many open issues), assignee ping-pong (to reduce personal accountability), and incorrect information in issues (since ageing tickets leads to loss of information).
Well-defined smells, with specific context factors, can be a remedy for these large uncertainties \xgls{ite} users face.

Finally, the results suggest that issue smell tools seem particularly useful for managers and product owners.
This can, however, be due to certain misconceptions or biases, such as developers being ``afraid'' of control, tool scatter, and overhead.
What makes smells particularly challenging is that their negative impact is often not immediate.
Future lab or observational studies with developers can control potential biases and clarify how smells should be presented and explained to different stakeholders in the \xgls{its}.
For instance, while issue structure visualisation could be helpful for navigating the knowledge graph around an issue, it may seem unnecessary or too complex for people who prefer list visualisation~\cite{Li_2012_REFSQ}.

\section{Summary}

In this chapter, I formed 40 Best Practices for \xglspl{ite} using an inductive approach.
The resulting catalogue can be used by practitioners to apply these recommendations in practice.
From a research perspective, researchers can now challenge and improve these Best Practices utilising a unified structure.
One finding from the catalogue-forming process is that some ontological attributes are discussed more than others in existing literature.
For example, the Violation section is heavily discussed in existing literature, while Context is quite rare.
Therefore, future work needs to fill this gap with case studies focused on investigating and enhancing existing Best Practices in industrial settings.

One core part of the Best Practices is their potential for automation.
The Violation section explicitly lists ``Algorithmic Detection'' as one of the dimensions.
Many of the formed Best Practices list either pseudocode or plain-text descriptions of how to algorithmically detect violations.
One way to further enhance the usability and application of these Best Practices in industrial settings, is to make these algorithms a reality.
In the next chapter, I implement these algorithms and apply them to my existing dataset of \xglspl{jr}.
The results showcase the prevalence of these violations in a real-world dataset of \xglspl{its}.

\chapter{Automation of Best Practices for Issue Tracking Ecosystems}  \label{ch:algorithms}

\epigraph{Automation is driving the decline of banal and repetitive tasks.}{Amber Rudd}

In this chapter, I demonstrate the application of algorithms to detect violations of Best Practices described in Chapter~\ref{ch:catalogue}.
Chapter~\ref{ch:ontology} introduced the ontology, thereby providing the structure necessary to describe \xgls{ite} Best Practices, and Chapter~\ref{ch:catalogue} then presented a catalogue of Best Practices formed into the ontology.
These chapters, however, don't go into detail regarding the algorithmic detection of associated violations.
While the ontology itself provides much value in structuring existing research and focusing future efforts, it does not require that all Best Practices have algorithmically detectable violations.
Simply by using the ontology as a guide, it is possible for an organisation to examine their \xglspl{ite}, reveal latent Best Practices, document them, and even start improving them by deploying manual processes.
However, it is much more desirable to have algorithmically enabled processes to automatically (or semi-automatically) detect violations of Best Practices.
Using these algorithms, practitioners can preemptively prevent issues from spreading within their \xglspl{its} and beyond.

To provide such algorithmic detection techniques, I conducted a mix of related work extraction and novel algorithm creation.
Some related work on \xgls{its} ``smells'' and ``Best Practices'' has provided algorithmic detection methods, in which case I cite and describe their work.
In other cases, I have developed novel detection algorithms.
I do not offer an exhaustive systematic list of algorithmic approaches, but rather an initial set of algorithms to show the feasibility and diversity of techniques available to support \xgls{ite} Best Practices.
As a result of this algorithmic exploration, I describe {\numItebpAlgs} algorithms that I designed or extended from the Best Practices.
Twelve of those algorithms are specific to Bug Reports, and 6 apply to all issue types.
This chapter contributes a concrete set of algorithms to detect violations of Best Practices, such that research can improve on them, and industry can apply them.

This chapter completes Part~\ref{part:solution} \nameref*{part:solution}, in combination with the previous Chapters~\ref{ch:ontology} and \ref{ch:catalogue}.
The next chapter starts Part~\ref{part:outlook} \nameref*{part:outlook}, which reflect on the previous chapters and further apply them.
In particular, the next chapter (Chapter~\ref{ch:tooling}) builds on the concepts of individual algorithms for violation detection, and looks at the greater picture of tooling for \xgls{ite} Best Practices.

\statementPublication{\cite{Montgomery_2025_BookChapter}}

\section{Research Methodology}  \label{sec:algorithms_methodology}

My primary objective with this chapter is to demonstrate the automation of \xgls{ite} Best Practices.
For this, I have a singular research question:
\begin{description}
    \item[RQ1] To what extent can \xgls{ite} Best Practice violations be automatically detected?
\end{description}

To investigate this research question, I explored the literature discussed in Chapters~\ref{ch:ontology} and \ref{ch:catalogue}, searching for algorithms.
Most of them describe detection techniques in at least pseudocode, while others say nothing.
As a result of this search, I implemented {\numItebpAlgs} algorithms.
To showcase the algorithmic detection of violations to \xgls{ite} Best Practices, I applied these algorithms to my Jira dataset presented in Chapter~\ref{ch:activities}.
This dataset has 16 Jiras, but I only use 13 of them in this chapter, for the same reason as described in Chapter~\ref{ch:evolution}: two of the original repos (MariaDB and Mindville) contain no comments, and Mojang only contains bugs.
To put that in a broader perspective for this chapter, presenting results from those repositories alongside the others would present a skewed perspective, since they are simply missing data that the others are not.
This would not be clear from the figures presented in this chapter, and would therefore give a false representation of the findings across the repositories.

For each \xgls{ite} Best Practice, I present the results of detecting violations in a consistent figure format.
This format can be seen in Figure~\ref{fig:bp_all_bug_reports}, which shows a summary of all \xgls{ite} Best Practice violations related to Bug Reports within the dataset.
Given the diversity of ways in which Jira is used, presenting a single value result (such as a global mean or median) would not provide meaningful insights into these ecosystems.
Accordingly, the figures split the results across Jiras and Activities (from Section~\ref{sec:artefacts_activities_findings}), presented as box plots across the distribution of projects at each intersection.
Each box plot represents the percentage of issues that violate the Best Practice, across all projects for a given Jira.

To describe Figure~\ref{fig:bp_all_bug_reports} in detail, I will use the first row as an example: ``\bp{SetBugReportAssignee}{UID}: \bp{SetBugReportAssignee}{NAM}''.
This row is plotting the percentage of violations to that Best Practice, which means that the ``Assignee'' field is not set on Bug Reports that are resolved and closed.
Each column of data represents a single \xgls{jr}, and the box plot visualises the percentage of violations to the total number of issues per project.
For example, the first column is visualising that the \xgls{jr} ``Apache'' has a median of 10\% violations to this Best Practice, across all projects in that \xgls{jr}.
In contrast, IntelDAOS has 0\% violations of this Best Practice.
Given the overall low violation rate, the y-scale is in symlog scale to highlight values in the 0--10\% range.
Other Best Practices, such as ``\bp{SetBugReportEnvironment}{UID}: \bp{SetBugReportEnvironment}{NAM}'', have their y-scale in linear scale, to better visualise the high violation rate.
The ``Overall'' plots presented on the right side of the figure show the results across all Jiras (still as a per-project distribution).

For the summary figures (Fig.~\ref{fig:bp_all_bug_reports} and Fig.~\ref{fig:bp_all_issue_types}): each row is a Best Practice being applied to all relevant issue types (i.e.~Bug Reports for Fig.~\ref{fig:bp_all_bug_reports} and all issue types for Fig.~\ref{fig:bp_all_issue_types}).
For the individual Best Practice figures (e.g., Fig.~\ref{fig:bp_\bp{SetBugReportAssignee}{REF}} or Fig.~\ref{fig:bp_\bp{SufficientDescription}{REF}}), each row is an issue type, or issue type theme.
This breakdown into issue type themes (where possible) allows an even more in-depth analysis of the violation rate for a given Best Practice.

For the {\numItebpAlgs} algorithms implemented in this chapter, there are two major categories: \xgls{ite} Best Practices related to Bug Reports and those that apply to all issue types.
This is not by design, but rather as a natural consequence of the \xgls{se} \xgls{its} research focus on Bug Reports over the last 20 years~\cite{Zhang_2016_CJ,Imran_2021_MSR}.
Since this thesis is primarily concerned with the collection and structuring of existing research, it makes sense that this is a major category of \xgls{ite} Best Practices.

\section{Algorithmic Detection of Violations: Bug Reports}

In this section, I implemented 12 algorithms that detect violations to \xgls{ite} Best Practices for Bug Reports.
These Best Practices are specifically designed for Bug Reports, whether for data structure reasons (e.g., custom priorities on Bug Reports), or process reasons (e.g., the inherent time pressure that exists for high-priority Bug Reports).
I discuss each of these algorithms in detail in the following subsections.
Figure~\ref{fig:bp_all_bug_reports} is a summary figure of all findings across eleven of the \xgls{ite} Best Practices for Bug Reports.\footnote{The twelfth algorithm was implemented outside the structure of this chapter, and therefore such a comparative figure cannot be produced.}
Each row represents a single Best Practice, where the violation detection algorithm was applied to the Bug Reports across the 13 Jira repos.

\afterpage{%
    \ActivateWarningFilters[largefigure]  
    \clearpage
    \begin{figure}[ht]
        \centering
        \vspace*{-10mm}  
        \makebox[0pt]{\includegraphics[width=1\textwidth]{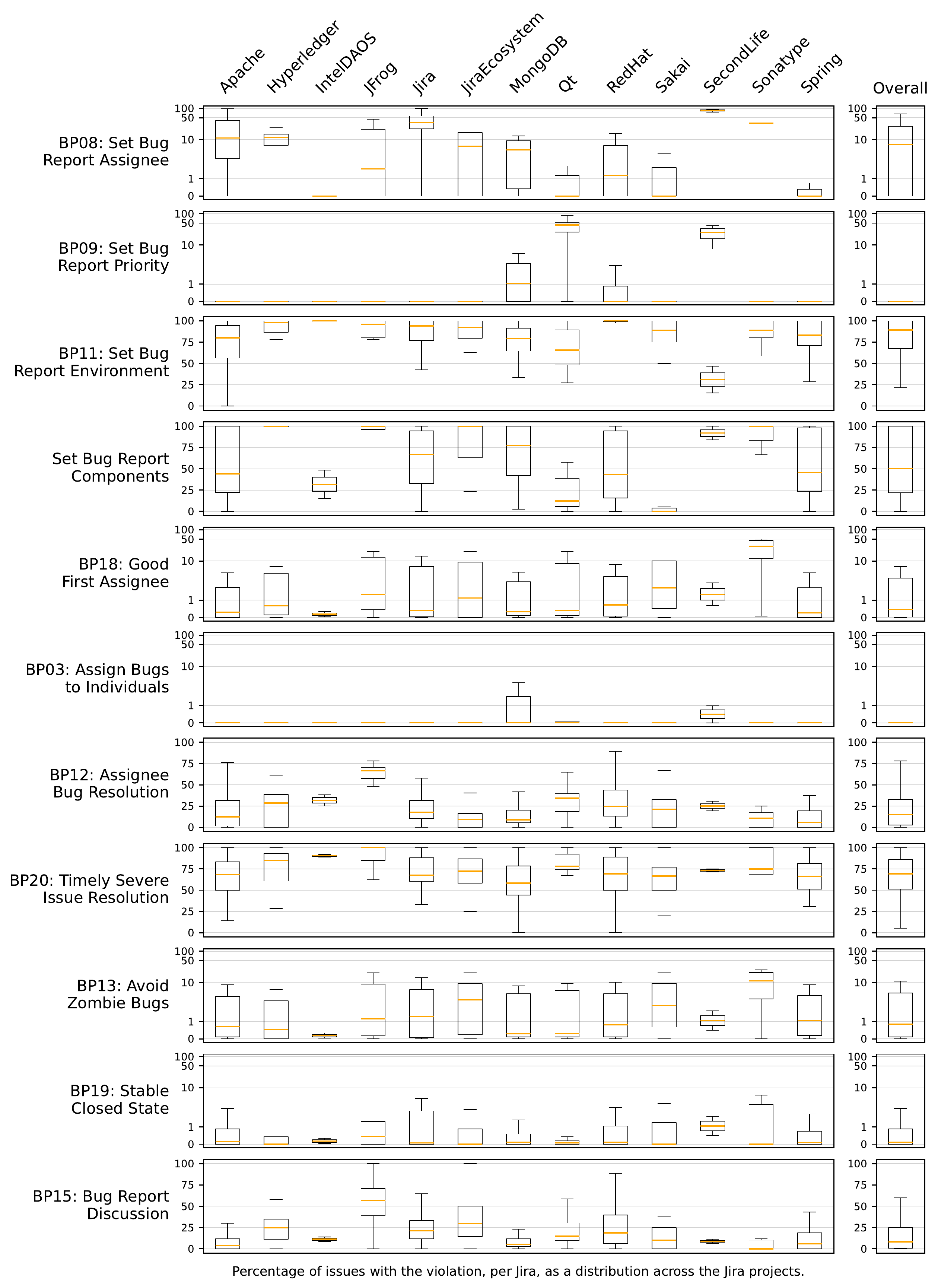}}
        \caption{Violations of Best Practices for Bug Reports.}
        \label{fig:bp_all_bug_reports}
    \end{figure}
    \clearpage
    \DeactivateWarningFilters[largefigure]  
}

These ``Bug Report'' algorithms \textit{can} also be applied to all other issue types, but the results should be interpreted with care.
They can be applied due to the shared nature of the underlying issue-type data models, but there are differences in which values are used for certain fields (such as priority).
Additionally, the recommendations of the \xgls{ite} Best Practice were designed specifically to apply to Bug Reports.
The result of applying these algorithms to other issue types can easily provide a false representation, or even a misleading one.
For comparison purposes, I have applied these algorithms across all issue type themes (Activities), but I only discuss some of those results in this chapter (the results with meaningful and correctly interpreted outcomes), and the figures are in Appendix~\ref{ch:appendix_additional_figures} to clearly separate and de-emphasise them.

Each row in Figure~\ref{fig:bp_all_bug_reports} is presented and discussed in greater detail in the following subsections.
Figure~\ref{fig:bp_all_bug_reports} provides an overview of the detailed discussions below, and as the only place these results can all be seen next to each other.
Six of the eleven sub-figures have their y-axis in symlog scale because the magnitude of those violation percentages are considerably low: generally between 1\% and 10\%.
The other five, however, are presented with a traditional linear scale.
This already says a lot about the differences across those violations, as some are much more prevalent than others.
Across the Best Practices, there are those with a median of 0\% violations (considering all projects from all Jiras), some at 50\%, and one at \mytilde90\%.
This shows the diversity of Best Practice violations, in terms of ``conformance'' or ``applicability'' to this generalised set of Jiras.

\FloatBarrier

\subsection{Set Bug Report Fields}

There is a group of Best Practices that are all related to setting Bug Report fields, including assignee (\bp{SetBugReportAssignee}{UID}), priority (\bp{SetBugReportPriority}{UID}), severity (\bp{SetBugReportSeverity}{UID}), environment (\bp{SetBugReportEnvironment}{UID}), and components.\footnote{The ``components'' version is not in my Best Practice catalogue (due to no one formally proposing it in the literature), but I have the data to apply these algorithms to it.}
These Best Practices, called ``Set Bug Report [Field]'' where [Field] is one of the above five fields, are all proposed by Qamar et al.~\cite{Qamar_2022_IST,Qamar_SEAA_2021}, and \textit{assignee} is also discussed by Prediger~\cite{Prediger_2023_MSc}.
These Best Practices are designed to encourage good Bug Report quality by having complete information on Bug Reports.
Without these fields, it is not possible to trace aspects such as who worked on Bug Reports (assignee), how important it was to the organisation (priority), how critical the problem itself was (severity), and where the problem was applicable (environment).

\begin{figure}[ht]
    \centering
    \includegraphics[width=\textwidth]{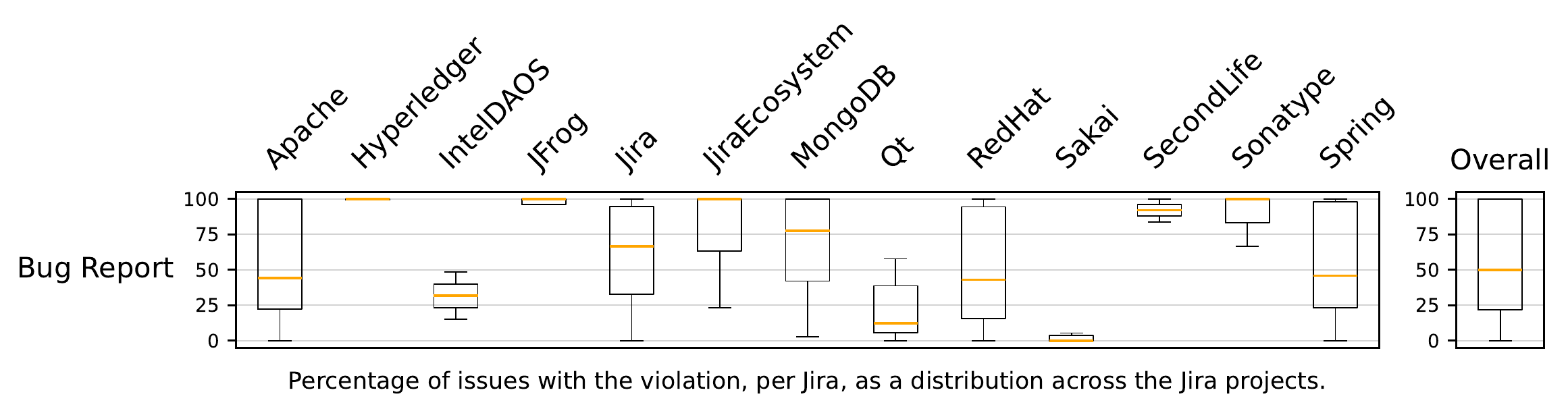}
    \caption{Violations to Set Bug Report Components (no catalogue item).}
    \label{fig:bp_set_bug_report_components}
\end{figure}

Qamar et al. describe an algorithmic approach in which the bug must be both \textit{fixed} (resolved) and \textit{closed}, and yet the fields are still unassigned (empty).
To capture the diverse ways in which Jira is used by different organisations, I performed an analysis of all potential statuses and resolutions, to produce a synonym list for what is ``fixed'' and what is ``closed''.
Using the results of that analysis, I implemented these algorithms, as described by Qamar et al.~\cite{Qamar_2022_IST}.
\begin{quote}
    ``First, we check whether the bug is fixed and closed. If so, we check whether the assignee field is empty or not. However, there are also some cases where the assignee field is not empty but an invalid email address is written. For instance, if the unassigned@gcc.gnu.org address is used as an assignee email, we consider this case a smell as well.''~\cite{Qamar_2022_IST}
\end{quote}
I present the results of these algorithmic violation detections in Figures~\ref{fig:bp_\bp{SetBugReportAssignee}{REF}}~(assignee), \ref{fig:bp_\bp{SetBugReportPriority}{REF}}~(priority), \ref{fig:bp_\bp{SetBugReportEnvironment}{REF}}~(environment), and \ref{fig:bp_set_bug_report_components}~(components).
Notably missing from these implementations is \textit{severity}, which doesn't exist as a field in Jira.\footnote{Online discussions regarding the severity field speculate that it is missing because it conceptually overlaps with \textit{priority} too much, creating more confusion than it \ltexdummy{helps~\cite{Jira_Missing_Severity_URL_1,Jira_Missing_Severity_URL_2,Jira_Missing_Severity_URL_3}}.}

\begin{figure}[ht]
    \centering
    \includegraphics[width=\textwidth]{2_figure/fig_bp_\bp{SetBugReportAssignee}{REF}.pdf}
    \caption{Violations to \bp{SetBugReportAssignee}{NAM} (\bp{SetBugReportAssignee}{UID}).}
    \label{fig:bp_\bp{SetBugReportAssignee}{REF}}
\end{figure}

First, I discuss the ``assignee'' field, displayed in Figure~\ref{fig:bp_\bp{SetBugReportAssignee}{REF}}.
Overall, \mytilde9\% of Bug Reports suffer from this violation.
Given the certainty of the detection method of this violation, this 9\% is rather concerning.
A fixed Bug Report must have been fixed by someone, and a closed state means that there is nothing more to do with this issue.
Therefore, with near-certainty, these 9\% of Bug Reports are missing information that should be there.
This 9\%, however, is not consistent across the Jiras.
Some Jiras have as much as 30\% of their Bug Reports missing the assignee (such as ``Jira''), and others have a 0\% violation rate, such as IntelDAOS.
Future research should investigate the rationale behind excluding the assignee from Bug Reports in organisations with a high violation rate.

\begin{figure}[ht]
    \centering
    \includegraphics[width=\textwidth]{2_figure/fig_bp_\bp{SetBugReportPriority}{REF}.pdf}
    \caption{Violations to \bp{SetBugReportPriority}{NAM} (\bp{SetBugReportPriority}{UID}).}
    \label{fig:bp_\bp{SetBugReportPriority}{REF}}
\end{figure}

Next, I discuss the ``priority'' field, displayed in Figure~\ref{fig:bp_\bp{SetBugReportPriority}{REF}}.
Across the Jiras, there is a very low violation rate, i.e., very few Bug Reports are missing the priority.
For the nine Jiras that have absolute 0\% violations, this field could be mandatory on Bug Reports (or all issue types).\footnote{Looking at the complete results across all issue types, presented in Figure~\ref{fig:bp_\bp{SetBugReportPriority}{REF}_all}, it indeed seems that ``priority'' is a required field across most issue types.}
For the remaining four Jiras, it seems MongoDB and RedHat (with less than 10\% violations) likely care about setting the priority, but for some reason do not enforce it, while Qt and SecondLife (with \mytilde40--50\% violations) do not use this field consistently.

\begin{figure}[ht]
    \centering
    \includegraphics[width=\textwidth]{2_figure/fig_bp_\bp{SetBugReportEnvironment}{REF}.pdf}
    \caption{Violations to \bp{SetBugReportEnvironment}{NAM} (\bp{SetBugReportEnvironment}{UID}).}
    \label{fig:bp_\bp{SetBugReportEnvironment}{REF}}
\end{figure}

Next, I discuss the ``environment'' field, displayed in Figure~\ref{fig:bp_\bp{SetBugReportEnvironment}{REF}}.
This Best Practice has the highest violation rate across all Bug Report Best Practices, with a median project violation rate of \mytilde90\% across all Jiras.
This violation rate is fairly similar across all Jiras, including some that have a median of 100\% (they do not use this field at all), and some that are closer to 75\% (barely use this field).
My interpretation of these results is that setting the environment field on Bug Reports is not a priority.\footnote{The violation rate on Requirements and Development issue types are even higher (see Fig.~\ref{fig:bp_\bp{SetBugReportEnvironment}{REF}_all}).}
An extension of this interpretation is that setting the environment field on Bug Reports is not a Best Practice.
However, it should also be considered that perhaps there are good reasons to have this Best Practice.
Future research needs to provide strong empirical evidence to the claim that setting the environment field on Bug Reports is indeed an \xgls{ite} Best Practice that should be followed.

Next, I discuss the ``components'' field, displayed in Figure~\ref{fig:bp_set_bug_report_components}.
The results show a varying degree of violations, from 100\%~(Hyperledger), to near 0\%~(Sakai).
The median across all projects is 50\%, which highlights the varying degree of violations.
It seems this Best Practice is important in some contexts, and not in others.
Further evidence for this context-specific assumption is the wide variance on the upper- and lower-quartiles, across multiple Jiras.
The evidence suggests that whether projects prioritise the environment field on Bug Reports is very project-dependent, with many at 100\% violation rate (they do not consider their absence to be a violation), and many near 0\% (they heavily use this field).
Apache, RedHat, and Spring, for example, have upper-quartiles near 100\%, and lower-quartiles near \mytilde25\%.
If an organisation wants to implement the Set Bug Report Components Best Practice, they should allow for project-specific opt-out to satisfy the needs of diverse project types.

As described in previous Best Practices, I also visualised the results for all issue type themes in figures available in the Appendix~\ref{ch:appendix_additional_figures}.
Interestingly, there is little difference across the issue type themes, compared to that of Bug Reports.
There is a near-identical trend in the percentage of violations, per Jira, across the issue type themes.
This trend holds largely true for all ``Set Bug Report Fields'' Best Practices, and across all organisations in my dataset.

\FloatBarrier

\subsection{\bp{GoodFirstAssignee}{NAM}}

\subsubsection{\bp{GoodFirstAssignee}{NAM} (\bp{GoodFirstAssignee}{UID})}

The ``\bp{GoodFirstAssignee}{NAM}'' Best Practice (see Table~\ref{tab:bp_\bp{GoodFirstAssignee}{REF}}) was proposed and later extended by \bp{GoodFirstAssignee}{AUT}.
The Best Practice is designed to decrease bug resolution time by ensuring a good first assignee on Bug Reports.
The idea being that performing multiple re-assignments increases the time to complete \ltexdummy{bugs~\cite{Qamar_SEAA_2021,Qamar_2022_IST}}.
The algorithmic approach is to look for evolutions of the assignee field, since every evolution is a re-assignment.
I don't count the ``creational'' evolutions, where the assignee is first set.
Qamar et al. also recommend not to count rapid reassignments: when an assignee is re-assigned within a 5-minute interval.
\begin{quote}
    \ltexignore{``The mining strategy for this smell is to look in the bug history for the assignee property. If the assignee field is changed more than twice, then we count it as a smell. Also, we observed that there are some multiple assignee changes done in a very short interval. We consider these multiple assignee changes as a mistake and do not count them as a smell. The interval duration is set as five minutes. Therefore, if multiple assignments are done in five minutes, they are counted as one assignment.''~\cite{Qamar_2022_IST}}
\end{quote}

While Qamar et al. count violations as ``more than twice'' (without justification), I count violations as more than once.
The first assignment is necessary, but the second assignment is already the first re-assignment.
I present the results of this implementation in Figure~\ref{fig:bp_\bp{GoodFirstAssignee}{REF}}.

\begin{figure}[ht]
    \centering
    \includegraphics[width=\textwidth]{2_figure/fig_bp_\bp{GoodFirstAssignee}{REF}.pdf}
    \caption{Violations to \bp{GoodFirstAssignee}{NAM} (\bp{GoodFirstAssignee}{UID}).}
    \label{fig:bp_\bp{GoodFirstAssignee}{REF}}
\end{figure}

The results show that this violation is quite rare, with a median occurrence rate of \mytilde0.5\%
This highlights the importance of this Best Practice to these organisations.
The log-scale figure shows that most \xglspl{jr} are at around 1\% violation rate, with only Sonatype having a high rate at \mytilde30\%.
Future work can unpack why Sonatype's violation rate is so high.

\FloatBarrier

\subsubsection{\bp{AssignBugsToIndividuals}{NAM} (\bp{AssignBugsToIndividuals}{UID})}

The ``\bp{AssignBugsToIndividuals}{NAM}'' Best Practice (see Table~\ref{tab:bp_\bp{AssignBugsToIndividuals}{REF}}) was proposed and later extended by \bp{AssignBugsToIndividuals}{AUT}.
The Best Practice is designed to foster traceability and ownership of Bug Reports by linking the individual assignee to the issue \textit{before it is resolved} (closed).
A not uncommon practice is to assign a team, group, or department to a Bug Report, which is then later assigned to one of the individuals within that team.
However, it is sometimes the case that someone will fix the bug and resolve the Bug Report, without assigning themselves to the Bug Report.
Without such a Best Practice, it becomes difficult to follow up on resolved bugs, particularly if they need to be reopened or investigated.

Qamar et al. describe a programmatic approach to detecting this smell~\cite{Qamar_SEAA_2021,Qamar_2022_IST}.
Search for Bug Reports that are resolved, and then check if the assignee is an individual.
I implemented their approach and applied it to my Jira dataset, producing the results displayed in Figure~\ref{fig:bp_\bp{AssignBugsToIndividuals}{REF}}.
\begin{quote}
    ``First, we check whether the bug is assigned. If so, we search for the selected keywords: `team', `group' and `backlog' in the assignee field. We found those keywords by manually inspecting the assignee names in each project. For example, in the MongoDB Core Server project history, there are some bugs assigned to \ltexignore{Backlog-Sharding} Team, which we consider it as a smell.''~\cite{Qamar_2022_IST}
\end{quote}

\begin{figure}[ht]
    \centering
    \includegraphics[width=\textwidth]{2_figure/fig_bp_\bp{AssignBugsToIndividuals}{REF}.pdf}
    \caption{Violations to \bp{AssignBugsToIndividuals}{NAM} (\bp{AssignBugsToIndividuals}{UID}).}
    \label{fig:bp_\bp{AssignBugsToIndividuals}{REF}}
\end{figure}

There are very few violations to this Best Practice.
MongoDB and SecondLife have the most violations.
MongoDB has an upper quartile at \mytilde2\%, and an upper whisker at \mytilde5\%.
SecondLife has the only median above 0\%, at \mytilde0.5\%, and a tight upper and lower quartile only \mytilde0.25\% above and below.
These results show that the Assign Bugs to Individuals Best Practice is rarely violated.

As described in the Best Practice table (Table~\ref{tab:bp_\bp{AssignBugsToIndividuals}{REF}}), this is a Best Practice that is only contextually relevant to Bug Reports.
Figure~\ref{fig:bp_\bp{AssignBugsToIndividuals}{REF}_all} also displays the results for the three issue type themes, Requirements, Development, and Maintenance.
MongoDB and SecondLife are again the only two Jiras that show any notable violations, and they continue to be low across the issue type themes.
Development is the only issue type that shows effectively no violations, which is an interesting area for future study.

Overall, the Assign Bugs to Individuals Best Practice is closely followed by the organisations under study.
Whether implicitly or explicitly, organisations seem to agree that Bug Reports (and other issue types) should be assigned to an individual before they are finally resolved.

\FloatBarrier

\subsubsection{\bp{AssigneeBugResolution}{NAM} (\bp{AssigneeBugResolution}{UID})}

The ``\bp{AssigneeBugResolution}{NAM}'' Best Practice (see Table~\ref{tab:bp_\bp{AssigneeBugResolution}{REF}}) was proposed and later extended by \bp{AssigneeBugResolution}{AUT}.
The Best Practice is designed to foster ownership and traceability between the assignee of a Bug Report, and the process of resolving it.
If anyone can resolve the report, then there is less ownership on the assignee to act on it.
Qamar et al. describe a programmatic approach to detecting this smell~\cite{Qamar_SEAA_2021,Qamar_2022_IST}.
Search for Bug Reports that are resolved, and then check the assignee against the resolver.
I implemented their approach, producing the following results displayed in Figure~\ref{fig:bp_\bp{AssigneeBugResolution}{REF}}.
Listed verbatim, their approach is:
\begin{quote}
    ``First, we check whether the bug is assigned and resolved. If so, we compare the assignee and the person who resolved the bug.''~\cite{Qamar_2022_IST}
\end{quote}

\begin{figure}[ht]
    \centering
    \includegraphics[width=\textwidth]{2_figure/fig_bp_\bp{AssigneeBugResolution}{REF}.pdf}
    \caption{Violations to \bp{AssigneeBugResolution}{NAM} (\bp{AssigneeBugResolution}{UID}).}
    \label{fig:bp_\bp{AssigneeBugResolution}{REF}}
\end{figure}

All \xglspl{jr} have some number of violations to this Best Practice, from \mytilde10\% (Spring) all the way up to \mytilde65\% (JFrog).
The \xglspl{iqr} are rather large, and the whiskers are also rather pronounced.
These results show that most projects in my dataset do not conform to this Best Practice.
Given the logical nature of this Best Practice, and the benefits provided, future work should investigate why this does not appear to be important within industrial projects.

\FloatBarrier

\subsection{Closing Bug Reports}

\subsubsection{\bp{TimelySevereIssueResolution}{NAM} (\bp{TimelySevereIssueResolution}{UID})}

The ``\bp{TimelySevereIssueResolution}{NAM}'' Best Practice (see Table~\ref{tab:bp_\bp{TimelySevereIssueResolution}{REF}}) was proposed by \bp{TimelySevereIssueResolution}{AUT}.
The Best Practice is designed to encourage quick resolution of Bug Reports.
In particular, quick resolution of Bug Reports that have been identified as ``severe'' in some way, either by the customer or the organisation.
The basic idea is that when a Bug Report is labelled as ``highly severe'' (a definition which differs from organisation to organisation), then the Bug Report should be closed (and hopefully resolved) within 48 hours.
The choice of 48 hours is simply stated as the default time chosen by one of the ``informants'' of Halverson et al.~\cite{Halverson_2006_CSCW}.
To detect violations to this Best Practice, we first have to identify what a ``high severity'' is within each \xgls{jr}.
I performed a manual analysis of all possible Priority values in my dataset, looking for those to label as ``high'' severity.\footnote{Jira does not have a ``Severity'' field; rather, they use the Priority field.}
For each potential value, I have the name of the Priority, as well as how many issues it was used in.
With these two pieces of information, here are the values I chose to mean ``high severity'':

\begin{itemize}
    \item \textbf{(> 1,000 issues each)}: \ltexignore{Critical, Blocker, P1: Critical, Highest, Critical - P2}
    \item \textbf{(< 1,000 issues each)}: \ltexignore{Urgent, Blocker - P1, P1, 2 - Critical, P1-Urgent, P0, 1 - Blocker}
    \item \textbf{(< 100 issues each)}: \ltexignore{P2-Critial, P1-Blocker, Blocking, Severe}
\end{itemize}

I then compare the closed date to the created date.
For my analysis, I selected a timeframe of 7 days.
My familiarity with the customer support domain~\cite{Montgomery_2017_RE,Montgomery_2017_RE_a,Montgomery_2018_REJ} has taught me that while 48 hours is commendable, it is not reasonable for many organisational contexts.
Instead, I selected a more relaxed one-week timeframe, allowing time for organisational units to connect and resolve the issue.
I present the findings of this algorithm in Figure~\ref{fig:bp_\bp{TimelySevereIssueResolution}{REF}}.

\begin{figure}[ht]
    \centering
    \includegraphics[width=\textwidth]{2_figure/fig_bp_\bp{TimelySevereIssueResolution}{REF}.pdf}
    \caption{Violations to \bp{TimelySevereIssueResolution}{NAM} (\bp{TimelySevereIssueResolution}{UID}).}
    \label{fig:bp_\bp{TimelySevereIssueResolution}{REF}}
\end{figure}

The results show that, despite the generous 7-day resolution window, a median of \mytilde50\% of issues across the projects violate this Best Practice.
The 50\% is fairly consistent across the \xglspl{jr}, except IntelDAOS and SecondLife at \mytilde75\%.
This result is difficult to interpret without diving deeper into the data, and cross-checking with stakeholders within these organisations.
They might conform to the idea that high-priority Bug Reports should be solved quickly, but only meet this 7-day threshold 50\% of the time.
It could also be that response and resolution times in \xgls{oss} projects are longer than in projects that normally cater to paying customers.
Regardless, it is clear that even at 7 days, this Best Practice is often violated in my dataset.

\FloatBarrier

\subsubsection{\bp{AvoidZombieBugs}{NAM} (\bp{AvoidZombieBugs}{UID}) \& \bp{ActiveBugReports}{NAM} (\bp{ActiveBugReports}{UID})}

The ``\bp{AvoidZombieBugs}{NAM}'' Best Practice (see Table~\ref{tab:bp_\bp{AvoidZombieBugs}{REF}}) was proposed by \bp{AvoidZombieBugs}{AUT}, and the ``\bp{ActiveBugReports}{NAM}'' Best Practice (see Table~\ref{tab:bp_\bp{ActiveBugReports}{REF}}) was proposed by \bp{ActiveBugReports}{AUT}.
These Best Practices were designed to foster a meaningful backlog by addressing Bug Reports before they get too old.
The idea being that after a certain period of inactivity, Bug Reports must either be closed, or worked on.
Of course, if organisations had the time to address all their Bug Reports, they would, so the problem is clearly resources, not desire.
However, this Best Practice emphasises a \textit{meaningful} backlog, which means that if something is in the backlog, it must still have meaning to the software.
Not all closed Bug Reports need to be resolved, and this Best Practice encourages closing Bug Reports that the organisation has no intention of addressing \textit{in the near future}.
How to define ``near future'' is the hard part.
Neither Halverson et al.~\cite{Halverson_2006_CSCW} nor Aranda and Venolia~\cite{Aranda_2009_ICSE} specify a timeline, whereas Qamar et al.~\cite{Qamar_2022_IST} specify a timeline of three months.
To detect violations to this Best Practice, I review all evolutions to every issue, and count the issues that have a gap in activity of three months or more.
I present the findings of this algorithm in Figure~\ref{fig:bp_\bp{AvoidZombieBugs}{REF}}.

\begin{figure}[ht]
    \centering
    \includegraphics[width=\textwidth]{2_figure/fig_bp_\bp{AvoidZombieBugs}{REF}.pdf}
    \caption{Violations to \bp{AvoidZombieBugs}{NAM} (\bp{AvoidZombieBugs}{UID}) \& \bp{ActiveBugReports}{NAM} (\bp{ActiveBugReports}{UID}).}
    \label{fig:bp_\bp{AvoidZombieBugs}{REF}}
\end{figure}

Across the organisations and projects, there is a median violation rate of \mytilde1\%.
This shows that practitioners are interested in keeping Bug Reports active and therefore meaningful.
Sonatype is an outlier, with a median violation rate of 10\%; however, this is still relatively small, and shows an overall commitment to active Bug Reports.
MongoDB and Qt show a particular commitment to this Best Practice, with a median at almost 0\%.

\FloatBarrier

\subsubsection{\bp{StableClosedState}{NAM} (\bp{StableClosedState}{UID})}

The ``\bp{StableClosedState}{NAM}'' Best Practice (see Table~\ref{tab:bp_\bp{StableClosedState}{REF}}) was proposed by \bp{StableClosedState}{AUT}.
The Best Practice is designed to improve stability of issues by discouraging the repeated opening and closing of Bug Reports.
Repeated opening and closing of bugs decreases software quality, increases maintenance \ltexdummy{costs}, and is frustrating for developers~\cite{Qamar_2022_IST}.
To detect these violations, I review all evolutions to the Status field, looking for issues that were re-opened.
In alignment with the recommendations from Qamar et al.~\cite{Qamar_2022_IST}, I do not count ``rapid'' cycles that occur in less than 5 minutes.
Any change in the Status that occurs less than 5 minutes before the last Status change is ignored.
I present the findings of this algorithm in Figure~\ref{fig:bp_\bp{StableClosedState}{REF}}.
\begin{quote}
    \ltexignore{``We check the history of the status field of the bug. Some projects explicitly use REOPENED value for the status field while others do not. In such cases, we check whether the status field is changed from Closed to another value, and we count it as a smell. \ltexdummy{Also}, we observed that there are some multiple status changes in a very short interval. We consider these changes a mistake and do not count them as a smell. The interval duration is set as five minutes. Therefore, if a bug status is changed multiple times in five minutes, it is counted as one change.''~\cite{Qamar_2022_IST}}
\end{quote}

\begin{figure}[ht]
    \centering
    \includegraphics[width=\textwidth]{2_figure/fig_bp_\bp{StableClosedState}{REF}.pdf}
    \caption{Violations to \bp{StableClosedState}{NAM} (\bp{StableClosedState}{UID}).}
    \label{fig:bp_\bp{StableClosedState}{REF}}
\end{figure}

As shown by the results, violations to this Best Practice are extremely rare, with a median violation rate of \mytilde0.01\%.
This is a clear sign that these organisations follow this Best Practice.
There is some variance in the \xglspl{iqr} and whiskers, but even then, the average is upper quartile is around 1\%.
Jira (the organisation) and Sonatype have the most violations to this Best Practice, with upper whiskers at \mytilde8\%.
Given the low medians across the \xglspl{jr}, it is likely they would prefer to avoid these violations altogether, and therefore would greatly benefit from processes or automations that support adherence to this Best Practice.

\FloatBarrier

\subsection{Bug Report Assorted}

\subsubsection{\bp{BugReportDiscussion}{NAM} (\bp{BugReportDiscussion}{UID})}

The ``\bp{BugReportDiscussion}{NAM}'' Best Practice (see Table~\ref{tab:bp_\bp{BugReportDiscussion}{REF}}) was proposed by \bp{BugReportDiscussion}{AUT}.
The Best Practice is designed to foster activity on Bug Reports through discussions in the comments.
The idea is that the comments on Bug Reports are part of a healthy \xgls{oss} community, and lack of comments is a sign that something is wrong.
To detect this violation, I review all closed Bug Reports, and check for those with no comments.
I present the findings of this algorithm in Figure~\ref{fig:bp_\bp{BugReportDiscussion}{REF}}.
\begin{quote}
    ``First, we check whether the bug is closed. If so, we check whether there is at least one comment in the bug.''~\cite{Qamar_2022_IST}
\end{quote}

\begin{figure}[ht]
    \centering
    \includegraphics[width=\textwidth]{2_figure/fig_bp_\bp{BugReportDiscussion}{REF}.pdf}
    \caption{Violations to \bp{BugReportDiscussion}{NAM} (\bp{BugReportDiscussion}{UID}).}
    \label{fig:bp_\bp{BugReportDiscussion}{REF}}
\end{figure}

The results overall show a median violation rate of \mytilde10\% across the \xglspl{jr}.
However, there is a mix of \xglspl{jr} with almost no violations (e.g., Sonatype with 0\%, and MongoDB with \mytilde5\%), and others such as JFrog with almost 60\% violation rate.
One thing to consider is the bots that some projects have installed, which may be offsetting their violations.
Another thing to consider is the popularity of the projects, and how that affects the outside discussion on their Bug Reports.
For a full and in-depth analysis, one could dig into whether the comments were bots, outsiders, or maintainers.

\FloatBarrier

\section{Algorithmic Detection Violations: All Issue Types}

In this section, I implement 6 algorithms that detect violations to \xgls{ite} Best Practices for all issue types.
I discuss each of these algorithms in the following subsections.
Figure~\ref{fig:bp_all_issue_types} is a summary of all findings across the 6 algorithms.
Each row represents a single Best Practice, where the violation detection algorithm was applied to all issues across the 13 \xglspl{jr}.

\afterpage{%
    \ActivateWarningFilters[largefigure]  
    \clearpage
    \begin{figure}[t]
        \centering
        \makebox[0pt]{\includegraphics[width=1\textwidth]{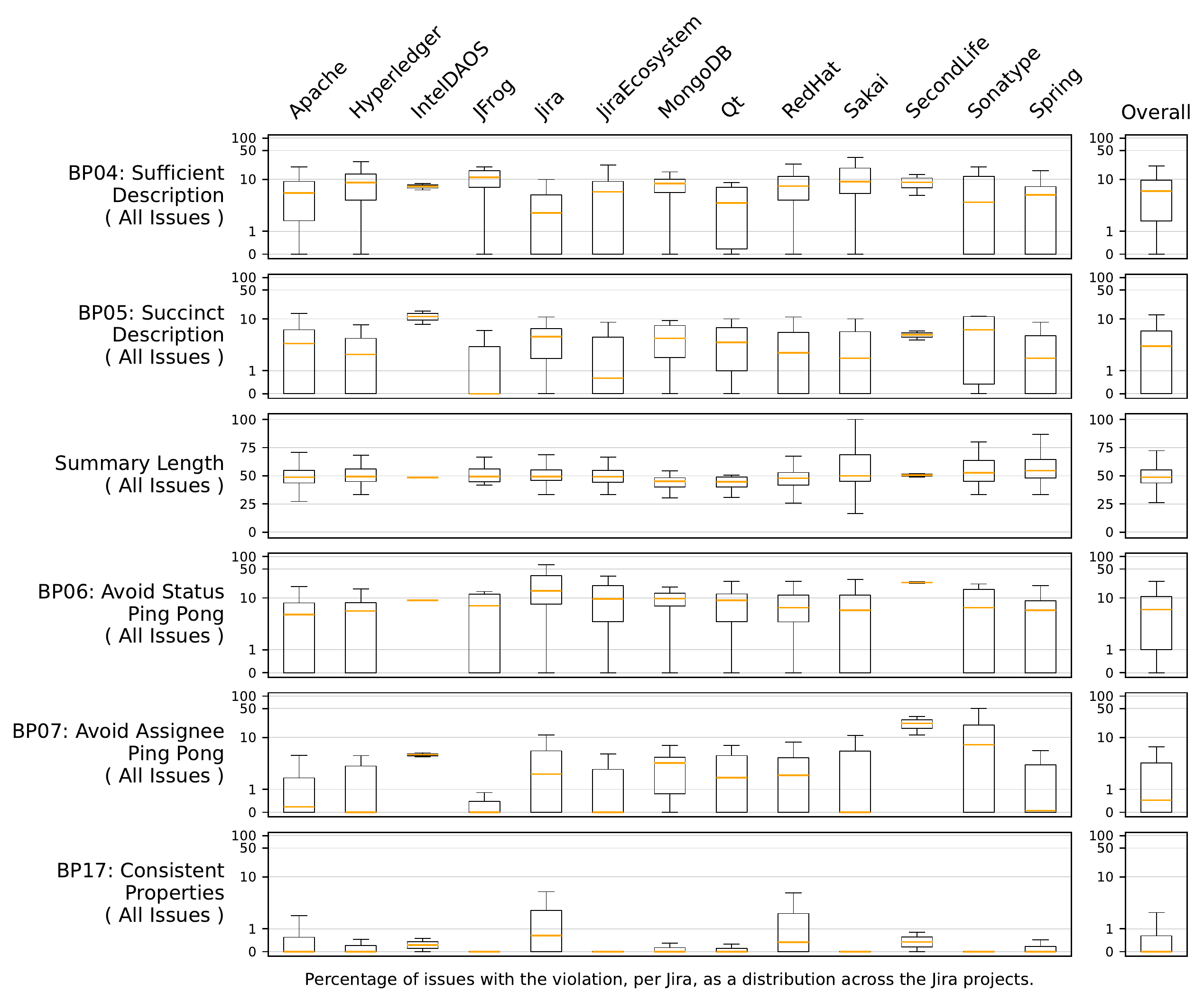}}
        \caption{Violations of Best Practices for all issue types.}
        \label{fig:bp_all_issue_types}
    \end{figure}
    \clearpage
    \DeactivateWarningFilters[largefigure]  
}

Five of the six sub-figures have their y-axis in symlog scale---highlighting the rarity of these violations, while only one is in linear scale.
Unlike the Bug Report Best Practices discussed presented above, these ``all issue types'' Best Practices have some trends across Jiras.
For example, Summary Length and \bp{AvoidStatusPingPong}{NAM} have roughly the same violation rate across the Jiras (50\% and 10\%, respectively).
Unlike the Bug Report Best Practices in Figure~\ref{fig:bp_all_bug_reports}, most of the Best Practices have quite low violation rates, except for Summary Length.
The following subsections go into detail regarding each of these Best Practices.

\FloatBarrier

\subsection{Summary and Description Length}

\subsubsection{\bp{SufficientDescription}{NAM} (\bp{SufficientDescription}{UID})}

The ``\bp{SufficientDescription}{NAM}'' Best Practice (see Table~\ref{tab:bp_\bp{AssignBugsToIndividuals}{REF}}) was proposed by \bp{SufficientDescription}{AUT}.
The Best Practice is designed to ensure that issue descriptions are up to a certain standard when it comes to length and basic information.
It is common to see issues with just a title and no description, or, if a description exists, it is just a short sentence.
This is likely due to the tacit knowledge the reporter assumes is known to everyone, but is likely not.

Prediger does not describe specifically how to detect this violation, although she discusses the concept of length (compared to that of, for example, semantic content).
Prediger does not specify the exact length that she considers ``sufficient'', likely because there is not a straightforward answer that applies universally to all issue types across all organisations.
That is a sign that this Best Practice needs much more research to understand the contextual factors that influence the application and tuning of this Best Practice.
Regardless, I implemented and applied an algorithm that automatically detects violations of this Best Practice, with a tunable threshold of length.
For the initial threshold, I set it to 10 words.
I think it is easy to argue that 10 words is not sufficiently long enough for a description to adequately describe an issue.
However, for the purpose of tuning a violation detection algorithm, I chose a number that is not likely to produce many false positives.
In other words, detected violations are indeed violations, thus not wasting the time of stakeholders who are being notified of this violation.
I present the results of this algorithmic violation detection in Figure~\ref{fig:bp_\bp{SufficientDescription}{REF}}.

\begin{figure}[ht]
    \centering
    \includegraphics[width=\textwidth]{2_figure/fig_bp_\bp{SufficientDescription}{REF}.pdf}
    \caption{Violations to \bp{SufficientDescription}{NAM} (\bp{SufficientDescription}{UID}).}
    \label{fig:bp_\bp{SufficientDescription}{REF}}
\end{figure}

Despite the generously low threshold of 10 words, most Jiras across all issue types suffer from 5--10\% violations per project (median).
Interestingly, Development issues appear to be struck the most by this violation (at \mytilde10\% median), although the difference between Requirements and Maintenance is only a few percentage points.
It is also worth noting that some of the Jiras have lower quartiles at 0\%, which means there is a substantial number of projects (\mytilde25\%) that do not suffer from this violation at all.

\FloatBarrier

\subsubsection{\bp{SuccinctDescription}{NAM} (\bp{SuccinctDescription}{UID})}

The ``\bp{SuccinctDescription}{NAM}'' Best Practice (see Table~\ref{tab:bp_\bp{SuccinctDescription}{REF}}) was proposed by \bp{SuccinctDescription}{AUT}.
The Best Practice is designed to ensure that issue descriptions are not too long, which adds unnecessary wasted time and potential misunderstandings from misinterpretations of the text.
While issue descriptions should be sufficiently long (``\bp{SufficientDescription}{NAM}'' Best Practice Table~\ref{fig:bp_\bp{SufficientDescription}{REF}}), this does not mean that longer is universally better.
As with many variable concepts, there is a trade-off with a meaningful compromise somewhere in the middle.

L\"uders does not describe specifically what threshold counts as ``succinct'' versus not, which is similar to the threshold of \bp{SufficientDescription}{NAM}.
While it may not be possible to pick a universal threshold, we simply do not know without further research investigating the contextual factors that influence this Best Practice.
Regardless, I implemented the detection of this violation, and set the initial threshold at 250 words.
I chose this threshold because that is the standard length of a page in a book, and represents multiple paragraphs of text.
At this point, it is unlikely that additional words are going to make the description any clearer, and perhaps editing is necessary if additional information is required.
I present the results of this algorithmic violation detection in Figure~\ref{fig:bp_\bp{SuccinctDescription}{REF}}.

\begin{figure}[ht]
    \centering
    \includegraphics[width=\textwidth]{2_figure/fig_bp_\bp{SuccinctDescription}{REF}.pdf}
    \caption{Violations to \bp{SuccinctDescription}{NAM} (\bp{SuccinctDescription}{UID}).}
    \label{fig:bp_\bp{SuccinctDescription}{REF}}
\end{figure}

The results show an interesting divide between Requirements, Development, and Maintenance issues: Development issues rarely suffer from this violation (\mytilde0\% median), Requirements minimally suffer from this violation (\mytilde1\% median), and Maintenance suffers from this violation much more (\mytilde8\% median).
This suggests that either Maintenance issues are of lower quality when it comes to succinct descriptions, or, Maintenance issues have a necessity for more information, and therefore longer descriptions.
Both of these potential phenomena require additional research to understand better.
For the purpose of this Best Practice, these results suggest that different issue types require different thresholds for what is considered a ``succinct description''.
Further research could uncover and prescribe what these thresholds are.

\FloatBarrier

\subsubsection{Summary Length}

\begin{figure}[ht]
    \centering
    \includegraphics[width=\textwidth]{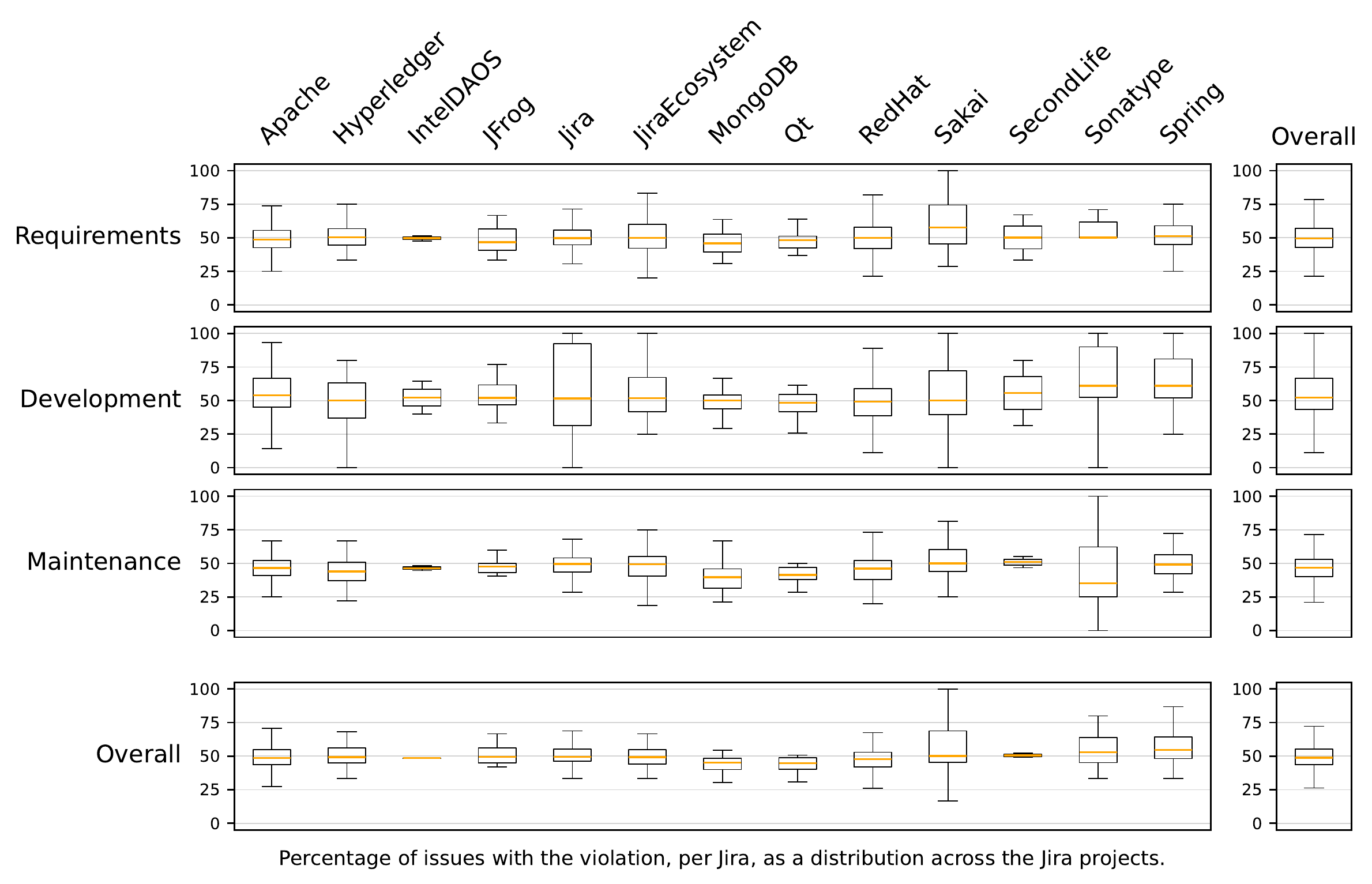}
    \caption{Violations to Summary Length.}
    \label{fig:bp_summary_length}
\end{figure}

The ``Summary Length'' Best Practice is not listed in the Best Practice catalogue, since it has not been formally proposed within the literature.
The reason for including it in this algorithms section is the transferability of the algorithms written for the Description field.
The Summary and Description field both share a similar role when it comes to describing the issue.
Perhaps more important for the Summary, however, is the exact length of the text.
While the Description can vary quite drastically depending on the type of issue being described, the Summary should always be a consistent length.
What defines ``consistent'' for a given project is not so clear.
Bohn describes an ideal range of 39 to 70 characters in his dissection of Summary length, citing literature~\cite{Zimmermann_2010_TSE}, company guidelines,\footnote{Bug Writing Guidelines by Bugzilla (\url{bugzilla.mozilla.org}) and eclipse (\url{bugs.eclipse.org}).} and his own personal investigations~\cite{Bohn_2024_MSc}.
Since the focus of my work is on implementation and application, and not the empirical investigation of what is the ideal length, this range will suffice.
Using this range of 39--70, I detect and flag all summaries that are outside this range as violations.
I present the results of this algorithmic violation detection in Figure~\ref{fig:bp_summary_length}.

With the current threshold, \mytilde50\% of issues across the projects are violating this Best Practice.
This is fairly consistent across the \xglspl{jr} and activity types, with varying degrees of variance visible in the \xglspl{iqr}.
Given the importance of the Summary field, and the emphasis on it as a central way to communicate and identify the issue, I don't think this \mytilde50\% violation rate is representative of the real impact of this Best Practice.
This number should be much lower.
While I believe keeping the Summary within a certain range is important, and likely indirectly enforced within many projects, I don't think it is as easy as a single threshold you can apply universally across projects.
Context likely plays an important role here, and without it, these results are lacking in realism.

\FloatBarrier

\subsection{Stable and Correct Fields}

\subsubsection{\bp{AvoidStatusPingPong}{NAM} (\bp{AvoidStatusPingPong}{UID})}

The ``\bp{AvoidStatusPingPong}{NAM}'' Best Practice (see Table~\ref{tab:bp_\bp{AvoidStatusPingPong}{REF}}) was proposed by \bp{AvoidStatusPingPong}{AUT}.
The Best Practice is designed to ensure that issues go through linear, intentional \textit{status} changes from open to closed.
If not adhering to this Best Practice, cycles can form within the status, which likely represents duplicate work and confused stakeholders.
To detect this Best Practice, create a state-transition graph from the history of all status changes in an \xgls{its}, and search those graphs for cycles.
If an organisation has ``allowed'' cycles (such as between ``\xgls{qa}'' and ``Dev''), then add these as exceptions within the algorithms.
I present the results of this algorithmic violation detection in Figure~\ref{fig:bp_\bp{AvoidStatusPingPong}{REF}}.

\begin{figure}[ht]
    \centering
    \includegraphics[width=\textwidth]{2_figure/fig_bp_\bp{AvoidStatusPingPong}{REF}.pdf}
    \caption{Violations to \bp{AvoidStatusPingPong}{NAM} (\bp{AvoidStatusPingPong}{UID}).}
    \label{fig:bp_\bp{AvoidStatusPingPong}{REF}}
\end{figure}

The results show that status cycles are actually quite common across the projects.
The median is \mytilde10\% for all Jira repositories, although repos such as IntelDAOS and SecondLife show medians between 25 and 50\%.
Given the prevalence across the Jira repos, this suggests that there are likely well-known and accepted cycles that occur.
To further refine this violation detection to remove what are likely false-positives, an analysis of the existing cycles needs to be done---per repo.
Then, once the researcher has a suitable understanding of \textit{what} is going on, they then need to approach the stakeholders within these repositories to understand \textit{why} these cycles are so common.
Some may be unintentional, but the more common and cross-repo cycle patterns are likely accepted forms of process for these issues.

\FloatBarrier

\subsubsection{\bp{AvoidAssigneePingPong}{NAM} (\bp{AvoidAssigneePingPong}{UID})}

The ``\bp{AvoidAssigneePingPong}{NAM}'' Best Practice (see Table~\ref{tab:bp_\bp{AvoidAssigneePingPong}{REF}}) was proposed by \bp{AvoidAssigneePingPong}{AUT}.
Similar to ``\bp{AvoidStatusPingPong}{NAM}'' above, this Best Practice is designed to ensure that issues go through linear, intentional \textit{assignee} changes.
When the assignee is re-assigned back and forth between different people, this is a sign that there is a disagreement in some external process or responsibility.
This can only be resolved through discussions regarding who should actually be working on it.
To detect this Best Practice, create a state-transition graph from the history of all assignee changes in an \xgls{its}, and search those graphs for cycles.
If an organisation has ``allowed'' assignee cycles, such as between people co-working on the issue, then add these as exceptions within the algorithms.
I present the results of this algorithmic violation detection in Figure~\ref{fig:bp_\bp{AvoidAssigneePingPong}{REF}}.

\begin{figure}[ht]
    \centering
    \includegraphics[width=\textwidth]{2_figure/fig_bp_\bp{AvoidAssigneePingPong}{REF}.pdf}
    \caption{Violations to \bp{AvoidAssigneePingPong}{NAM} (\bp{AvoidAssigneePingPong}{UID}).}
    \label{fig:bp_\bp{AvoidAssigneePingPong}{REF}}
\end{figure}

The results show that assignee cycles are quite rare, with an overall median of \mytilde0.5\%; however, roughly half of the repos have an assignee cycle median between \mytilde5--10\%.
The SecondLife (\mytilde25\%) repo has a high count of assignee cycles.

\FloatBarrier

\subsubsection{\bp{ConsistentProperties}{NAM} (\bp{ConsistentProperties}{UID})}

The ``\bp{ConsistentProperties}{NAM}'' Best Practice (see Table~\ref{tab:bp_\bp{ConsistentProperties}{REF}}) was proposed by \bp{ConsistentProperties}{AUT}.
The Best Practice is designed to keep the fields within issues as up-to-date as possible, considering information that is sometimes written in the Description or Comments fields.
Despite fields that are designed to hold specific information, stakeholders sometimes write things such as ``this issue is now resolved'', without setting the ``Resolution'' field.
This Best Practice states that no one should state information else where on the issue when the fields could instead be updated.

L{\"u}ders does not \ltexdummy{propose} a specific algorithmic approach to identifying violations of this Best Practice.
Given that these violations occur in natural language (in the Description or Comments fields), detection becomes an \xgls{nlp} problem.
I chose to develop a preliminary, naive implementation of this detection algorithm using context-dependent information and text searching.
First, I collect a list of all fields that exist within issues (which I already conducted in Chapter~\ref{ch:evolution}).
Then, I search the full evolution history of all issues for values those fields were ever set to, separated per Jira.
Finally, I search the Description and Comments for any instance of both a field,\footnote{For computation reasons, I only search for the fields IssueType, Status, Priority, and Resolution. These fields have historical values in the range of \textless 200, whereas the other fields have upwards of 10,000 potential values for each field, per Jira. Future research can investigate why this is, and how to reduce the computational complexity to make this naive approach feasible for these fields.} and one of its possible values (within the context of the Jira).
Like most \xgls{nlp} tasks, the results are fraught with false positives, but the recall will be nearly 100\%, except in cases of spelling mistakes.
As Dan Berry describes, recall is often preferred over precision, since a human can quickly assess whether a result is correct or not, but finding missed results requires an extraordinary amount of time~\cite{Berry_2021_EMSE}.
Accordingly, I left my preliminary implementation as is.
I present the results of this algorithmic violation detection in Figure~\ref{fig:bp_\bp{ConsistentProperties}{REF}}.

\begin{figure}[ht]
    \centering
    \includegraphics[width=\textwidth]{2_figure/fig_bp_\bp{ConsistentProperties}{REF}.pdf}
    \caption{Violations to \bp{ConsistentProperties}{NAM} (\bp{ConsistentProperties}{UID}).}
    \label{fig:bp_\bp{ConsistentProperties}{REF}}
\end{figure}

There are almost no violations to this Best Practice, across all Jiras, despite the admittedly high false positive rate.
The median across all projects is 0\%, and only four of the Jiras show a median between 0 and 1\%.
Inconsistent properties appear to be more of a problem in Maintenance issues, which makes sense given the increased number of stakeholders and stakeholder groups who are involved with Maintenance.
With these increased roles, knowing when and how to update the fields would become less certain, and perhaps not even possible for some stakeholder groups (such as people external to the organisation).
Analytically, it would seem that inconsistent properties are a problem when they occur, but empirically we can see that this occurs very rarely.
I also assume that once my naive preliminary script is updated to be more intelligent, the number of detected violations will be even lower.

\FloatBarrier

\section{Summary}

In this chapter, I implemented {\numItebpAlgs} different algorithms to assist in the automated detection of violations of Best Practices for \xglspl{ite}.
These algorithms are the starting place for \xgls{its} plugins that support practitioners in real-time and within their working context.
With these algorithms, researchers can now challenge and improve the state-of-the-art for these Best Practices.
Additionally, the results presented in this chapter can serve as a starting place for comparative studies looking to either improve on the algorithms, as well as inspiration for areas where case studies could investigate these phenomena.

With the algorithms constructed and applied to a real dataset, the next step is to consider the tooling they would be a part of.
The logic has been written, but product design and integration in practitioner settings is a difficult task.
How can these algorithms be integrated?
When should this information be presented to practitioners?
How much control should be given to users?
I address these questions with tooling recommendations in the next chapter.

    \part{Outlook Recommendation}  \label{part:outlook}
    
\chapter{Tool Support for Best Practices in Issue Tracking Ecosystems}  \label{ch:tooling}

\epigraph{We become what we behold. We shape our tools and then our tools shape us.}{Marshall McLuhan}

In this chapter, I present a number of recommendations and guidelines for creating recommender systems for \xgls{ite} Best Practices.
In Chapters~\ref{ch:catalogue} and \ref{ch:algorithms}, I presented a catalogue of \xgls{ite} Best Practices and algorithms to detect violations of those Best Practices, but those are all in isolation of each other and in isolation from their solution space.
One of the findings from previous chapters is that \xgls{ite} Best Practices (and the problems that they address) are context-dependent, which is captured in the ontology presented in Chapter~\ref{ch:ontology}.
This leads to the logical conclusion that the solution space for recommender systems for these Best Practices is also context-dependent.
Recommender systems for \xgls{ite} Best Practices need to consider and allow for the context factors to guide their implementation and usage.

In this chapter, I focus on overall tool design for such a complete system that addresses the context-dependent nature of \xgls{ite} Best Practices.
I present descriptions of desired attributes of recommender systems, and in some cases I have already developed preliminary tools and demos to showcase some of this functionality.
I then present these designs to industry participants to get their feedback on visual design and proposed usage of such tools.
The primary feedback from the interviewed industry participants is confirmation of the usefulness of the context-dependent features.
This includes the configuration screen, which allows individuals, teams, and organisations to set different preferences for their Best Practices, including the tuning of certain variables (not just binary ``on'' or ``off'' for each Best Practice).
Another finding is that certain context factors play a larger role.
For example, the role a stakeholder has within the organisation is an important context factor for not only which Best Practices they prefer, but also entire tool design choices.
For example, I found that developers prefer integrated feedback directly within \xglspl{its} while creating or editing issues, whereas managers prefer summary tools that give reports across the \xgls{its}.

This chapter contributes both design aspects of recommender systems for \xgls{ite} Best Practices, and empirical feedback on these design aspects.
Researchers can take this work and directly build on it with their own designs and future studies on our designs.
Practitioners can already implement these designs within their own systems to report, understand, and enforce Best Practices within their organisations.

\statementPublication{\cite{Prediger_2023_MSc}}

\section{Research Methodology}

My primary objective with this chapter is to prescribe and validate recommendations for tooling that supports \xgls{ite} Best Practices.
To prescribe recommendations, I will use a deductive approach, applying the collective knowledge presented in this thesis.
I'm not seeking to provide empirical evidence, but rather to act as a knowledge expert on the topics presented in this thesis, and therefore provide value in the form of final recommendations.
To validate the recommendations, I sought to gather insights about the potential usefulness of the recommendations.
For this, I interviewed 26 practitioners who use \xglspl{its}, with a focus on showcasing the recommendations, features, and screenshots from this chapter.
I first asked about tooling in general to collect ideas without the bias of a specific tool or feature.
I then presented four screenshots to guide the discussion and stimulate in-depth answers: (1) a configuration screen for the detection of individual smells, (2) a dashboard summarising the detected smells across the \xgls{its}, (3) a detailed view showing detected smells for the currently viewed issue, and (4) an issue-link graph visualisation highlighting link smells~\cite{OpenReq_2023_Online,OpenReqGitHub_2023_Online}.
\ltexdummy{I} also asked if the shown features could address the problems and smells discussed earlier in the interview.
\ltexdummy{I} encouraged participants to share all thoughts on software tooling to support addressing \xgls{ite} problems and smells, particularly when manifested as smells.

\section{Recommendations for Tool Support}

\begin{mybox}{Nested Configurations}
    Use nested layers of configuration files to accommodate the layers of context that affect the applicability of \xgls{ite} Best Practices in industrial settings.
\end{mybox}

\xgls{ite} Best Practices are context dependent (see Chapter~\ref{ch:ontology}), and some of those contexts are naturally layered.
For example, organisations are composed of teams, and teams are composed of people.
Each of those contexts may affect whether to apply a Best Practice or not, and each of them are likely to affect the tuning of the Best Practices.
For example, an organisation may adopt the ``Assignee Bug Resolution'' Best Practice (the bug assignee should be the one to resolve the bug; see Table~\ref{tab:bp_assignee_bug_resolution}), but they have a contractor team based in another country, and they prefer that someone internal always resolves the Bug Reports, regardless of who is assigned.
In this case, the context of that particular team is in disagreement with the context of the organisation for the application of this Best Practice.
However, instead of viewing it as a disagreement, it would be better to imagine that the team context is a sub-layer within the organisation, and so just this one team needs an exception to the organisational decision.
This is where one can apply the power of nested layers of configuration for \xgls{ite} Best Practices.

Configurations outline whether a Best Practice applies (should be automatically checked for violations) as well as the tuning of the Best Practices where that is relevant (e.g., ``minimum number of words'' for the Best Practice ``Sufficient Description'').
A basic configuration file could be in a simple format such as JSON, with a ``key: value'' system.
The nesting of contexts can be applied by having different configuration files that set the same parameters.
Then, once an organisation has decided on what constitutes a context layer, and which layers are nested within each other, the configuration files can hierarchically override each other as the nested layers of context are applied.
In our example above regarding the Assignee Bug Resolution, the organisation would have a organisational-wide configuration with the Assignee Bug Resolution Best Practice \textit{enabled}, and the specific contractor team would have their own configuration where the Assignee Bug Resolution Best Practice is \textit{disabled}.
Since the \textit{team context} is more specific than the \textit{organisational context}, the team configuration overrides the organisational one.
However, the team configuration file in this example would only set this single configuration, and the rest would be left blank, such that they would adopt whatever configurations apply from the organisation.
The team context is not independent of the organisation, it is just more specific and therefore \textit{can override} specific recommendations, \textit{where needed}.

Conceptually, what constitutes a context layer, and which layers are nested within each other (or perhaps only partially overlapping), is a complex problem in and of itself.
From a technology perspective, it is easy to implement these configuration files, and to write an override method that consumes all files and outputs the specific configuration for a given a context.
What an organisation must decide, as a process decision, is which contexts are allowed to have their own configuration files, and what is the hierarchical layering of these contexts within their given organisation.
As a starting point, I recommend the following context configuration files, listed in order of increasing precedence: \textbf{organisation, team, project, sprint, individual}.
However, there are many good reasons to use additional configuration files, as well as different precedence orders.
In many customer-centric organisations, the customer might have a say in organisational processes and the quality therein.
For example, an organisation could tune the Timely Severe Issue Resolution Best Practice to four days, but a particular customer might need a timely severe issue resolution of just one day.
In this case, the customer should have a configuration file that is likely just after ``team'' in the precedence.

\begin{figure}[ht]
    \centering
    \includegraphics[width=\textwidth]{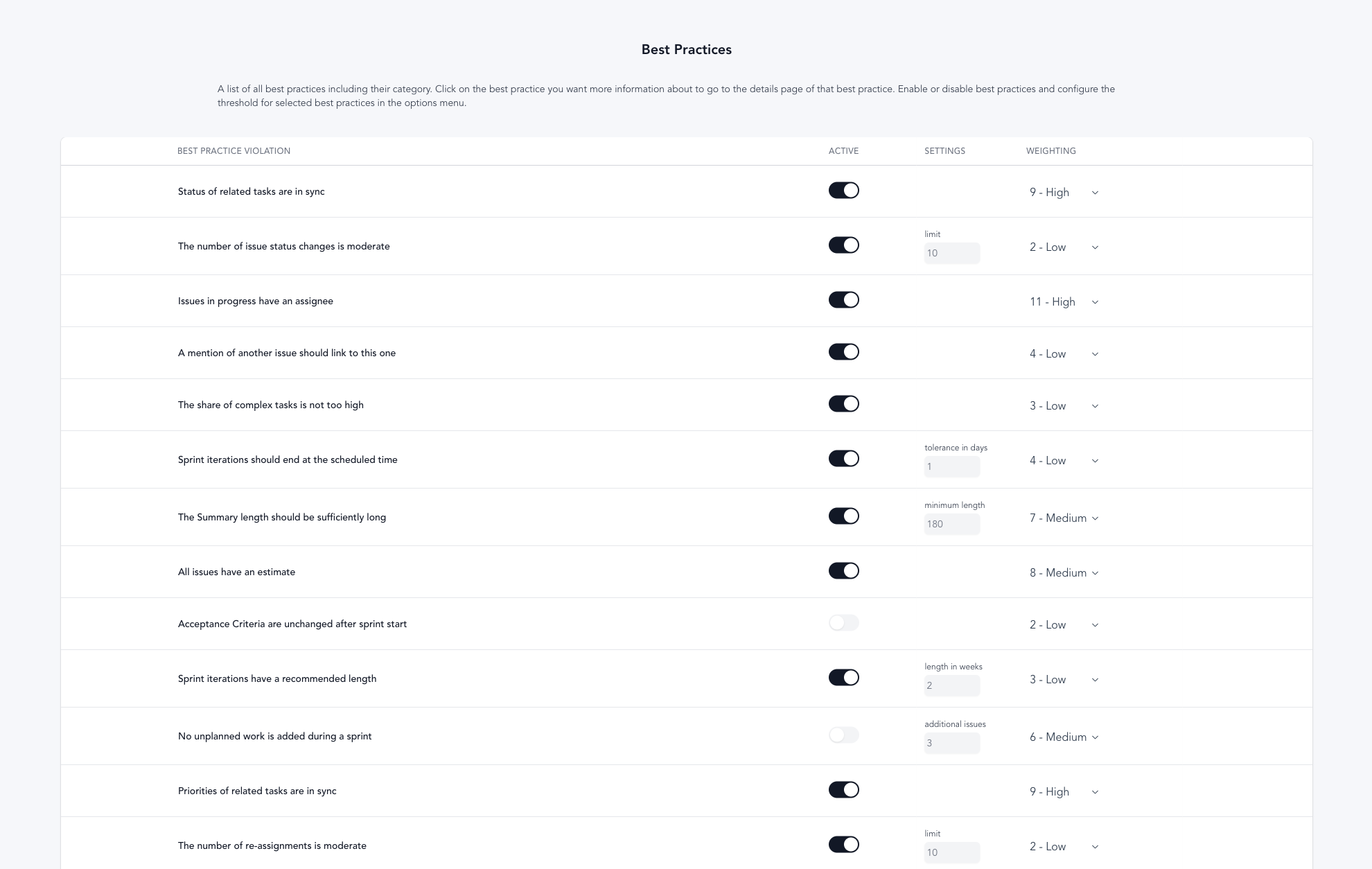}
    \caption{ITE Best Practice Configuration}
    \label{fig:tooling_bp_config}
\end{figure}

Figure~\ref{fig:tooling_bp_config} shows a prototype of the Best Practice configuration screen, as developed by Nina Prediger~\cite{Prediger_2023_MSc}.
For each Best Practice, the user (or team, organisation) can enable or disable it, tune it with particular best-practice-specific settings, and even give it a weight (importance).
This configuration screen does not show the layering of configurations, but that would likely be a separate interface only designed to managed by one person within the organisation.
The configuration screen presented in Figure~\ref{fig:tooling_bp_config}, however, would be utilised by every person in the organisation who wishes to adapt their own configurations, or the configurations of their project, team, or customer, on behalf of those entities.

\begin{mybox}{Cold Start using \xgls{ite} Best Practice Historical Analysis}
    Use historical analysis of your \xgls{its} to understand which \xgls{ite} Best Practices to address within your organisation.
\end{mybox}

The initial catalogue of \xgls{ite} Best Practices that I presented in Chapter~\ref{ch:catalogue} already has 40 Best Practices, and the hope would be to grow this list to account for all quality-based methods for improving \xglspl{ite}.
With so many potential catalogue items, it may be confusing for an organisation to chose and start implementing these Best Practices.
A data-driven approach to this decision would be for the organisation to first run the algorithms I presented in Chapter~\ref{ch:algorithms} on their \xgls{its}.
These results would inform them which Best Practices are being most violated within their organisation, and therefore which to address first.
It could also be the case that certain Best Practices have many reported violations, but the organisation could decide that these are in fact not violations within their context and defined processes.
Either way, this would ground their understanding of \xgls{ite} Best Practices in their specific context, and be the starting point for organisation-wide decisions.

\begin{mybox}{Just-in-Time Feedback for \xgls{its} Users}
    Deliver feedback on \xgls{ite} Best Practices immediately through \xgls{its} plugins that annotate issues.
\end{mybox}

Remembering all the Best Practices implemented in a \textit{given context} is not feasible for the average \xgls{ite} stakeholder, let alone \textit{all contexts}.
To train their memory and provide immediate feedback, I recommend utilising \xgls{its} plugins to prompt \xgls{its} stakeholders at the exact moment they have violated a Best Practice implemented by their organisation.
Here are some plugin ideas that could provide immediate just-in-time feedback on \xgls{ite} Best Practices:
\begin{description}
    \item[Sufficient Description] When creating or editing an issue, check the description length when the user tries to ``save'' the issue, and prompt them with a warning if the description is too short. At the extreme, it could also block them entirely from saving the issue until the description is sufficiently long.
    \item[Set Bug Report Assignee] When creating or editing an issue, check if the assignee field is empty when the user attempts to ``save'' the issue.
    \item[Set Bug Report X] Same as ``Set Bug Report Assignee'' above, but now applies to the other Best Practices of this structure.
    \item[Consistent Properties] When editing the description or adding a new comment, check the text when the user tries to ``save'' their work, and prompt them with a warning if any issue field names are found in their text (which is a sign that they are adding information to the description or comment that should instead be changed directly in the properties).
    \item[Bug-to-Commit Linking] When a user attempts to change the resolution to ``fixed'' on a Bug Report, check if the Bug Report has the commit linked in the description or custom field for commit links. If not, warn or stop them.
\end{description}

Those are just a few examples of just-in-time feedback that can be delivered to \xgls{its} stakeholders through native \xgls{its} plugins.
With the use of \xgls{nlp} techniques, more just-in-time checks can be performed on the textual fields.
Research applying \xgls{nlp} techniques has produced promising results in the last couple of decades~\cite{Zhao_2022_ACMCompSurvey,Montgomery_2025_BookChapter,Gemkow_2018_RE,Schlutter_2018_REFSQ,Pudlitz_2019_RE}.
Overall, there are many opportunities to support \xgls{its} stakeholders and guide them towards Best Practice conformance.

\begin{mybox}{Issue Report for \xgls{its} Users}
    Deliver feedback on \xgls{ite} Best Practices through an \xgls{its} plugin that annotates individual issues with a report of which Best Practices are violated.
\end{mybox}

There are many reasons Best Practices could be violated, including when an organisation adopts a new Best Practice, which may already be violated by thousands of existing issues.
It is not reasonable to assume that stakeholders will immediately update all issues to conform to the new Best Practices.
Instead, an \xgls{its} plugin can be used to create a small issue report in the properties section of each issue.
Many of the Best Practices are applied to a single issue, which means it is possible to give a report on the quality of that issue, regarding how many violations it has.
When someone is working on that issue, be that creation or editing, this report would be immediately visible to them, in the context of their work.

\begin{figure}[ht]
    \centering
    \includegraphics[width=\textwidth]{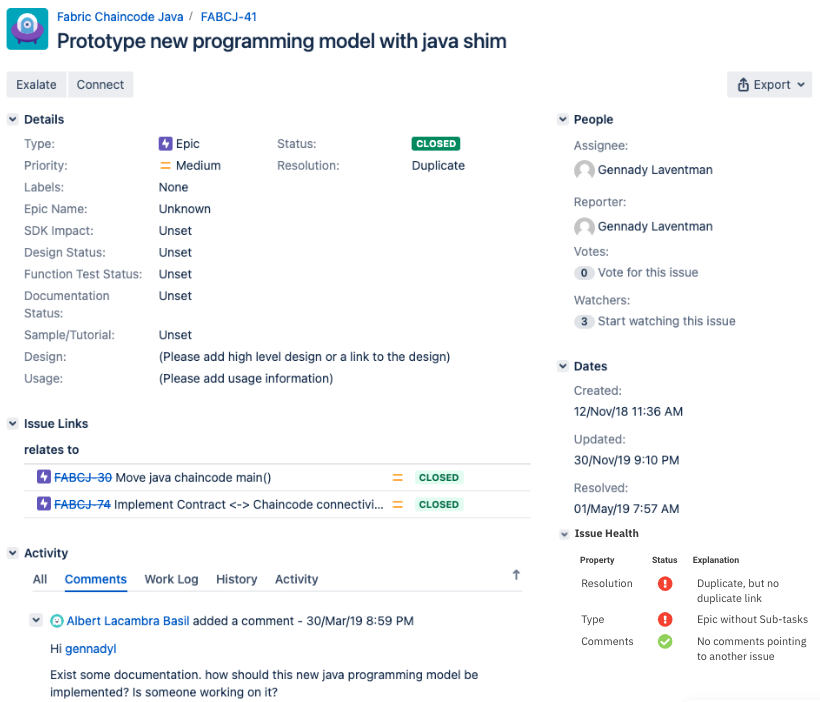}
    \caption{ITE Best Practice Issue Report}
    \label{fig:tooling_bp_issue_report}
\end{figure}

Using this issue report, relevant stakeholders who have navigated to a particular issue will see the issue report, and they can make an in-the-moment decision regarding whether to improve the issue or not.
This has the benefit that stakeholders will have the context of the issue in mind when they see the reported Best Practice violations.
Figure~\ref{fig:tooling_bp_issue_report} shows an example of what this could look like, in the bottom right of an issue, labelled here as ``Issue Health''.

\begin{mybox}{Dashboard for Managers}
    Summarise the conformance to \xgls{ite} Best Practices across the \xgls{its} by creating a dashboard for managers.
\end{mybox}

\begin{figure}[ht]
    \centering
    \includegraphics[width=\textwidth]{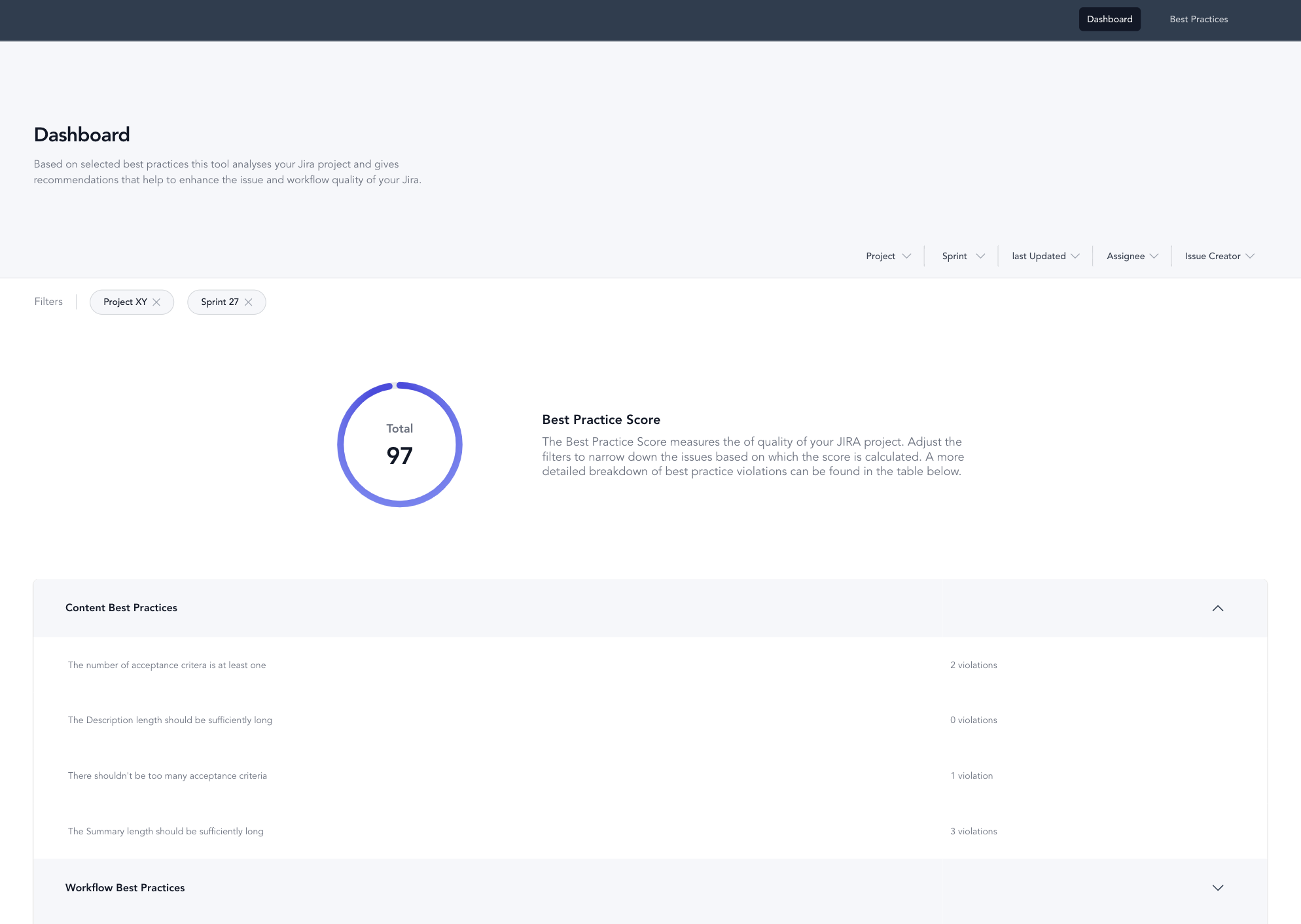}
    \caption{ITE Best Practice Dashboard}
    \label{fig:tooling_bp_dashboard}
\end{figure}

While just-in-time feedback and issue reports for \xgls{its} users are appropriate to help guide them, it is missing an \xgls{its}-wide perspective on the status and progress of Best Practice conformance.
Over time, an organisation should see progress towards a higher-quality \xgls{its} through more conformance to the Best Practices they choose.
For this, I recommend implementing a dashboard that summarises the conformance to Best Practices across the \xgls{its}.
Figure~\ref{fig:tooling_bp_dashboard} shows a prototype of the Best Practice dashboard, as developed by Nina Prediger~\cite{Prediger_2023_MSc}.

There are three key features within this dashboard: summarisation, filtering, and historical records.
It is important to summarise the conformance in a single value, such that a quick judgement can be made on the current progress.
\ltexignore{It is essential that the dashboard supports filtering to specific contexts (e.g., projects and sprints) so that managers can dive into the data, to contextualise and investigate the specifics behind the summarisation that is reported.}
Finally, keeping historical records of the summarisations presented allows managers to see progress over time, instead of just a single current summarised value.
For the \xgls{its} Jira, where historical records of all issue changes are already stored, it is possible to see what the summarisation value would be at any point in the past.
Combined, these three features would allow a manager to see the current status, set goals, and see progress over time.

\section{Practitioner Feedback on Tool Support}

I first asked participants about any \textit{general tooling insights} they have, and then showed them the screenshots of our prototype tools to get feedback on specific ways of \textit{enhancing \xglspl{its}}.
The participants described that tool support must be easy to use, integrated well into their workflow, and provide reliable results.
By meeting these requirements, they believe a tool could help improve the efficiency and effectiveness of the \xgls{ite} process.

\subsection{General Tooling Insights}

\textbf{Smell Detection Tooling to Support Existing Meetings.}
Participants mentioned that smell detection tooling could be used to support the quality of their \xgls{its} in existing meetings such as triage and sprint planning (P01, P08, P11, P13, P14, P20, P25).
Multiple participants mentioned that triage meetings are essential to organising \xglspl{its} and ensuring that issue reports have the necessary information for developers to act on them.
These meetings are already a way to ensure that issues in the sprint adhere to a quality standard and can be resolved.
Participants mentioned that smell detection could support these meetings by highlighting potential problems.
P14 also mentioned the potential to support retro meetings: ``it would be nice to look at a dashboard and say we were 50\% worse in this smell''.

\textbf{Prevention over Detection.}
Participants said that instead of detecting the smells, preventing them from happening might be more useful; otherwise there would still be an overhead of fixing smells after the fact (P03, P05, P08, P09, P13, P16, P17, P24, P26).
Furthermore, one participant noted that a post-mortem of the issue reports might not be the correct way to fix them, but that the user should instead be alerted during the creation or editing.
For example, P05 said that ``smells should be prevented during issue creation, for example: if you want to close an issue as duplicate, you get a pop-up if there is no link''.
P08 agreed: ``it would have to be directly integrated into [the \xgls{its}]; a separate tool would not make sense''.
The participants noted that an essential requirement for adequate tool support in \xglspl{ite} is the direct integration into their \xgls{its} workflow.
This means the tool should be easy to use and not require too much overhead while creating or modifying issues; otherwise, it will not be used.
P25 said, \ltexdummy{``if you are not able to really address these smells, you have to re-think whether these kinds of tracking systems, properties, and linking are working for you. It reminds you to rethink different things''}.

\textbf{Detecting Link Smells.}
21 out of the 26 participants believe that link detection would be helpful, P18 and P20 were unsure about the usefulness and P04, P24, and P25 did not consider link detection helpful.
P01, P02, P07, and P16 voiced that link detection would help locate issues or relevant issue reports in the \xgls{its}.
A generated graph or tree representation was also suggested by some participants to achieve a high-level view of the connections between projects or gain an overview or understanding of the context.
P24 said, \ltexdummy{``suggested links need to be reviewed and when it turns out that they are not actually related, [extra work was created]''. They would ``rather miss a link than have to inspect unnecessary ones''}.

\textbf{Tooling Pitfalls to Avoid.}
Participants specifically mentioned that smell detection would be more useful for managers and not developers (P02, P05, P07, and P09).
They also mentioned that the tool should not generate too many false positives because the additional noise is ``annoying'' and users do not want to ``sift through a large collection of smells''.
They expressed concern that the user might receive too many notifications (P06, P07, P12, P14, P15, P21, P22, P24, P26).
P07 stated the system \ltexdummy{``should not spam too much; if the information is not too obvious then it would be very helpful''}.
Notifications can be annoying, especially outside the \xgls{its}, e.g., e-mails~\cite{Meyer_2021_TSE}.
P14 also cautioned that there might be social impacts of such a tool, making users feel bad instead of helping them improve.
Additionally, participants mentioned that the tool should not obstruct the workflow, as it creates more overhead.
P13 said that ``enforced processes are not good'' and P22 said ``it should not completely block [your workflow]''.
Several participants voiced their scepticism toward automated processes (P07, P08, P12, P15, P18, P21, P23, P24).
They said that the results need to be transparent, interpretable, and manually reviewed.
However, P24 reported from experience that ``suggested links need to be reviewed, then it turns out that they are not actually related'', which led to the creation of extra work. They would ``rather miss a link than have to inspect unnecessary ones''.
There also must be a general acceptance of the tool in the team for it to be used.
Three participants (P05, P15, P23) were unsure about the usefulness of smells and participant P09 stated that it would not be useful for them.
It was noted that some kinds of smell detection would likely be more useful for managers than developers.
Additionally, one participant noted that smell detection can be counter-productive for developers.

\subsection{Enhancing Issue Trackers}

\textbf{(1) Configuration.}
We already saw that most smells are not perceived universally as problematic, and some are indeed part of the good practice.
I showed the participants a screenshot of a smell configuration page, containing a toggle for each smell (and some had a configurable value).
Almost all participants said that a configuration is a must-have, a tool ``must be customisable because some smells would not work or would constantly [be detected]'' (P03).
P10 mentioned that the tool could be misused and hinder more work than help if it is not configured properly.
P13 cautioned that ``deactivating smells is a nice [feature], but [they] need to understand and document why smells are deactivated''.
P26 also added that a \textit{default} configuration is needed.
Finally, P05 mentioned that they would like ``syntax to write their own smells''.

\textbf{(2) Dashboard.}
I showed the participants a dashboard presenting the overall ``health'' of the \xgls{its}.
Multiple participants commented that a dashboard has more value for managers, with little value for developers.
It was also noted that it should not be the absolute values but the trends over time.
P14 said, ``I don't care about the actual amount, I care about the trend''.
P21 said, \ltexdummy{``it would be nice to see the change over time with the health, so it is not only the static view that is important, but the development over time''}.
P24 said that tools like this should be used with a goal in mind because ```optimising for green isn't always good'' and that the metrics shouldn't become the targets as they can ``distract from what the customer wants''.
P14 cautioned that ``metrics can get in the way of non-tracked benefits people bring; reducing people to metrics might distract from the important social dimensions of the workplace''.
``As a manager, I feel this could be a great way to connect with the developers'' (P20).

\textbf{(3) Detailed Issue View.}
I showed the participants an issue view with integrated feedback on detected smells for this individual issue.
Developers seemed to prefer smell detection for individual issues rather than in an \xgls{its} dashboard.
P17 said that ``[the information] needs to be pre-digested into actions being requested of the developers; they can then decide to do it, or not'' and P02 said ``but they don't need the stats, just the lists of smells and where/how to fix it''.
P06 said that ``issue page hints could be very helpful''.

\textbf{(4) Link Graph Visualisation.}
I showed the participants a graph view of interlinked issues, including which link smells each issue has.
Smell lists and visualisations can help detect violations or identify relevant links, but finding the right balance between the two is essential.
Participants mentioned that lists are sufficient and visualisation is not needed for managing links.
Visualisations may be preferred at higher levels (e.g., epics) but may be overkill at lower levels (e.g., subtasks).
A high-level abstraction from lower-level links may be sufficient, such as by extracting ``related'' epics based on the presence of related user stories, tasks, or bugs.
P22 said that this also depends on the usage of links: ``depends how you work with it, what kind of project you are working in, and if a team uses a lot of issue links: then this would be helpful''.

\section{Discussion}

Users are interested in software frameworks that can detect violations of Best Practices.
These tools may be particularly useful for managers, but they must be integrated into users' workflows.
While visualisations are nice, it may not be necessary for the detectors to be valuable.
The users interviewed had different and sometimes conflicting priorities and needs regarding the tools' functionalities.
Existing links and larger issue graphs are more important for the R\&D team lead and product managers, who may benefit from visualisation, confirming the results of Li and Maalej~\cite{Li_2012_REFSQ}.
Tooling could unearth underlying process problems, for instance through violation detection.
Often missing or changing properties or too many options might lead users to rethink their processes and finally lessen the bloat by revamping their workflow.

For efficient \xgls{its} usage, finding the right balance between restriction and freedom is essential.
Restricting users too much can reduce efficiency if they become frustrated with the process, but if the degree of freedom is too high and things are not checked, many mistakes can occur that lead to problems in the future.
Restriction and freedom may depend on the specific needs and processes of the team or organisation using the \xgls{its}.
Herraiz et al.~\cite{Herraiz_2008_MSR}, for example, argued for simplifying the issue report form in Eclipse.
Too many options to choose from make it hard for reporters to choose the correct one and should, thus, be limited.

In large \xglspl{its}, automation is necessary due to the overhead involved in tasks such as reminding someone to update properties (e.g., status, due date, estimation) or updating properties based on the status of linked issues (e.g., if all sub-tasks are closed, the epic should be closed).
While Jira automations exist~\cite{Jira_2024_automation}, they seem not to be widely used.
One reason could be the missing awareness of practitioners of these features.
Another reason could be the overhead of configuring and updating these rules.
Supporting or recommending specific automation rules based on \xgls{ite} usage patterns need to be investigated.
Overall, there appears to be a need for a (smart) guide on how to use \xglspl{its} efficiently.

\section{Summary}

In this chapter, I recommended five core features for tooling support for Best Practices in \xglspl{ite}.
I then interviewed practitioners to get their feedback on these features.
Practitioners agreed with the benefits these features bring to \xglspl{ite}.
Depending on the context of the practitioner, whether manager, developer, or large company, they had slightly different perspectives on the usefulness of these features.
Overall, these findings showcase the potential for improvement in \xglspl{ite} through specific tooling features.
In this chapter, I started the discussion regarding recommendations for future work for this research, with a particular focus on practitioner impact through tooling support.
In the next chapter, I summarise the contributions of this thesis, and outline future work areas specifically aimed at researchers.

\chapter{Conclusion}  \label{ch:conclusion}

\epigraph{Nach dem Spiel ist vor dem Spiel.}{German Expression}

In this thesis, I addressed three primary gaps through four parts, across ten chapters, producing nine central contributions (see Fig.~\ref{fig:thesis_overview}).
I investigated the difficulties practitioners experience with \xglspl{its} by interviewing 26 practitioners.
I collected and analysed 16 Jira repositories to better understand the complexities that exist within \xglspl{its}.
\ltexdummy{I empirically} formed the Best Practices Ontology for \xgls{ite}, created an initial catalogue of {\numItebp} Best Practices, and implemented {\numItebpAlgs} algorithms to automatically detect Best Practice violations.
Through these empirically driven investigations, I have made progress towards addressing the three primary gaps.
We now know considerably more about the difficulties encountered by practitioners using \xglspl{its}, as well as the complexities within \xglspl{its}.
We also now have context-dependent recommendations, and a theoretical structure to further improve this area of research and practice.

\section{Threats to Validity}  \label{sec:conclusion_threats}

In this thesis, I conducted various data gathering and analysis techniques.
In this section, I will summarise the threats to validity across the techniques in aggregate.
Where necessary, I also discuss specific concerns with certain techniques applied in particular chapters.
This thesis collects and processes both quantitative and qualitative data, with a particular focus on understanding and summarising large amounts of textual data.

The collection of the Jira dataset introduced a potential sampling bias into the results, as I did not get a ``random sample'' of all Jira repos to conduct this research on.
However, I feel this bias is outweighed by the benefit this large dataset provides, as I do not know of any larger set to which I could apply stratified representative sampling.
In the end, the dataset is across 16 Jiras (including world-leading software organisations such as Apache, RedHat, and Spring), with 2.7 million issues and 30 million evolutions.
Given the large dataset, I believe the claims made in this thesis indeed represent a larger understanding of public \xglspl{its} in practice.
Additionally, other \xglspl{its} may have other features and be used differently, but as noted in Chapter~\ref{ch:activities}: ``Jira is by far the most popular tool in the \xgls{ite} and agile project management markets''.

Observer bias represents one main risk for interview studies, where the interviewers elicit the statements they are expecting or hoping for.
I took multiple measures to mitigate this risk, as discussed in Chapter~\ref{ch:challenges}.
Notably, I explicitly encouraged participants to disagree and share their personal opinions.
There were also two interviewers in each session, we posed the questions as neutral as possible, and we exposed participants to information only when they had to see it.
Due to participants' disagreements as well as the heterogeneous perceptions gathered, this bias appears minor.
I might have misunderstood the interviewees or missed important points in my reporting, particularly as I did not record the sessions (for the criticality of the discussion and for better spontaneous engagement).
To mitigate this risk, both interviewers took notes and asked for clarifications whenever needed.
We also discussed each interview session afterwards to mitigate the threat of memory recall.
Finally, our analysis of the notes was conducted by three authors (for each interview), is fully documented, and each sentence traced to the findings.

The interview sample includes only 26 experienced software practitioners.
I think that this is a rather large sample, comparable to seminal qualitative studies in \xgls{se}.
The sample provided enough diversity and redundancy to derive the findings.
However, I refrain from claiming generalisability of the results to software practitioners, and this never was the goal.
Nevertheless, the diversity of the sample, involved companies, and redundancy of observation give me reason to believe that the overall trends hold true.
To achieve quantifiable and generalisable results, follow-up studies (such as surveys or experiments with practitioners) would be needed.
The qualitative findings serve as starting hypothesis and variables to measure for such studies.
Similarly, the results are based on the subjective statements of practitioners.
In theory, what people do might diverge from what they say.
Therefore, triangulation with observation studies or artefact analysis will likely lead to stronger evidence and more insights.

This thesis relies heavily on the use of qualitative methods to interpret and categorise real-world phenomena
(e.g., issue type themes in Chapter~\ref{ch:activities} and the information themes in Chapter~\ref{ch:evolution}), which can result in threats to construct validity.
To mitigate these threats, I followed strict qualitative guidelines designed by prominent methods researchers and analysed thousands of data points to achieve strong resonance and saturation.
Where possible, I released the analysed datasets, including intermediate steps.
Additionally, I referenced outside materials where available, triangulating my interpretations with explicit descriptions of these constructs.
By following these methodological guidelines, I am confident in the reliability of the findings.

\section{Summary of Contributions}

\textbf{Cross-Study Finding that Context is Key.}
My findings across the interview studies and historical analysis highlight that ``context is key'', confirming that context plays a central role in shaping \xglspl{ite}, including problems and solutions.
It is no surprise to \xgls{se} researchers that context plays an important role in a \xgls{se} concept.
However, \xglspl{ite} have largely been studied and reported without the context playing a central role in any of the experimental design, the reported factors for case studies, or the statements made about findings.
This means that studies of \xglspl{ite} are limited to the exact (often unreported) context in which they are studied, and findings are not contextualised for practitioners to know when and how to apply them in their own context.
The interview studies revealed that problems experienced by practitioners are context-sensitive (occurring under some conditions and not under others), and the impact of these problems was also context-dependent.
Some \xgls{ite} problems and solutions were also in conflict with each other, and their prioritisation was dependent on the relevant context factors.
The results of the historical analyses in Chapters~\ref{ch:activities} and~\ref{ch:evolution} show a diverse usage of Jira, with different \xgls{se} activities and information types emphasised in different contexts.
To treat \xglspl{ite} as universal would be a mistake, and for specific empirical investigations it would introduce major flaws to construct and conclusion validity.
My findings highlight the importance of understanding, declaring, and investigating context factors when researching \xglspl{ite}.

\textbf{``\xglsfirst{ite}'' as a Fundamental Concept.}
I introduced the concept of an \xglsfirst{ite}: ``an \xgls{its} tool, and all surrounding contextual factors that affect the \xgls{its}, including the stakeholders who interact with it, the company they work for, the team they work with, and the project they are working on.''
There currently exists many terms to describe tools such as ``Jira'', and they are both manifold and too narrow.
Existing terms focus on the tool itself, rather than the larger context that surrounds, affects, and is affected by the tool.
The concept of an \xgls{ite} includes the environment around the tool, thus encapsulating and unifying what the term means.
The term unifies and expands on the existing term usage regarding the rhetoric around \xglspl{its}, issue tracking systems, and bug trackers.
The benefits of ``\xgls{ite}'' as a concept goes beyond just addressing these limitations, and include drawing attention to the surrounding ecosystem, and encouraging an explanation when a researcher does not account for these ecosystem factors in their research.

\textbf{An Understanding of Practitioner Problems within \xglspl{ite}.}
I interviewed 26 practitioners who regularly interact with \xglspl{ite} and produced a list of commonly reported problems.
\xglspl{its} are commonly reported in pop-culture as difficult to use, but the studied details of such difficulties are lacking.
We have lots of research regarding specific problems within \xglspl{ite}, such as Bug Report quality~\cite{Bettenburg_2008_FSE,Zimmermann_2010_TSE} and the correctness of ``requirements'' issue types~\cite{VanCan_2024_REFSQ,Winkler_2018_REFSQ}.
However, these pinpoint analyses are often not focused on the problem itself (e.g., its existence and the impact on industry), but rather trying to solve the problem.
I found a wide variety of problems across three major categories: \xgls{ite} information, \xgls{ite} workflows, and \xgls{ite} organisational aspects.
Of particular interest was the trade-offs that occurred between potential solutions to one problem, that would then increase the likelihood or impact of a different problem.
For example, one of the most common problems reported is ``workflow bloat'', in which the \xgls{its} workflow involves so many mandatory steps, that practitioners felt either overwhelmed or frustrated in the amount of time they had to dedicate to just interfacing with the \xgls{its}.
\ltexignore{However, an equally prevalent problem was people not using the \xgls{its} properly, and many practitioners recommended that the \xgls{its} should enforce \textit{more} on the user to solve improper \xgls{its} usage---which would then also increase the amount of workflow bloat.}
Overall, the findings highlight that even today, after decades of research into \xglspl{ite}, there are still many problems to address.

\textbf{Dataset of Issue Tracking Systems.}
I collected and published data from 16 public Jira repositories containing {\mytilde}2000 projects, 2.7 million issues, and 30 million evolutions~\cite{Montgomery_2022_MSR}.
While GitHub is a popular, public, and well-studied resource for \xglspl{its}, it is also a code-first platform, and lacks certain customisation features available in Jira.
According to 6Sense~\cite{6SenseJira_2024_Online}, Datanyze~\cite{DatanyzeJira_2024_Online}, and Enlyft~\cite{EnlyftJira_2024_Online}, Jira is by far the most popular tool in the \xgls{its} and agile project management markets.
However, finding \xglspl{jr} to analyse is difficult because Jira is a private tool hosted by individual organisations.
This Jira dataset that I published is the single largest public collection of Jira repositories that exists.
Other datasets do exist, but they are either subsets of my dataset\footnote{My dataset has the full Apache Jira Repository, while it is quite common to find Jira datasets that have a subset of projects from that Jira.} or considerably small (1--10k issues) custom sets that have some other special data attribute (such as custom labelling).
This dataset has been viewed 10,000 times and downloaded over 2,700 times,\footnote{\url{https://doi.org/10.5281/zenodo.5882881}} showing the interest that exists for such data.

\textbf{Data-Driven Characterisation of \xglspl{its}.}
I investigated the Jira dataset and revealed key information about practitioner usage of Jira, including the activities that are conducted within \xglspl{its}, the information that is managed, and the evolution that occurs.
\xglspl{its} have been studied for many decades, but largely from specifics angles, without a holistic perspective.
While there are many specific problems that that should and have been studied, a holistic perspective provides a comparative summary that allows for a complete picture of the system under study.
My investigation revealed three high-level activities in \xglspl{its}: Requirements, Development, and Maintenance.
For each, a subset of artefacts can be found, including Epics and User Stories for Requirements, Tasks for Development, and Bug Reports for Maintenance.
Requirements issues account for \mytilde30\% of all issues (median \% issues per project), Development accounts for \mytilde20\%, and Maintenance accounts for \mytilde50\%.
My investigation also revealed different information types within \xglspl{its}: Content, MetaContent, RepoStructure, Workflow, and Community.
My analysis then revealed how often these information types evolve, including the average of \mytilde8 evolutions per issue, and most evolutions occurring to Workflow (\mytilde40\%) and Community (\mytilde30\%).
Overall, my analyses reveals and details many intricate inner workings of \xglspl{its} that confirm some existing assumptions about \xglspl{its}, and challenge others.

\textbf{Best Practice Ontology for \xglspl{ite}.}
I created an ontological structure for \xgls{ite} Best Practices that collects and re-frames decades of research into quality aspects for \xgls{ite}.
The ontology provides a unified structure to research, discuss, and communicate quality attributes for \xglspl{its}.
The ontology, while not proposed as theory, has many important constructs to be framed as theory by future work.
I also contributed five propositions which should be scrutinised by future work looking to contribute theory in this area.
The ontology has five sections and 16 dimensions, with the sections being: Meta, Summary, Recommendation, Context, and Violation.
The ontology acts as a guide for those creating \xgls{ite} Best Practices, but also for those transforming or challenging existing \xgls{ite} Best Practices.

\textbf{Catalogue of Best Practices for \xglspl{ite}.}
I formed {\numItebp} Best Practices from existing research into quality factors for \xglspl{ite}.
The catalogue acts as a starting place for both researchers and practitioners to build on.
Future research can refute and add to the knowledge that they contain, in a structured way.
Practitioners can review and apply the Best Practices directly in their \xglspl{ite}.
The catalogue was constructed using techniques similar to a secondary study.
Primary articles were collected, and then key information was extracted and summarised in the ontological structure, producing the catalogue.
The catalogue is incomplete, and requires future work to fill in the missing unknowns.

\textbf{Algorithms Detecting Violations of Best Practices for \xglspl{ite}.}
I collected and created {\numItebpAlgs} algorithms to automate the detection of violations to \xgls{ite} Best Practices.
These algorithms allow the immediate application of these techniques in industrial settings.
They are also a starting place for cross-analysis research, applying the findings I presented with my dataset.
The algorithms are all built for Jira data, but their application only requires data that can be found in most \xglspl{its}.
My findings from applying them to my Jira dataset revealed a mix of well-followed Best Practices (low violation rate), and ignored or unimportant Best Practices.

\textbf{Nested Configurations to Satisfy Contextual Needs.}
I introduced the concept of Nested Configurations for tooling solutions that address \xgls{ite} Best Practices as a direct recommendation to deal with the importance and complexity of context factors in \xglspl{ite}.
\xgls{ite} research needs to acknowledge, specify, and investigate \xgls{ite} phenomena with context in mind, and once those context factors have been identified, practitioners need a way to apply them into their \xgls{its} tooling.
Practitioners also require a way to adapt the tooling to their context-specific needs.
The solution to these situations is nested configurations, such that layers of settings can adapt and conform to contexts---whether identified in research or practice.
In Chapter~\ref{ch:tooling}, I offered a recommended structure for nested configurations, listed in order of increasing precedence: organisation, team, project, sprint, and individual.
This means that an organisation should define their preference for Best Practices in their \xgls{ite}, but a team should be able to override particular Best Practices if their context demands it.
There are also many good reasons to use additional configuration files, as well as different precedence orders.
Overall, the nested configuration concept can adapt to many---and perhaps all---organisation contexts.

\section{Future Work}

Future work in the area of quality factors for \xglspl{its} and \xglspl{ite} can take many potential directions.
Given the momentum created by this thesis, I am biased towards the further investigation, improvement, and dissemination of \xgls{ite} Best Practices.
However, there are other potential directions as well, including an entirely different conceptualisation of \xgls{ite} or the creation of a new (or transformed) structure similar to \xgls{ite} Best Practices.
For the sake of a focused perspective, I will describe the future work I see for \xgls{ite} Best Practices.

\subsubsection{Investigate and Characterise Specific Patterns of ITS Usage}

We need research investigating and characterising specific patterns of \xgls{its} usage.
My investigation of \xglspl{its} in Chapters~\ref{ch:activities} and~\ref{ch:evolution} surfaced many findings, but the landscape of holistic information about \xglspl{its} and \xglspl{ite} is still very limited.
Specific \xgls{se} process models tend to view \xglspl{its} as supporting Agile \xgls{se} (see Section~\ref{sec:background_agile_re}) or a Change-based form of \xgls{re} (see Section~\ref{sec:background_change_based_re}), however, my intuition after working with \xglspl{its} for more than a decade~\cite{Montgomery_2017_RE,Montgomery_2017_RE_a,Montgomery_2017_MScThesis} is that \xglspl{its} support notably specific and alternative forms of \xgls{se} across all phases of \xgls{se}.
Much of the continued research in this area treats these systems as homogeneous, making broad statements about the findings and recommending simplified solutions for practitioners.
Additionally, much of the tangential research views \xglspl{its} as merely simple tools to house more complex processes.
All these assumptions, as shown by my results, are simply not true.

Regarding specific directions for future work, I recommend exploratory empirical research that is grounded in both historical \xgls{its} data and interviews or surveys for follow-up member checking.
These exploratory investigations need to apply rigorous content analysis methods such as \xgls{ta} to form a coherent and consistent perspective on the data.
I recommend applying an \textit{inductive} \xgls{ta} with \textit{rich overview} as the level of analysis, working to surface \textit{semantic} themes.
My assumption is that specific patterns of usage will emerge, which we can then member-check with industry participants, and eventually begin to generalise to other industrial contexts that apply similar processes.
My expectation is that a clear set of patterns will emerge, which can be found across types of \xglspl{its}, in different \xglspl{ite}.
The work of \ltexdummy{van Can} and Dalpiaz~\cite{VanCan_2024_REFSQ} is already conducting research in this direction: starting bottom-up, and investigating what exists in \xglspl{its} (inductively), instead of presupposing models of usage.
I believe that similar continued efforts is the best way to fully understand and therefore offer the best recommendations for these industrial \xglspl{ite}.

\subsubsection{Investigate and Characterise Context Factors for ITE Best Practices}

We need research investigating and characterising context factors for \xgls{ite} Best Practices, to make these recommendations more specific, applicable, and adaptable.
Related work in the area of context for tool support has highlighted that context is often dynamic and learned~\cite{Happel_2008_RSSE,Maalej_2009_ASE}.
My investigation into practitioner challenges with \xglspl{its} in Chapter~\ref{ch:challenges} and structuring of existing research into \xgls{ite} Best Practices in Chapter~\ref{ch:catalogue} show the importance and lacking of known context factors in \xglspl{ite}.
In addition to the knowledge we need on ``specific patterns of \xgls{its} usage'', we also need to consider the context factors that affect those patterns of usage.
\xgls{ite} Best Practice recommendations that work consistently for a given \xgls{its} usage pattern may not apply if specific context factors change.
\ltexignore{As discussed at length in Section~\ref{sec:ontology_context}, I believe it is possible to remove all notions of subjectiveness when it comes to the application of \xgls{ite} Best Practices because the ``it depends'' exceptions are just the result of hidden factors at play.}
While ``smells'' still have a valuable role in the \xgls{se} landscape (both in research and practice), I strongly believe that a mature understanding of quality factors for \xglspl{ite} involves elevating ourselves from smells to Best Practices---which account for context factors.

Regarding specific directions for future work, I recommend multiple-case study research with an embedded design (see \ltexdummy{Runeson et al.~\cite{Runeson_2012_Wiley} Section 3.2.3}).
The cases should be different project settings within a largely shared context, and the units of analysis should be the different \xgls{ite} Best Practices.
The largely shared contexts grants similarities that simplify the investigation (e.g., same company and maybe even same team), while the different project settings (e.g., different project, sprint, or customer) grants experimental-design-like differences to focus the analysis on.
The number of \xgls{ite} Best Practices under study should be kept low to focus the analysis (max 3--5).
The data collection methods should include data analysis of the live repositories, but more importantly, interviews, observations, and surveys of the people involved to understand the success (or not) of the Best Practices on their \xgls{ite}.
\ltexignore{After considerable data collection, the data analysis should involve in-depth cross-case analyses to compare which shared context factors had similar impacts on the different Best Practices, and which separate context factors likely played a role in their respective successes or failures.}
Within the scope of a single embedded multiple-case study, it is reasonable to assume that meaningful results can emerge that outline the relevance and impact of multiple context factors for multiple \xgls{ite} Best Practices.
I believe that this form of targetted research would reveal many meaningful insights that further our understanding of both \xglspl{ite} and the Best Practices that apply to them.

\subsubsection{Validate the Initial Catalogue of ITE Best Practices}

We need research validating the initial catalogue of \xgls{ite} Best Practices, to increase the confidence in the extracted information and discover flaws in the ontological design.
My creation of the \xgls{ite} Best Practice Ontology in Chapter~\ref{ch:ontology}, followed by my creation of the initial catalogue in Chapter~\ref{ch:catalogue}, produced an empirically grounded starting place for this conceptual structure.
However, there is still much work to be done before any reasonable trust should be put in the results.
The \xgls{ite} Best Practice Ontology is designed to structure, support, and encourage well-formed theory in the area of quality factors for \xgls{ite}.
The catalogue is designed to offer a starting place for what these Best Practices could look like, and offer some amount of robustness testing for the ontology itself.
However, the nature of these initial conceptual frameworks, including the design of my work, is such that follow-up research is needed to validate the results.
In my case, there are two possible avenues for validation: either the ontology itself from a purely theoretical perspective, or any of the \xgls{ite} Best Practices from either theoretical and practical perspectives.
I will only discuss the latter avenue: validation of specific \xgls{ite} Best Practices.

Regarding specific directions for future work, I recommend a mix of two techniques.
First, holistic single-case studies for practitioner feedback and investigation of loose casual effects of the Best Practice on meaningful outcomes.
Second, historical data analysis for systematic investigation of Best Practice application, implementation, and intervention.
By the nature of the initial formation of the Best Practices, it is not known which parts of each Best Practice need to be validated the most.
For this reason, I recommend a more open and exploratory case-study-like investigation of the application of these Best Practices in industrial settings.
I believe the quickest and most direct way to challenge (and therefore learn about) the Best Practices is to explore their full application.
These lessons learned can then be directly applied to the Best Practices to confirm or update the Best Practices.
I also recommend historical data analysis due to the rich nature of \xgls{its} data, including historical analysis of pre- and post-intervention outcomes.
Many of the desired benefits of \xgls{ite} Best Practices can be directly measured within \xglspl{its}, and therefore it is possible to algorithmically and systematically study the impact of Best Practices quantitatively.
Quantitative analysis of sociotechnical outcomes has been repeatedly challenged in \xgls{se} research~\cite{Razzaq_2025_CS}, including the creation of productivity frameworks that consider a wide range of context factors and desired sociotechnical outcomes~\cite{Forsgren_2021_Queue}.
This should be considered when investigating Best Practices quantitatively, since what is ``desirable'' and what is quantitatively ``measurable'' may not intersect for a given \xgls{ite} Best Practice.
I believe a combination of these techniques will lead to more robust, validated \xgls{ite} Best Practices.

\subsubsection{Engineer a Complete Tooling Solution to Integrate ITE Best Practices into ITSs}

We need to engineer a complete tooling solution to integrate \xgls{ite} Best Practices into \xglspl{its} to streamline and support practitioner adoption of the Best Practices.
My work has provided much of the foundational pieces required for a complete tooling solution.
The Best Practices are provided by Chapter~\ref{ch:catalogue}, the algorithms are provided in Chapter~\ref{ch:algorithms}, and the tooling solutions are provided in Chapter~\ref{ch:tooling}.
However, there is still much engineering work to be done.
Product development requires additional work, including analysis of which product features need to exist, which user groups should be targeted, how to integrate into the various \xglspl{its}, which \xgls{its} to begin with, and which \xgls{ite} Best Practices to start with.
Of notable importance, is to consider how the tooling can integrate, communicate, and support the adoption and maintenance of the Best Practices.
The integration of the violation detection algorithms is the most straightforward aspect, and the visualisation of the results of these algorithms is similarly approachable.
Other aspects, however, can be much harder to integrate, and can easily add to the existing complexity and feature bloat that already plagues \xglspl{its}.
For example, it is not clear how to support process suggestions made in the ``Recommendation'' section of \xgls{ite} Best Practices.

Regarding specific directions for future work, I recommend to first engineer the algorithmic detection and display of violations, then build the system of nested configurations, then investigate classes of process recommendations and implement them one at a time---with extreme sensitivity to visual and process clutter.
The algorithmic detection work has already begun, with my algorithms presented and published in Chapter~\ref{ch:algorithms}.
All that is missing is the integration of these into a live system with real-time access to the \xgls{its} database, which requires custom connectors for each \xgls{its} data model.
Once real-time algorithmic detection is built, then a custom display of these findings need to be built into the individual display of issues.
To support the management perspective (as discussed in Chapter~\ref{ch:tooling}), a dashboard should be built to visualise the results across larger sets of issues.
The implementation of a nested configuration system should be fairly straightforward, given the design description in Chapter~\ref{ch:tooling}.
At any given point, the system needs to be aware of both the nested configurations, and the current context within the system.
The result is a decision whether to apply the algorithms or not.
Finally, the process recommendations are indeed unique, and require intense investigation.
\ltexignore{Examples of process recommendations include real-time suggestions while writing or editing issues, pop-ups that suggest things to users while using other parts of the \xgls{its}, required reading of agreements before performing certain actions within the \xgls{its}, and role-based restrictions on actions.}
To begin this process, my recommendation is to focus on in-the-flow and in-the-moment suggestions for \xgls{its} users.
This brings information to users in the exact context it is intended for.
Users can then make in-the-moment decisions whether they want to act.
Other process recommendations can be considered and integrated over time.
I believe that these combinations of tooling integrations will create meaningful and useful improvements to \xglspl{ite}.

\subsubsection{Investigate the Full Application of ITE Best Practices in Industrial Settings}

We need to investigate the full application of \xgls{ite} Best Practices in industrial settings to understand the true impact of applying such recommendations in industry.
Considering the full culmination of my thesis work, the ultimate long-term goal is to improve the quality of \xglspl{ite} overall.
This goal requires the application of all future work items mentioned above, and then the application of the finalised and complete system in industrial contexts.
For meaningful improvements to real use cases, we need a broad characterisation of specific \xgls{its} usage patterns.
For repeated success in new environments, we need a broad characterisation of context factors for \xgls{ite} Best Practices.
To obtain confidence in the desired impact of individual \xgls{ite} Best Practices, we need evidence of the success of the Best Practices from multiple case studies.
For easy and repeatable integration of \xgls{ite} Best Practices, we need tooling that supports and adapts to many \xglspl{its} and \xglspl{ite}.
Regarding specific directions for future work, I wish all the best skill and luck to the researcher that undertakes such a large, complex, and long-term mission.
I believe that such a mission is possible, and would bring many needed improvements to the ever-growing world of \xglspl{ite}.


    \part{Appendices}  \label{part:appendices}
    \appendix

\chapter{Full Catalogue of ITE Best Practices}  \label{ch:appendix_catalogue}

In this appendix, I show all 40 Best Practices for \xgls{ite}.
First, I show the table of contents again, to act as a guide.
Then, I list the Best Practices one at a time.

\bpToC{appendix}

\bpTable{GoodBugReport}
\bpTable{AtomicFeatureRequests}
\bpTable{AssignBugsToIndividuals}
\bpTable{SufficientDescription}
\bpTable{SuccinctDescription}
\bpTable{AvoidStatusPingPong}
\bpTable{AvoidAssigneePingPong}
\bpTable{SetBugReportAssignee}
\bpTable{SetBugReportPriority}
\bpTable{SetBugReportSeverity}
\bpTable{SetBugReportEnvironment}
\bpTable{AssigneeBugResolution}
\bpTable{AvoidZombieBugs}
\bpTable{ActiveBugReports}
\bpTable{BugReportDiscussion}
\bpTable{RespectfulCommunication}
\bpTable{ConsistentProperties}
\bpTable{GoodFirstAssignee}
\bpTable{StableClosedState}
\bpTable{TimelySevereIssueResolution}
\bpTable{IssueCreationGuidelines}
\bpTable{OnTopicDiscussions}

\bpTable{BugToCommitLinking}
\bpTable{LinkDuplicates}
\bpTable{MinimalLinkTypes}
\bpTable{RecordLinks}
\bpTable{RealisticDependencies}
\bpTable{SingularRelationships}
\bpTable{ConnectedHierarchies}
\bpTable{HighDependencyBugsFirst}
\bpTable{SearchReminders}

\bpTable{OrderedProductBacklog}
\bpTable{TeamProducedWorkEstimates}
\bpTable{StoryPointsOverHours}
\bpTable{EstimateAllItems}
\bpTable{AvoidUnplannedWork}
\bpTable{RecommendedSprintLength}
\bpTable{ConsistentSprintLength}
\bpTable{UseAcceptanceCriteria}
\bpTable{LimitAcceptanceCriteria}

\chapter{Additional Figures}  \label{ch:appendix_additional_figures}

Many of the Best Practice figures presented in Chapter~\ref{ch:algorithms} are showing data limited to just Bug Reports.
Here, I present those same figures, but applied across all issue type themes.
They are presented here to provide some comparative value to see how similar the results are across the issue type themes, when compared to Bug Reports.

\begin{figure}[ht]
    \centering
    \includegraphics[width=\textwidth]{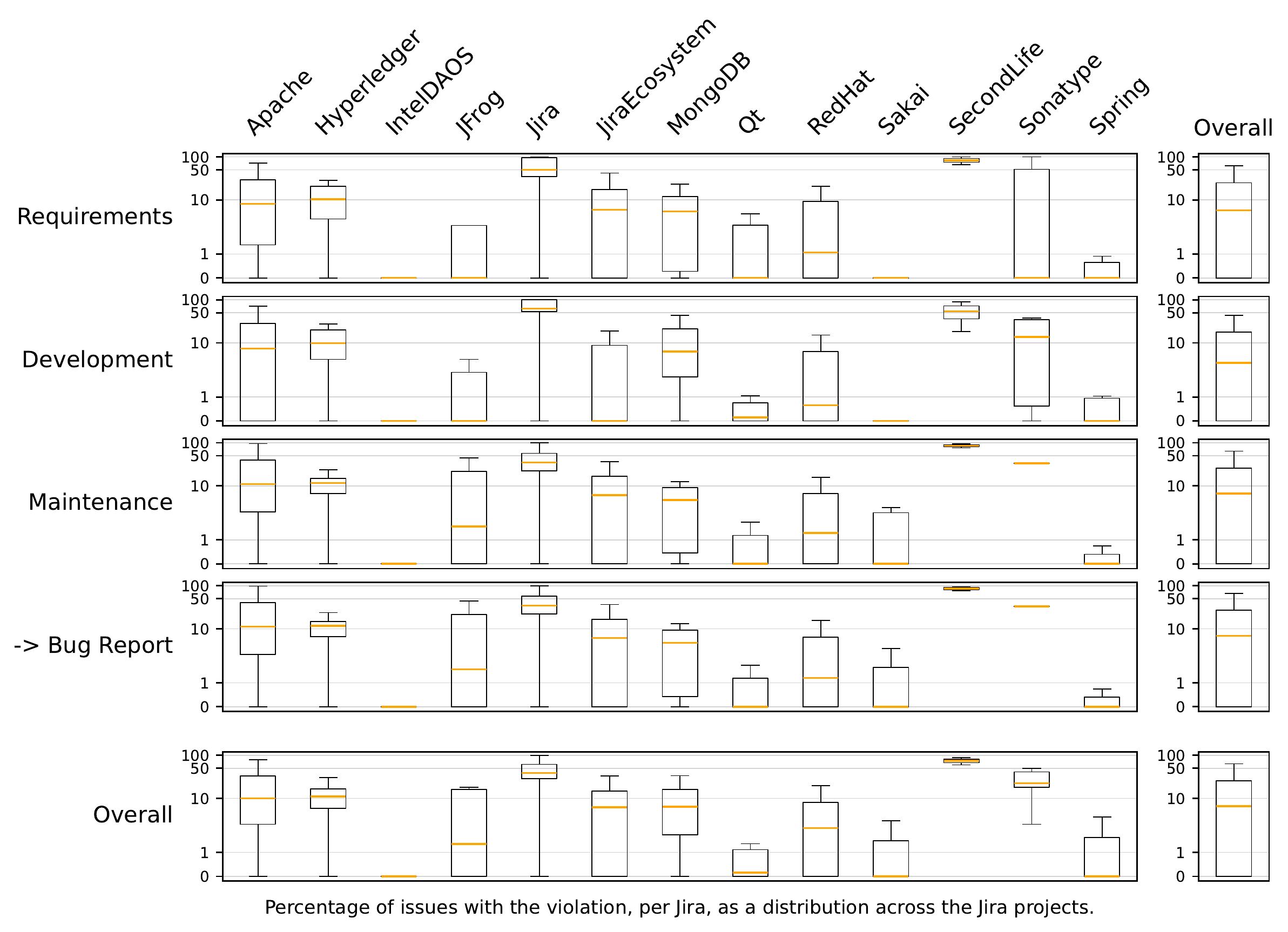}
    \caption{Violations to Set Bug Report Assignee (All Themes).}
    \label{fig:bp_set_bug_report_assignee_all}
\end{figure}

\begin{figure}[ht]
    \centering
    \includegraphics[width=\textwidth]{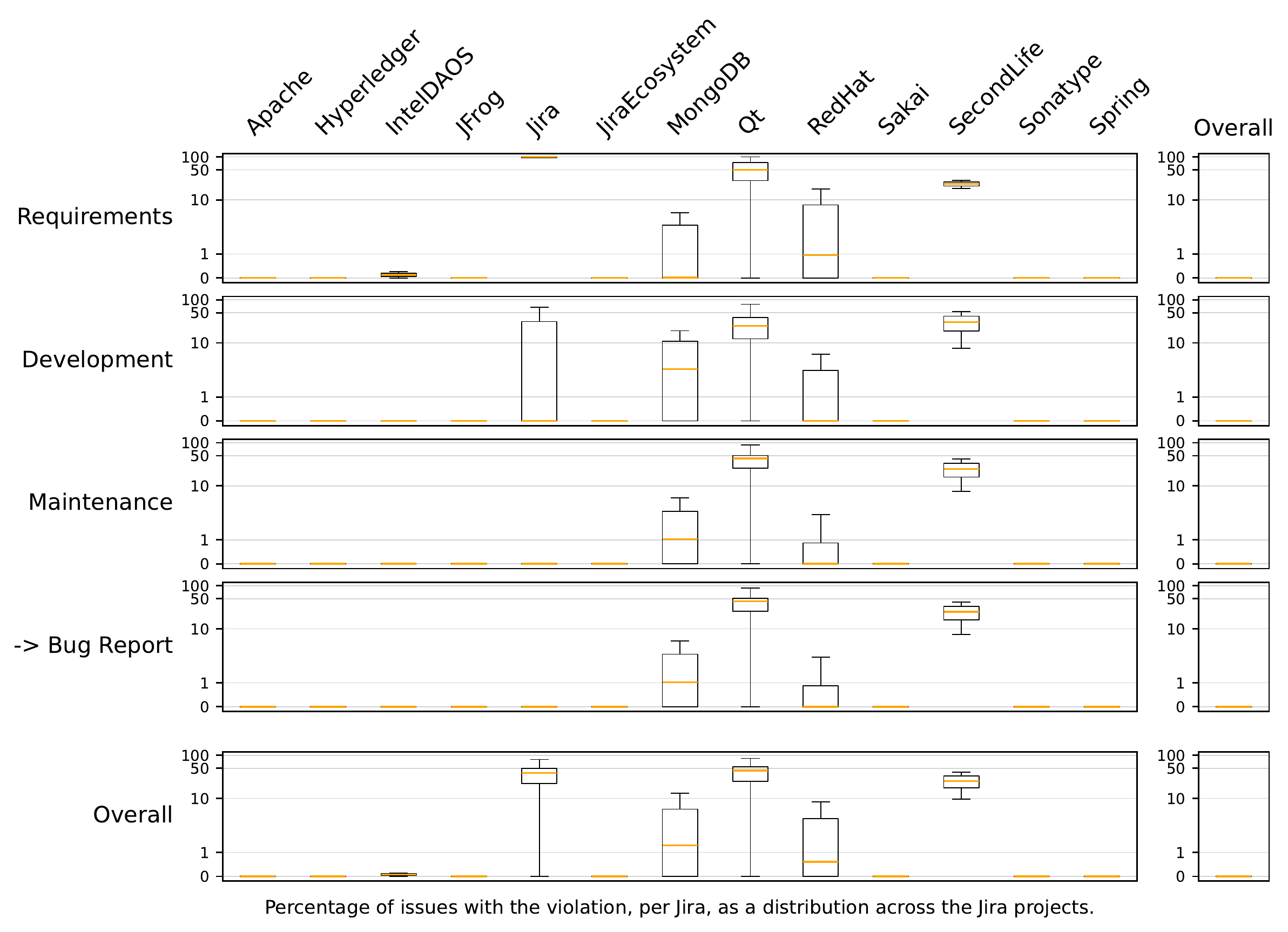}
    \caption{Violations to Set Bug Report Priority (All Themes).}
    \label{fig:bp_set_bug_report_priority_all}
\end{figure}

\begin{figure}[ht]
    \centering
    \includegraphics[width=\textwidth]{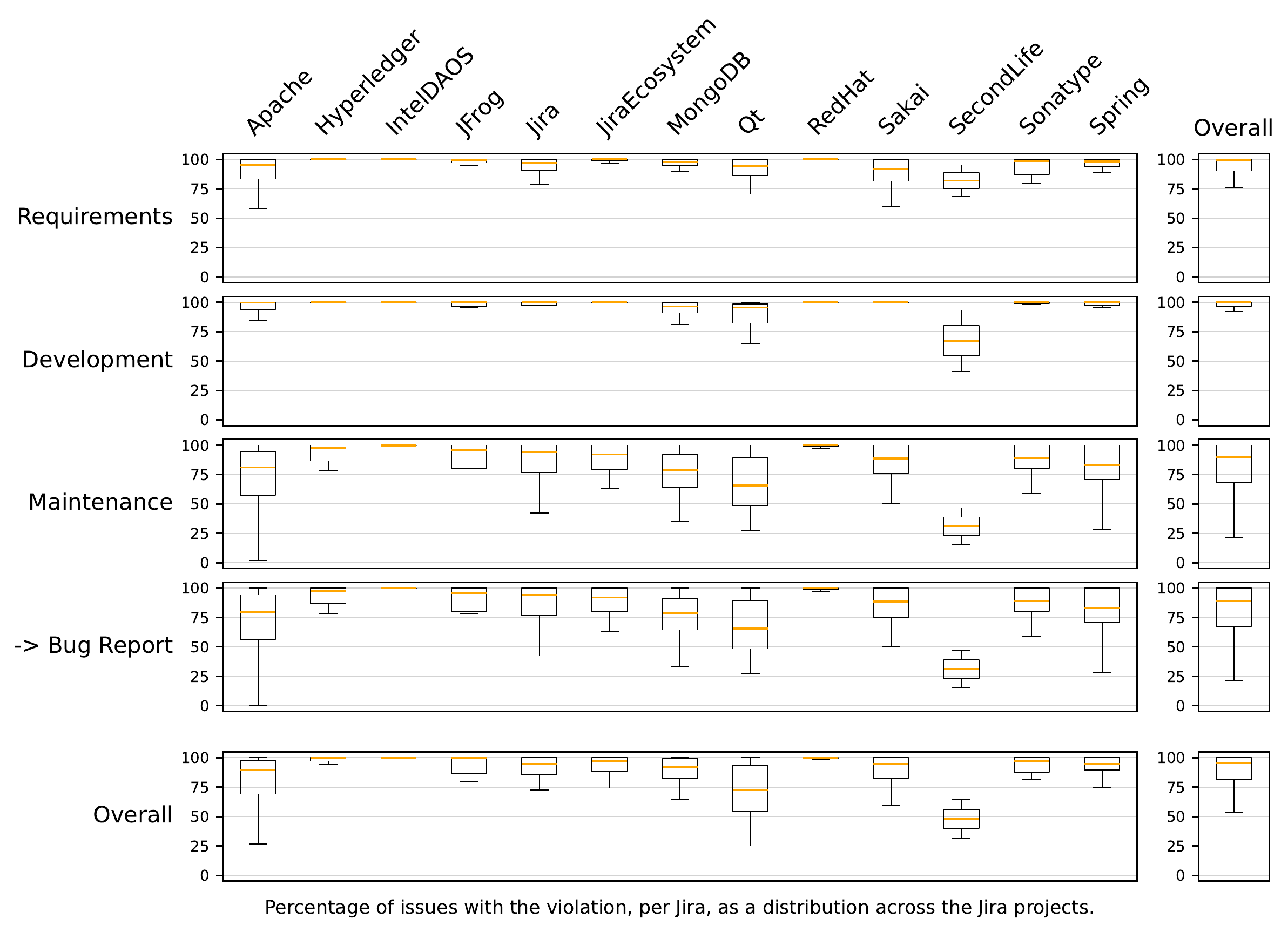}
    \caption{Violations to Set Bug Report Environment (All Themes).}
    \label{fig:bp_set_bug_report_environment_all}
\end{figure}

\begin{figure}[ht]
    \centering
    \includegraphics[width=\textwidth]{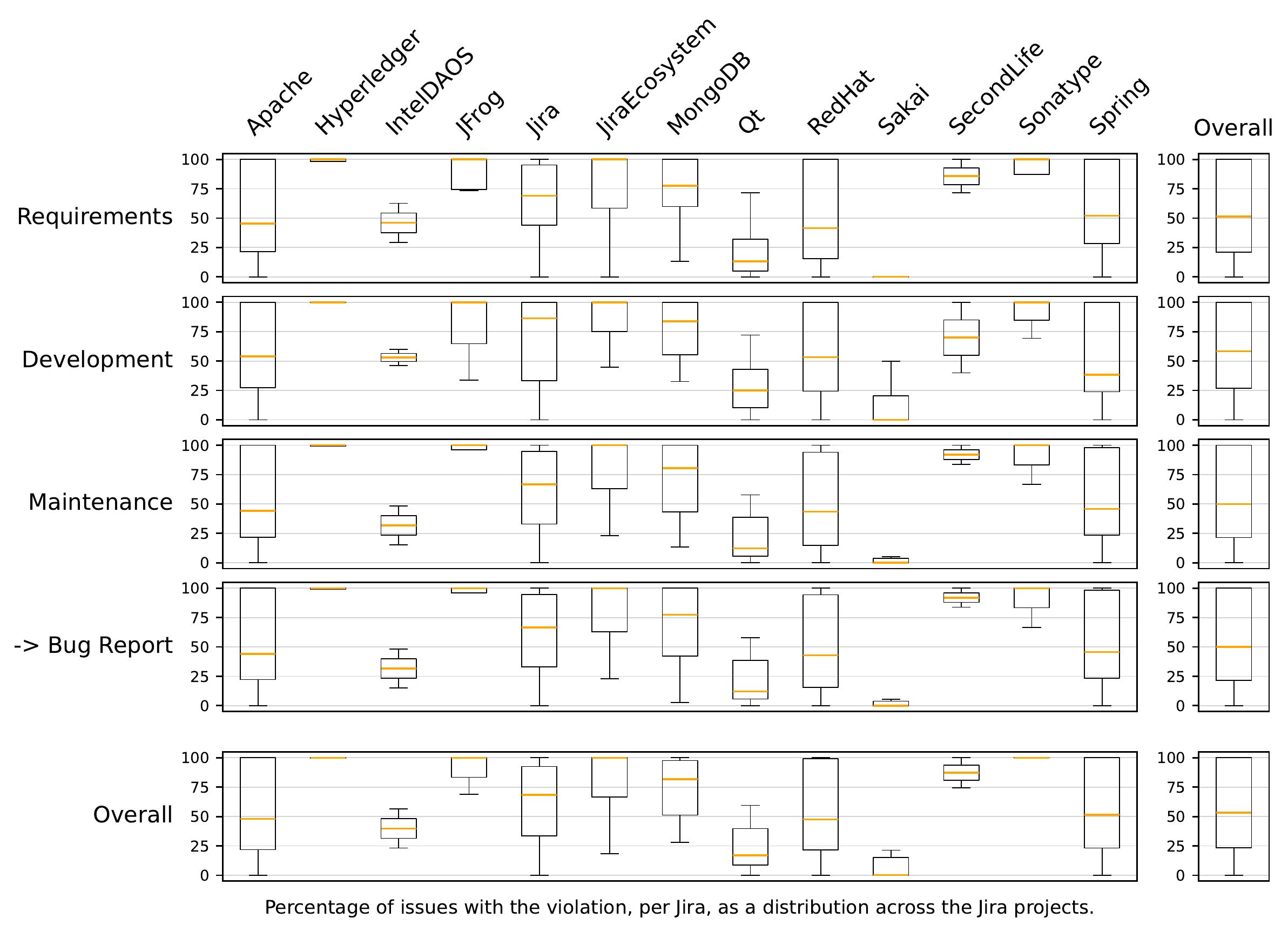}
    \caption{Violations to Set Bug Report Components (All Themes).}
    \label{fig:bp_set_bug_report_components_all}
\end{figure}

\begin{figure}[ht]
    \centering
    \includegraphics[width=\textwidth]{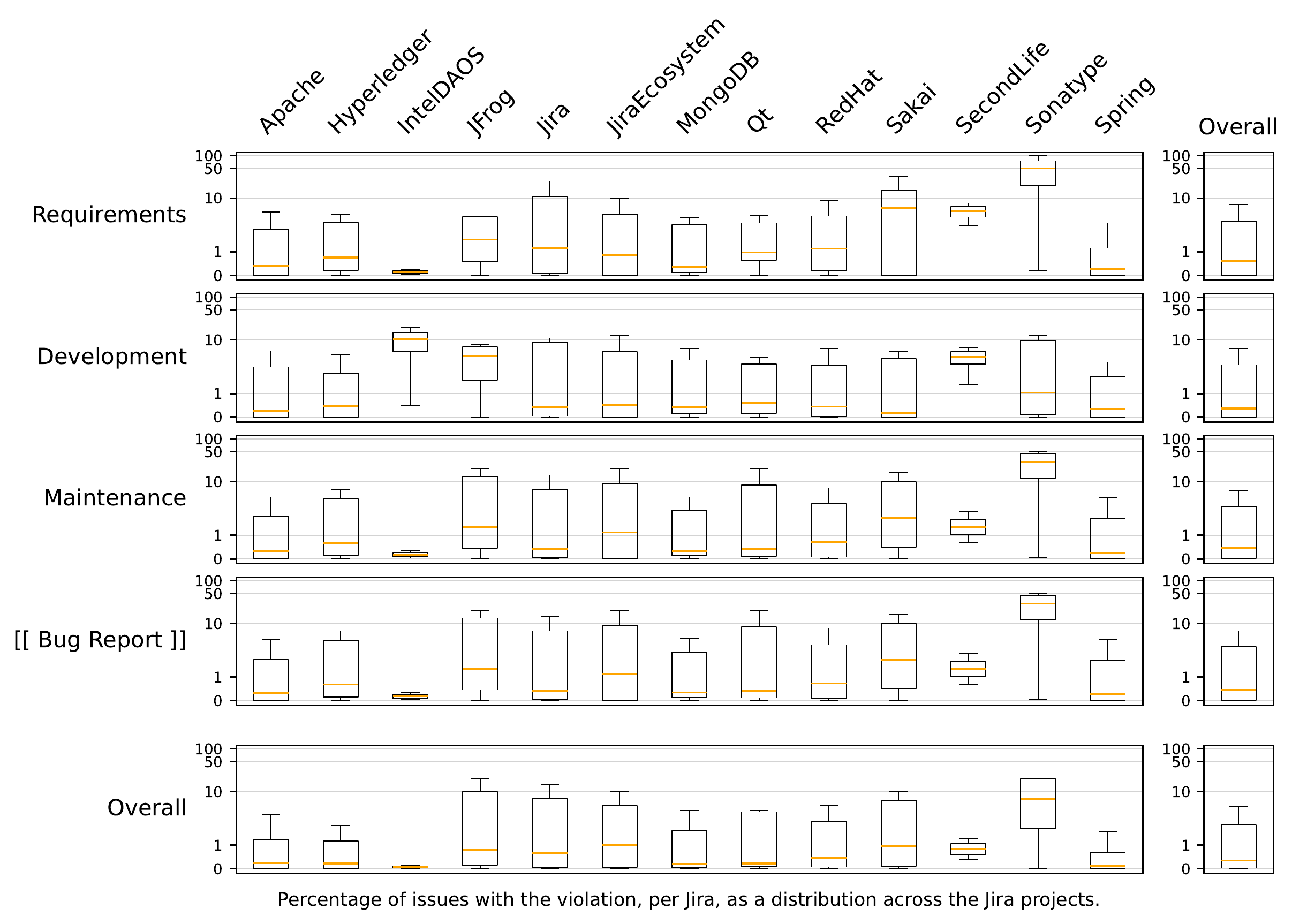}
    \caption{Violations to Good First Assignee (All Themes).}
    \label{fig:bp_good_first_assignee_all}
\end{figure}

\begin{figure}[ht]
    \centering
    \includegraphics[width=\textwidth]{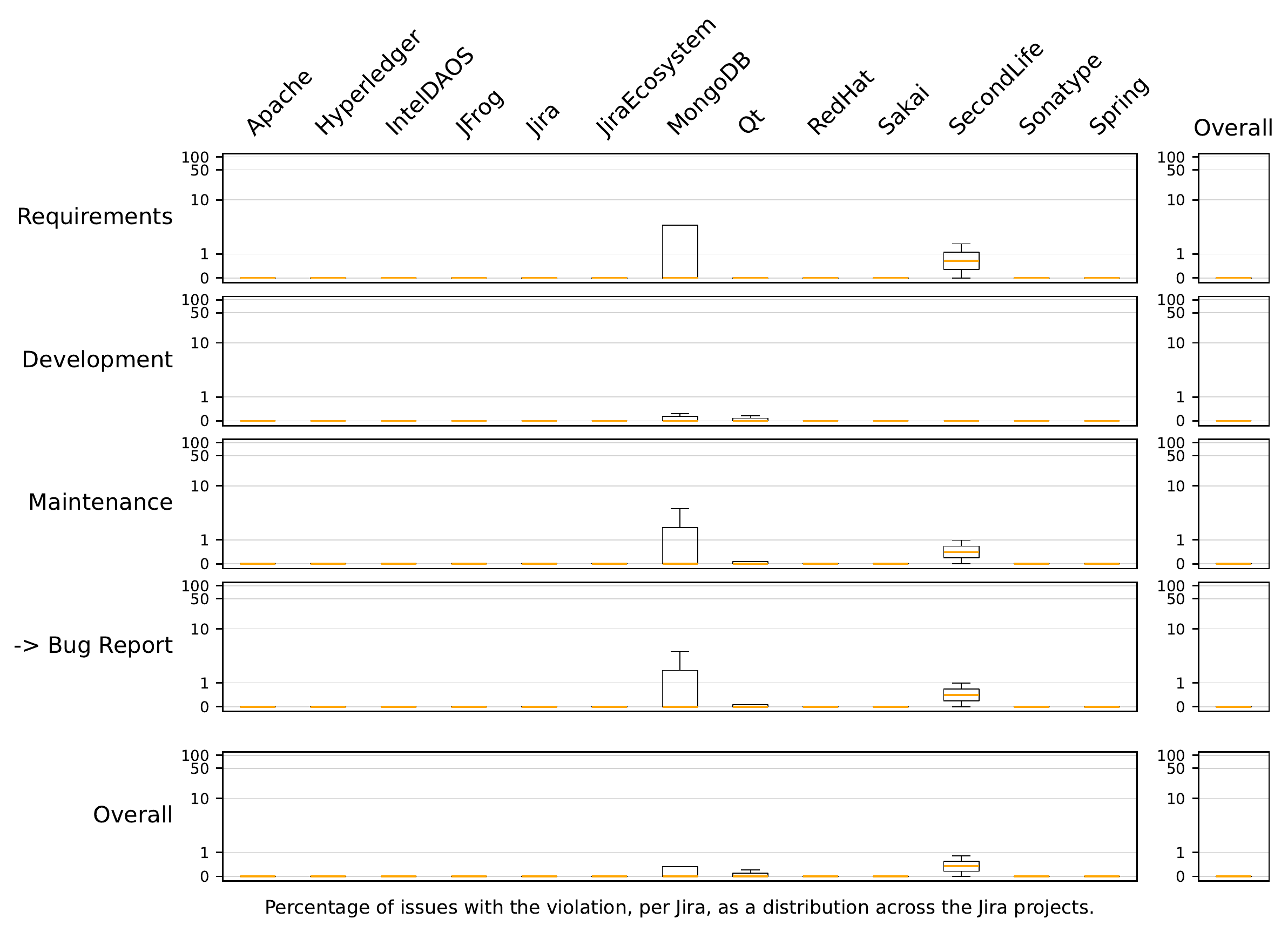}
    \caption{Violations to Assign Bugs to Individuals (All Themes).}
    \label{fig:bp_assign_bugs_to_individuals_all}
\end{figure}

\begin{figure}[ht]
    \centering
    \includegraphics[width=\textwidth]{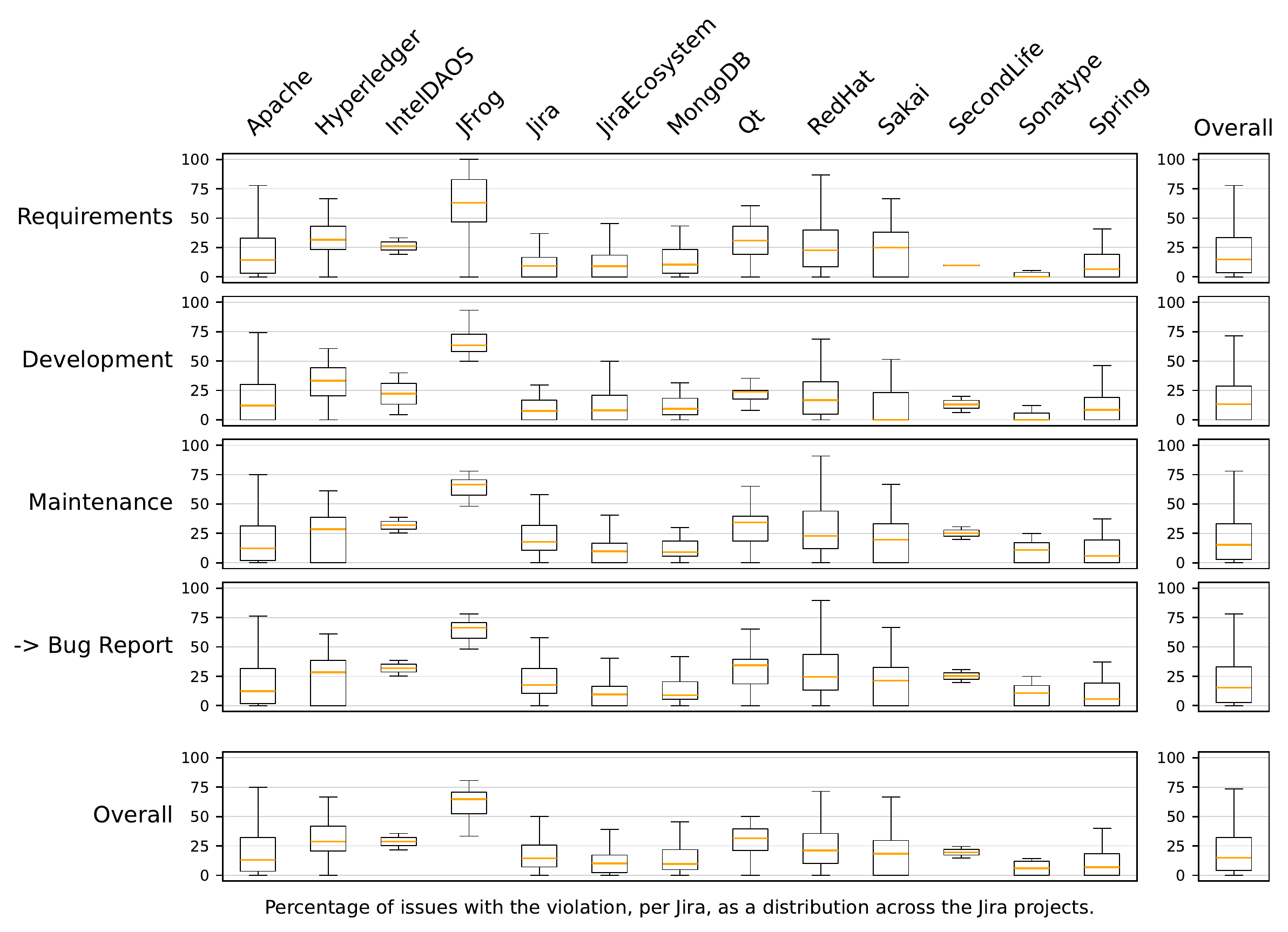}
    \caption{Violations to Assignee Bug Resolution (All Themes).}
    \label{fig:bp_assignee_bug_resolution_all}
\end{figure}

\begin{figure}[ht]
    \centering
    \includegraphics[width=\textwidth]{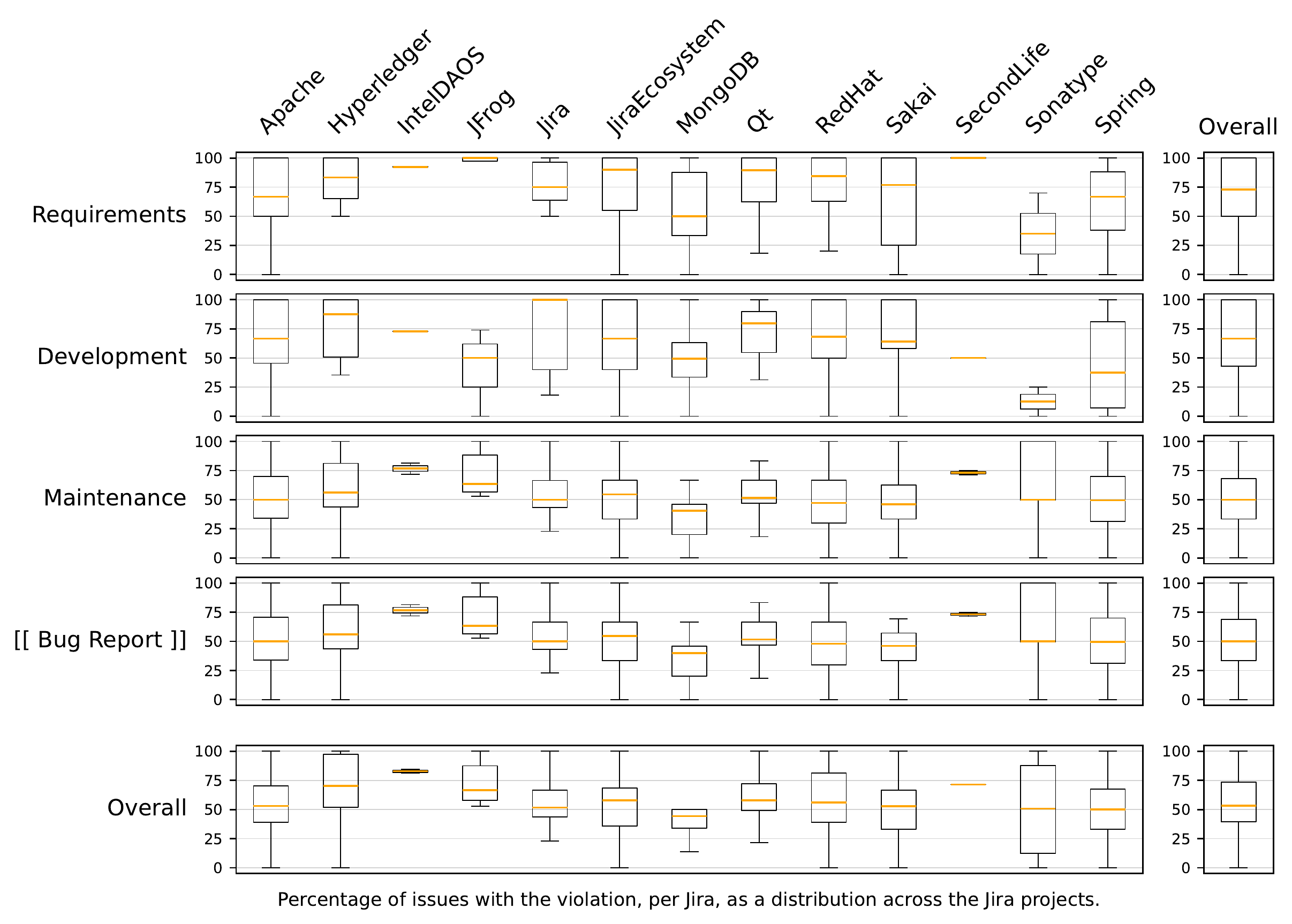}
    \caption{Violations to Timely Severe Issue Resolution (All Themes).}
    \label{fig:bp_timely_severe_issue_resolution_all}
\end{figure}

\begin{figure}[ht]
    \centering
    \includegraphics[width=\textwidth]{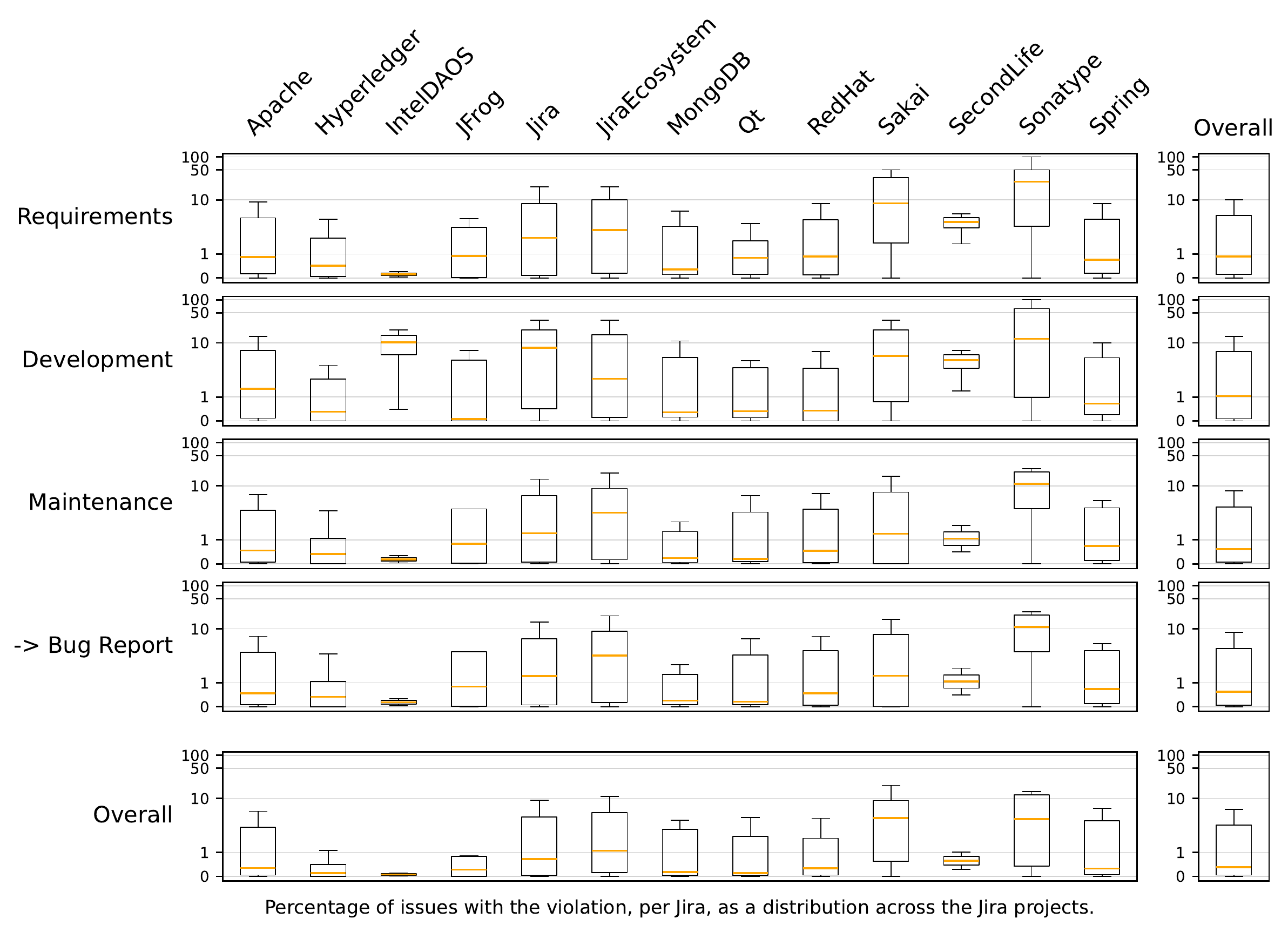}
    \caption{Violations to Avoid Zombie Bugs (All Themes).}
    \label{fig:bp_avoid_zombie_bugs_all}
\end{figure}

\begin{figure}[ht]
    \centering
    \includegraphics[width=\textwidth]{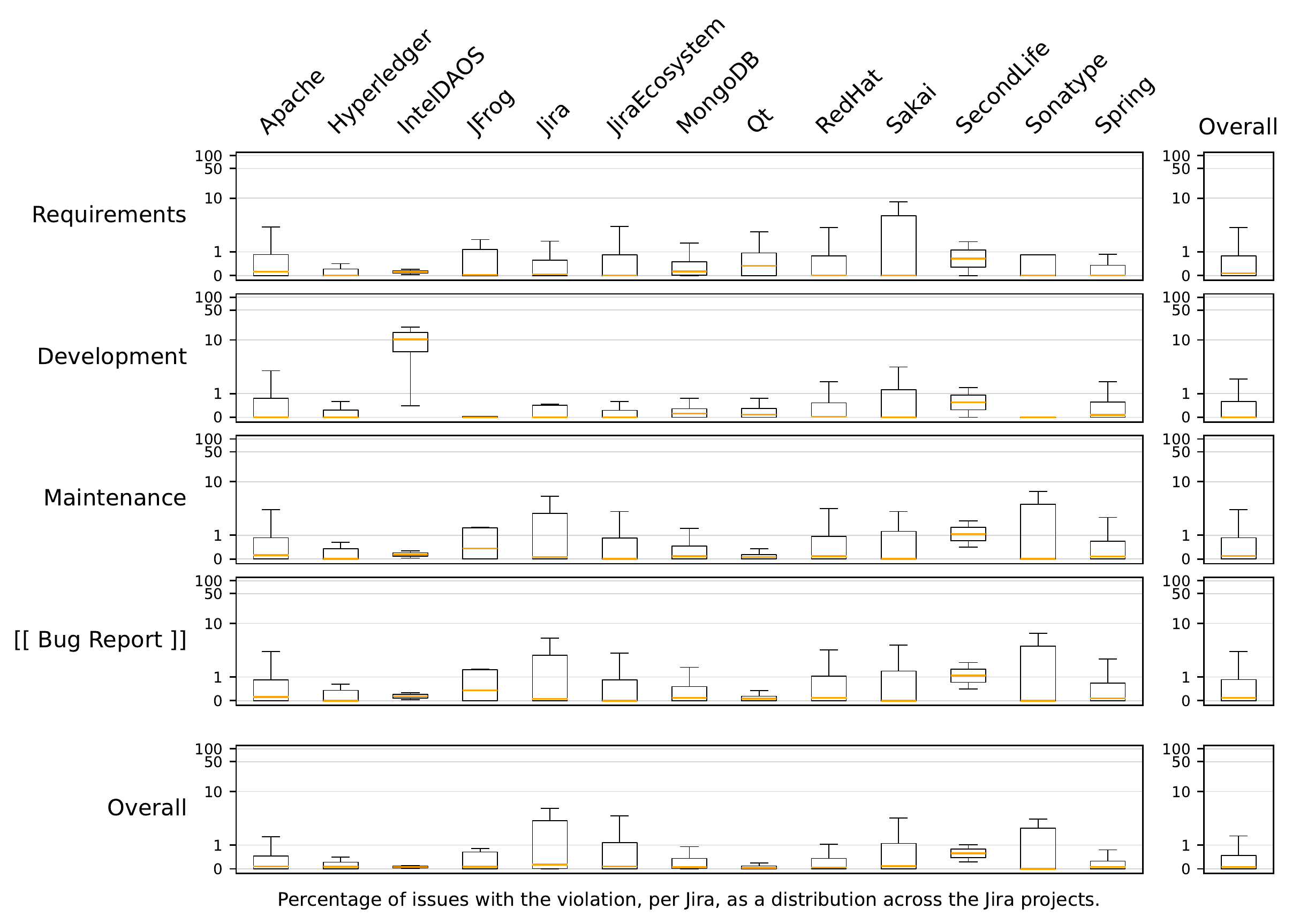}
    \caption{Violations to Stable Closed State (All Themes).}
    \label{fig:bp_stable_closed_state_all}
\end{figure}

\begin{figure}[ht]
    \centering
    \includegraphics[width=\textwidth]{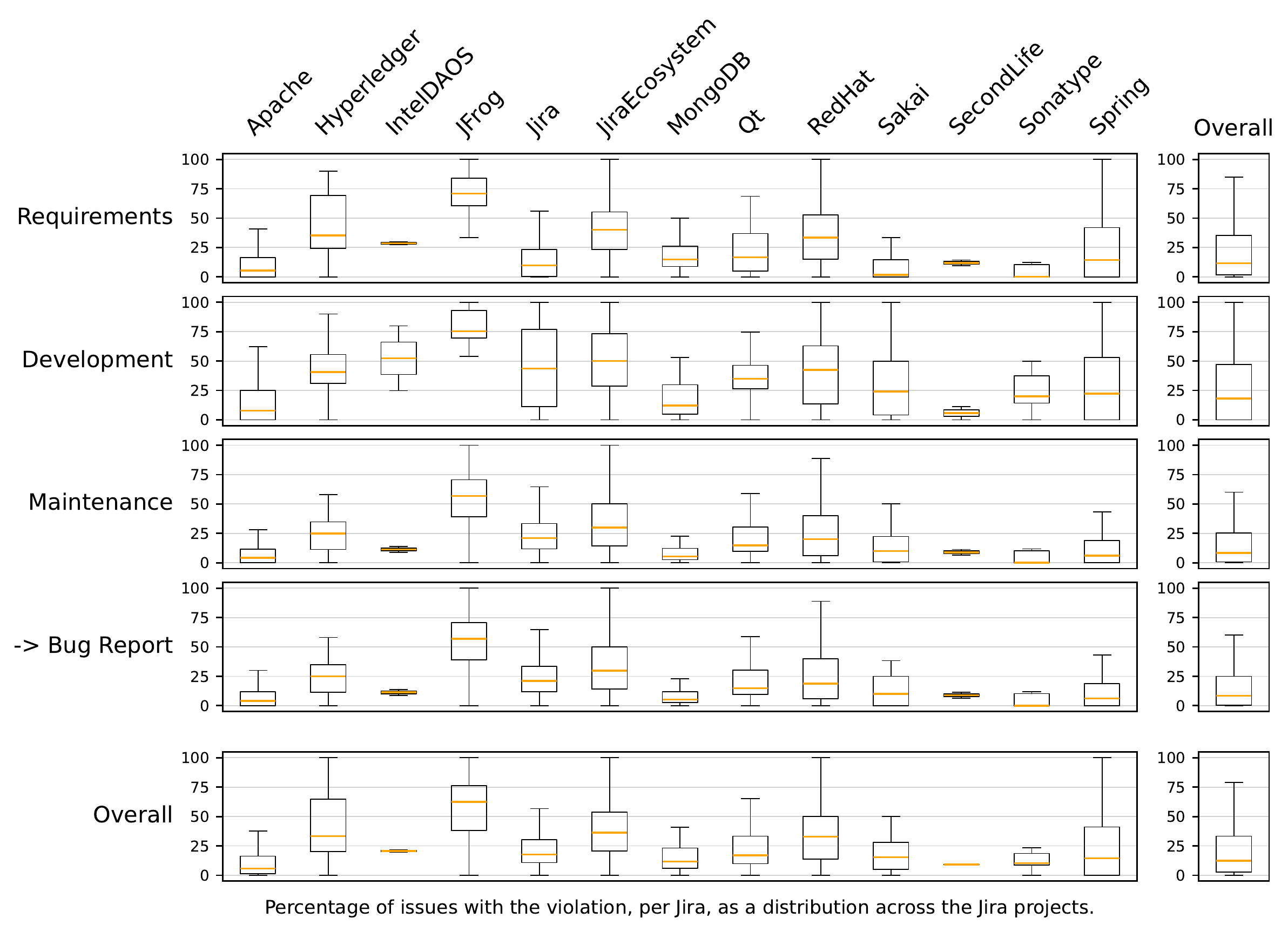}
    \caption{Violations to Bug Report Discussion (All Themes).}
    \label{fig:bp_bug_report_discussion_all}
\end{figure}

\tocAddFauxChapter{List of Figures}
\listoffigures

\tocAddFauxChapter{List of Tables}
\listoftables

\tocAddFauxChapter{List of Publications}
\chapter*{List of Publications}

\newcommand*{\listPub}[1]{\small\fullcite{#1}~\cite{#1}}

\noindent
My publications, central to the thesis, included in this work (partly verbatim):
\begin{itemize}
    \item Montgomery, L. and Maalej, W. `Beyond Bug Reports and Feature Requests: How Do Software Projects Manage Different Issue Types?' In: To be submitted\dots (2025)
    \item Montgomery, L., Lüders, C., Rahe, C. and Maalej, W. `Smells? It depends! An Interview Study of Issue Tracking Problems in Practice'. In: To be submitted\dots (2025)
    \item \listPub{Montgomery_2025_BookChapter}
    \item \listPub{Montgomery_2022_MSR}
    \item \listPub{Montgomery_2022_REJ}
\end{itemize}

\noindent
My publications, not central to the thesis, produced during my PhD:
\begin{itemize}
    \item \listPub{Frattini_2024_JSS}
    \item \listPub{Frattini_2024_EMSE}
    \item \listPub{Frattini_2023_REJ}
    \item \listPub{Frattini_2023_REFSQ}
    \item \listPub{Frattini_2022_RE}
    \item \listPub{Puhlfürß_2022_ICSME}
    \item \listPub{Pham_2019_RE}
    \item \listPub{Stanik_2018_ICSME}
    \item \listPub{Fucci_2018_RESFQ}
\end{itemize}

\printglossary[
    title=Abbreviations,
    type=abbreviations,
    style=mystyle,
    nogroupskip=true,       
    nonumberlist=true,      
]
    
\printglossary[
    type=glossary,
    title=Glossary,
    nogroupskip=true,   
    nonumberlist=true   
]

\tocAddFauxChapter{Bibliography}

\nocite{Bretting_2024_Intelligenz}

{
    \emergencystretch 3em
    \printbibliography      

@string{computing = {Computing}}

@string{computer = {{IEEE Computer Society} Computer}}

@string{springer = {Springer-Verlag}}

@online{6SenseJira_2024_Online,
  title   = {6Sense Jira Market Share},
  url     = {https://6sense.com/tech/productivity/jira-software-market-share},
  urldate = {2024-07-03}
}

@article{Adhya_2015_TC,
  author    = {Adhya, Esha},
  journal   = {Technical Communication},
  number    = {3},
  pages     = {183--192},
  publisher = {Society for Technical Communication},
  title     = {{K}ey elements of an effective style guide in the new age},
  volume    = {62},
  year      = {2015}
}

@article{AgileManifesto_2001_SD,
  author    = {Fowler, Martin and Highsmith, Jim and others},
  journal   = {Software development},
  number    = {8},
  pages     = {28--35},
  publisher = {San Francisco, CA: Miller Freeman, Inc., 1993},
  title     = {{T}he agile manifesto},
  volume    = {9},
  year      = {2001}
}

@article{AlDallal_2017_TSE,
  author    = {Jehad Al Dallal and
               Anas Abdin},
  bibsource = {dblp computer science bibliography, https://dblp.org},
  doi       = {10.1109/TSE.2017.2658573},
  journal   = {{IEEE} Trans. Software Eng.},
  number    = {1},
  pages     = {44--69},
  title     = {{E}mpirical {E}valuation of the {I}mpact of {O}bject-Oriented {C}ode {R}efactoring on {Q}uality {A}ttributes: {A} {S}ystematic {L}iterature {R}eview},
  volume    = {44},
  year      = {2018}
}

@book{Alexander_1997_Book,
  author    = {Christopher Alexander and
               Sara Ishikawa and
               Murray Silverstein and
               Max Jacobson and
               Ingrid Fiksdahl{-}King and
               Shlomo Angel},
  bibsource = {dblp computer science bibliography, https://dblp.org},
  isbn      = {978-0-19-501919-3},
  publisher = {Oxford University Press},
  title     = {{A} {P}attern {L}anguage - {T}owns, {B}uildings, {C}onstruction},
  year      = {1977}
}

@article{Alsaqqa_2020_iJIM,
  author    = {Samar Alsaqqa and
               Samer Sawalha and
               Heba Abdel{-}Nabi},
  title     = {Agile Software Development: Methodologies and Trends},
  journal   = {Int. J. Interact. Mob. Technol.},
  volume    = {14},
  number    = {11},
  pages     = {246--270},
  year      = {2020},
  doi       = {10.3991/IJIM.V14I11.13269},
  bibsource = {dblp computer science bibliography, https://dblp.org}
}

@article{AmoakoGyampah_1993_IM,
  author    = {Kwasi Amoako{-}Gyampah and
               Kathy B. White},
  bibsource = {dblp computer science bibliography, https://dblp.org},
  doi       = {10.1016/0378-7206(93)90021-K},
  journal   = {Information \& Management},
  number    = {1},
  pages     = {1--10},
  title     = {{U}ser involvement and user satisfaction: {A}n exploratory contingency model},
  volume    = {25},
  year      = {1993}
}

@inproceedings{Anvik_2005_OOPSLA,
  author    = {John Anvik and
               Lyndon Hiew and
               Gail C. Murphy},
  bibsource = {dblp computer science bibliography, https://dblp.org},
  booktitle = {Proceedings of the 2005 {OOPSLA} workshop on Eclipse Technology eXchange,
               {ETX} 2005, San Diego, California, USA, October 16-17, 2005},
  doi       = {10.1145/1117696.1117704},
  editor    = {Margaret{-}Anne D. Storey and
               Michael G. Burke and
               Li{-}Te Cheng and
               Andr{\'{e}} van der Hoek},
  pages     = {35--39},
  publisher = {{ACM}},
  title     = {{C}oping with an open bug repository},
  year      = {2005}
}

@inproceedings{Anvik_2006_ICSE,
  author    = {John Anvik and
               Lyndon Hiew and
               Gail C. Murphy},
  bibsource = {dblp computer science bibliography, https://dblp.org},
  booktitle = {28th International Conference on Software Engineering {(ICSE} 2006),
               Shanghai, China, May 20-28, 2006},
  doi       = {10.1145/1134285.1134336},
  editor    = {Leon J. Osterweil and
               H. Dieter Rombach and
               Mary Lou Soffa},
  pages     = {361--370},
  publisher = {{ACM}},
  title     = {{W}ho should fix this bug?},
  year      = {2006}
}

@inproceedings{Aranda_2009_ICSE,
  author    = {Jorge Aranda and
               Gina Venolia},
  bibsource = {dblp computer science bibliography, https://dblp.org},
  booktitle = {31st International Conference on Software Engineering, {ICSE} 2009,
               May 16-24, 2009, Vancouver, Canada, Proceedings},
  doi       = {10.1109/ICSE.2009.5070530},
  pages     = {298--308},
  publisher = {{IEEE}},
  title     = {{T}he secret life of bugs: {G}oing past the errors and omissions in software repositories},
  year      = {2009}
}

@article{Arora_2019_ESE,
  author    = {Chetan Arora and
               Mehrdad Sabetzadeh and
               Lionel C. Briand},
  bibsource = {dblp computer science bibliography, https://dblp.org},
  doi       = {10.1007/S10664-019-09693-X},
  journal   = {Empir. Softw. Eng.},
  number    = {4},
  pages     = {2509--2539},
  title     = {{A}n empirical study on the potential usefulness of domain models for completeness checking of requirements},
  volume    = {24},
  year      = {2019}
}

@inproceedings{Arya_2019_ICSE,
  author    = {Deeksha M. Arya and
               Wenting Wang and
               Jin L. C. Guo and
               Jinghui Cheng},
  bibsource = {dblp computer science bibliography, https://dblp.org},
  booktitle = {Proceedings of the 41st International Conference on Software Engineering,
               {ICSE} 2019, Montreal, QC, Canada, May 25-31, 2019},
  doi       = {10.1109/ICSE.2019.00058},
  editor    = {Joanne M. Atlee and
               Tevfik Bultan and
               Jon Whittle},
  pages     = {454--464},
  publisher = {{IEEE} / {ACM}},
  title     = {{A}nalysis and detection of information types of open source software issue discussions},
  year      = {2019}
}

@online{asana_2024_Online,
  title   = {asana},
  url     = {https://asana.com/},
  urldate = {2024-07-03}
}

@online{Azure_2024_Online,
  title   = {Azure},
  url     = {https://azure.microsoft.com/en-us/products/devops/boards},
  urldate = {2024-07-03}
}

@inproceedings{Bachmann_2010_FSE,
  author    = {Adrian Bachmann and
               Christian Bird and
               Foyzur Rahman and
               Premkumar T. Devanbu and
               Abraham Bernstein},
  bibsource = {dblp computer science bibliography, https://dblp.org},
  booktitle = {Proceedings of the 18th {ACM} {SIGSOFT} International Symposium on
               Foundations of Software Engineering, 2010, Santa Fe, NM, USA, November
               7-11, 2010},
  doi       = {10.1145/1882291.1882308},
  editor    = {Gruia{-}Catalin Roman and
               Andr{\'{e}} van der Hoek},
  pages     = {97--106},
  publisher = {{ACM}},
  title     = {{T}he missing links: bugs and bug-fix commits},
  year      = {2010}
}

@inproceedings{Banerjee_2015_BIGDSE,
  author    = {Sean Banerjee and
               Bojan Cukic},
  bibsource = {dblp computer science bibliography, https://dblp.org},
  booktitle = {1st {IEEE/ACM} International Workshop on Big Data Software Engineering,
               {BIGDSE} 2015, Florence, Italy, May 23, 2015},
  doi       = {10.1109/BIGDSE.2015.16},
  editor    = {Luciano Baresi and
               Tim Menzies and
               Andreas Metzger and
               Thomas Zimmermann},
  pages     = {37--43},
  publisher = {{IEEE} Computer Society},
  title     = {{O}n the {C}ost of {M}ining {V}ery {L}arge {O}pen {S}ource {R}epositories},
  year      = {2015}
}

@article{Bano_2017_EMSE,
  author    = {Muneera Bano and
               Didar Zowghi and
               Francesca {da Rimini}},
  bibsource = {dblp computer science bibliography, https://dblp.org},
  doi       = {10.1007/S10664-016-9465-1},
  journal   = {Empir. Softw. Eng.},
  number    = {5},
  pages     = {2339--2372},
  title     = {{U}ser satisfaction and system success: an empirical exploration of user involvement in software development},
  volume    = {22},
  year      = {2017}
}

@online{BaseCamp_2024_Online,
  title   = {BaseCamp},
  url     = {https://basecamp.com/},
  urldate = {2024-07-03}
}

@inproceedings{Bavota_2016_MSR,
  author    = {Gabriele Bavota and
               Barbara Russo},
  bibsource = {dblp computer science bibliography, https://dblp.org},
  booktitle = {Proceedings of the 13th International Conference on Mining Software
               Repositories, {MSR} 2016, Austin, TX, USA, May 14-22, 2016},
  doi       = {10.1145/2901739.2901742},
  editor    = {Miryung Kim and
               Romain Robbes and
               Christian Bird},
  pages     = {315--326},
  publisher = {{ACM}},
  title     = {{A} large-scale empirical study on self-admitted technical debt},
  year      = {2016}
}

@inproceedings{Baysal_2013_ICSE,
  author    = {Olga Baysal},
  bibsource = {dblp computer science bibliography, https://dblp.org},
  booktitle = {35th International Conference on Software Engineering, {ICSE} '13,
               San Francisco, CA, USA, May 18-26, 2013},
  doi       = {10.1109/ICSE.2013.6606729},
  editor    = {David Notkin and
               Betty H. C. Cheng and
               Klaus Pohl},
  pages     = {1407--1410},
  publisher = {{IEEE} Computer Society},
  title     = {{I}nforming development decisions: from data to information},
  year      = {2013}
}

@inproceedings{Baysal_2013_ICSE_2,
  author    = {Olga Baysal and
               Reid Holmes and
               Michael W. Godfrey},
  bibsource = {dblp computer science bibliography, https://dblp.org},
  booktitle = {35th International Conference on Software Engineering, {ICSE} '13,
               San Francisco, CA, USA, May 18-26, 2013},
  doi       = {10.1109/ICSE.2013.6606674},
  editor    = {David Notkin and
               Betty H. C. Cheng and
               Klaus Pohl},
  pages     = {1185--1188},
  publisher = {{IEEE} Computer Society},
  title     = {{S}ituational awareness: personalizing issue tracking systems},
  year      = {2013}
}

@inproceedings{Beck_1996_ICSE,
  author    = {Kent L. Beck and
               James Coplien and
               Ron Crocker and
               Lutz Dominick and
               Gerard Meszaros and
               Frances Paulisch and
               John M. Vlissides},
  bibsource = {dblp computer science bibliography, https://dblp.org},
  booktitle = {18th International Conference on Software Engineering, Berlin, Germany,
               March 25-29, 1996, Proceedings},
  editor    = {H. Dieter Rombach and
               T. S. E. Maibaum and
               Marvin V. Zelkowitz},
  pages     = {103--114},
  publisher = {{IEEE} Computer Society},
  title     = {{I}ndustrial {E}xperience with {D}esign {P}atterns},
  url       = {http://portal.acm.org/citation.cfm?id=227726.227747},
  year      = {1996}
}

@article{Bernhardt_2012_ARN,
  author  = {B. Bernhardt and T. Singer},
  doi     = {10.1146/annurev-neuro-062111-150536},
  journal = {Annual review of neuroscience},
  pages   = { 1-23 },
  title   = {{T}he neural basis of empathy},
  volume  = {35},
  year    = {2012}
}

@article{Berntzen_2023_EMSE,
  author    = {Marthe Nordengen Berntzen and
               Viktoria Stray and
               Nils Brede Moe and
               Rashina Hoda},
  bibsource = {dblp computer science bibliography, https://dblp.org},
  doi       = {10.1007/S10664-023-10349-0},
  journal   = {Empir. Softw. Eng.},
  number    = {5},
  pages     = {114},
  title     = {{R}esponding to change over time: {A} longitudinal case study on changes in coordination mechanisms in large-scale agile},
  volume    = {28},
  year      = {2023}
}

@article{Berry_2021_EMSE,
  author    = {Berry, Daniel M},
  journal   = {Empirical Software Engineering},
  number    = {6},
  pages     = {111},
  publisher = {Springer},
  title     = {{E}mpirical evaluation of tools for hairy requirements engineering tasks},
  volume    = {26},
  year      = {2021}
}

@inproceedings{Bertram_2010_CSCW,
  author    = {Dane Bertram and
               Amy Voida and
               Saul Greenberg and
               Robert J. Walker},
  bibsource = {dblp computer science bibliography, https://dblp.org},
  booktitle = {Proceedings of the 2010 {ACM} Conference on Computer Supported Cooperative
               Work, {CSCW} 2010, Savannah, Georgia, USA, February 6-10, 2010},
  doi       = {10.1145/1718918.1718972},
  editor    = {Kori Inkpen and
               Carl Gutwin and
               John C. Tang},
  pages     = {291--300},
  publisher = {{ACM}},
  title     = {{C}ommunication, collaboration, and bugs: the social nature of issue tracking in small, collocated teams},
  year      = {2010}
}

@inproceedings{Bettenburg_2008_FSE,
  address   = {Atlanta Georgia, USA},
  author    = {Nicolas Bettenburg and Sascha Just and Adrian Schr{\"o}ter and Cathrin Weiss and Rahul Premraj and Thomas Zimmermann},
  booktitle = {SIGSOFT '08/FSE-16: Proceedings of the 16th ACM SIGSOFT International Symposium on Foundations of software engineering},
  doi       = {10.1145/1453101.1453146},
  pages     = {208--318},
  publisher = {Association for Computing Machinery},
  title     = {What Makes a Good Bug Report?},
  year      = {2008}
}

@inproceedings{Bissyande_2013_CSMR,
  author    = {Tegawend{\'{e}} F. Bissyand{\'{e}} and
               Ferdian Thung and
               Shaowei Wang and
               David Lo and
               Lingxiao Jiang and
               Laurent R{\'{e}}veill{\`{e}}re},
  bibsource = {dblp computer science bibliography, https://dblp.org},
  booktitle = {17th European Conference on Software Maintenance and Reengineering,
               {CSMR} 2013, Genova, Italy, March 5-8, 2013},
  doi       = {10.1109/CSMR.2013.19},
  editor    = {Anthony Cleve and
               Filippo Ricca and
               Maura Cerioli},
  pages     = {89--98},
  publisher = {{IEEE} Computer Society},
  title     = {{E}mpirical {E}valuation of {B}ug {L}inking},
  year      = {2013}
}

@phdthesis{Bohn_2024_MSc,
  author = {Piet Bohn},
  school = {University of Hamburg},
  title  = {{E}xploring {L}{L}{M}s as a {T}ool for {A}utomated {D}etection and {C}orrection of {I}ssue {T}racker {S}mells},
  type   = {MSc Thesis},
  year   = {2024}
}

@inproceedings{Borg_2013_CSMR,
  author    = {Markus Borg and
               Dietmar Pfahl and
               Per Runeson},
  bibsource = {dblp computer science bibliography, https://dblp.org},
  booktitle = {17th European Conference on Software Maintenance and Reengineering,
               {CSMR} 2013, Genova, Italy, March 5-8, 2013},
  doi       = {10.1109/CSMR.2013.18},
  editor    = {Anthony Cleve and
               Filippo Ricca and
               Maura Cerioli},
  pages     = {79--88},
  publisher = {{IEEE} Computer Society},
  title     = {{A}nalyzing {N}etworks of {I}ssue {R}eports},
  year      = {2013}
}

@phdthesis{Borg_PhDThesis_2015,
  author    = {Markus Borg},
  bibsource = {dblp computer science bibliography, https://dblp.org},
  school    = {Lund University, Sweden},
  title     = {{F}rom {B}ugs to {D}ecision {S}upport - {L}everaging {H}istorical {I}ssue {R}eports in {S}oftware {E}volution},
  url       = {http://lup.lub.lu.se/record/5268091},
  year      = {2015}
}

@inproceedings{Bortis_2011_CHASE,
  author    = {Gerald Bortis and
               Andr{\'{e}} van der Hoek},
  bibsource = {dblp computer science bibliography, https://dblp.org},
  booktitle = {Proceedings of the 4th International Workshop on Cooperative and Human
               Aspects of Software Engineering, {CHASE} 2011, Waikiki, Honolulu,
               HI, USA, May 21, 2011},
  doi       = {10.1145/1984642.1984659},
  editor    = {Marcelo Cataldo and
               Cleidson R. B. de Souza and
               Yvonne Dittrich and
               Rashina Hoda and
               Helen Sharp},
  pages     = {69--71},
  publisher = {{ACM}},
  title     = {{T}eambugs: a collaborative bug tracking tool},
  year      = {2011}
}

@article{Braun_2006_QRP,
  author    = {Braun, Virginia and Clarke, Victoria},
  journal   = {Qualitative research in psychology},
  number    = {2},
  pages     = {77--101},
  publisher = {Taylor \& Francis},
  title     = {{U}sing thematic analysis in psychology},
  volume    = {3},
  year      = {2006}
}

@book{Braun_2021_Book,
  address   = {California, USA},
  author    = {Braun, Virginia and Clarke, Victoria},
  isbn      = {9781526417305},
  publisher = {SAGE Publications},
  title     = {{T}hematic {A}nalysis: {A} {P}ractical {G}uide},
  year      = {2021}
}

@misc{Bretting_2024_Intelligenz,
  author = {Sara Bretting},
  year   = {2024},
  title  = {Intelligenz},
  note   = {Licence Agreement: This work is licensed to Lloyd Montgomery for use in his dissertation (contract is private). This artwork may not be used outside the context of this dissertation. The creator of the artwork, Sara Bretting, acknowledges that the dissertation is licensed under CC BY 4.0 and therefore her artwork can be distributed under the same terms while imbedded in the dissertation.}
}

@inproceedings{Breu_2010_CSCW,
  author    = {Silvia Breu and
               Rahul Premraj and
               Jonathan Sillito and
               Thomas Zimmermann},
  bibsource = {dblp computer science bibliography, https://dblp.org},
  booktitle = {Proceedings of the 2010 {ACM} Conference on Computer Supported Cooperative
               Work, {CSCW} 2010, Savannah, Georgia, USA, February 6-10, 2010},
  doi       = {10.1145/1718918.1718973},
  editor    = {Kori Inkpen and
               Carl Gutwin and
               John C. Tang},
  pages     = {301--310},
  publisher = {{ACM}},
  title     = {{I}nformation needs in bug reports: improving cooperation between developers and users},
  year      = {2010}
}

@book{Brown_1998_Book,
  author    = {Brown, William H and Malveau, Raphael C and McCormick, Hays and Mowbray, Thomas J},
  publisher = {John Wiley \& Sons, Inc.},
  title     = {{A}nti{P}atterns: refactoring software, architectures, and projects in crisis},
  year      = {1998}
}

@online{Bugzilla_2024_Online,
  title   = {Bugzilla},
  url     = {https://www.bugzilla.org/},
  urldate = {2024-07-03}
}

@inproceedings{Carver_2016_ESEM,
  author    = {Jeffrey C. Carver and
               Oscar Dieste and
               Nicholas A. Kraft and
               David Lo and
               Thomas Zimmermann},
  bibsource = {dblp computer science bibliography, https://dblp.org},
  booktitle = {Proceedings of the 10th {ACM/IEEE} International Symposium on Empirical
               Software Engineering and Measurement, {ESEM} 2016, Ciudad Real, Spain,
               September 8-9, 2016},
  doi       = {10.1145/2961111.2962597},
  pages     = {56:1--56:10},
  publisher = {{ACM}},
  title     = {{H}ow {P}ractitioners {P}erceive the {R}elevance of {ESEM} {R}esearch},
  year      = {2016}
}

@inproceedings{Catolino_ICSE_2021,
  author    = {Gemma Catolino and
               Fabio Palomba and
               Damian Tamburri and
               Alexander Serebrenik},
  bibsource = {dblp computer science bibliography, https://dblp.org},
  booktitle = {43rd {IEEE/ACM} International Conference on Software Engineering:
               Software Engineering in Society, {ICSE} {(SEIS)} 2021, Madrid, Spain,
               May 25-28, 2021},
  doi       = {10.1109/ICSE-SEIS52602.2021.00017},
  pages     = {77--86},
  publisher = {{IEEE}},
  title     = {{U}nderstanding {C}ommunity {S}mells {V}ariability: {A} {S}tatistical {A}pproach},
  year      = {2021}
}

@article{Cavalcanti_2014_JSEP,
  author    = {Yguarat{\~{a}} Cerqueira Cavalcanti and
               Paulo Anselmo da Mota Silveira Neto and
               Ivan do Carmo Machado and
               Tassio Vale and
               Eduardo Santana de Almeida and
               Silvio Romero de Lemos Meira},
  bibsource = {dblp computer science bibliography, https://dblp.org},
  doi       = {10.1002/SMR.1639},
  journal   = {J. Softw. Evol. Process.},
  number    = {7},
  pages     = {620--653},
  title     = {{C}hallenges and opportunities for software change request repositories: a systematic mapping study},
  volume    = {26},
  year      = {2014}
}

@inproceedings{Chaparro_2019_ESECFSE,
  author    = {Oscar Chaparro and
               Carlos Bernal{-}C{\'{a}}rdenas and
               Jing Lu and
               Kevin Moran and
               Andrian Marcus and
               Massimiliano {Di Penta} and
               Denys Poshyvanyk and
               Vincent Ng},
  bibsource = {dblp computer science bibliography, https://dblp.org},
  booktitle = {Proceedings of the {ACM} Joint Meeting on European Software Engineering
               Conference and Symposium on the Foundations of Software Engineering,
               {ESEC/SIGSOFT} {FSE} 2019, Tallinn, Estonia, August 26-30, 2019},
  doi       = {10.1145/3338906.3338947},
  editor    = {Marlon Dumas and
               Dietmar Pfahl and
               Sven Apel and
               Alessandra Russo},
  pages     = {86--96},
  publisher = {{ACM}},
  title     = {{A}ssessing the quality of the steps to reproduce in bug reports},
  year      = {2019}
}

@article{Clercq_2019_PR,
  author  = {Dirk {De Clercq} and Inam Ul Haq and Muhammad Umer Azeem},
  doi     = {10.1108/PR-02-2018-0052},
  journal = {Personnel Review},
  title   = {{W}hy happy employees help},
  year    = {2019}
}

@inproceedings{Cruzes_2011_ESEM,
  address   = {Banff, Alberta, Canada},
  author    = {Cruzes, Daniela S. and Dyba, Tore},
  booktitle = {2011 {{International Symposium}} on {{Empirical Software Engineering}} and {{Measurement}}},
  doi       = {10.1109/ESEM.2011.36},
  issn      = {1949-3789},
  month     = sep,
  pages     = {275--284},
  publisher = {IEEE Computer Society},
  title     = {{R}ecommended {{Steps}} for {{Thematic {S}ynthesis}} in {{Software {E}ngineering}}},
  year      = {2011}
}

@techreport{Dalpiaz_2018_DS,
  author    = {Fabiano Dalpiaz},
  doi       = {10.17632/7zbk8zsd8y.1},
  publisher = {Mendeley Data},
  title     = {{R}equirements data sets (user stories)},
  year      = {2018}
}

@online{DatanyzeJira_2024_Online,
  title   = {Datanyze Jira Market Share},
  url     = {https://www.datanyze.com/market-share/project-management--217/jira-market-share},
  urldate = {2024-07-03}
}

@inproceedings{Davies_2014_ESEM,
  author    = {Steven Davies and
               Marc Roper},
  bibsource = {dblp computer science bibliography, https://dblp.org},
  booktitle = {2014 {ACM-IEEE} International Symposium on Empirical Software Engineering
               and Measurement, {ESEM} '14, Torino, Italy, September 18-19, 2014},
  doi       = {10.1145/2652524.2652541},
  pages     = {26:1--26:10},
  publisher = {{ACM}},
  title     = {{W}hat's in a bug report?},
  year      = {2014}
}

@inproceedings{Davis_1993_CS,
  author    = {Alan M. Davis and Scott Overmyer and Kathleen Jordan and Joseph Caruso and Fatma Dandashi and Anhtuan Dinh and Gary Kincaid and Glen Ledeboer and Patricia Reynolds and Pradip Sitaram and Anh Ta and Mary Theofanos},
  title     = {Identifying and measuring quality in a software requirements specification},
  booktitle = {Proceedings of the First International Software Metrics Symposium,
               {METRICS} 1993, May 21-22, 1993, Balimore, Maryland, {USA}},
  pages     = {141--152},
  publisher = {{IEEE} Computer Society},
  year      = {1993},
  doi       = {10.1109/METRIC.1993.263792}
}

@inproceedings{Deshmukh_2017_ICSME,
  author    = {Jayati Deshmukh and
               K. M. Annervaz and
               Sanjay Podder and
               Shubhashis Sengupta and
               Neville Dubash},
  bibsource = {dblp computer science bibliography, https://dblp.org},
  booktitle = {2017 {IEEE} International Conference on Software Maintenance and Evolution,
               {ICSME} 2017, Shanghai, China, September 17-22, 2017},
  doi       = {10.1109/ICSME.2017.69},
  pages     = {115--124},
  publisher = {{IEEE} Computer Society},
  title     = {{T}owards {A}ccurate {D}uplicate {B}ug {R}etrieval {U}sing {D}eep {L}earning {T}echniques},
  year      = {2017}
}

@inproceedings{Eckhardt_2016_ICSE,
  author    = {Jonas Eckhardt and
               Andreas Vogelsang and
               Daniel Mendez},
  bibsource = {dblp computer science bibliography, https://dblp.org},
  booktitle = {Proceedings of the 38th International Conference on Software Engineering,
               {ICSE} 2016, Austin, TX, USA, May 14-22, 2016},
  doi       = {10.1145/2884781.2884788},
  editor    = {Laura K. Dillon and
               Willem Visser and
               Laurie A. Williams},
  pages     = {832--842},
  publisher = {{ACM}},
  title     = {{A}re "non-functional" requirements really non-functional?: an investigation of non-functional requirements in practice},
  year      = {2016}
}

@inproceedings{Eckhardt_2016_RE,
  author    = {Jonas Eckhardt and
               Andreas Vogelsang and
               Henning Femmer and
               Philipp Mager},
  bibsource = {dblp computer science bibliography, https://dblp.org},
  booktitle = {24th {IEEE} International Requirements Engineering Conference, {RE}
               2016, Beijing, China, September 12-16, 2016},
  doi       = {10.1109/RE.2016.24},
  pages     = {46--55},
  publisher = {{IEEE} Computer Society},
  title     = {{C}hallenging {I}ncompleteness of {P}erformance {R}equirements by {S}entence {P}atterns},
  year      = {2016}
}

@inproceedings{Eloranta_2013_APSEC,
  author    = {Veli{-}Pekka Eloranta and
               Kai Koskimies and
               Tommi Mikkonen and
               Jyri Vuorinen},
  bibsource = {dblp computer science bibliography, https://dblp.org},
  booktitle = {20th Asia-Pacific Software Engineering Conference, {APSEC} 2013, Ratchathewi,
               Bangkok, Thailand, December 2-5, 2013 - Volume 1},
  doi       = {10.1109/APSEC.2013.72},
  editor    = {Pornsiri Muenchaisri and
               Gregg Rothermel},
  pages     = {503--510},
  publisher = {{IEEE} Computer Society},
  title     = {{S}crum {A}nti-Patterns - {A}n {E}mpirical {S}tudy},
  year      = {2013}
}

@article{Eloranta_2016_IST,
  author    = {Veli{-}Pekka Eloranta and
               Kai Koskimies and
               Tommi Mikkonen},
  bibsource = {dblp computer science bibliography, https://dblp.org},
  doi       = {10.1016/J.INFSOF.2015.12.003},
  journal   = {Inf. Softw. Technol.},
  pages     = {194--203},
  title     = {{E}xploring {S}crum{B}ut - {A}n empirical study of {S}crum anti-patterns},
  volume    = {74},
  year      = {2016}
}

@online{EnlyftJira_2024_Online,
  title   = {Enlyft Jira Market Share},
  url     = {https://enlyft.com/tech/products/atlassian-jira},
  urldate = {2024-07-03}
}

@misc{EntSize_2024_Online,
  howpublished = {\url{https://ec.europa.eu/eurostat/statistics-explained/index.php?title=Glossary:Enterprise_size}},
  note         = {Accessed: 2024-03-18},
  title        = {Eurostat: Statistics Explained - Enterprise size}
}

@inproceedings{Ernst_2012_empiRE,
  address   = {{Chicago, IL, USA}},
  author    = {Ernst, Neil A. and Murphy, Gail C.},
  booktitle = {2012 {{Second IEEE International Workshop}} on {{Empirical Requirements Engineering}} ({{EmpiRE}})},
  doi       = {10.1109/EmpiRE.2012.6347678},
  issn      = {2329-6356},
  month     = {9},
  pages     = {25--32},
  publisher = {{IEEE}},
  title     = {{C}ase {S}tudies in {J}ust-in-Time {R}equirements {A}nalysis},
  year      = {2012}
}

@inproceedings{Feller_2000_ICIS,
  author    = {Joseph Feller and
               Brian Fitzgerald},
  bibsource = {dblp computer science bibliography, https://dblp.org},
  booktitle = {Proceedings of the Twenty-First International Conference on Information
               Systems, {ICIS} 2000, Brisbane, Australia, December 10-13, 2000},
  editor    = {Soon Ang and
               Helmut Krcmar and
               Wanda J. Orlikowski and
               Peter Weill and
               Janice I. DeGross},
  pages     = {58--69},
  publisher = {Association for Information Systems},
  title     = {{A} framework analysis of the open source software development paradigm},
  url       = {http://aisel.aisnet.org/icis2000/7},
  year      = {2000}
}

@article{Femmer_2017_JSS,
  author    = {Henning Femmer and
               Daniel Mendez and
               Stefan Wagner and
               Sebastian Eder},
  bibsource = {dblp computer science bibliography, https://dblp.org},
  doi       = {10.1016/J.JSS.2016.02.047},
  journal   = {J. Syst. Softw.},
  pages     = {190--213},
  title     = {{R}apid quality assurance with {R}equirements {S}mells},
  volume    = {123},
  year      = {2017}
}

@inproceedings{Femmer_2017_REW,
  author    = {Henning Femmer and
               Michael Unterkalmsteiner and
               Tony Gorschek},
  bibsource = {dblp computer science bibliography, https://dblp.org},
  booktitle = {{IEEE} 25th International Requirements Engineering Conference Workshops,
               {RE} 2017 Workshops, Lisbon, Portugal, September 4-8, 2017},
  doi       = {10.1109/REW.2017.18},
  pages     = {400--406},
  publisher = {{IEEE} Computer Society},
  title     = {{W}hich {R}equirements {A}rtifact {Q}uality {D}efects are {A}utomatically {D}etectable? {A} {C}ase {S}tudy},
  year      = {2017}
}

@article{Femmer_2019_IEEES,
  author    = {Henning Femmer and
               Andreas Vogelsang},
  bibsource = {dblp computer science bibliography, https://dblp.org},
  doi       = {10.1109/MS.2018.110161823},
  journal   = {{IEEE} Softw.},
  number    = {3},
  pages     = {83--91},
  title     = {{R}equirements {Q}uality {I}s {Q}uality in {U}se},
  volume    = {36},
  year      = {2019}
}

@article{Ferrari_2019_ASE,
  author    = {Alessio Ferrari and
               Andrea Esuli},
  bibsource = {dblp computer science bibliography, https://dblp.org},
  doi       = {10.1007/S10515-019-00261-7},
  journal   = {Autom. Softw. Eng.},
  number    = {3},
  pages     = {559--598},
  title     = {{A}n {NLP} approach for cross-domain ambiguity detection in requirements engineering},
  volume    = {26},
  year      = {2019}
}

@inproceedings{Fischbach_2020_ESEM,
  author    = {Jannik Fischbach and
               Henning Femmer and
               Daniel Mendez and
               Davide Fucci and
               Andreas Vogelsang},
  bibsource = {dblp computer science bibliography, https://dblp.org},
  booktitle = {{ESEM} '20: {ACM} / {IEEE} International Symposium on Empirical Software
               Engineering and Measurement, Bari, Italy, October 5-7, 2020},
  doi       = {10.1145/3382494.3421462},
  editor    = {Maria Teresa Baldassarre and
               Filippo Lanubile and
               Marcos Kalinowski and
               Federica Sarro},
  pages     = {41:1--41:10},
  publisher = {{ACM}},
  title     = {{W}hat {M}akes {A}gile {T}est {A}rtifacts {U}seful?: {A}n {A}ctivity-Based {Q}uality {M}odel from a {P}ractitioners' {P}erspective},
  year      = {2020}
}

@inproceedings{Fischbach_2020_RE,
  author    = {Jannik Fischbach and
               Benedikt Hauptmann and
               Lukas Konwitschny and
               Dominik Spies and
               Andreas Vogelsang},
  bibsource = {dblp computer science bibliography, https://dblp.org},
  booktitle = {28th {IEEE} International Requirements Engineering Conference, {RE}
               2020, Zurich, Switzerland, August 31 - September 4, 2020},
  doi       = {10.1109/RE48521.2020.00053},
  editor    = {Travis D. Breaux and
               Andrea Zisman and
               Samuel Fricker and
               Martin Glinz},
  pages     = {388--393},
  publisher = {{IEEE}},
  title     = {{T}owards {C}ausality {E}xtraction from {R}equirements},
  year      = {2020}
}

@inproceedings{Fischbach_2021_PROFES,
  author    = {Jannik Fischbach and
               Julian Frattini and
               Daniel Mendez and
               Michael Unterkalmsteiner and
               Henning Femmer and
               Andreas Vogelsang},
  bibsource = {dblp computer science bibliography, https://dblp.org},
  booktitle = {Product-Focused Software Process Improvement - 22nd International
               Conference, {PROFES} 2021, Turin, Italy, November 26, 2021, Proceedings},
  doi       = {10.1007/978-3-030-91452-3\_6},
  editor    = {Luca Ardito and
               Andreas Jedlitschka and
               Maurizio Morisio and
               Marco Torchiano},
  pages     = {85--102},
  publisher = {Springer},
  series    = {Lecture Notes in Computer Science},
  title     = {{H}ow {D}o {P}ractitioners {I}nterpret {C}onditionals in {R}equirements?},
  volume    = {13126},
  year      = {2021}
}

@inproceedings{Fischbach_2021_REFSQ,
  author    = {Jannik Fischbach and
               Julian Frattini and
               Arjen Spaans and
               Maximilian Kummeth and
               Andreas Vogelsang and
               Daniel Mendez and
               Michael Unterkalmsteiner},
  bibsource = {dblp computer science bibliography, https://dblp.org},
  booktitle = {Requirements Engineering: Foundation for Software Quality - 27th International
               Working Conference, {REFSQ} 2021, Essen, Germany, April 12-15, 2021,
               Proceedings},
  doi       = {10.1007/978-3-030-73128-1\_2},
  editor    = {Fabiano Dalpiaz and
               Paola Spoletini},
  pages     = {19--36},
  publisher = {Springer},
  series    = {Lecture Notes in Computer Science},
  title     = {{A}utomatic {D}etection of {C}ausality in {R}equirement {A}rtifacts: {T}he {C}i{R}{A} {A}pproach},
  volume    = {12685},
  year      = {2021}
}

@inproceedings{Fitzgerald_2011_RE,
  author    = {Camilo Fitzgerald and
               Emmanuel Letier and
               Anthony Finkelstein},
  bibsource = {dblp computer science bibliography, https://dblp.org},
  booktitle = {{RE} 2011, 19th {IEEE} International Requirements Engineering Conference,
               Trento, Italy, August 29 2011 - September 2, 2011},
  doi       = {10.1109/RE.2011.6051658},
  pages     = {229--238},
  publisher = {{IEEE} Computer Society},
  title     = {{E}arly failure prediction in feature request management systems},
  year      = {2011}
}

@article{Forsgren_2021_Queue,
  author    = {Nicole Forsgren and
               Margaret{-}Anne D. Storey and
               Chandra Shekhar Maddila and
               Thomas Zimmermann and
               Brian Houck and
               Jenna L. Butler},
  bibsource = {dblp computer science bibliography, https://dblp.org},
  doi       = {10.1145/3454122.3454124},
  journal   = {{ACM} Queue},
  number    = {1},
  pages     = {20--48},
  title     = {{T}he {SPACE} of {D}eveloper {P}roductivity: {T}here's more to it than you think},
  volume    = {19},
  year      = {2021}
}

@book{Fowler_2018_Book,
  author    = {Martin Fowler},
  bibsource = {dblp computer science bibliography, https://dblp.org},
  isbn      = {978-0-201-48567-7},
  publisher = {Addison-Wesley},
  series    = {Addison Wesley object technology series},
  title     = {{R}efactoring - {I}mproving the {D}esign of {E}xisting {C}ode},
  url       = {http://martinfowler.com/books/refactoring.html},
  year      = {1999}
}

@inproceedings{Franch_2017_RE,
  author    = {Xavier Franch and
               Daniel Mendez and
               Marc Oriol and
               Andreas Vogelsang and
               Rogardt Heldal and
               Eric Knauss and
               Guilherme H. Travassos and
               Jeffrey C. Carver and
               Oscar Dieste and
               Thomas Zimmermann},
  bibsource = {dblp computer science bibliography, https://dblp.org},
  booktitle = {25th {IEEE} International Requirements Engineering Conference, {RE}
               2017, Lisbon, Portugal, September 4-8, 2017},
  doi       = {10.1109/RE.2017.17},
  editor    = {Ana Moreira and
               Jo{\~{a}}o Ara{\'{u}}jo and
               Jane Hayes and
               Barbara Paech},
  pages     = {382--387},
  publisher = {{IEEE} Computer Society},
  title     = {{H}ow do {P}ractitioners {P}erceive the {R}elevance of {R}equirements {E}ngineering {R}esearch? {A}n {O}ngoing {S}tudy},
  year      = {2017}
}

@article{Franch_2020_TSE,
  title     = {How do practitioners perceive the relevance of requirements engineering research?},
  author    = {Franch, Xavier and Mendez, Daniel and Vogelsang, Andreas and Heldal, Rogardt and Knauss, Eric and Oriol, Marc and Travassos, Guilherme H. and Carver, Jeffrey C. and Zimmermann, Thomas},
  journal   = {IEEE Transactions on Software Engineering},
  volume    = {48},
  number    = {6},
  pages     = {1947--1964},
  year      = {2020},
  publisher = {IEEE}
}

@article{Franch_2022_TSE,
  author    = {Xavier Franch and
               Daniel Mendez and
               Andreas Vogelsang and
               Rogardt Heldal and
               Eric Knauss and
               Marc Oriol and
               Guilherme H. Travassos and
               Jeffrey C. Carver and
               Thomas Zimmermann},
  bibsource = {dblp computer science bibliography, https://dblp.org},
  doi       = {10.1109/TSE.2020.3042747},
  journal   = {{IEEE} Trans. Software Eng.},
  number    = {6},
  pages     = {1947--1964},
  title     = {{H}ow do {P}ractitioners {P}erceive the {R}elevance of {R}equirements {E}ngineering {R}esearch?},
  volume    = {48},
  year      = {2022}
}

@inproceedings{Frattini_2022_RE,
  author    = {Julian Frattini and
               Lloyd Montgomery and
               Jannik Fischbach and
               Michael Unterkalmsteiner and
               Daniel Mendez and
               Davide Fucci},
  bibsource = {dblp computer science bibliography, https://dblp.org},
  booktitle = {30th {IEEE} International Requirements Engineering Conference, {RE}
               2022, Melbourne, Australia, August 15-19, 2022},
  doi       = {10.1109/RE54965.2022.00041},
  pages     = {274--280},
  publisher = {{IEEE}},
  title     = {{A} {L}ive {E}xtensible {O}ntology of {Q}uality {F}actors for {T}extual {R}equirements},
  year      = {2022}
}

@inproceedings{Frattini_2023_REFSQ,
  author    = {Julian Frattini and
               Lloyd Montgomery and
               Davide Fucci and
               Jannik Fischbach and
               Michael Unterkalmsteiner and
               Daniel Mendez},
  bibsource = {dblp computer science bibliography, https://dblp.org},
  booktitle = {Joint Proceedings of {REFSQ-2023} Workshops, Doctoral Symposium, Posters
               {\&} Tools Track and Journal Early Feedback co-located with the 28th
               International Conference on Requirements Engineering: Foundation for
               Software Quality {(REFSQ} 2023), Barcelona, Catalunya, Spain, April
               17-20, 2023},
  editor    = {Alessio Ferrari},
  publisher = {CEUR-WS.org},
  series    = {{CEUR} Workshop Proceedings},
  title     = {{L}et's {S}top {B}uilding at the {F}eet of {G}iants: {R}ecovering unavailable {R}equirements {Q}uality {A}rtifacts},
  url       = {https://ceur-ws.org/Vol-3378/NLP4RE-paper3.pdf},
  volume    = {3378},
  year      = {2023}
}

@article{Frattini_2023_REJ,
  author    = {Julian Frattini and
               Lloyd Montgomery and
               Jannik Fischbach and
               Daniel Mendez and
               Davide Fucci and
               Michael Unterkalmsteiner},
  bibsource = {dblp computer science bibliography, https://dblp.org},
  doi       = {10.1007/S00766-023-00405-Y},
  journal   = {Requir. Eng.},
  number    = {4},
  pages     = {507--520},
  title     = {{R}equirements quality research: a harmonized theory, evaluation, and roadmap},
  volume    = {28},
  year      = {2023}
}

@article{Frattini_2024_EMSE,
  author     = {Julian Frattini and
                Davide Fucci and
                Richard Torkar and
                Lloyd Montgomery and
                Michael Unterkalmsteiner and
                Jannik Fischbach and
                Daniel Mendez},
  bibsource  = {dblp computer science bibliography, https://dblp.org},
  doi        = {10.48550/ARXIV.2401.01154},
  eprint     = {2401.01154},
  eprinttype = {arXiv},
  journal    = {CoRR},
  title      = {{A}pplying {B}ayesian {D}ata {A}nalysis for {C}ausal {I}nference about {R}equirements {Q}uality: {A} {R}eplicated {E}xperiment},
  volume     = {abs/2401.01154},
  year       = {2024}
}

@article{Frattini_2024_JSS,
  author    = {Julian Frattini and
               Lloyd Montgomery and
               Davide Fucci and
               Michael Unterkalmsteiner and
               Daniel Mendez and
               Jannik Fischbach},
  bibsource = {dblp computer science bibliography, https://dblp.org},
  doi       = {10.1016/J.JSS.2024.112120},
  journal   = {J. Syst. Softw.},
  pages     = {112120},
  title     = {{R}equirements quality research artifacts: {R}ecovery, analysis, and management guideline},
  volume    = {216},
  year      = {2024}
}

@inproceedings{Fucci_2018_ESEM,
  author    = {Davide Fucci and
               Cristina Palomares and
               Xavier Franch and
               Dolors Costal and
               Mikko Raatikainen and
               Martin Stettinger and
               Zijad Kurtanovic and
               Tero Kojo and
               Lars Koenig and
               Andreas A. Falkner and
               Gottfried Schenner and
               Fabrizio Brasca and
               Tomi M{\"{a}}nnist{\"{o}} and
               Alexander Felfernig and
               Walid Maalej},
  bibsource = {dblp computer science bibliography, https://dblp.org},
  booktitle = {Proceedings of the 12th {ACM/IEEE} International Symposium on Empirical
               Software Engineering and Measurement, {ESEM} 2018, Oulu, Finland,
               October 11-12, 2018},
  doi       = {10.1145/3239235.3240498},
  editor    = {Markku Oivo and
               Daniel Mendez and
               Audris Mockus},
  pages     = {19:1--19:10},
  publisher = {{ACM}},
  title     = {{N}eeds and challenges for a platform to support large-scale requirements engineering: a multiple-case study},
  year      = {2018}
}

@inproceedings{Fucci_2018_RESFQ,
  author    = {Davide Fucci and
               Christoph Stanik and
               Lloyd Montgomery and
               Zijad Kurtanovic and
               Timo Johann and
               Walid Maalej},
  bibsource = {dblp computer science bibliography, https://dblp.org},
  booktitle = {Joint Proceedings of {REFSQ-2018} Workshops, Doctoral Symposium, Live
               Studies Track, and Poster Track co-located with the 23rd International
               Conference on Requirements Engineering: Foundation for Software Quality
               {(REFSQ} 2018), Utrecht, The Netherlands, March 19, 2018},
  editor    = {Klaus Schmid},
  publisher = {CEUR-WS.org},
  series    = {{CEUR} Workshop Proceedings},
  title     = {{R}esearch on {NLP} for {RE} at the {U}niversity of {H}amburg: {A} {R}eport},
  url       = {https://ceur-ws.org/Vol-2075/NLP4RE\_paper8.pdf},
  volume    = {2075},
  year      = {2018}
}

@article{Garshol_2004_JIS,
  author    = {Lars Marius Garshol},
  bibsource = {dblp computer science bibliography, https://dblp.org},
  doi       = {10.1177/0165551504045856},
  journal   = {J. Inf. Sci.},
  number    = {4},
  pages     = {378--391},
  title     = {{M}etadata? {T}hesauri? {T}axonomies? {T}opic {M}aps! {M}aking {S}ense of it all},
  volume    = {30},
  year      = {2004}
}

@inproceedings{Gemkow_2018_RE,
  author    = {Tim Gemkow and
               Miro Conzelmann and
               Kerstin Hartig and
               Andreas Vogelsang},
  bibsource = {dblp computer science bibliography, https://dblp.org},
  booktitle = {26th {IEEE} International Requirements Engineering Conference, {RE}
               2018, Banff, AB, Canada, August 20-24, 2018},
  doi       = {10.1109/RE.2018.00052},
  editor    = {G{\"{u}}nther Ruhe and
               Walid Maalej and
               Daniel Amyot},
  pages     = {412--417},
  publisher = {{IEEE} Computer Society},
  title     = {{A}utomatic {G}lossary {T}erm {E}xtraction from {L}arge-Scale {R}equirements {S}pecifications},
  year      = {2018}
}

@online{GitHub_2024_Online,
  title   = {GitHub},
  url     = {https://github.com/},
  urldate = {2024-07-03}
}

@online{GitLab_2024_Online,
  title   = {GitLab},
  url     = {https://about.gitlab.com/},
  urldate = {2024-07-03}
}

@inproceedings{Giuliano_2018_CASCON,
  author    = {Giuliano Antoniol and
               Kamel Ayari and
               Massimiliano {Di Penta} and
               Foutse Khomh and
               Yann{-}Ga{\"{e}}l Gu{\'{e}}h{\'{e}}neuc},
  bibsource = {dblp computer science bibliography, https://dblp.org},
  booktitle = {Proceedings of the 2008 conference of the Centre for Advanced Studies
               on Collaborative Research, October 27-30, 2008, Richmond Hill, Ontario,
               Canada},
  doi       = {10.1145/1463788.1463819},
  editor    = {Marsha Chechik and
               Mark R. Vigder and
               Darlene A. Stewart},
  pages     = {23},
  publisher = {{IBM}},
  title     = {{I}s it a bug or an enhancement?: a text-based approach to classify change requests},
  year      = {2008}
}

@incollection{Gotel_2012_BOOK,
  author    = {Orlena Gotel and
               Jane Cleland{-}Huang and
               Jane Huffman Hayes and
               Andrea Zisman and
               Alexander Egyed and
               Paul Gr{\"{u}}nbacher and
               Alex Dekhtyar and
               Giuliano Antoniol and
               Jonathan I. Maletic and
               Patrick M{\"{a}}der},
  bibsource = {dblp computer science bibliography, https://dblp.org},
  booktitle = {Software and Systems Traceability},
  doi       = {10.1007/978-1-4471-2239-5\_1},
  editor    = {Jane Cleland{-}Huang and
               Olly Gotel and
               Andrea Zisman},
  pages     = {3--22},
  publisher = {Springer},
  title     = {{T}raceability {F}undamentals},
  year      = {2012}
}

@article{Gregor_2006_MIS,
  author    = {Shirley Gregor},
  bibsource = {dblp computer science bibliography, https://dblp.org},
  journal   = {{MIS} Q.},
  number    = {3},
  pages     = {611--642},
  title     = {{T}he {N}ature of {T}heory in {I}nformation {S}ystems},
  url       = {http://misq.org/the-nature-of-theory-in-information-systems.html},
  volume    = {30},
  year      = {2006}
}

@article{Groen_2017_IEEESoftware,
  author    = {Eduard C. Groen and
               Norbert Seyff and
               Raian Ali and
               Fabiano Dalpiaz and
               J{\"{o}}rg D{\"{o}}rr and
               Emitza Guzman and
               Mahmood Hosseini and
               Jordi Marco and
               Marc Oriol and
               Anna Perini and
               Melanie J. C. Stade},
  bibsource = {dblp computer science bibliography, https://dblp.org},
  doi       = {10.1109/MS.2017.33},
  journal   = {{IEEE} Softw.},
  number    = {2},
  pages     = {44--52},
  title     = {{T}he {C}rowd in {R}equirements {E}ngineering: {T}he {L}andscape and {C}hallenges},
  volume    = {34},
  year      = {2017}
}

@article{Gruber_1993_KA,
  author    = {Thomas Gruber},
  journal   = {Knowledge acquisition},
  number    = {2},
  pages     = {199--220},
  publisher = {Elsevier},
  title     = {{A} translation approach to portable ontology specifications},
  volume    = {5},
  year      = {1993}
}

@article{Gruber_1995_IJHCS,
  author    = {Thomas Gruber},
  journal   = {International journal of human-computer studies},
  number    = {5-6},
  pages     = {907--928},
  publisher = {Elsevier},
  title     = {{T}oward principles for the design of ontologies used for knowledge sharing?},
  volume    = {43},
  year      = {1995}
}

@article{Gruber_1995_Online,
  author       = {Thomas Gruber},
  howpublished = {\url{https://web.archive.org/web/20100716004426/http://www-ksl.stanford.edu/kst/what-is-an-ontology.html}},
  note         = {Accessed: 2024-08-27},
  title        = {{W}hat is an {O}ntology?},
  year         = {1995}
}

@article{Gruber_2009_Online,
  author       = {Thomas Gruber},
  howpublished = {\url{https://web.archive.org/web/20100715224852/http://tomgruber.org/writing/ontology-definition-2007.htm}},
  note         = {Accessed: 2024-08-27},
  title        = {Ontology},
  year         = {2009}
}

@inproceedings{Guzman_2014_RE,
  author    = {Emitza Guzman and
               Walid Maalej},
  bibsource = {dblp computer science bibliography, https://dblp.org},
  booktitle = {{IEEE} 22nd International Requirements Engineering Conference, {RE}
               2014, Karlskrona, Sweden, August 25-29, 2014},
  doi       = {10.1109/RE.2014.6912257},
  editor    = {Tony Gorschek and
               Robyn R. Lutz},
  pages     = {153--162},
  publisher = {{IEEE} Computer Society},
  title     = {{H}ow {D}o {U}sers {L}ike {T}his {F}eature? {A} {F}ine {G}rained {S}entiment {A}nalysis of {A}pp {R}eviews},
  year      = {2014}
}

@inproceedings{Halverson_2006_CSCW,
  author    = {Christine A. Halverson and
               Jason B. Ellis and
               Catalina Danis and
               Wendy A. Kellogg},
  bibsource = {dblp computer science bibliography, https://dblp.org},
  booktitle = {Proceedings of the 2006 {ACM} Conference on Computer Supported Cooperative
               Work, {CSCW} 2006, Banff, Alberta, Canada, November 4-8, 2006},
  doi       = {10.1145/1180875.1180883},
  editor    = {Pamela J. Hinds and
               David Martin},
  pages     = {39--48},
  publisher = {{ACM}},
  title     = {{D}esigning task visualizations to support the coordination of work in software development},
  year      = {2006}
}

@inproceedings{Happel_2008_RSSE,
  author    = {Hans{-}J{\"{o}}rg Happel and
               Walid Maalej},
  title     = {Potentials and challenges of recommendation systems for software development},
  booktitle = {Proceedings of the 2008 International Workshop on Recommendation Systems
               for Software Engineering, {RSSE} 2008, Atlanta, GA, USA, November
               9, 2008},
  pages     = {11--15},
  publisher = {{ACM}},
  year      = {2008},
  doi       = {10.1145/1454247.1454251},
  bibsource = {dblp computer science bibliography, https://dblp.org}
}

@inproceedings{Hassan_2008_FSM,
  author       = {Hassan, Ahmed E.},
  booktitle    = {2008 frontiers of software maintenance},
  organization = {IEEE},
  pages        = {48--57},
  title        = {{T}he road ahead for mining software repositories},
  year         = {2008}
}

@inproceedings{He_2020_ICPC,
  author    = {Jianjun He and
               Ling Xu and
               Meng Yan and
               Xin Xia and
               Yan Lei},
  bibsource = {dblp computer science bibliography, https://dblp.org},
  booktitle = {{ICPC} '20: 28th International Conference on Program Comprehension,
               Seoul, Republic of Korea, July 13-15, 2020},
  doi       = {10.1145/3387904.3389263},
  pages     = {117--127},
  publisher = {{ACM}},
  title     = {{D}uplicate {B}ug {R}eport {D}etection {U}sing {D}ual-Channel {C}onvolutional {N}eural {N}etworks},
  year      = {2020}
}

@inproceedings{Heck_2013_IWPSE,
  author    = {Petra Heck and
               Andy Zaidman},
  bibsource = {dblp computer science bibliography, https://dblp.org},
  booktitle = {13th International Workshop on Principles of Software Evolution, {IWPSE}
               2013, Proceedings, August 19-20, 2013, Saint Petersburg, Russia},
  doi       = {10.1145/2501543.2501550},
  editor    = {Romain Robbes and
               Gregorio Robles},
  pages     = {43--52},
  publisher = {{ACM}},
  title     = {{A}n analysis of requirements evolution in open source projects: recommendations for issue trackers},
  year      = {2013}
}

@article{Heck_2014_JSEP,
  author    = {Petra Heck and
               Andy Zaidman},
  bibsource = {dblp computer science bibliography, https://dblp.org},
  doi       = {10.1002/SMR.1678},
  journal   = {J. Softw. Evol. Process.},
  number    = {12},
  pages     = {1280--1296},
  title     = {{H}orizontal traceability for just-in-time requirements: the case for open source feature requests},
  volume    = {26},
  year      = {2014}
}

@article{Heck_2016_REJ,
  author    = {Petra Heck and
               Andy Zaidman},
  bibsource = {dblp computer science bibliography, https://dblp.org},
  doi       = {10.1007/S00766-016-0247-5},
  journal   = {Requir. Eng.},
  number    = {4},
  pages     = {453--473},
  title     = {{A} framework for quality assessment of just-in-time requirements: the case of open source feature requests},
  volume    = {22},
  year      = {2017}
}

@inproceedings{Herraiz_2008_MSR,
  author    = {Israel Herraiz and
               Daniel M. Germ{\'{a}}n and
               Jes{\'{u}}s M. Gonz{\'{a}}lez{-}Barahona and
               Gregorio Robles},
  bibsource = {dblp computer science bibliography, https://dblp.org},
  booktitle = {Proceedings of the 2008 International Working Conference on Mining
               Software Repositories, {MSR} 2008 (Co-located with ICSE), Leipzig,
               Germany, May 10-11, 2008, Proceedings},
  doi       = {10.1145/1370750.1370786},
  editor    = {Ahmed E. Hassan and
               Michele Lanza and
               Michael W. Godfrey},
  pages     = {145--148},
  publisher = {{ACM}},
  title     = {{T}owards a simplification of the bug report form in eclipse},
  year      = {2008}
}

@inproceedings{Herzig_2013_ICSE,
  author    = {Kim Herzig and
               Sascha Just and
               Andreas Zeller},
  bibsource = {dblp computer science bibliography, https://dblp.org},
  booktitle = {35th International Conference on Software Engineering, {ICSE} '13,
               San Francisco, CA, USA, May 18-26, 2013},
  doi       = {10.1109/ICSE.2013.6606585},
  editor    = {David Notkin and
               Betty H. C. Cheng and
               Klaus Pohl},
  pages     = {392--401},
  publisher = {{IEEE} Computer Society},
  title     = {{I}t's not a bug, it's a feature: how misclassification impacts bug prediction},
  year      = {2013}
}

@article{Hesse_2016_IST,
  author    = {Tom{-}Michael Hesse and
               Veronika Lerche and
               Marcus Seiler and
               Konstantin Kn{\"{o}}{\ss} and
               Barbara Paech},
  bibsource = {dblp computer science bibliography, https://dblp.org},
  doi       = {10.1016/J.INFSOF.2016.06.003},
  journal   = {Inf. Softw. Technol.},
  pages     = {36--51},
  title     = {{D}ocumented decision-making strategies and decision knowledge in open source projects: {A}n empirical study on {F}irefox issue reports},
  volume    = {79},
  year      = {2016}
}

@inproceedings{Huo_2014_ICSME,
  author    = {Da Huo and
               Tao Ding and
               Collin McMillan and
               Malcom Gethers},
  bibsource = {dblp computer science bibliography, https://dblp.org},
  booktitle = {30th {IEEE} International Conference on Software Maintenance and Evolution,
               Victoria, BC, Canada, September 29 - October 3, 2014},
  doi       = {10.1109/ICSME.2014.22},
  pages     = {1--10},
  publisher = {{IEEE} Computer Society},
  title     = {{A}n {E}mpirical {S}tudy of the {E}ffects of {E}xpert {K}nowledge on {B}ug {R}eports},
  year      = {2014}
}

@article{Iftikhar_2024_IST,
  author    = {Umar Iftikhar and
               Nauman Bin Ali and
               J{\"{u}}rgen B{\"{o}}rstler and
               Muhammad Usman},
  bibsource = {dblp computer science bibliography, https://dblp.org},
  doi       = {10.1016/J.INFSOF.2023.107348},
  journal   = {Inf. Softw. Technol.},
  pages     = {107348},
  title     = {{A} tertiary study on links between source code metrics and external quality attributes},
  volume    = {165},
  year      = {2024}
}

@inproceedings{Imran_2021_MSR,
  author    = {Mia Mohammad Imran and
               Agnieszka Ciborowska and
               Kostadin Damevski},
  bibsource = {dblp computer science bibliography, https://dblp.org},
  booktitle = {18th {IEEE/ACM} International Conference on Mining Software Repositories,
               {MSR} 2021, Madrid, Spain, May 17-19, 2021},
  doi       = {10.1109/MSR52588.2021.00029},
  pages     = {167--178},
  publisher = {{IEEE}},
  title     = {{A}utomatically {S}electing {F}ollow-up {Q}uestions for {D}eficient {B}ug {R}eports},
  year      = {2021}
}

@article{Izadi_2022_EMSE,
  author    = {Maliheh Izadi and
               Kiana Akbari and
               Abbas Heydarnoori},
  bibsource = {dblp computer science bibliography, https://dblp.org},
  doi       = {10.1007/S10664-021-10085-3},
  journal   = {Empir. Softw. Eng.},
  number    = {2},
  pages     = {50},
  title     = {{P}redicting the objective and priority of issue reports in software repositories},
  volume    = {27},
  year      = {2022}
}

@phdthesis{Janak_2009_PhDThesis,
  author = {Jan{\'a}k, Ji{\v{r}}{\'\i}},
  school = {Masarykova univerzita, Fakulta informatiky},
  title  = {{I}ssue tracking systems},
  year   = {2009}
}

@article{Jayatilleke_2018_IST,
  author    = {Shalinka Jayatilleke and
               Richard Lai},
  bibsource = {dblp computer science bibliography, https://dblp.org},
  doi       = {10.1016/J.INFSOF.2017.09.004},
  journal   = {Inf. Softw. Technol.},
  pages     = {163--185},
  title     = {{A} systematic review of requirements change management},
  volume    = {93},
  year      = {2018}
}

@inproceedings{Jeong_2009_FSE,
  author    = {Gaeul Jeong and
               Sunghun Kim and
               Thomas Zimmermann},
  bibsource = {dblp computer science bibliography, https://dblp.org},
  booktitle = {Proceedings of the 7th joint meeting of the European Software Engineering
               Conference and the {ACM} {SIGSOFT} International Symposium on Foundations
               of Software Engineering, 2009, Amsterdam, The Netherlands, August
               24-28, 2009},
  doi       = {10.1145/1595696.1595715},
  editor    = {Hans van Vliet and
               Val{\'{e}}rie Issarny},
  pages     = {111--120},
  publisher = {{ACM}},
  title     = {{I}mproving bug triage with bug tossing graphs},
  year      = {2009}
}

@online{Jira_2024_automation,
  title   = {Jira Automations},
  url     = {https://www.atlassian.com/software/jira/guides/expand-jira/automation-use-cases},
  urldate = {2024-07-03}
}

@online{Jira_2024_Online,
  title   = {Jira},
  url     = {https://www.atlassian.com/software/jira},
  urldate = {2024-07-03}
}

@online{Jira_Missing_Severity_URL_1,
  title   = {Online Jira discussion about missing severity field.},
  url     = {https://community.atlassian.com/t5/Jira-questions/Remind-me-why-Jira-removed-the-Severity-field/qaq-p/773901},
  urldate = {2024-11-15}
}

@online{Jira_Missing_Severity_URL_2,
  title   = {Online Jira discussion about missing severity field.},
  url     = {https://community.atlassian.com/t5/Jira-questions/Why-there-is-no-Severity-Field-for-Issue-Type-Bug/qaq-p/1117643},
  urldate = {2024-11-15}
}

@online{Jira_Missing_Severity_URL_3,
  title   = {Online Jira discussion about missing severity field.},
  url     = {https://community.atlassian.com/t5/Jira-questions/where-to-find-jira-issue-severity/qaq-p/1696032},
  urldate = {2024-11-15}
}

@online{JiraJoke1_2024_Online,
  title   = {Jira Joke on Reddit},
  url     = {https://www.reddit.com/r/ProgrammerHumor/comments/qt28x3/a_different_level_of_hate/},
  urldate = {2024-11-20}
}

@online{JiraJoke2_2024_Online,
  title   = {Jira Jokes on the Atlassian Forum},
  url     = {https://community.atlassian.com/t5/Watercooler-discussions/Friday-Fun-JIRA-memes/td-p/816534},
  urldate = {2024-11-20}
}

@online{JiraJoke3_2024_Online,
  title   = {Jira Joke on Reddit: ``Happy teams start with Jira, \dots and then they are sad.''},
  url     = {https://www.reddit.com/r/ProgrammerHumor/comments/1ffpe7f/andthentheyaresad/},
  urldate = {2024-11-20}
}

@online{JiraJoke4_2024_Online,
  title   = {A website dedicated entirely to quotes and opinions about how they dislike Jira.},
  url     = {https://ifuckinghatejira.com/},
  urldate = {2024-11-20}
}

@inproceedings{Johann_2017_RE,
  author    = {Timo Johann and
               Christoph Stanik and
               Alireza Alizadeh and
               Walid Maalej},
  bibsource = {dblp computer science bibliography, https://dblp.org},
  booktitle = {25th {IEEE} International Requirements Engineering Conference, {RE}
               2017, Lisbon, Portugal, September 4-8, 2017},
  doi       = {10.1109/RE.2017.71},
  editor    = {Ana Moreira and
               Jo{\~{a}}o Ara{\'{u}}jo and
               Jane Hayes and
               Barbara Paech},
  pages     = {21--30},
  publisher = {{IEEE} Computer Society},
  title     = {{SAFE:} {A} {S}imple {A}pproach for {F}eature {E}xtraction from {A}pp {D}escriptions and {A}pp {R}eviews},
  year      = {2017}
}

@article{Johnson_2003_CSE,
  author    = {Johnson, Jeffrey N and Dubois, Paul F},
  journal   = {Computing in Science \& Engineering},
  number    = {6},
  pages     = {71--77},
  publisher = {IEEE},
  title     = {{I}ssue tracking},
  volume    = {5},
  year      = {2003}
}

@inproceedings{Just_2008_IEEESymposium,
  author    = {Sascha Just and
               Rahul Premraj and
               Thomas Zimmermann},
  bibsource = {dblp computer science bibliography, https://dblp.org},
  booktitle = {{IEEE} Symposium on Visual Languages and Human-Centric Computing,
               {VL/HCC} 2008, Herrsching am Ammersee, Germany, 15-19 September 2008,
               Proceedings},
  doi       = {10.1109/VLHCC.2008.4639063},
  pages     = {82--85},
  publisher = {{IEEE} Computer Society},
  title     = {{T}owards the next generation of bug tracking systems},
  year      = {2008}
}

@inproceedings{Kabbedijk_2009_RE,
  author    = {Jaap Kabbedijk and
               Sjaak Brinkkemper and
               Slinger Jansen and
               Bas van der Veldt},
  bibsource = {dblp computer science bibliography, https://dblp.org},
  booktitle = {{RE} 2009, 17th {IEEE} International Requirements Engineering Conference,
               Atlanta, Georgia, USA, August 31 - September 4, 2009},
  doi       = {10.1109/RE.2009.28},
  pages     = {281--286},
  publisher = {{IEEE} Computer Society},
  title     = {{C}ustomer {I}nvolvement in {R}equirements {M}anagement: {L}essons from {M}ass {M}arket {S}oftware {D}evelopment},
  year      = {2009}
}

@inproceedings{Kamata_Tamai_2007_CS,
  author    = {Mayumi Itakura Kamata and
               Tetsuo Tamai},
  title     = {How Does Requirements Quality Relate to Project Success or Failure?},
  booktitle = {15th {IEEE} International Requirements Engineering Conference, {RE}
               2007, October 15-19th, 2007, New Delhi, India},
  pages     = {69--78},
  publisher = {{IEEE} Computer Society},
  year      = {2007},
  doi       = {10.1109/RE.2007.31},
  bibsource = {dblp computer science bibliography, https://dblp.org}
}

@article{KarrWisniewski_2010_CHB,
  author    = {Pamela Karr Wisniewski and
               Ying Lu},
  bibsource = {dblp computer science bibliography, https://dblp.org},
  doi       = {10.1016/J.CHB.2010.03.008},
  journal   = {Comput. Hum. Behav.},
  number    = {5},
  pages     = {1061--1072},
  title     = {{W}hen more is too much: {O}perationalizing technology overload and exploring its impact on knowledge worker productivity},
  volume    = {26},
  year      = {2010}
}

@inproceedings{Kassab_2018_IPCC,
  author    = {Mohamad Kassab and
               Joanna F. DeFranco and
               Valdemar Vicente Graciano Neto},
  title     = {An Empirical Investigation on the Satisfaction Levels with the Requirements
               Engineering Practices: Agile vs. Waterfall},
  booktitle = {{IEEE} International Professional Communication Conference, ProComm
               2018, Toronto, ON, Canada, July 22-25, 2018},
  pages     = {118--124},
  publisher = {{IEEE}},
  year      = {2018},
  doi       = {10.1109/PROCOMM.2018.00033},
  bibsource = {dblp computer science bibliography, https://dblp.org}
}

@techreport{Kitchenham_2007_TR,
  author      = {Kitchenham, Barbara and Charters, Stuart},
  institution = {{Keele University}},
  month       = jul,
  number      = {EBSE-2007-01},
  pages       = {65},
  title       = {{G}uidelines for {P}erforming {{Systematic {L}iterature {R}eviews}} in {{Software {E}ngineering}}},
  year        = {2007}
}

@article{Kitchenham_2013_IST,
  author    = {Barbara Kitchenham and
               Pearl Brereton},
  bibsource = {dblp computer science bibliography, https://dblp.org},
  doi       = {10.1016/J.INFSOF.2013.07.010},
  journal   = {Inf. Softw. Technol.},
  number    = {12},
  pages     = {2049--2075},
  title     = {{A} systematic review of systematic review process research in software engineering},
  volume    = {55},
  year      = {2013}
}

@book{Kuchana_2004_Book,
  author    = {Kuchana, Partha},
  publisher = {Auerbach Publications},
  title     = {{S}oftware architecture design patterns in {J}ava},
  year      = {2004}
}

@article{Kundisch_2021_BISE,
  author    = {Kundisch, Dennis and Muntermann, Jan and Oberl{\"a}nder, Anna Maria and Rau, Daniel and R{\"o}glinger, Maximilian and Schoormann, Thorsten and Szopinski, Daniel},
  journal   = {Business \& Information Systems Engineering},
  pages     = {1--19},
  publisher = {Springer},
  title     = {{A}n update for taxonomy designers: methodological guidance from information systems research},
  year      = {2021}
}

@inproceedings{Lamkanfi_2010_MSR,
  author    = {Ahmed Lamkanfi and
               Serge Demeyer and
               Emanuel Giger and
               Bart Goethals},
  bibsource = {dblp computer science bibliography, https://dblp.org},
  booktitle = {Proceedings of the 7th International Working Conference on Mining
               Software Repositories, {MSR} 2010 (Co-located with ICSE), Cape Town,
               South Africa, May 2-3, 2010, Proceedings},
  doi       = {10.1109/MSR.2010.5463284},
  editor    = {Jim Whitehead and
               Thomas Zimmermann},
  pages     = {1--10},
  publisher = {{IEEE} Computer Society},
  title     = {{P}redicting the severity of a reported bug},
  year      = {2010}
}

@inproceedings{Lamkanfi_2011_CSMR,
  author    = {Ahmed Lamkanfi and
               Serge Demeyer and
               Quinten David Soetens and
               Tim Verdonck},
  bibsource = {dblp computer science bibliography, https://dblp.org},
  booktitle = {15th European Conference on Software Maintenance and Reengineering,
               {CSMR} 2011, 1-4 March 2011, Oldenburg, Germany},
  doi       = {10.1109/CSMR.2011.31},
  editor    = {Tom Mens and
               Yiannis Kanellopoulos and
               Andreas Winter},
  pages     = {249--258},
  publisher = {{IEEE} Computer Society},
  title     = {{C}omparing {M}ining {A}lgorithms for {P}redicting the {S}everity of a {R}eported {B}ug},
  year      = {2011}
}

@book{Laplante_2022_Book,
  title     = {Requirements engineering for software and systems},
  author    = {Laplante, Phillip A and Kassab, Mohamad},
  year      = {2022},
  publisher = {Auerbach Publications}
}

@inproceedings{Lazar_2014_MSR,
  author    = {Alina Lazar and
               Sarah Ritchey and
               Bonita Sharif},
  bibsource = {dblp computer science bibliography, https://dblp.org},
  booktitle = {11th Working Conference on Mining Software Repositories, {MSR} 2014,
               Proceedings, May 31 - June 1, 2014, Hyderabad, India},
  doi       = {10.1145/2597073.2597128},
  editor    = {Premkumar T. Devanbu and
               Sung Kim and
               Martin Pinzger},
  pages     = {392--395},
  publisher = {{ACM}},
  title     = {{G}enerating duplicate bug datasets},
  year      = {2014}
}

@inproceedings{Li_2012_EASE,
  author    = {Juan Li and
               He Zhang and
               Liming Zhu and
               D. Ross Jeffery and
               Qing Wang and
               Mingshu Li},
  bibsource = {dblp computer science bibliography, https://dblp.org},
  booktitle = {16th International Conference on Evaluation {\&} Assessment in Software
               Engineering, {EASE} 2012, Ciudad Real, Spain, May 14-15, 2012. Proceedings},
  doi       = {10.1049/IC.2012.0002},
  editor    = {Maria Teresa Baldassarre and
               Marcela Genero and
               Emilia Mendes and
               Mario Piattini},
  pages     = {12--21},
  publisher = {{IET} - The Institute of Engineering and Technology / {IEEE} Xplore},
  title     = {{P}reliminary results of a systematic review on requirements evolution},
  year      = {2012}
}

@inproceedings{Li_2012_REFSQ,
  author    = {Yang Li and
               Walid Maalej},
  bibsource = {dblp computer science bibliography, https://dblp.org},
  booktitle = {Requirements Engineering: Foundation for Software Quality - 18th International
               Working Conference, {REFSQ} 2012, Essen, Germany, March 19-22, 2012.
               Proceedings},
  doi       = {10.1007/978-3-642-28714-5\_17},
  editor    = {Bj{\"{o}}rn Regnell and
               Daniela E. Damian},
  pages     = {194--210},
  publisher = {Springer},
  series    = {Lecture Notes in Computer Science},
  title     = {{W}hich {T}raceability {V}isualization {I}s {S}uitable in {T}his {C}ontext? {A} {C}omparative {S}tudy},
  volume    = {7195},
  year      = {2012}
}

@inproceedings{Li_2016_CAiSE,
  author    = {Feng{-}Lin Li and
               Jennifer Horkoff and
               Lin Liu and
               Alexander Borgida and
               Giancarlo Guizzardi and
               John Mylopoulos},
  bibsource = {dblp computer science bibliography, https://dblp.org},
  booktitle = {Advanced Information Systems Engineering - 28th International Conference,
               CAiSE 2016, Ljubljana, Slovenia, June 13-17, 2016. Proceedings},
  doi       = {10.1007/978-3-319-39696-5\_14},
  editor    = {Selmin Nurcan and
               Pnina Soffer and
               Marko Bajec and
               Johann Eder},
  pages     = {221--238},
  publisher = {Springer},
  series    = {Lecture Notes in Computer Science},
  title     = {{E}ngineering {R}equirements with {D}esiree: {A}n {E}mpirical {E}valuation},
  volume    = {9694},
  year      = {2016}
}

@inproceedings{Li_2018_APSEC,
  author    = {Lisha Li and
               Zhilei Ren and
               Xiaochen Li and
               Weiqin Zou and
               He Jiang},
  bibsource = {dblp computer science bibliography, https://dblp.org},
  booktitle = {25th Asia-Pacific Software Engineering Conference, {APSEC} 2018, Nara,
               Japan, December 4-7, 2018},
  doi       = {10.1109/APSEC.2018.00053},
  pages     = {386--395},
  publisher = {{IEEE}},
  title     = {{H}ow {A}re {I}ssue {U}nits {L}inked? {E}mpirical {S}tudy on the {L}inking {B}ehavior in {G}it{H}ub},
  year      = {2018}
}

@inproceedings{Li_2022_ESEM,
  author    = {Yingling Li and
               Xing Che and
               Yuekai Huang and
               Junjie Wang and
               Song Wang and
               Yawen Wang and
               Qing Wang},
  bibsource = {dblp computer science bibliography, https://dblp.org},
  booktitle = {{ESEM} '22: {ACM} / {IEEE} International Symposium on Empirical Software
               Engineering and Measurement, Helsinki, Finland, September 19 - 23,
               2022},
  doi       = {10.1145/3544902.3546257},
  editor    = {Fernanda Madeiral and
               Casper Lassenius and
               Tayana Conte and
               Tomi M{\"{a}}nnist{\"{o}}},
  pages     = {1--11},
  publisher = {{ACM}},
  title     = {{A} {T}ale of {T}wo {T}asks: {A}utomated {I}ssue {P}riority {P}rediction with {D}eep {M}ulti-task {L}earning},
  year      = {2022}
}

@inproceedings{Lo_2015_ESECFSE,
  author    = {David Lo and
               Nachiappan Nagappan and
               Thomas Zimmermann},
  bibsource = {dblp computer science bibliography, https://dblp.org},
  booktitle = {Proceedings of the 2015 10th Joint Meeting on Foundations of Software
               Engineering, {ESEC/FSE} 2015, Bergamo, Italy, August 30 - September
               4, 2015},
  doi       = {10.1145/2786805.2786809},
  editor    = {Elisabetta Di Nitto and
               Mark Harman and
               Patrick Heymans},
  pages     = {415--425},
  publisher = {{ACM}},
  title     = {{H}ow practitioners perceive the relevance of software engineering research},
  year      = {2015}
}

@inproceedings{Lucassen_2016_REFSQ,
  author    = {Garm Lucassen and
               Fabiano Dalpiaz and
               Jan Martijn E. M. van der Werf and
               Sjaak Brinkkemper},
  bibsource = {dblp computer science bibliography, https://dblp.org},
  booktitle = {Requirements Engineering: Foundation for Software Quality - 22nd International
               Working Conference, {REFSQ} 2016, Gothenburg, Sweden, March 14-17,
               2016, Proceedings},
  doi       = {10.1007/978-3-319-30282-9\_14},
  editor    = {Maya Daneva and
               Oscar Pastor},
  pages     = {205--222},
  publisher = {Springer},
  series    = {Lecture Notes in Computer Science},
  title     = {{T}he {U}se and {E}ffectiveness of {U}ser {S}tories in {P}ractice},
  volume    = {9619},
  year      = {2016}
}

@inproceedings{Lüders_2022_MSR,
  author    = {Clara L{\"{u}}ders and
               Abir Bouraffa and
               Walid Maalej},
  bibsource = {dblp computer science bibliography, https://dblp.org},
  booktitle = {19th {IEEE/ACM} International Conference on Mining Software Repositories,
               {MSR} 2022, Pittsburgh, PA, USA, May 23-24, 2022},
  doi       = {10.1145/3524842.3528457},
  pages     = {48--60},
  publisher = {{ACM}},
  title     = {{B}eyond {D}uplicates: {T}owards {U}nderstanding and {P}redicting {L}ink {T}ypes in {I}ssue {T}racking {S}ystems},
  year      = {2022}
}

@inproceedings{Lüders_2022_RE,
  author    = {Clara L{\"{u}}ders and
               Tim Pietz and
               Walid Maalej},
  bibsource = {dblp computer science bibliography, https://dblp.org},
  booktitle = {30th {IEEE} International Requirements Engineering Conference, {RE}
               2022, Melbourne, Australia, August 15-19, 2022},
  doi       = {10.1109/RE54965.2022.00010},
  pages     = {26--38},
  publisher = {{IEEE}},
  title     = {{A}utomated {D}etection of {T}yped {L}inks in {I}ssue {T}rackers},
  year      = {2022}
}

@phdthesis{Lüders_2023_PhDThesis,
  author = {Clara Lüders},
  school = {University of Hamburg, Germany},
  title  = {{M}ining and {U}nderstanding {I}ssue {L}inks {T}owards a {B}etter {I}ssue {M}anagement},
  type   = {PhD Thesis},
  url    = {https://ediss.sub.uni-hamburg.de/handle/ediss/10310},
  year   = {2023}
}

@inproceedings{Lueders_2019_RE,
  author    = {Clara L{\"{u}}ders and
               Mikko Raatikainen and
               Joaquim Motger and
               Walid Maalej},
  bibsource = {dblp computer science bibliography, https://dblp.org},
  booktitle = {27th {IEEE} International Requirements Engineering Conference, {RE}
               2019, Jeju Island, Korea (South), September 23-27, 2019},
  doi       = {10.1109/RE.2019.00070},
  editor    = {Daniela E. Damian and
               Anna Perini and
               Seok{-}Won Lee},
  pages     = {492--493},
  publisher = {{IEEE}},
  title     = {{O}pen{R}eq {I}ssue {L}ink {M}ap: {A} {T}ool to {V}isualize {I}ssue {L}inks in {J}ira},
  year      = {2019}
}

@inproceedings{Maalej_2009_ASE,
  author    = {Walid Maalej},
  title     = {Task-First or Context-First? Tool Integration Revisited},
  booktitle = {{ASE} 2009, 24th {IEEE/ACM} International Conference on Automated
               Software Engineering, Auckland, New Zealand, November 16-20, 2009},
  pages     = {344--355},
  publisher = {{IEEE} Computer Society},
  year      = {2009},
  doi       = {10.1109/ASE.2009.36},
  bibsource = {dblp computer science bibliography, https://dblp.org}
}

@book{Maalej_2013_MARKBook,
  editor    = {Walid Maalej and Anil Kumar Thurimella},
  title     = {Managing Requirements Knowledge},
  publisher = {Springer},
  year      = {2013},
  doi       = {10.1007/978-3-642-34419-0},
  isbn      = {978-3-642-34418-3},
  bibsource = {dblp computer science bibliography, https://dblp.org}
}

@inproceedings{Maalej_2014_EmpiRE,
  author    = {Walid Maalej and
               Zijad Kurtanovic and
               Alexander Felfernig},
  bibsource = {dblp computer science bibliography, https://dblp.org},
  booktitle = {4th {IEEE} International Workshop on Empirical Requirements Engineering,
               EmpiRE 2014, Karlskrona, Sweden, August 25, 2014},
  doi       = {10.1109/EMPIRE.2014.6890118},
  editor    = {Maya Daneva and
               Richard Berntsson{-}Svensson and
               Xavier Franch and
               Nazim H. Madhavji and
               Sabrina Marczak},
  pages     = {64--71},
  publisher = {{IEEE} Computer Society},
  title     = {{W}hat stakeholders need to know about requirements},
  year      = {2014}
}

@article{Maalej_2015_IEEESoftware,
  author    = {Walid Maalej and
               Maleknaz Nayebi and
               Timo Johann and
               G{\"{u}}nther Ruhe},
  bibsource = {dblp computer science bibliography, https://dblp.org},
  doi       = {10.1109/MS.2015.153},
  journal   = {{IEEE} Softw.},
  number    = {1},
  pages     = {48--54},
  title     = {{T}oward {D}ata-Driven {R}equirements {E}ngineering},
  volume    = {33},
  year      = {2016}
}

@inproceedings{Maalej_2015_RE,
  author    = {Walid Maalej and
               Hadeer Nabil},
  bibsource = {dblp computer science bibliography, https://dblp.org},
  booktitle = {23rd {IEEE} International Requirements Engineering Conference, {RE}
               2015, Ottawa, ON, Canada, August 24-28, 2015},
  doi       = {10.1109/RE.2015.7320414},
  editor    = {Didar Zowghi and
               Vincenzo Gervasi and
               Daniel Amyot},
  pages     = {116--125},
  publisher = {{IEEE} Computer Society},
  title     = {{B}ug report, feature request, or simply praise? {O}n automatically classifying app reviews},
  year      = {2015}
}

@article{Maalej_2016_RE,
  author    = {Walid Maalej and
               Zijad Kurtanovic and
               Hadeer Nabil and
               Christoph Stanik},
  bibsource = {dblp computer science bibliography, https://dblp.org},
  doi       = {10.1007/S00766-016-0251-9},
  journal   = {Requir. Eng.},
  number    = {3},
  pages     = {311--331},
  title     = {{O}n the automatic classification of app reviews},
  volume    = {21},
  year      = {2016}
}

@article{Madampe_2022_TSE,
  author    = {Kashumi Madampe and
               Rashina Hoda and
               John C. Grundy},
  bibsource = {dblp computer science bibliography, https://dblp.org},
  doi       = {10.1109/TSE.2021.3104732},
  journal   = {{IEEE} Trans. Software Eng.},
  number    = {10},
  pages     = {3737--3752},
  title     = {{A} {F}aceted {T}axonomy of {R}equirements {C}hanges in {A}gile {C}ontexts},
  volume    = {48},
  year      = {2022}
}

@online{Mantis_2024_Online,
  title   = {Mantis},
  url     = {https://mantisbt.org/},
  urldate = {2024-07-03}
}

@article{Mendez_2017_EMSE,
  author    = {Daniel Mendez and
               Stefan Wagner and
               Marcos Kalinowski and
               Michael Felderer and
               Priscilla Mafra and
               Antonio Vetr{\`{o}} and
               Tayana Conte and
               Marie{-}Therese Christiansson and
               Desmond Greer and
               Casper Lassenius and
               Tomi M{\"{a}}nnist{\"{o}} and
               Maleknaz Nayebi and
               Markku Oivo and
               Birgit Penzenstadler and
               Dietmar Pfahl and
               Rafael Prikladnicki and
               G{\"{u}}nther Ruhe and
               Andr{\'{e}} Schekelmann and
               Sagar Sen and
               Rodrigo O. Sp{\'{\i}}nola and
               Ahmet Tuzcu and
               Jose Luis de la Vara and
               Roel J. Wieringa},
  title     = {Naming the pain in requirements engineering - Contemporary problems, causes, and effects in practice},
  journal   = {Empirical Sofware Engineering},
  volume    = {22},
  number    = {5},
  pages     = {2298--2338},
  year      = {2017},
  doi       = {10.1007/S10664-016-9451-7},
  bibsource = {dblp computer science bibliography, https://dblp.org}
}

@misc{Mendez_2021_Lecture,
  author = {Daniel Mendez},
  doi    = {10.6084/m9.figshare.17128613.v1},
  title  = {{R}equirements {E}ngineering {L}ecture {S}eries},
  year   = {2021}
}

@inproceedings{Menzies_2008_ICSM,
  author    = {Tim Menzies and
               Andrian Marcus},
  bibsource = {dblp computer science bibliography, https://dblp.org},
  booktitle = {24th {IEEE} International Conference on Software Maintenance {(ICSM}
               2008), September 28 - October 4, 2008, Beijing, China},
  doi       = {10.1109/ICSM.2008.4658083},
  pages     = {346--355},
  publisher = {{IEEE} Computer Society},
  title     = {{A}utomated severity assessment of software defect reports},
  year      = {2008}
}

@inproceedings{Merten_2016_RE,
  author    = {Thorsten Merten and
               Mat{\'{u}}s Falis and
               Paul H{\"{u}}bner and
               Thomas Quirchmayr and
               Simone B{\"{u}}rsner and
               Barbara Paech},
  bibsource = {dblp computer science bibliography, https://dblp.org},
  booktitle = {24th {IEEE} International Requirements Engineering Conference, {RE}
               2016, Beijing, China, September 12-16, 2016},
  doi       = {10.1109/RE.2016.8},
  pages     = {166--175},
  publisher = {{IEEE} Computer Society},
  title     = {{S}oftware {F}eature {R}equest {D}etection in {I}ssue {T}racking {S}ystems},
  year      = {2016}
}

@inproceedings{Merten_2016_REFSQ,
  author    = {Thorsten Merten and
               Daniel Kr{\"{a}}mer and
               Bastian Mager and
               Paul Schell and
               Simone B{\"{u}}rsner and
               Barbara Paech},
  bibsource = {dblp computer science bibliography, https://dblp.org},
  booktitle = {Requirements Engineering: Foundation for Software Quality - 22nd International
               Working Conference, {REFSQ} 2016, Gothenburg, Sweden, March 14-17,
               2016, Proceedings},
  doi       = {10.1007/978-3-319-30282-9\_4},
  editor    = {Maya Daneva and
               Oscar Pastor},
  pages     = {45--62},
  publisher = {Springer},
  series    = {Lecture Notes in Computer Science},
  title     = {{D}o {I}nformation {R}etrieval {A}lgorithms for {A}utomated {T}raceability {P}erform {E}ffectively on {I}ssue {T}racking {S}ystem {D}ata?},
  volume    = {9619},
  year      = {2016}
}

@inproceedings{Meyer_2014_FSE,
  author    = {Andr{\'{e}} N. Meyer and
               Thomas Fritz and
               Gail C. Murphy and
               Thomas Zimmermann},
  bibsource = {dblp computer science bibliography, https://dblp.org},
  booktitle = {Proceedings of the 22nd {ACM} {SIGSOFT} International Symposium on
               Foundations of Software Engineering, (FSE-22), Hong Kong, China, November
               16 - 22, 2014},
  doi       = {10.1145/2635868.2635892},
  editor    = {Shing{-}Chi Cheung and
               Alessandro Orso and
               Margaret{-}Anne D. Storey},
  pages     = {19--29},
  publisher = {{ACM}},
  title     = {{S}oftware developers' perceptions of productivity},
  year      = {2014}
}

@article{Meyer_2021_TSE,
  author    = {Andr{\'{e}} N. Meyer and
               Earl T. Barr and
               Christian Bird and
               Thomas Zimmermann},
  bibsource = {dblp computer science bibliography, https://dblp.org},
  doi       = {10.1109/TSE.2019.2904957},
  journal   = {{IEEE} Trans. Software Eng.},
  number    = {5},
  pages     = {863--880},
  title     = {{T}oday {W}as a {G}ood {D}ay: {T}he {D}aily {L}ife of {S}oftware {D}evelopers},
  volume    = {47},
  year      = {2021}
}

@online{Miro_2024_Online,
  title   = {Miro},
  url     = {https://Miro.com/},
  urldate = {2024-07-03}
}

@article{Mockus_2002_TOSEM,
  author    = {Audris Mockus and
               Roy T. Fielding and
               James D. Herbsleb},
  bibsource = {dblp computer science bibliography, https://dblp.org},
  doi       = {10.1145/567793.567795},
  journal   = {{ACM} Trans. Softw. Eng. Methodol.},
  number    = {3},
  pages     = {309--346},
  title     = {{T}wo case studies of open source software development: {A}pache and {M}ozilla},
  volume    = {11},
  year      = {2002}
}

@online{Monday_2024_Online,
  title   = {Monday tool.},
  url     = {https://monday.com},
  urldate = {2024-11-20}
}

@phdthesis{Montgomery_2017_MScThesis,
  author = {Lloyd Montgomery},
  school = {University of Victoria},
  title  = {Escalation prediction using feature engineering: addressing support ticket escalations within IBM's ecosystem},
  type   = {MSc Thesis},
  year   = {2017}
}

@inproceedings{Montgomery_2017_RE,
  author    = {Lloyd Montgomery and
               Daniela E. Damian},
  bibsource = {dblp computer science bibliography, https://dblp.org},
  booktitle = {25th {IEEE} International Requirements Engineering Conference, {RE}
               2017, Lisbon, Portugal, September 4-8, 2017},
  doi       = {10.1109/RE.2017.61},
  editor    = {Ana Moreira and
               Jo{\~{a}}o Ara{\'{u}}jo and
               Jane Hayes and
               Barbara Paech},
  pages     = {362--371},
  publisher = {{IEEE} Computer Society},
  title     = {{W}hat do {S}upport {A}nalysts {K}now {A}bout {T}heir {C}ustomers? {O}n the {S}tudy and {P}rediction of {S}upport {T}icket {E}scalations in {L}arge {S}oftware {O}rganizations},
  year      = {2017}
}

@inproceedings{Montgomery_2017_RE_a,
  author    = {Lloyd Montgomery and
               Emma Reading and
               Daniela E. Damian},
  bibsource = {dblp computer science bibliography, https://dblp.org},
  booktitle = {25th {IEEE} International Requirements Engineering Conference, {RE}
               2017, Lisbon, Portugal, September 4-8, 2017},
  doi       = {10.1109/RE.2017.62},
  editor    = {Ana Moreira and
               Jo{\~{a}}o Ara{\'{u}}jo and
               Jane Hayes and
               Barbara Paech},
  pages     = {452--455},
  publisher = {{IEEE} Computer Society},
  title     = {{E}{C}rits - {V}isualizing {S}upport {T}icket {E}scalation {R}isk},
  year      = {2017}
}

@article{Montgomery_2018_REJ,
  author    = {Lloyd Montgomery and
               Daniela E. Damian and
               Tyson Bulmer and
               Shaikh Quader},
  bibsource = {dblp computer science bibliography, https://dblp.org},
  doi       = {10.1007/S00766-018-0292-3},
  journal   = {Requir. Eng.},
  number    = {3},
  pages     = {333--355},
  title     = {{C}ustomer support ticket escalation prediction using feature engineering},
  volume    = {23},
  year      = {2018}
}

@inproceedings{Montgomery_2022_MSR,
  author    = {Lloyd Montgomery and
               Clara L{\"{u}}ders and
               Walid Maalej},
  bibsource = {dblp computer science bibliography, https://dblp.org},
  booktitle = {19th {IEEE/ACM} International Conference on Mining Software Repositories,
               {MSR} 2022, Pittsburgh, PA, USA, May 23-24, 2022},
  doi       = {10.1145/3524842.3528486},
  pages     = {73--77},
  publisher = {{ACM}},
  title     = {{A}n {A}lternative {I}ssue {T}racking {D}ataset of {P}ublic {J}ira {R}epositories},
  year      = {2022}
}

@article{Montgomery_2022_REJ,
  author    = {Lloyd Montgomery and
               Davide Fucci and
               Abir Bouraffa and
               Lisa Scholz and
               Walid Maalej},
  bibsource = {dblp computer science bibliography, https://dblp.org},
  doi       = {10.1007/S00766-021-00367-Z},
  journal   = {Requir. Eng.},
  number    = {2},
  pages     = {183--209},
  title     = {{E}mpirical research on requirements quality: a systematic mapping study},
  volume    = {27},
  year      = {2022}
}

@incollection{Montgomery_2025_BookChapter,
  address   = {Cham, Switzerland},
  author    = {Montgomery, Lloyd and L{\"u}ders, Clara and Maalej, Walid},
  booktitle = {Handbook of Natural Language Processing for Requirements Engineering},
  chapter   = {11},
  editor    = {A. Ferrari and G. Deshpande},
  pages     = {???--???},
  publisher = {Springer Nature Switzerland AG},
  title     = {Mining Issue Trackers: Concepts and Techniques},
  year      = {2024}
}

@online{Montgomery_2025_PhDThesisReplicationPackage,
  title   = {Replication Package for PhD Dissertation},
  author  = {Lloyd Montgomery},
  url     = {https://doi.org/10.5281/zenodo.14669551},
  urldate = {2025-01-17}
}

@article{Nickerson_2013_EJIS,
  author    = {Robert C. Nickerson and
               Upkar Varshney and
               Jan Muntermann},
  bibsource = {dblp computer science bibliography, https://dblp.org},
  doi       = {10.1057/EJIS.2012.26},
  journal   = {Eur. J. Inf. Syst.},
  number    = {3},
  pages     = {336--359},
  title     = {{A} method for taxonomy development and its application in information systems},
  volume    = {22},
  year      = {2013}
}

@misc{OpenReq_2023_Online,
  author  = {OpenReq},
  title   = {{H}omepage},
  url     = {https://openreq.eu/},
  urldate = {2023-01-03}
}

@misc{OpenReqGitHub_2023_Online,
  author  = {OpenReq},
  title   = {{P}roject {G}it{H}ub {S}ite},
  url     = {https://github.com/openreqeu},
  urldate = {2023-01-03}
}

@inproceedings{Ortu_2015_PROMISE,
  author    = {Marco Ortu and
               Giuseppe Destefanis and
               Bram Adams and
               Alessandro Murgia and
               Michele Marchesi and
               Roberto Tonelli},
  bibsource = {dblp computer science bibliography, https://dblp.org},
  booktitle = {Proceedings of the 11th International Conference on Predictive Models
               and Data Analytics in Software Engineering, {PROMISE} 2015, Beijing,
               China, October 21, 2015},
  doi       = {10.1145/2810146.2810147},
  editor    = {Ayse Bener and
               Leandro L. Minku and
               Burak Turhan},
  pages     = {1:1--1:4},
  publisher = {{ACM}},
  title     = {{T}he {JIRA} {R}epository {D}ataset: {U}nderstanding {S}ocial {A}spects of {S}oftware {D}evelopment},
  year      = {2015}
}

@inproceedings{Pagano_2013_RE,
  author    = {Dennis Pagano and
               Walid Maalej},
  bibsource = {dblp computer science bibliography, https://dblp.org},
  booktitle = {21st {IEEE} International Requirements Engineering Conference, {RE}
               2013, Rio de Janeiro-RJ, Brazil, July 15-19, 2013},
  doi       = {10.1109/RE.2013.6636712},
  pages     = {125--134},
  publisher = {{IEEE} Computer Society},
  title     = {{U}ser feedback in the appstore: {A}n empirical study},
  year      = {2013}
}

@article{Palomba_2021_TSE,
  author    = {Fabio Palomba and
               Damian Tamburri and
               Francesca Arcelli Fontana and
               Rocco Oliveto and
               Andy Zaidman and
               Alexander Serebrenik},
  bibsource = {dblp computer science bibliography, https://dblp.org},
  doi       = {10.1109/TSE.2018.2883603},
  journal   = {{IEEE} Trans. Software Eng.},
  number    = {1},
  pages     = {108--129},
  title     = {{B}eyond {T}echnical {A}spects: {H}ow {D}o {C}ommunity {S}mells {I}nfluence the {I}ntensity of {C}ode {S}mells?},
  volume    = {47},
  year      = {2021}
}

@inproceedings{Parra_2018_ICSECP,
  author    = {Eugenio Parra and
               Jose Luis de la Vara and
               Luis Alonso},
  bibsource = {dblp computer science bibliography, https://dblp.org},
  booktitle = {Proceedings of the 40th International Conference on Software Engineering:
               Companion Proceeedings, {ICSE} 2018, Gothenburg, Sweden, May 27 -
               June 03, 2018},
  doi       = {10.1145/3183440.3195095},
  editor    = {Michel Chaudron and
               Ivica Crnkovic and
               Marsha Chechik and
               Mark Harman},
  pages     = {199--200},
  publisher = {{ACM}},
  title     = {{A}nalysis of requirements quality evolution},
  year      = {2018}
}

@inproceedings{Parra_2019_ISSREW,
  author    = {Parra, Eugenio and Alonso, Luis and Mendieta, Roy and de la Vara, Jose Luis},
  booktitle = {2019 {{IEEE International Symposium}} on {{Software Reliability Engineering Workshops}} ({{ISSREW}})},
  doi       = {10.1109/ISSREW.2019.00047},
  month     = oct,
  pages     = {79--84},
  title     = {{A}dvances in {{Artefact {Q}uality {A}nalysis}} for {{Safety}}-{{Critical {S}ystems}}},
  year      = {2019}
}

@techreport{Pena_2020_DS,
  author    = {Francisco Javier Peña Veitía},
  doi       = {10.17632/bw9md35c29.2},
  publisher = {Mendeley Data},
  title     = {{I}dentifying {U}ser {S}tories in {I}ssues records},
  year      = {2020}
}

@inproceedings{Perez_2021_ICPC,
  author    = {Quentin Perez and
               Pierre{-}Antoine Jean and
               Christelle Urtado and
               Sylvain Vauttier},
  bibsource = {dblp computer science bibliography, https://dblp.org},
  booktitle = {29th {IEEE/ACM} International Conference on Program Comprehension,
               {ICPC} 2021, Madrid, Spain, May 20-21, 2021},
  doi       = {10.1109/ICPC52881.2021.00014},
  pages     = {47--58},
  publisher = {{IEEE}},
  title     = {{B}ug or not bug? {T}hat is the question},
  year      = {2021}
}

@inproceedings{Petersen_2008_EASE,
  author    = {Kai Petersen and
               Robert Feldt and
               Shahid Mujtaba and
               Michael Mattsson},
  bibsource = {dblp computer science bibliography, https://dblp.org},
  booktitle = {12th International Conference on Evaluation and Assessment in Software
               Engineering, {EASE} 2008, University of Bari, Italy, 26-27 June 2008},
  editor    = {Giuseppe Visaggio and
               Maria Teresa Baldassarre and
               Stephen G. Linkman and
               Mark Turner},
  publisher = {{BCS}},
  series    = {Workshops in Computing},
  title     = {{S}ystematic {M}apping {S}tudies in {S}oftware {E}ngineering},
  url       = {http://ewic.bcs.org/content/ConWebDoc/19543},
  year      = {2008}
}

@inproceedings{Petersen_2009_PROFES,
  author    = {Kai Petersen and
               Claes Wohlin and
               Dejan Baca},
  editor    = {Frank Bomarius and
               Markku Oivo and
               P{\"{a}}ivi Jaring and
               Pekka Abrahamsson},
  title     = {The Waterfall Model in Large-Scale Development},
  booktitle = {Product-Focused Software Process Improvement, 10th International Conference,
               {PROFES} 2009, Oulu, Finland, June 15-17, 2009. Proceedings},
  series    = {Lecture Notes in Business Information Processing},
  volume    = {32},
  pages     = {386--400},
  publisher = {Springer},
  year      = {2009},
  doi       = {10.1007/978-3-642-02152-7\_29},
  bibsource = {dblp computer science bibliography, https://dblp.org}
}

@article{Petersen_2015_IST,
  author    = {Kai Petersen and
               Sairam Vakkalanka and
               Ludwik Kuzniarz},
  bibsource = {dblp computer science bibliography, https://dblp.org},
  doi       = {10.1016/J.INFSOF.2015.03.007},
  journal   = {Inf. Softw. Technol.},
  pages     = {1--18},
  title     = {{G}uidelines for conducting systematic mapping studies in software engineering: {A}n update},
  volume    = {64},
  year      = {2015}
}

@inproceedings{Pham_2019_RE,
  author    = {Yen Dieu Pham and
               Lloyd Montgomery and
               Walid Maalej},
  bibsource = {dblp computer science bibliography, https://dblp.org},
  booktitle = {27th {IEEE} International Requirements Engineering Conference Workshops,
               {RE} 2019 Workshops, Jeju Island, Korea (South), September 23-27,
               2019},
  doi       = {10.1109/REW.2019.00008},
  pages     = {7--11},
  publisher = {{IEEE}},
  title     = {{R}enovating {R}equirements {E}ngineering: {F}irst {T}houghts to {S}hape {R}equirements {E}ngineering as a {P}rofession},
  year      = {2019}
}

@book{Pohl_2016_Book,
  author    = {Pohl, Klaus},
  publisher = {Rocky Nook, Inc.},
  title     = {Requirements engineering fundamentals: a study guide for the certified professional for requirements engineering exam-foundation level-IREB compliant},
  year      = {2016}
}

@phdthesis{Prediger_2023_MSc,
  author = {Nina Prediger},
  school = {University of Hamburg},
  title  = {{V}isualising {D}ata and {B}est {P}ractices in {J}ira {I}ssue {R}epositories},
  type   = {MSc Thesis},
  year   = {2023}
}

@inproceedings{Pudlitz_2019_RE,
  author    = {Florian Pudlitz and
               Florian Brokhausen and
               Andreas Vogelsang},
  bibsource = {dblp computer science bibliography, https://dblp.org},
  booktitle = {27th {IEEE} International Requirements Engineering Conference, {RE}
               2019, Jeju Island, Korea (South), September 23-27, 2019},
  doi       = {10.1109/RE.2019.00031},
  editor    = {Daniela E. Damian and
               Anna Perini and
               Seok{-}Won Lee},
  pages     = {211--222},
  publisher = {{IEEE}},
  title     = {{E}xtraction of {S}ystem {S}tates from {N}atural {L}anguage {R}equirements},
  year      = {2019}
}

@inproceedings{Puhlfürß_2022_ICSME,
  author    = {Tim Puhlf{\"{u}}r{\ss} and
               Lloyd Montgomery and
               Walid Maalej},
  bibsource = {dblp computer science bibliography, https://dblp.org},
  booktitle = {{IEEE} International Conference on Software Maintenance and Evolution,
               {ICSME} 2022, Limassol, Cyprus, October 3-7, 2022},
  doi       = {10.1109/ICSME55016.2022.00043},
  pages     = {379--383},
  publisher = {{IEEE}},
  title     = {{A}n {E}xploratory {S}tudy of {D}ocumentation {S}trategies for {P}roduct {F}eatures in {P}opular {G}it{H}ub {P}rojects},
  year      = {2022}
}

@article{Qamar_2022_IST,
  author    = {Khushbakht Ali Qamar and
               Emre S{\"{u}}l{\"{u}}n and
               Eray T{\"{u}}z{\"{u}}n},
  bibsource = {dblp computer science bibliography, https://dblp.org},
  doi       = {10.1016/J.INFSOF.2022.106972},
  journal   = {Inf. Softw. Technol.},
  pages     = {106972},
  title     = {{T}axonomy of bug tracking process smells: {P}erceptions of practitioners and an empirical analysis},
  volume    = {150},
  year      = {2022}
}

@inproceedings{Qamar_SEAA_2021,
  author    = {Khushbakht Ali Qamar and
               Emre S{\"{u}}l{\"{u}}n and
               Eray T{\"{u}}z{\"{u}}n},
  bibsource = {dblp computer science bibliography, https://dblp.org},
  booktitle = {47th Euromicro Conference on Software Engineering and Advanced Applications,
               {SEAA} 2021, Palermo, Italy, September 1-3, 2021},
  doi       = {10.1109/SEAA53835.2021.00026},
  editor    = {Maria Teresa Baldassarre and
               Giuseppe Scanniello and
               Amund Skavhaug},
  pages     = {138--147},
  publisher = {{IEEE}},
  title     = {{T}owards a {T}axonomy of {B}ug {T}racking {P}rocess {S}mells: {A} {Q}uantitative {A}nalysis},
  year      = {2021}
}

@article{Ralph_2018_TSE,
  author    = {Paul Ralph},
  bibsource = {dblp computer science bibliography, https://dblp.org},
  doi       = {10.1109/TSE.2018.2796554},
  journal   = {{IEEE} Trans. Software Eng.},
  number    = {7},
  pages     = {712--735},
  title     = {{T}oward {M}ethodological {G}uidelines for {P}rocess {T}heories and {T}axonomies in {S}oftware {E}ngineering},
  volume    = {45},
  year      = {2019}
}

@article{Razzaq_2025_CS,
  author    = {Abdul Razzaq and Jim Buckley and Qin Lai and Tingting Yu and Goetz Botterweck},
  title     = {A Systematic Literature Review on the Influence of Enhanced Developer
               Experience on Developers' Productivity: Factors, Practices, and
               Recommendations},
  journal   = {{ACM} Comput. Surv.},
  volume    = {57},
  number    = {1},
  pages     = {13:1--13:46},
  year      = {2025},
  doi       = {10.1145/3687299}
}

@online{RedMine_2024_Online,
  title   = {RedMine},
  url     = {https://www.redmine.org/},
  urldate = {2024-07-03}
}

@inproceedings{Regnell_2008_REFSQ,
  author    = {Bj{\"{o}}rn Regnell and
               Richard Berntsson{-}Svensson and
               Krzysztof Wnuk},
  bibsource = {dblp computer science bibliography, https://dblp.org},
  booktitle = {Requirements Engineering: Foundation for Software Quality, 14th International
               Working Conference, {REFSQ} 2008, Montpellier, France, June 16-17,
               2008, Proceedings},
  doi       = {10.1007/978-3-540-69062-7\_11},
  editor    = {Barbara Paech and
               Colette Rolland},
  pages     = {123--128},
  publisher = {Springer},
  series    = {Lecture Notes in Computer Science},
  title     = {{C}an {W}e {B}eat the {C}omplexity of {V}ery {L}arge-Scale {R}equirements {E}ngineering?},
  volume    = {5025},
  year      = {2008}
}

@article{Reinartz_2004_JMR,
  author    = {Reinartz, Werner and Krafft, Manfred and Hoyer, Wayne D},
  journal   = {Journal of marketing research},
  number    = {3},
  pages     = {293--305},
  publisher = {SAGE Publications Sage CA: Los Angeles, CA},
  title     = {{T}he customer relationship management process: {I}ts measurement and impact on performance},
  volume    = {41},
  year      = {2004}
}

@inproceedings{Rocha_2016_SANER,
  author    = {Henrique Rocha and
               Marco T{\'{u}}lio Valente and
               Humberto Marques{-}Neto and
               Gail C. Murphy},
  bibsource = {dblp computer science bibliography, https://dblp.org},
  booktitle = {{IEEE} 23rd International Conference on Software Analysis, Evolution,
               and Reengineering, {SANER} 2016, Suita, Osaka, Japan, March 14-18,
               2016 - Volume 1},
  doi       = {10.1109/SANER.2016.87},
  pages     = {46--56},
  publisher = {{IEEE} Computer Society},
  title     = {{A}n {E}mpirical {S}tudy on {R}ecommendations of {S}imilar {B}ugs},
  year      = {2016}
}

@book{Runeson_2012_Wiley,
  author    = {Per Runeson and
               Martin H{\"{o}}st and
               Austen Rainer and
               Bj{\"{o}}rn Regnell},
  bibsource = {dblp computer science bibliography, https://dblp.org},
  isbn      = {978-1-118-10435-4},
  publisher = {Wiley},
  title     = {{C}ase {S}tudy {R}esearch in {S}oftware {E}ngineering - {G}uidelines and {E}xamples},
  url       = {http://eu.wiley.com/WileyCDA/WileyTitle/productCd-1118104358.html},
  year      = {2012}
}

@article{Ruparelia_2010_SEN,
  author    = {Ruparelia, Nayan B},
  journal   = {ACM SIGSOFT Software Engineering Notes},
  number    = {3},
  pages     = {8--13},
  publisher = {ACM New York, NY, USA},
  title     = {{S}oftware development lifecycle models},
  volume    = {35},
  year      = {2010}
}

@inproceedings{Saha_2015_MSR,
  author    = {Ripon K. Saha and
               Julia Lawall and
               Sarfraz Khurshid and
               Dewayne E. Perry},
  bibsource = {dblp computer science bibliography, https://dblp.org},
  booktitle = {12th {IEEE/ACM} Working Conference on Mining Software Repositories,
               {MSR} 2015, Florence, Italy, May 16-17, 2015},
  doi       = {10.1109/MSR.2015.31},
  editor    = {Massimiliano {Di Penta} and
               Martin Pinzger and
               Romain Robbes},
  pages     = {258--268},
  publisher = {{IEEE} Computer Society},
  title     = {{A}re {T}hese {B}ugs {R}eally "{N}ormal"?},
  year      = {2015}
}

@inproceedings{Saito_2017_RE,
  author    = {Shinobu Saito and
               Yukako Iimura and
               Aaron K. Massey and
               Annie I. Ant{\'{o}}n},
  bibsource = {dblp computer science bibliography, https://dblp.org},
  booktitle = {25th {IEEE} International Requirements Engineering Conference, {RE}
               2017, Lisbon, Portugal, September 4-8, 2017},
  doi       = {10.1109/RE.2017.33},
  editor    = {Ana Moreira and
               Jo{\~{a}}o Ara{\'{u}}jo and
               Jane Hayes and
               Barbara Paech},
  pages     = {194--203},
  publisher = {{IEEE} Computer Society},
  title     = {{H}ow {M}uch {U}ndocumented {K}nowledge is there in {A}gile {S}oftware {D}evelopment?: {C}ase {S}tudy on {I}ndustrial {P}roject {U}sing {I}ssue {T}racking {S}ystem and {V}ersion {C}ontrol {S}ystem},
  year      = {2017}
}

@inproceedings{Sandusky_2004_MSR,
  author    = {Robert J. Sandusky and
               Les Gasser and
               Gabriel Ripoche},
  bibsource = {dblp computer science bibliography, https://dblp.org},
  booktitle = {Proceedings of the 1st International Workshop on Mining Software Repositories,
               MSR at ICSE 2004, Edinburgh, Scotland, UK, 25th May 2004},
  editor    = {Ahmed E. Hassan and
               Richard C. Holt and
               Audris Mockus},
  pages     = {80--84},
  title     = {{B}ug {R}eport {N}etworks: {V}arieties, {S}trategies, and {I}mpacts in a {F/OSS} {D}evelopment {C}ommunity},
  year      = {2004}
}

@online{Savanna_2024_Online,
  title   = {Savanna},
  url     = {https://savannah.gnu.org},
  urldate = {2025-01-16}
}

@article{Scacchi_2002_IEEPS,
  author    = {Walt Scacchi},
  bibsource = {dblp computer science bibliography, https://dblp.org},
  doi       = {10.1049/IP-SEN:20020202},
  journal   = {{IEE} Proc. Softw.},
  number    = {1},
  pages     = {24--39},
  title     = {{U}nderstanding the requirements for developing open source software systems},
  volume    = {149},
  year      = {2002}
}

@article{Scacchi_2006_SPIP,
  author    = {Walt Scacchi and
               Joseph Feller and
               Brian Fitzgerald and
               Scott A. Hissam and
               Karim R. Lakhani},
  bibsource = {dblp computer science bibliography, https://dblp.org},
  doi       = {10.1002/SPIP.255},
  journal   = {Softw. Process. Improv. Pract.},
  number    = {2},
  pages     = {95--105},
  title     = {{U}nderstanding {F}ree/{O}pen {S}ource {S}oftware {D}evelopment {P}rocesses},
  volume    = {11},
  year      = {2006}
}

@article{Scacchi_2007_AC,
  author    = {Walt Scacchi},
  bibsource = {dblp computer science bibliography, https://dblp.org},
  doi       = {10.1016/S0065-2458(06)69005-0},
  journal   = {Advances in Computers},
  pages     = {243--295},
  title     = {{F}ree/{O}pen {S}ource {S}oftware {D}evelopment: {R}ecent {R}esearch {R}esults and {M}ethods},
  volume    = {69},
  year      = {2007}
}

@inproceedings{Scacchi_2007_FSE,
  author    = {Walt Scacchi},
  bibsource = {dblp computer science bibliography, https://dblp.org},
  booktitle = {Proceedings of the 6th joint meeting of the European Software Engineering
               Conference and the {ACM} {SIGSOFT} International Symposium on Foundations
               of Software Engineering, 2007, Dubrovnik, Croatia, September 3-7,
               2007},
  doi       = {10.1145/1287624.1287689},
  editor    = {Ivica Crnkovic and
               Antonia Bertolino},
  pages     = {459--468},
  publisher = {{ACM}},
  title     = {{F}ree/open source software development},
  year      = {2007}
}

@inproceedings{Scacchi_2010_FoSER,
  author    = {Walt Scacchi},
  bibsource = {dblp computer science bibliography, https://dblp.org},
  booktitle = {Proceedings of the Workshop on Future of Software Engineering Research,
               FoSER 2010, at the 18th {ACM} {SIGSOFT} International Symposium on
               Foundations of Software Engineering, 2010, Santa Fe, NM, USA, November
               7-11, 2010},
  doi       = {10.1145/1882362.1882427},
  editor    = {Gruia{-}Catalin Roman and
               Kevin J. Sullivan},
  pages     = {315--320},
  publisher = {{ACM}},
  title     = {{T}he future of research in free/open source software development},
  year      = {2010}
}

@inproceedings{Schlutter_2018_REFSQ,
  author    = {Aaron Schlutter and
               Andreas Vogelsang},
  bibsource = {dblp computer science bibliography, https://dblp.org},
  booktitle = {Joint Proceedings of {REFSQ-2018} Workshops, Doctoral Symposium, Live
               Studies Track, and Poster Track co-located with the 23rd International
               Conference on Requirements Engineering: Foundation for Software Quality
               ({REFSQ} 2018), Utrecht, The Netherlands, March 19, 2018},
  editor    = {Klaus Schmid},
  publisher = {CEUR-WS.org},
  series    = {{CEUR} Workshop Proceedings},
  title     = {{K}nowledge {R}epresentation of {R}equirements {D}ocuments {U}sing {N}atural {L}anguage {P}rocessing},
  url       = {https://ceur-ws.org/Vol-2075/NLP4RE\_paper9.pdf},
  volume    = {2075},
  year      = {2018}
}

@inproceedings{Seiler_2017_REFSQ,
  address   = {Cham},
  author    = {Seiler, Marcus and Paech, Barbara},
  booktitle = {Requirements Engineering: Foundation for Software Quality},
  editor    = {Gr{\"u}nbacher, Paul and Perini, Anna},
  isbn      = {978-3-319-54045-0},
  pages     = {174--180},
  publisher = {Springer International Publishing},
  title     = {{U}sing {T}ags to {S}upport {F}eature {M}anagement {A}cross {I}ssue {T}racking {S}ystems and {V}ersion {C}ontrol {S}ystems},
  year      = {2017}
}

@article{Shanks_2002_AJIS,
  author    = {Shanks, Graeme and others},
  journal   = {Australasian Journal of Information Systems},
  number    = {1},
  publisher = {Australian Computer Society},
  title     = {{G}uidelines for conducting positivist case study research in information systems},
  volume    = {10},
  year      = {2002}
}

@inproceedings{Shi_2013_RE,
  author    = {Lin Shi and
               Qing Wang and
               Mingshu Li},
  bibsource = {dblp computer science bibliography, https://dblp.org},
  booktitle = {21st {IEEE} International Requirements Engineering Conference, {RE}
               2013, Rio de Janeiro-RJ, Brazil, July 15-19, 2013},
  doi       = {10.1109/RE.2013.6636713},
  pages     = {135--144},
  publisher = {{IEEE} Computer Society},
  title     = {{L}earning from evolution history to predict future requirement changes},
  year      = {2013}
}

@article{Singer_2009_ANYAS,
  author  = {T. Singer and C. Lamm},
  doi     = {10.1111/j.1749-6632.2009.04418.x},
  journal = {Annals of the New York Academy of Sciences},
  title   = {{T}he {S}ocial {N}euroscience of {E}mpathy},
  volume  = {1156},
  year    = {2009}
}

@article{Sjøberg_2008_Book,
  author    = {Sjøberg, Dag IK and Dyba, Tore and Anda, Bente CD and Hannay, Jo E},
  journal   = {Guide to advanced empirical software engineering},
  pages     = {312--336},
  publisher = {Springer},
  title     = {{B}uilding theories in software engineering},
  year      = {2008}
}

@online{Smartsheet_2024_Online,
  title   = {smartsheet tool.},
  url     = {https://www.smartsheet.com/},
  urldate = {2024-11-20}
}

@article{Sommerville_2011_Book,
  title   = {Software engineering (ed.)},
  author  = {Sommerville, Ian},
  journal = {America: Pearson Education Inc},
  year    = {2011}
}

@online{SpiraTest_2024_Online,
  title   = {SpiraTest},
  url     = {https://www.inflectra.com/Products/SpiraTest/Highlights/Bug-Tracking.aspx},
  urldate = {2024-07-03}
}

@article{Standish_1994_ChaosReport,
  title   = {The chaos report},
  author  = {Standish Group},
  journal = {The Standish Group},
  pages   = {1--16},
  year    = {1994}
}

@inproceedings{Stanik_2018_ICSME,
  author    = {Christoph Stanik and
               Lloyd Montgomery and
               Daniel Martens and
               Davide Fucci and
               Walid Maalej},
  bibsource = {dblp computer science bibliography, https://dblp.org},
  booktitle = {2018 {IEEE} International Conference on Software Maintenance and Evolution,
               {ICSME} 2018, Madrid, Spain, September 23-29, 2018},
  doi       = {10.1109/ICSME.2018.00027},
  pages     = {172--182},
  publisher = {{IEEE} Computer Society},
  title     = {{A} {S}imple {N}{L}{P}-Based {A}pproach to {S}upport {O}nboarding and {R}etention in {O}pen {S}ource {C}ommunities},
  year      = {2018}
}

@article{Stol_2018_TOSEM,
  author    = {Klaas{-}Jan Stol and
               Brian Fitzgerald},
  bibsource = {dblp computer science bibliography, https://dblp.org},
  doi       = {10.1145/3241743},
  journal   = {{ACM} Trans. Softw. Eng. Methodol.},
  number    = {3},
  pages     = {11:1--11:51},
  title     = {{T}he {ABC} of {S}oftware {E}ngineering {R}esearch},
  volume    = {27},
  year      = {2018}
}

@online{Sutherland_2020_ScrumGuide,
  author  = {Sutherland J., Schwaber K.},
  title   = {The Scrum Guide—The Definitive Guide to Scrum: The Rules of the Game.},
  url     = {https://scrumguides.org/docs/scrumguide/v2020/2020-Scrum-Guide-US.pdf},
  urldate = {2024-09-09}
}

@article{Tamburri_2014_JISA,
  author    = {Damian Tamburri and
               Philippe Kruchten and
               Patricia Lago and
               Hans van Vliet},
  bibsource = {dblp computer science bibliography, https://dblp.org},
  doi       = {10.1186/S13174-015-0024-6},
  journal   = {J. Internet Serv. Appl.},
  number    = {1},
  pages     = {10:1--10:17},
  title     = {{S}ocial debt in software engineering: insights from industry},
  volume    = {6},
  year      = {2015}
}

@article{Tamburri_2016_IEEESoftware,
  author    = {Damian Tamburri and
               Rick Kazman and
               Hamed Fahimi},
  bibsource = {dblp computer science bibliography, https://dblp.org},
  doi       = {10.1109/MS.2016.144},
  journal   = {{IEEE} Softw.},
  number    = {6},
  pages     = {70--79},
  title     = {{T}he {A}rchitect's {R}ole in {C}ommunity {S}hepherding},
  volume    = {33},
  year      = {2016}
}

@article{Tamburri_2019_TSE,
  author    = {Damian Tamburri and
               Fabio Palomba and
               Rick Kazman},
  bibsource = {dblp computer science bibliography, https://dblp.org},
  doi       = {10.1109/TSE.2019.2901490},
  journal   = {{IEEE} Trans. Software Eng.},
  number    = {3},
  pages     = {630--652},
  title     = {{E}xploring {C}ommunity {S}mells in {O}pen-{S}ource: {A}n {A}utomated {A}pproach},
  volume    = {47},
  year      = {2021}
}

@inproceedings{Telemaco_2019_CIbSE,
  author    = {Ulisses Telemaco and
               Toacy C. Oliveira and
               Paulo S. C. Alencar and
               Donald Cowan},
  bibsource = {dblp computer science bibliography, https://dblp.org},
  booktitle = {Proceedings of the {XXII} Iberoamerican Conference on Software Engineering,
               CIbSE 2019, La Habana, Cuba, April 22-26, 2019},
  editor    = {Beatriz Mar{\'{\i}}n},
  pages     = {30--43},
  publisher = {Curran Associates},
  title     = {{A} {C}atalog of {B}ad {A}gile {S}mells for {A}gility {A}ssessment},
  year      = {2019}
}

@article{Telemaco_2020_IEEEAccess,
  author    = {Ulisses Telemaco and
               Toacy C. Oliveira and
               Paulo S. C. Alencar and
               Donald Cowan},
  bibsource = {dblp computer science bibliography, https://dblp.org},
  doi       = {10.1109/ACCESS.2020.2989106},
  journal   = {{IEEE} Access},
  pages     = {79239--79259},
  title     = {{A} {C}atalogue of {A}gile {S}mells for {A}gility {A}ssessment},
  volume    = {8},
  year      = {2020}
}

@inproceedings{Terdchanakul_2017_ICSME,
  author    = {Pannavat Terdchanakul and
               Hideaki Hata and
               Passakorn Phannachitta and
               Kenichi Matsumoto},
  bibsource = {dblp computer science bibliography, https://dblp.org},
  booktitle = {2017 {IEEE} International Conference on Software Maintenance and Evolution,
               {ICSME} 2017, Shanghai, China, September 17-22, 2017},
  doi       = {10.1109/ICSME.2017.14},
  pages     = {534--538},
  publisher = {{IEEE} Computer Society},
  title     = {{B}ug or {N}ot? {B}ug {R}eport {C}lassification {U}sing {N}-{G}ram {IDF}},
  year      = {2017}
}

@article{Thayer_1997_Book,
  title     = {Software engineering project management},
  author    = {Thayer, Richard H and Yourdon, E},
  journal   = {Software engineering project management},
  pages     = {72--104},
  year      = {1997},
  publisher = {IEEE Computer Society Press}
}

@inproceedings{Thompson_2016_MSR,
  author    = {C. Albert Thompson and
               Gail C. Murphy and
               Marc Palyart and
               Marko Gasparic},
  bibsource = {dblp computer science bibliography, https://dblp.org},
  booktitle = {Proceedings of the 13th International Conference on Mining Software
               Repositories, {MSR} 2016, Austin, TX, USA, May 14-22, 2016},
  doi       = {10.1145/2901739.2901779},
  editor    = {Miryung Kim and
               Romain Robbes and
               Christian Bird},
  pages     = {281--285},
  publisher = {{ACM}},
  title     = {{H}ow software developers use work breakdown relationships in issue repositories},
  year      = {2016}
}

@inproceedings{Tian_2012_WCRE,
  author    = {Yuan Tian and
               David Lo and
               Chengnian Sun},
  bibsource = {dblp computer science bibliography, https://dblp.org},
  booktitle = {19th Working Conference on Reverse Engineering, {WCRE} 2012, Kingston,
               ON, Canada, October 15-18, 2012},
  doi       = {10.1109/WCRE.2012.31},
  pages     = {215--224},
  publisher = {{IEEE} Computer Society},
  title     = {{I}nformation {R}etrieval {B}ased {N}earest {N}eighbor {C}lassification for {F}ine-Grained {B}ug {S}everity {P}rediction},
  year      = {2012}
}

@inproceedings{Tian_2013_ICSM,
  author    = {Yuan Tian and
               David Lo and
               Chengnian Sun},
  bibsource = {dblp computer science bibliography, https://dblp.org},
  booktitle = {2013 {IEEE} International Conference on Software Maintenance, Eindhoven,
               The Netherlands, September 22-28, 2013},
  doi       = {10.1109/ICSM.2013.31},
  pages     = {200--209},
  publisher = {{IEEE} Computer Society},
  title     = {{DRONE:} {P}redicting {P}riority of {R}eported {B}ugs by {M}ulti-factor {A}nalysis},
  year      = {2013}
}

@inproceedings{Tomova_2018_TSE,
  author    = {Mihaela Todorova Tomova and
               Michael Rath and
               Patrick M{\"{a}}der},
  bibsource = {dblp computer science bibliography, https://dblp.org},
  booktitle = {Proceedings of the 40th International Conference on Software Engineering:
               Companion Proceeedings, {ICSE} 2018, Gothenburg, Sweden, May 27 -
               June 03, 2018},
  doi       = {10.1145/3183440.3195086},
  editor    = {Michel Chaudron and
               Ivica Crnkovic and
               Marsha Chechik and
               Mark Harman},
  pages     = {181--182},
  publisher = {{ACM}},
  title     = {{U}se of trace link types in issue tracking systems},
  year      = {2018}
}

@online{Trac_2024_Online,
  title   = {Trac},
  url     = {https://trac.edgewall.org/},
  urldate = {2024-07-03}
}

@online{Trello_2024_Online,
  title   = {Trello},
  url     = {https://trello.com/},
  urldate = {2024-07-03}
}

@inproceedings{Tufano_2016_ASE,
  author    = {Michele Tufano and
               Fabio Palomba and
               Gabriele Bavota and
               Massimiliano {Di Penta} and
               Rocco Oliveto and
               Andrea De Lucia and
               Denys Poshyvanyk},
  bibsource = {dblp computer science bibliography, https://dblp.org},
  booktitle = {Proceedings of the 31st {IEEE/ACM} International Conference on Automated
               Software Engineering, {ASE} 2016, Singapore, September 3-7, 2016},
  doi       = {10.1145/2970276.2970340},
  editor    = {David Lo and
               Sven Apel and
               Sarfraz Khurshid},
  pages     = {4--15},
  publisher = {{ACM}},
  title     = {{A}n empirical investigation into the nature of test smells},
  year      = {2016}
}

@inproceedings{Tuna_2022_ICSESEIP,
  author    = {Erdem Tuna and
               Vladimir Kovalenko and
               Eray T{\"{u}}z{\"{u}}n},
  bibsource = {dblp computer science bibliography, https://dblp.org},
  booktitle = {44th {IEEE/ACM} International Conference on Software Engineering:
               Software Engineering in Practice, {ICSE} {(SEIP)} 2022, Pittsburgh,
               PA, USA, May 22-24, 2022},
  doi       = {10.1109/ICSE-SEIP55303.2022.9793952},
  pages     = {77--86},
  publisher = {{IEEE}},
  title     = {{B}ug {T}racking {P}rocess {S}mells {I}n {P}ractice},
  year      = {2022}
}

@inproceedings{VanCan_2024_REFSQ,
  author    = {Ashley {van Can} and
               Fabiano Dalpiaz},
  bibsource = {dblp computer science bibliography, https://dblp.org},
  booktitle = {Requirements Engineering: Foundation for Software Quality - 30th International
               Working Conference, {REFSQ} 2024, Winterthur, Switzerland, April 8-11,
               2024, Proceedings},
  doi       = {10.1007/978-3-031-57327-9\_19},
  editor    = {Daniel Mendez and
               Ana Moreira},
  pages     = {305--321},
  publisher = {Springer},
  series    = {Lecture Notes in Computer Science},
  title     = {{R}equirements {I}nformation in {B}acklog {I}tems: {C}ontent {A}nalysis},
  volume    = {14588},
  year      = {2024}
}

@article{VanCan_2025_IST,
  title     = {Locating requirements in backlog items: Content analysis and experiments with large language models},
  author    = {Ashley {van Can} and Fabiano Dalpiaz},
  journal   = {Information and Software Technology},
  volume    = {179},
  pages     = {107644},
  year      = {2025},
  publisher = {Elsevier}
}

@article{Vogelsang_2019_ST,
  author    = {Andreas Vogelsang and
               Daniel Mendez and
               Xavier Franch},
  bibsource = {dblp computer science bibliography, https://dblp.org},
  journal   = {Softwaretechnik-Trends},
  number    = {1},
  pages     = {9--10},
  title     = {{I}s {RE} {R}esearch {R}elevant for {P}ractitioners? {F}irst {R}esults from the {R}{E}-pract {S}tudy},
  url       = {https://fb-swt.gi.de/fileadmin/FB/SWT/Softwaretechnik-Trends/Verzeichnis/Band\_39\_Heft\_1/04-Vogelsang\_et\_al.pdf},
  volume    = {39},
  year      = {2019}
}

@article{Wagner_2019_TOSEM,
  author    = {Stefan Wagner and
               Daniel Mendez and
               Michael Felderer and
               Antonio Vetr{\`{o}} and
               Marcos Kalinowski and
               Roel J. Wieringa and
               Dietmar Pfahl and
               Tayana Conte and
               Marie{-}Therese Christiansson and
               Desmond Greer and
               Casper Lassenius and
               Tomi M{\"{a}}nnist{\"{o}} and
               Maleknaz Nayebi and
               Markku Oivo and
               Birgit Penzenstadler and
               Rafael Prikladnicki and
               G{\"{u}}nther Ruhe and
               Andr{\'{e}} Schekelmann and
               Sagar Sen and
               Rodrigo O. Sp{\'{\i}}nola and
               Ahmet Tuzcu and
               Jose Luis de la Vara and
               Dietmar Winkler},
  title     = {Status Quo in Requirements Engineering: {A} Theory and a Global Family
               of Surveys},
  journal   = {{ACM} Trans. Softw. Eng. Methodol.},
  volume    = {28},
  number    = {2},
  pages     = {9:1--9:48},
  year      = {2019},
  doi       = {10.1145/3306607},
  bibsource = {dblp computer science bibliography, https://dblp.org}
}

@inproceedings{Wang_2008_ICSE,
  author    = {Xiaoyin Wang and
               Lu Zhang and
               Tao Xie and
               John Anvik and
               Jiasu Sun},
  bibsource = {dblp computer science bibliography, https://dblp.org},
  booktitle = {30th International Conference on Software Engineering {(ICSE} 2008),
               Leipzig, Germany, May 10-18, 2008},
  doi       = {10.1145/1368088.1368151},
  editor    = {Wilhelm Sch{\"{a}}fer and
               Matthew B. Dwyer and
               Volker Gruhn},
  pages     = {461--470},
  publisher = {{ACM}},
  title     = {{A}n approach to detecting duplicate bug reports using natural language and execution information},
  year      = {2008}
}

@article{Wei_2014_AJIBM,
  author  = {L. Wei and R. Yazdanifard},
  doi     = {10.4236/AJIBM.2014.41002},
  journal = {American Journal of Industrial and Business Management},
  pages   = {9--12},
  title   = {{T}he impact of {P}ositive {R}einforcement on {E}mployees’ {P}erformance in {O}rganizations},
  volume  = {4},
  year    = {2014}
}

@inproceedings{Winkler_2018_REFSQ,
  author    = {Jonas Paul Winkler and
               Andreas Vogelsang},
  bibsource = {dblp computer science bibliography, https://dblp.org},
  booktitle = {Requirements Engineering: Foundation for Software Quality - 24th International
               Working Conference, {REFSQ} 2018, Utrecht, The Netherlands, March
               19-22, 2018, Proceedings},
  doi       = {10.1007/978-3-319-77243-1\_4},
  editor    = {Erik Kamsties and
               Jennifer Horkoff and
               Fabiano Dalpiaz},
  pages     = {57--71},
  publisher = {Springer},
  series    = {Lecture Notes in Computer Science},
  title     = {{U}sing {T}ools to {A}ssist {I}dentification of {N}on-requirements in {R}equirements {S}pecifications - {A} {C}ontrolled {E}xperiment},
  volume    = {10753},
  year      = {2018}
}

@inproceedings{Winter_2020_REFSQ,
  author    = {Katharina Winter and
               Henning Femmer and
               Andreas Vogelsang},
  bibsource = {dblp computer science bibliography, https://dblp.org},
  booktitle = {Requirements Engineering: Foundation for Software Quality - 26th International
               Working Conference, {REFSQ} 2020, Pisa, Italy, March 24-27, 2020,
               Proceedings [{REFSQ} 2020 was postponed]},
  doi       = {10.1007/978-3-030-44429-7\_1},
  editor    = {Nazim H. Madhavji and
               Liliana Pasquale and
               Alessio Ferrari and
               Stefania Gnesi},
  pages     = {3--18},
  publisher = {Springer},
  series    = {Lecture Notes in Computer Science},
  title     = {{H}ow {D}o {Q}uantifiers {A}ffect the {Q}uality of {R}equirements?},
  volume    = {12045},
  year      = {2020}
}

@inproceedings{Xavier_2020_MSR,
  author    = {Laerte Xavier and
               Fabio Ferreira and
               Rodrigo Brito and
               Marco T{\'{u}}lio Valente},
  bibsource = {dblp computer science bibliography, https://dblp.org},
  booktitle = {{MSR} '20: 17th International Conference on Mining Software Repositories,
               Seoul, Republic of Korea, 29-30 June, 2020},
  doi       = {10.1145/3379597.3387459},
  editor    = {Sunghun Kim and
               Georgios Gousios and
               Sarah Nadi and
               Joseph Hejderup},
  pages     = {137--146},
  publisher = {{ACM}},
  title     = {{B}eyond the {C}ode: {M}ining {S}elf-Admitted {T}echnical {D}ebt in {I}ssue {T}racker {S}ystems},
  year      = {2020}
}

@inproceedings{Xia_2013_WCRE,
  author    = {Xin Xia and
               David Lo and
               Xinyu Wang and
               Bo Zhou},
  bibsource = {dblp computer science bibliography, https://dblp.org},
  booktitle = {20th Working Conference on Reverse Engineering, {WCRE} 2013, Koblenz,
               Germany, October 14-17, 2013},
  doi       = {10.1109/WCRE.2013.6671282},
  editor    = {Ralf L{\"{a}}mmel and
               Rocco Oliveto and
               Romain Robbes},
  pages     = {72--81},
  publisher = {{IEEE} Computer Society},
  title     = {{A}ccurate developer recommendation for bug resolution},
  year      = {2013}
}

@inproceedings{Xuan_2012_ICSE,
  author    = {Jifeng Xuan and
               He Jiang and
               Zhilei Ren and
               Weiqin Zou},
  bibsource = {dblp computer science bibliography, https://dblp.org},
  booktitle = {34th International Conference on Software Engineering, {ICSE} 2012,
               June 2-9, 2012, Zurich, Switzerland},
  doi       = {10.1109/ICSE.2012.6227209},
  editor    = {Martin Glinz and
               Gail C. Murphy and
               Mauro Pezz{\`{e}}},
  pages     = {25--35},
  publisher = {{IEEE} Computer Society},
  title     = {{D}eveloper prioritization in bug repositories},
  year      = {2012}
}

@article{Zhang_2015_SCIS,
  author    = {Jie Zhang and
               Xiaoyin Wang and
               Dan Hao and
               Bing Xie and
               Lu Zhang and
               Hong Mei},
  bibsource = {dblp computer science bibliography, https://dblp.org},
  doi       = {10.1007/S11432-014-5241-2},
  journal   = {Sci. China Inf. Sci.},
  number    = {2},
  pages     = {1--24},
  title     = {{A} survey on bug-report analysis},
  volume    = {58},
  year      = {2015}
}

@article{Zhang_2016_CJ,
  author    = {Tao Zhang and
               He Jiang and
               Xiapu Luo and
               Alvin T. S. Chan},
  bibsource = {dblp computer science bibliography, https://dblp.org},
  doi       = {10.1093/COMJNL/BXV114},
  journal   = {Comput. J.},
  number    = {5},
  pages     = {741--773},
  title     = {{A} {L}iterature {R}eview of {R}esearch in {B}ug {R}esolution: {T}asks, {C}hallenges and {F}uture {D}irections},
  volume    = {59},
  year      = {2016}
}

@article{Zhao_2022_ACMCompSurvey,
  author    = {Liping Zhao and
               Waad Alhoshan and
               Alessio Ferrari and
               Keletso J. Letsholo and
               Muideen A. Ajagbe and
               Erol{-}Valeriu Chioasca and
               Riza Theresa Batista{-}Navarro},
  bibsource = {dblp computer science bibliography, https://dblp.org},
  doi       = {10.1145/3444689},
  journal   = {{ACM} Comput. Surv.},
  number    = {3},
  pages     = {55:1--55:41},
  title     = {{N}atural {L}anguage {P}rocessing for {R}equirements {E}ngineering: {A} {S}ystematic {M}apping {S}tudy},
  volume    = {54},
  year      = {2022}
}

@article{Zhou_2016_JSE,
  author    = {Yu Zhou and
               Yanxiang Tong and
               Ruihang Gu and
               Harald C. Gall},
  bibsource = {dblp computer science bibliography, https://dblp.org},
  doi       = {10.1002/SMR.1770},
  journal   = {J. Softw. Evol. Process.},
  number    = {3},
  pages     = {150--176},
  title     = {{C}ombining text mining and data mining for bug report classification},
  volume    = {28},
  year      = {2016}
}

@article{Zimmermann_2010_TSE,
  author    = {Thomas Zimmermann and
               Rahul Premraj and
               Nicolas Bettenburg and
               Sascha Just and
               Adrian Schr{\"o}ter and
               Cathrin Weiss},
  bibsource = {dblp computer science bibliography, https://dblp.org},
  doi       = {10.1109/TSE.2010.63},
  journal   = {{IEEE} Trans. Software Eng.},
  number    = {5},
  pages     = {618--643},
  title     = {{W}hat {M}akes a {G}ood {B}ug {R}eport?},
  volume    = {36},
  year      = {2010}
}

@article{Zou_2020_TSE,
  author    = {Weiqin Zou and
               David Lo and
               Zhenyu Chen and
               Xin Xia and
               Yang Feng and
               Baowen Xu},
  bibsource = {dblp computer science bibliography, https://dblp.org},
  doi       = {10.1109/TSE.2018.2870414},
  journal   = {{IEEE} Trans. Software Eng.},
  number    = {8},
  pages     = {836--862},
  title     = {{H}ow {P}ractitioners {P}erceive {A}utomated {B}ug {R}eport {M}anagement {T}echniques},
  volume    = {46},
  year      = {2020}
}

@inproceedings{Zowghi_2018_AffectRE,
  author    = {Didar Zowghi},
  bibsource = {dblp computer science bibliography, https://dblp.org},
  booktitle = {1st International Workshop on Affective Computing for Requirements
               Engineering, AffectRE at RE 2018, Banff, AB, Canada, August 21, 2018},
  doi       = {10.1109/AFFECTRE.2018.00008},
  editor    = {Davide Fucci and
               Nicole Novielli and
               Emitza Guzman},
  pages     = {13},
  publisher = {{IEEE}},
  title     = {"{A}ffects" of {U}ser {I}nvolvement in {S}oftware {D}evelopment},
  year      = {2018}
}
}
    
\tocAddFauxChapter{Term Index}
\printindex[termindex]

\tocAddFauxChapter{Author Index}
\printindex
    
\clearpage                  
\pagestyle{empty}	        
\cleardoubleevenpage        

\end{document}